\documentclass[a4paper,11pt]{report}
\usepackage{a4wide}
\usepackage{amsmath,amsfonts,bbm,mathrsfs,amsthm}
\usepackage[dvips]{graphics,graphicx,psfrag}
\usepackage{epsfig}
\usepackage{color}
\usepackage[PostScript=dvips]{diagrams}
\usepackage{cancel}
\usepackage{algorithmic}

\newtheorem*{thm}{Theorem}
\newtheorem*{prop}{Proposition}
\newtheorem*{definition}{Definition}

\newcommand{\vm}{\vec{m}}
\newcommand{\vn}{\vec{n}}

\psfrag{rg0}{$r\gg0$}
\psfrag{rl0}{$r\ll0$}
\psfrag{critloc}{$\frac{\partial\mathcal{W}_{eff}}{\partial u_i}=f_i(u_i,t_i)=0$}
\psfrag{l2}{$\mathcal{L}_2$}
\psfrag{bs1}{$\bar{\mathcal{S}_1}$}
\psfrag{bl1}{$\bar{\mathcal{L}_1}$}
\psfrag{bs2}{$\bar{\mathcal{S}}_2$}
\psfrag{bl2}{$\bar{\mathcal{L}}_2$}
\psfrag{s1}{${\mathcal{S}}_1$}
\psfrag{s2}{$\mathcal{S}_2$}
\psfrag{l1}{${\mathcal{L}}_1$}
\psfrag{psil2bl1}{$\psi_{\mathcal{L}_2\bar{\mathcal{L}_1}}$}
\psfrag{psibl1s1}{$\psi_{\bar{\mathcal{L}}_1\mathcal{S}_1}$}
\psfrag{psis1l2}{$\psi_{\mathcal{S}_1\mathcal{L}_2}$}
\psfrag{beta3}{$\beta_3$}
\psfrag{beta2}{$\beta_2$}
\psfrag{beta1}{$\frac{\beta_1}{2}$}
\psfrag{bpsil2bl1}{$\bar{\psi}_{\mathcal{L}_2\bar{\mathcal{L}}_1}$}
\psfrag{phibl1bs1}{$\phi_{\bar{\mathcal{L}}_1\bar{\mathcal{S}}_1}$}
\psfrag{phibs1l2}{$\phi_{\bar{\mathcal{S}}_1\mathcal{L}_2}$}
\psfrag{phibl2bl1}{$\phi_{\bar{\mathcal{L}}_2\bar{\mathcal{L}}_1}$}
\psfrag{phis1bl2}{$\phi_{\mathcal{S}_1\bar{\mathcal{L}}_2}$}
\psfrag{phil2l1}{$\phi_{\mathcal{L}_2\mathcal{L}_1}$}
\psfrag{phil1s1}{$\phi_{\mathcal{L}_1\mathcal{S}_1}$}

\begin{document}
\thispagestyle{empty}
\begin{center}
\includegraphics{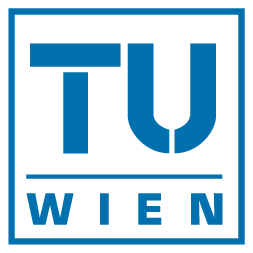}
\end{center}
\vspace{3mm}
\begin{center}
{\LARGE DISSERTATION}
\end{center}
\vspace{10mm}
\begin{center}
{\LARGE\bf D--Branes in Topological String Theory}
\end{center}
\vspace{12mm}
\begin{center}
ausgef\"uhrt zum Zwecke der Erlangung des akademischen Grades eines\\
{\em Doktor der Naturwissenschaften}
\end{center}
\vspace{7mm}
\begin{center}
unter der Leitung von
\end{center}
\begin{center}
Ao. Univ. Prof. Dipl.-Ing. Dr. techn. Maximilian Kreuzer\\
E136\\
Institut f\"ur Theoretische Physik
\end{center}
\begin{center}
und
\end{center}
\begin{center}
Prof. Wolfgang Lerche\\
CERN
\end{center}
\vspace{7mm}
\begin{center}
eingereicht an der Technischen Universit\"at Wien\\
Fakult\"at f\"ur Physik
\end{center}
\vspace{4mm}
\begin{center}
von
\end{center}
\vspace{2mm}
\begin{center}
{\bf Dipl.-Ing. Johanna Knapp}\\
{\tt johanna.knapp''at''cern.ch}
\end{center}
\vspace{2cm}
Wien, am 14. August 2007
\newpage
\thispagestyle{empty}
\mbox{}
\newpage
\thispagestyle{empty}
\begin{center}
\end{center}
\vspace{7cm}
\begin{flushright}
 to my family and my partner
\end{flushright}
\chapter*{Kurzfassung}
\pagenumbering{roman}
Das Hauptthema dieser Doktorarbeit sind D--branes in topologischer Stringtheorie. Topologische Stringtheorie beschreibt einen Untersektor der vollen Stringtheorie, der sich auf die Nullmoden der physikalischen Felder beschr\"ankt. Man erh\"alt eine solche Theorie durch den so genannten topologischen Twist, welcher, aus dem Blickwinkel der Weltfl\"achenwirkung, einer mit der Theorie vertr\"aglichen Redefinition der Spins der Felder entspricht. Der topologische Twist kann auf zwei Arten durchgef\"uhrt werden, woraus zwei unterschiedliche und a priori unabh\"angige Theorien resultieren, die als A-- und B--Modell bezeichnet werden.\\
F\"ur offene Strings kann man Randbedingungen definieren, welche mit dem topologischen Twist konsistent sind. Diese werden als A-- bzw. B--branes bezeichnet. \\Vor Kurzem wurde eine neue Beschreibung von topologischen D--branes im B--Modell, welches durch ein Landau--Ginzburg Modell realisiert werden kann, gefunden. Solche Landau--Ginzburg Theorien sind durch ein Superpotential charakterisiert und B--branes werden durch Matrixfaktorisierungen dieses Superpotentials beschrieben. Diese Doktorarbeit besch\"aftigt sich mit den Eigenschaften und Anwendungen dieses Formalismus. Eine der Problemstellungen befasst sich mit der Berechnung des effektiven Superpotentials $\mathcal{W}_{eff}$. Dieses Objekt ist von ph\"anomenologischem Interesse, da man es als vierdimensionales Superpotential einer $N=1$ supersymmetrischen Calabi--Yau Kompaktifizierung interpretieren kann. Es wird gezeigt, wie man diese Gr\"o\ss e f\"ur minimale Modelle berechnen kann. Diese Modelle sind n\"utzliche Spielzeugmodelle f\"ur Stringkompaktifizierungen. Als eines der Hauptresultate dieser Doktorarbeit zeigen wir, dass es m\"oglich ist $\mathcal{W}_{eff}$ zu berechnen, indem man die allgemeinsten nicht--lineare Deformationen von Matrixfaktorisierungen berechnet. Ein bemerkenswertes Resultat in der Stringtheorie besagt, dass A-Modell und B--Modell durch eine Dualit\"at miteinander verbunden sind, die sogenannte Mirrorsymmetrie. Diese besagt, dass das A--Modell auf einer bestimmten Calabi--Yau Mannigfalitgkeit \"aquivalent ist zum B--Modell auf der Mirror--Calabi--Yau. \\
Mirrorsymmetrie kann erweitert werden um auch offene Strings zu beschreiben. Diese Dualit\"at ist bis jetzt noch nicht ausreichend verstanden, insbesondere f\"ur kompakte Calabi--Yau Mannigfalitgkeiten. In dieser Doktorarbeit werden wir uns ausserdem mit Mirrorsymmetrie f\"ur offene Strings f\"ur die einfachste kompakte Calabi--Yau befassen, f\"ur den Torus. Unter der Verwendung von Matrixfaktorisierungen werden Dreipunktfunktionen im B--Modell berechnet und es wird gezeigt, dass diese mit den entsprechenden Gr\"o\ss en im A--Modell \"ubereinstimmen.
\chapter*{Abstract}
The main focus of this thesis are D--branes in topological string theory. \\
Topological string theory describes a subsector of the full string theory, that only captures the zero--modes of the physical fields. It can be obtained by an operation called topological twist, which is, at the level of the worldsheet action, a consistent redefinition of the spins of the fields. The topological twist can be done in two ways. This leads to two different, a priori independent, models, called the A--model and the B--model. If we consider open string theory, there are boundary conditions which are consistent with the topological twist. These boundary conditions are called A--type and B--type D--branes, respectively.\\
Recently, a new description of D--branes in the topological B--model has been found. The B--model can be realized in terms of a Landau--Ginzburg theory, which is characterized by a superpotential. B--type D--branes are described by matrix factorizations of this superpotential. This thesis is devoted to exploring the properties and applications of this formalism.\\
One of the central challenges of this thesis is the calculation of the effective superpotential $\mathcal{W}_{eff}$. This quantity is of phenomenological interest since it has an interpretation as a four--dimensional space--time superpotential in $N=1$ supersymmetric string compactifications. We show how to calculate it for the minimal models, a subclass of Landau--Ginzburg theories, which serve as useful toy models for string compactifications. One of the main results of this thesis is that $\mathcal{W}_{eff}$ can be obtained by calculating the most general, non--linear deformation of a matrix factorization. \\
A remarkable result in string theory states that the A--model and the B--model are related by a duality known as mirror symmetry. The statement is that the A--model on a Calabi--Yau manifold is equivalent to the B--model on the mirror Calabi--Yau. \\
Mirror symmetry can be extended to open strings. Open string mirror symmetry is by now poorly understood, in particular for compact Calabi--Yau spaces. In this thesis we will verify open string mirror symmetry for the simplest compact Calabi--Yau, the torus. Using matrix factorization techniques we calculate three--point correlators in the B--model and show that they match with the correlators we compute in the A--model. 
\chapter*{Acknowledgements}
I want to thank my supervisor at CERN, Wolfgang Lerche, for accepting me as his student and for suggesting research problems which are well suited for making one's first steps in scientific work. I appreciate that I was left the time and the freedom to come to terms with the subject in my own way. Furthermore I want to thank Wolfgang for giving me insight into the beautiful field of topological string theory and for various short discussions which immediately cured my confusion when I was stuck in some calculation.\\
It is a pleasure to thank my supervisor in Vienna, Maximilian Kreuzer, for constant support all through my theoretical physics career, and in particular for suggesting to me to do my PhD at CERN which I never would have dared without his encouragement. I also want to thank him for pleasant and instructive discussions during my stay in Vienna in February 2007. \\
I want to express my gratitude to Hans Jockers for expertly answering many of my questions and for discussions which helped me see the things I was working on in a wider context.  
Furthermore I want to thank Emanuel Scheidegger for helpful discussions on topological string theory and modular forms. Thanks also to Marcos Mari\~no for organizing such a great student seminar.\\
As promised, Marlene Weiss gets her own paragraph in this acknowledgements section. It is hard to find such a great office mate. I really enjoyed the many discussions we had on a broad variety of topics, sometimes even physics. I want to thank her for invaluable company and support during the last two years.\\
Furthermore I want to thank James Bedford, Cedric Delaunay, I\~naki Garcia--Etxebarria, Tanju Gleisberg, Stefan Hohenegger, Tristan Maillard, Are Rachlow, Fouad Saad, John Ward, Jan Winter and all the other students at CERN for many pleasant encounters. \\
I am deeply grateful to my boyfriend for more support than I could possibly ask for. I want to thank my friends and family for their continuing support, in particular Gige for taking care of all the administrative stuff in Austria. Thanks also to Mrs. M\"ossmer for help with the paperwork related to registering my thesis. Finally, I want to thank all the people who ever cooked for me, and Guide Michelin and Gault Millau for leading the way to most enjoyable decadence.  \\\\
I also want to acknowledge the recently deceased Professor Wolfgang Kummer, who taught me a great share of my theoretical physics general eductation and who has positively influenced my decision to focus on theory.
\setcounter{tocdepth}{1}
\tableofcontents
\newpage
\pagenumbering{arabic}
\chapter{Introduction}
\section{Motivation}
String theory has been a challenge to theoretical physicists for over thirty years. Originally introduced in the late sixties to describe hadronic resonances, it was soon realized that string theory could be a possible candidate for a 'theory of everything'. This is due to the fact that the physical spectrum of closed string theory contains a massless spin $2$ excitation, which has a natural interpretation as a graviton. It is also possible to realize gauge theories in string theory, which can be achieved, for instance, by taking into account the open string sector. This lead string theorists to the insight that string theory could be a theory which not only is a consistent theory of quantum gravity but also unifies the fundamental interactions at the string scale. The initial enthusiasm subsided when physicist recognized the conceptual complexity of this theory, but was rekindled again by the 'string revolutions' which revealed new exciting and unexpected aspects of string theory. String theory is often accused of being a field in mathematics rather than a physical theory because experimentally testable predictions of physical phenomena are difficult to find. Phenomena like gravitational waves and supersymmetry, which may be measured in the near future at gravitational experiments like LIGO or LISA or at the LHC, respectively, may point towards string theory but they do not imply that string theory is {\em the} theory of nature. Despite this criticism string theory has proven to be a consistent theory in the most amazing ways and has by no means been falsified. By now there exists no other candidate for a theory of everything, which is better understood and easier to handle than string theory. \\\\
Let us now point out some characteristic features of string theory, which make it difficult to relate it to the 'real world' but also imply some intriguing fundamental concepts which may be realized in nature. \\
First, we should mention that supersymmetry is crucial for a string theory to be consistent. Since we, obviously, do not live in a supersymmetric world, supersymmetry has to be broken. Phenomenology implies that one should have at most $N=1$ supersymmetry at the string scale, which is then completely broken at low energies. The mechanisms which break supersymmetry may go to work well below the string scale at energies which are sufficiently low to be measured at the LHC. This implies that string theories with $N=1$ supersymmetry are by far the most promising candidates for realistic models.\\
Another characteristic property of string theory is that, perturbatively, it can only be consistently defined in ten dimensions. From a string theory point of view, our four--dimensional space time is an effective low--energy realization of this ten--dimensional theory. So what happens to the other dimensions? The standard picture is that, unlike four--dimensional space--time, the extra dimensions are small and compact and have therefore not been observed. The structure of the internal dimensions affects the observables in the four--dimensional theory. The most prominent candidates for these compact manifolds are the Calabi--Yau manifolds and generalizations thereof. A Calabi--Yau manifold is a Ricci flat K\"ahler manifold, where the K\"ahler property is implied by supersymmetry and the Ricci flatness constraint comes from the requirement that the theory compactified on the Calabi--Yau is consistent at the quantum level. Unfortunately, there are millions of possible candidates for Calabi--Yau compactifications, all leading to different string vacua. There immediately arises the question which Calabi--Yau manifold is the right one to describe our world. This problem has come into focus recently with the introduction of flux compactifications, which yield a vast number of vacua which are actually quite similar. This is known as the string landscape. It suggests that our universe may be a (possibly metastable) point in the landscape of string vacua which might as well have been realized in a slightly different way. In order to shed light on this issue, statistical methods paired with anthropic\footnote{Such considerations and the lack of physical predictions have brought string theory the criticism of being some kind of religion. See \cite{ausgang} for a reaction to this.} and 'entropic' considerations have been applied to study the structure of the landscape. \\
At the beginnings of string theory it was hoped that there is a unique string theory which yields a unique vacuum which contains our universe. As we have just seen the hope to find a unique vacuum out of string theory in a straight forward manner has been thoroughly shattered. But also string theory itself is not unique, at least not in the way it was expected in the beginning. String theory in ten dimensions comes in five incarnations: Type I string theory is an open/closed string theory of unoriented strings. The heterotic string is a string theory which has fermions only in the left--moving sector. It can be realized with two gauge groups, $E_{8}\times E_8$ and $SO(32)$. The $E_8\times E_8$ heterotic string was the first string theory for which $N=1$ supersymmetric theories in four dimensions could be realized. The heterotic string has received new attention lately when it was found that one can obtain the Standard Model from the $E_8\times E_8$ theory. There are two string theories which have $N=2$ supersymmetry, the type IIA and the type IIB string. Furthermore, at the non--perturbative level, there is M--theory in eleven dimensions and F--theory in twelve dimensions.\\
The existence of so many different string theories may look like a drawback concerning the search of a unique theory of nature but actually one of the most fascinating aspects of string theory comes to rescue. It turns out that all these theories are related by certain dualities. The existence and consistency of these dualities may be viewed as one of the most convincing arguments in favor of string theory. We can make a distinction between two types of dualities: there are the strong/weak coupling dualities which relate the perturbative sector of one theory to the strong--coupling regime of the dual theory. The second type are dualities which relate the perturbative sectors of two theories, where one of the two is, in some sense, easier to handle than the other one. \\
One of the most prominent examples of a strong/weak duality is the AdS/CFT correspondence. This is a gauge--gravity duality, which relates a gravity theory with matter, i.e. a closed string theory, in a certain space--time, to a supersymmetric gauge theory without gravity at the boundary of this space--time. One of the current achievements in this field is that it provides analytic methods to investigate the strong coupling regimes of gauge theories. Quite recently, string theory methods have been used to calculate the properties of quark--gluon plasma. It is quite amusing that after more than twenty years, string theory has found back to its roots, being, once again, successful in describing phenomena in QCD. At the moment it looks like this is a promising field in physics where string theory may actually be able to make predictions that could be tested in the near future.\\
An impressive example of a string duality which relates the perturbative sectors of two string theories is mirror symmetry. Mirror symmetry identifies the type IIA string compactified on a Calabi--Yau manifold with type IIB string theory compactified on a different Calabi--Yau, which is the 'mirror' of the first manifold. Mirror symmetry for closed strings is very well understood, but looking only at the closed string sector poses two problems: As we have mentioned before, type II compactifications have $N=2$ supersymmetry, whereas the phenomenologically relevant models only have $N=1$ supersymmetry. Furthermore, if we only consider closed strings, the theory we describe is a pure theory of gravity. If we want to realize gauge theories in type II string theories, we also have to include the open string sector. This brings us close to the topic of this thesis.\\\\
If one wants to consider open string theory, it is necessary to specify boundary conditions at the ends of the string. There are two types of boundary conditions: those where the ends of the open string move freely (Neumann conditions) and those where the string ends are fixed (Dirichlet conditions). Dirichlet boundary conditions have been ignored in string theory for a long time since there were no suitable objects in the theory for the open string to end on. In 1995 such objects were found by Polchinski. It was shown that open strings can end on D--branes. These are non--perturbative, dynamical, extended objects -- they are solitons in string theory.  D--branes explicitly break Lorentz invariance. This makes them look unphysical at first sight, but we can only be sure that Lorentz invariance is unbroken in four dimensions, this may however not be the case in the compact dimensions. Thus, we can safely introduce D--branes into our string theory as long as we choose Neumann boundary conditions in the four--dimensional space--time. \\
An open string cannot move in the directions normal to the D--brane but it can move freely on the worldvolume of the D--brane. Therefore the open string has additional degrees of freedom as compared to the closed string. These degrees of freedom happen to be gauge degrees of freedom. Thus, we can realize gauge theories on the worldvolumes of D--branes. In certain limits, we can picture D--branes as infinitely extended planes. If we consider $N$ coincident D--branes this gives rise to a $U(N)$ gauge theory. \\
There are even more benefits of D--branes. Since they break translation invariance, D--branes also break supersymmetry\footnote{One can see this from the supersymmetry algebra $\{Q,Q\}\sim P$, where $Q$ is the supersymmetry generator and $P$ generates the translations.}. Generic D--brane configurations will break supersymmetry completely, but there is a particular kind of (BPS) D--branes, which actually preserve half of the supersymmetry. Thus, D--branes provide a means to break the the $N=2$ supersymmetry of type II string theories down to $N=1$. \\\\
It is of central interest to study D--branes and mirror symmetry in type II string compactifications but it turns out to be a very difficult task in many cases. So we have to find some simplifications. This leads us to the topological string. Topological string theory describes a subsector of the physical string theory. In principle, topological string theories are exactly solvable theories. They only deal with a subsector of the observables of the full string theory. This sector captures the topological properties of the theory. One can obtain a topological theory out of an $N=2$ supersymmetric theory by an operation on the $N=2$ algebra, called the topological twist. The topological twist can be performed in two different ways, which leads to two a priori different theories. These are usually referred to as the A--model and the B--model. \\
The A--model and the B--model are related by mirror symmetry. The B--model is in this case the 'easy' theory, where 'easy' means in particular that it does not receive any quantum corrections. The A--model, on the contrary, obtains instanton corrections and this makes it hard to calculate correlation functions in this model. By mirror symmetry, we can however calculate the A--model quantities by considering the B--model on the mirror Calabi--Yau.\\
One benefit of topological string theory is that the moduli dependence of the theories simplifies a lot. On a generic Calabi--Yau compactification all the physical quantities will depend on the moduli of the Calabi--Yau. The moduli are parameters, which can be divided into two classes: roughly speaking, the K\"ahler moduli determine the size of the Calabi--Yau whereas the the complex structure moduli parameterize its shape. It turns out that the A--model only depends on the K\"ahler moduli whereas the B--model only depends on the complex structure moduli.\\
We have already mentioned that topological string theory only describes a subsector of the full theory. It is thus natural to ask which physical quantities, if any, the topological string computes. It can be shown that topological string theory computes certain terms in the action of the four--dimensional effective theory one obtains by compactifying type II string theory on a Calabi--Yau. In the closed string case the relevant quantity is the free energy of the topological string. In the effective action this quantity enters as the coefficient of the term responsible for the gravitational correction to the scattering of graviphotons. Quite recently, it was found that the topological string also counts the microstates of extremal black holes. This fascinating result indicates that string theory gives a microscopic description of black hole entropy, as one would expect from a good theory of quantum gravity. \\
It is possible to introduce D--branes in topological string theory. In $N=1$ compactifications with open strings the topological string computes the superpotential of the $N=1$ theory in four dimensions. This quantity will be of particular interest in this thesis. Mirror symmetry also works for open topological strings. Open string mirror symmetry is called homological mirror symmetry and is by now not sufficiently well understood. In this thesis we will discuss homological mirror symmetry for a toy model. \\\\
The main focus of this thesis will be on D--branes in B--type topological string theories. As we will discuss in the following chapter the topological B--model can be realized in terms of a supersymmetric Landau--Ginzburg model. Such models are characterized by a Landau--Ginzburg superpotential. D--branes in such a theory are characterized by matrix factorizations on this superpotential. The aim of this thesis is to investigate this rather new description of D--branes. 
\section{Summary and Outline}
Let us now give the general plan of this thesis. In chapter \ref{chap-topstring} we review some aspects of topological strings and D--branes. In chapter \ref{chap-mf} we focus on B--type topological Landau--Ginzburg models with boundaries. We show that D--branes in these models are realized in terms of matrix factorizations of the Landau--Ginzburg superpotential and discuss their properties. Chapter \ref{chap-effsupo} is devoted to the effective superpotential. In particular we will be concerned with two interpretations of the effective superpotential, namely as the generating function of open string disk amplitudes and as the quantity which encodes the obstructions to deformations of D--branes. These two distinct interpretations lead to two methods for calculating the effective superpotential. We test these methods in chapter \ref{chap-minmod} for a special class of models, the minimal models. These models serve as toy models for Calabi--Yau compactifications. Chapter \ref{chap-torus} is devoted to the simplest Calabi--Yau, the torus $T^2$. We discuss D--branes in the A--model and the B--model and verify homological mirror symmetry by comparing three--point functions, which we compute independently in both models. In chapter \ref{chap-outlook} we point out some open problems. Furthermore we give additional results and details on certain calculations in three appendices.\\\\
The results which have been obtained during the production of this thesis have been published in \cite{Knapp:2006rd,Knapp:2007kq}.

\chapter{Topological Strings and D--branes}
\label{chap-topstring}
In this chapter we review some aspects of topological strings and D--branes. The intention is to explain some of the general background which is useful for understanding D--branes in topological Landau--Ginzburg models. We also want to show how Landau--Ginzburg models appear in the setup of topological string theory and mirror symmetry. The contents of this chapter can be summarized in the following picture:
\begin{center}
\begin{figure}[h]
\includegraphics{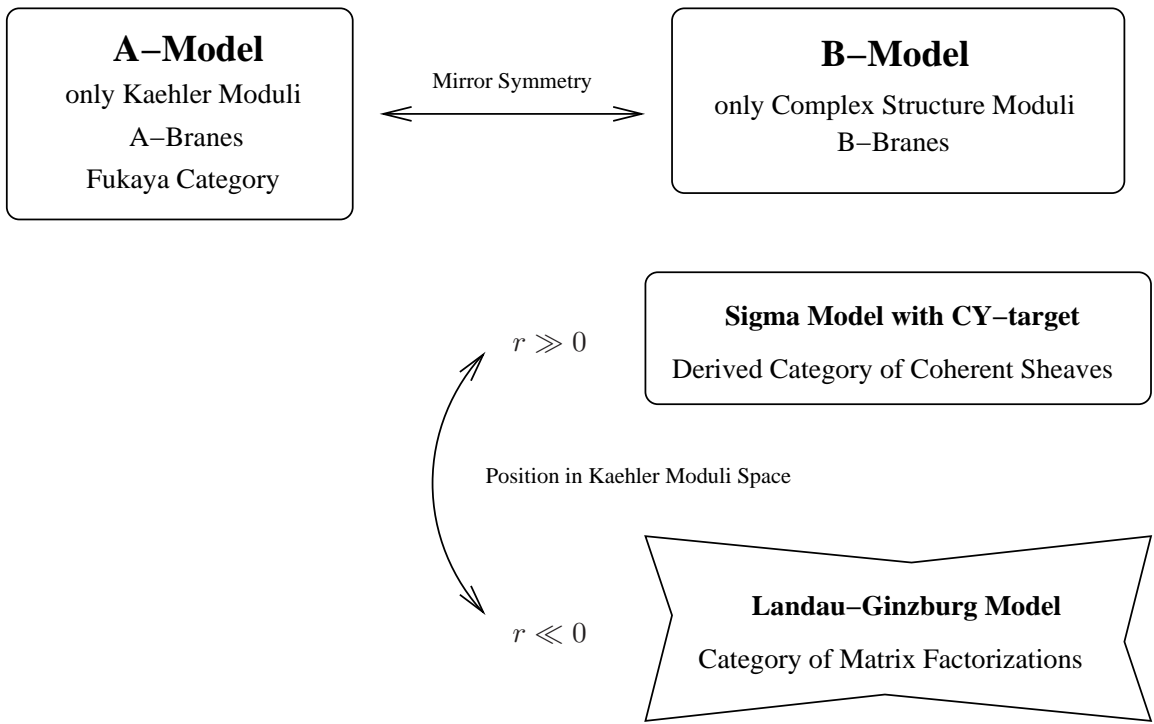}
\caption{Matrix factorizations in the 'big picture' of Topological String Theory.}
\end{figure}
\end{center}
Here we only have drawn the details on the B--model, which will be our focus.\\
In section \ref{sec-n=2} we discuss $N=2$ supersymmetric theories without boundary and review the basics of topological string theory. We point out that the topological B--model can be realized in two ways, depending on the position in K\"ahler moduli space. Section \ref{sec-branes} is devoted to D--banes. We summarize the properties of D--branes in the A--model and the B--model and their description in terms of categories.\\ 
There are many good books and review articles which discuss the subjects we cover here. An exhaustive discussion of mirror symmetry and the mathematical and physical background necessary to understand it is given in \cite{mirrorsymmetry}. A classic review article on string theory and Calabi--Yau manifolds is \cite{Greene:1996cy}. Two recent reviews on topological string theory are \cite{Neitzke:2004ni,Vonk:2005yv}. The very basic facts on D--branes can be found in Polchinski's books \cite{polchinski1,polchinski2}. D--branes in the context of categories are reviewed e. g. in \cite{Sharpe:2003dr,Aspinwall:2004jr}.

\section{Aspects of $N=2$ Theories}
\label{sec-n=2}
In this section we summarize some aspects of $N=2$ supersymmetric theories, which will be relevant for this thesis. We start by discussing the $N=2$ superconformal algebra and the ring of chiral primary fields \cite{Lerche:1989uy}. As examples for the concrete realization of these theories we discuss the non--linear sigma model and the Landau--Ginzburg model. We go on to review how one can obtain topological field theories from non--linear sigma models, following \cite{Witten:1991zz}. We define the A--model and the B--model, which are related by mirror symmetry. Finally we discuss the Calabi--Yau/Landau--Ginzburg correspondence, summarizing the results of \cite{Witten:1993yc}. In this paper is shown that the non--linear sigma model and the Landau--Ginzburg model can be viewed as 'phases' of a supersymmetric gauge theory. These phases correspond to different points in K\"ahler moduli space. This implies that the B--model, which is independent of the K\"ahler moduli, has two equivalent realizations in terms of a non--linear sigma model and a Landau--Ginzburg theory. 

\subsection{$N=2$ Superconformal Theories}
An $N=2$ superconformal algebra is generated by an energy--momentum tensor $T(z)$, two supercharges $G^{\pm}(z)$ of conformal charge $\frac{3}{2}$ and a $U(1)$ R--current $J(z)$. The algebra is determined by the following operator products:
\begin{align}
\label{eq-n2sca}
T(z)T(w)&\sim \frac{\frac{c}{2}}{(z-w)^4}+\frac{2T(w)}{(z-w)^2}+\frac{\partial_w T(w)}{z-w}\nonumber\\
T(z)G^{\pm}(w)&\sim\frac{\frac{3}{2}G^{\pm}(w)}{(z-w)^2}+\frac{\partial_wG^{\pm}(w)}{z-w}\nonumber\\
T(z)J(w)&\sim \frac{J(w)}{(z-w)^2}+\frac{\partial_wJ(w)}{z-w}\nonumber\\
G^+(z)G^{-}(w)&\sim \frac{\frac{2c}{3}}{(z-w)^3}+\frac{2J(w)}{(z-w)^2}+\frac{2T(w)+\partial_w J(w)}{z-w}\nonumber\\
J(z)G^{\pm}(w)&\sim\pm\frac{G^{\pm}(w)}{z-w}\nonumber\\
J(z)J(w)&\sim\frac{\frac{c}{3}}{(z-w)^2}
\end{align}
Here $c$ is the central charge of the theory.
These operators have a mode expansion:
\begin{align}
&T(z)=\sum_nL_nz^{-n-2}\qquad J(z)=\sum_nJ_nz^{-n-1}\qquad G^{\pm}(z)=\sum_nG^{\pm}_{n\pm a}z^{-(n\pm a)-\frac{3}{2}}
\end{align}
The parameter $a$, $0\leq a\leq 1$ controls the boundary conditions of the fermions: for integral $a$, we are in the Ramond sector, half--integral values of $a$ imply periodic boundary conditions, corresponding to the Neveu--Schwarz sector.\\
One particular property of $N=2$ superconformal theories is the existence of the chiral ring of primary fields \cite{Lerche:1989uy}. A {\em chiral primary} field $\phi$ is defined as a field whose operator product with $G^+$ does not contain any singular terms. We can associate a state $|\phi\rangle$ to $\phi$. A (anti--)chiral primary state is then defined by the condition:
\begin{align}
G^{+}_{-\frac{1}{2}}|\phi\rangle=0&\qquad\qquad G^{-}_{\frac{1}{2}}|\phi\rangle=0
\end{align}
The second equation defines the antichiral primary. The superconformal algebra tells us that
\begin{align}
\{G^-_{\frac{1}{2}},G^{+}_{-\frac{1}{2}}\}&=2L_0-J_0.
\end{align}
For a chiral primary field $\phi$ we have:
\begin{align}
&\langle\phi|\{G^-_{\frac{1}{2}},G^{+}_{-\frac{1}{2}}\}|\phi\rangle=\langle\phi|2L_0-J_0|\phi\rangle=\langle \phi|2h_{\phi}-q_{\phi}|\phi\rangle=0,
\end{align} 
where we introduced the conformal weight $h_{\phi}$ and the $R$--charge $q_{\phi}$. Thus, chiral primary fields satisfy $h_{\phi}=\frac{q_{\phi}}{2}$. Using $(G^+_{-\frac{1}{2}})^{\dag}=G^-_{\frac{1}{2}}$, we can write
\begin{align}
\langle\phi|\{G^-_{\frac{1}{2}},G^{+}_{-\frac{1}{2}}\}|\phi\rangle&=\big|G^-_{\frac{1}{2}}|\phi\rangle\big|^2+\big|G^+_{-\frac{1}{2}}|\phi\rangle\big|^2,
\end{align}
which is positive definite. Thus, for any state $|\psi\rangle$ we have $h_{\psi}\geq\frac{q_{\psi}}{2}$. \\
We now consider the operator product between two chiral primary fields $\phi$ and $\chi$:
\begin{align}
\label{eq-chirope}
\phi(z)\chi(w)=\sum_i(z-w)^{h_{\psi_i}-h_{\phi}-h_{\chi}}\psi_i(w)
\end{align}
Note that $R$--charges add up in the operator product so that one has $q_{\psi_i}=q_{\phi}+q_{\chi}$. The relation to the conformal weights then implies that $h_{\psi_i}\geq h_{\phi}+h_{\chi}$. This entails that there are no singular terms in the operator product (\ref{eq-chirope}). So, taking the limit $z\rightarrow w$, the operator product is non--zero only if $\psi$ is a chiral primary field. Thus, the chiral primary fields form a closed, non--singular ring under the operator product. This is called the {\em chiral ring}.\\
We make one further observation. The $N=2$ algebra implies:
\begin{align}
\{G^-_{\frac{3}{2}},G^{+}_{-\frac{3}{2}}\}&=2L_0-3J_0+\frac{2}{3}c
\end{align}
From this it follows that the conformal weights of chiral primary fields are bounded from above by $\frac{c}{6}$. We thus reach the conclusion that there is only a finite number of chiral primary fields. \\
An analogous discussion can be made for antichiral states. In $N=(2,2)$ superconformal theories we can have chiral and anti--chiral primaries in the holomorphic and the antiholomorphic sector, yielding four different chiral rings, where two are complex conjugate of the other two. We denote them by $(c,c)$, $(c,a)$, $(a,c)$ and $(a,a)$.\\\\
Models with $N=2$ superconformal symmetry can be realized in various ways. We will now discuss the non--linear sigma models and the supersymmetric Landau--Ginzburg models. A non--linear sigma model in two dimensions governs maps $\phi:\Sigma\rightarrow X$ from a worldsheet $\Sigma$ to a target manifold $X$. In general, $X$ is a Riemannian manifold. $N=2$ supersymmetry constrains it to be K\"ahler. We define coordinates $z$, $\bar{z}$ on $\Sigma$ and $\phi^i$ on $X$. $\phi$ can be described {\em locally} by $\phi^i(z,\bar{z})$. The action has the following form:
\begin{align}
\label{eq-glsm}
S=&2t\int_{\Sigma}d^2z\bigg(g_{i\bar{i}}(\partial_z\phi^i\partial_{\bar{z}}\phi^{\bar{i}}+\partial_{\bar{z}}\phi^i\partial_z\phi^{\bar{i}})
+i\psi_-^{\bar{i}}D_z\psi_-^ig_{\bar{i}i}+i\psi_+^{\bar{i}}D_{\bar{z}}\psi_+^ig_{\bar{i}i}+R_{i\bar{i}j\bar{j}}\psi_+^i\psi_+^{\bar{i}}\psi_-^j\psi_-^{\bar{j}} \bigg)
\end{align}
Here $t$ is a coupling constant which will become important when we consider topological amplitudes, $g_{i\bar{j}}=\partial_i\partial_{\bar{j}} K(\phi,\bar{\phi})$ is the K\"ahler metric and $R_{i\bar{i}j\bar{j}}$ is the curvature constructed from the connection $\Gamma^i_{\phantom{i}jk}=g^{i\bar{k}}\partial_jg_{k\bar{k}}$. 
It is also possible to turn on a $B$--field $B_{i\bar{j}}$, which we will set to 0 here.
The covariant derivative is defined as follows:
\begin{align}
D_{\bar{z}}\psi_+^i&=\frac{\partial}{\partial\bar{z}}\psi_+^i+\frac{\partial\phi^j}{\partial\bar{z}}\Gamma^i_{\phantom{i}jk}\psi_+^k
\end{align}
Let $K,\bar{K}$ be the (anti--)canonical bundles in $\Sigma$, then the fermions are sections of the following bundles:
\begin{align}
\psi_+^i\in\Gamma(K^{\frac{1}{2}}\otimes\phi^{\ast}(T^{1,0}X))\qquad\qquad&\psi_+^{\bar{i}}\in\Gamma(K^{\frac{1}{2}}\otimes\phi^{\ast}(T^{0,1}X))\nonumber\\
\psi_-^i\in\Gamma(\bar{K}^{\frac{1}{2}}\otimes\phi^{\ast}(T^{1,0}X))\qquad\qquad&\psi_-^{\bar{i}}\in\Gamma(\bar{K}^{\frac{1}{2}}\otimes\phi^{\ast}(T^{0,1}X))
\end{align}
Assigning charges $Q_{\pm},\bar{Q}_{\pm}$ to the supersymmetry generators, the supersymmetry transformation laws defined by $\delta=\epsilon_+Q_--\epsilon_-Q_+-\bar{\epsilon}_+\bar{Q}_-+\bar{\epsilon}_-\bar{Q}_+$ are:
\begin{align}
\begin{array}{lllllll}
\delta\phi^i&=&\epsilon_-\psi^i_+-\epsilon_+\psi^i_- &\qquad&\delta\phi^{\bar{i}}&=&-\bar{\epsilon}_-\psi_+^{\bar{i}}+\bar{\epsilon}_+\psi_-^{\bar{i}}\\
\delta\psi_+^i&=&-i\bar{\epsilon}_-\partial_z\phi^i+\epsilon_+\psi_-^j\Gamma^i_{\phantom{i}jm}\psi^m_+&\qquad&\delta\psi_+^{\bar{i}}&=&i\epsilon_-\partial_z\phi^{\bar{i}}-\bar{\epsilon}_+\psi_-^{\bar{j}}\Gamma^{\bar{i}}_{\phantom{i}\bar{j}\bar{m}}\psi_+^{\bar{m}}\\
\delta\psi_-^i&=&i\bar{\epsilon}_+\partial_{\bar{z}}\phi^i-\epsilon_-\psi_+^j\Gamma^{i}_{\phantom{i}jm}\psi_-^m&\qquad&\delta\psi_-^{\bar{i}}&=&-i\epsilon_-\partial_{\bar{z}}\phi^{\bar{i}}+\bar{\epsilon}_-\psi_+^{\bar{j}}\Gamma^{\bar{i}}_{\phantom{i}\bar{j}\bar{m}}\psi_+^{\bar{m}}
\end{array}
\end{align}
The parameters $\epsilon_-,\bar{\epsilon}_-$ are sections of $K^{\frac{1}{2}}$ and $\epsilon_+,\bar{\epsilon}_+$ are sections of $\bar{K}^{\frac{1}{2}}$. Note that the action of the non--linear sigma model cannot be globally defined on a worldsheet for genus $g\neq 1$. At genus $g=1$ the canonical budle is trivial and thus the fermions are scalars which can be globally defined. For any other genus this is not the case. \\
A second important example of an $N=2$ supersymmetric theory is the Landau--Ginzburg model:
\begin{align}
\label{eq-lg}
S_{\Sigma}&=\int_{\Sigma}d^2xd^4\theta K(\Phi,\bar{\Phi})+\int_{\Sigma}d^2xd^2\theta W(\Phi)+\:c.c.,
\end{align}
where $K$ is the K\"ahler potential and $W$ is the holomorphic superpotential. $\Phi$ is a chiral superfield. A Landau--Ginzburg model is entirely characterized by its superpotential. The model itself is not a conformal field theory but it is believed to flow to a unique conformal fixed point in the infrared. The chiral ring $\mathcal{R}$ is isomorphic to the Jacobi ring \cite{Lerche:1989uy}, i.e. given a superpotential $W(\Phi_i)$ we have:
\begin{align}
\mathcal{R}&=\frac{\mathbbm{C}[\Phi_i]}{(\partial_jW(\Phi_i))}
\end{align}
These models will be of central interest in this thesis. We will give an exhaustive discussion of the supersymmetry variations and boundary conditions in section \ref{sec-bdrylg}.

\subsection{The Topological Twist and Mirror Symmetry}
In this section we discuss how to get a topological field theory out of an $N=2$ superconformal theory. A topological field theory is characterized by the requirement that correlators of physical operators are independent of the metric $g_{ij}$ of the manifold the theory lives on. Physical states are defined by the cohomology of a fermionic nilpotent operator $Q$, the BRST operator. Topological field theories can be classified according to the form of their actions \cite{Birmingham:1991ty}: A topological field theory of Schwarz type is of the form $S_S=\{Q,V\}$ where $V$ may also depend on the metric. A topological field theory of Witten type has an action $S_W=S_c+\{Q,V\}$, where $S_c$ is independent of the metric. Topological field theories have the crucial propoerty that the energy--momentum tensor $T_{ab}$ is $Q$--exact:
\begin{align}
T_{ab}=\{Q,B_{ab}\}
\end{align}
Restricting to the topological sector means that we restrict to a subsector of the physical theory. In particular, we will discuss that the topological theory localizes on the zero--modes of the physical fields and thus gives information about the vacuum sector of the full theory. There are two ways to get a topological field theory out of an $N=2$ superconformal field theory, the corresponding models are called the A--model and the B--model\footnote{Actually, there is another possibility which yields the {\em half--twisted} models \cite{Witten:1991zz}.}. A remarkable property is that these two models are related by mirror symmetry. 

\subsubsection{Twisting the $N=2$ Algebra}
At the level of the superconformal algebra, we can get the algebra of the topological theory by an operation called the {\em topological twist}, which is defined as follows:
\begin{align}
\label{eq-twist}
T(z)&\rightarrow T(z)\pm\frac{1}{2}\partial J(z)\nonumber\\
J(z)&\rightarrow\pm J(z)
\end{align}
With these redefinitions the algebra becomes:
\begin{align}
T(z)T(w)&\sim \frac{2T(w)}{(z-w)^2}+\frac{\partial_w T(w)}{z-w}\nonumber\\
T(z)G(w)&\sim\frac{2G(w)}{(z-w)^2}+\frac{\partial_wG(w)}{z-w}\nonumber\\
T(z)Q(w)&\sim\frac{Q(w)}{(z-w)^2}+\frac{\partial_wQ(w)}{z-w}\nonumber\\
T(z)J(w)&\sim -\frac{\hat{c}}{(z-w)^3}+\frac{J(w)}{(z-w)^2}+\frac{\partial_wJ(w)}{z-w}\nonumber\\
Q(z)G(w)&\sim \frac{\hat{c}}{(z-w)^3}+\frac{J(w)}{(z-w)^2}+\frac{T(w)}{z-w}\nonumber\\
J(z)G(w)&\sim-\frac{G(w)}{z-w}\nonumber\\
J(z)Q(w)&\sim\frac{Q(w)}{z-w}\nonumber\\
J(z)J(w)&\sim\frac{\hat{c}}{(z-w)^2}
\end{align}
Instead of the supersymmetry generators $G^{\pm}$ we have introduced two new quantities:
\begin{align}
G(z)=\sum_nG_nz^{n-2}\qquad&Q(z)=\sum_nQ_nz^{n-1}
\end{align}
$G(z)$ has conformal weight $2$ and $Q(z)$ has weight $1$. This implies that $\int Q$ is a scalar and can be defined globally on worldsheets of arbitrary genus. Furthermore $Q$ is fermionic and thus has the right properties to define the BRST operator of the topological field theory. We observe that the conformal anomaly in the operator product of the energy momentum tensor with itself has vanished. An important consequence is that the dimension in which a topological string theory can be consistently defined is not constrained to the critical dimension $D=10$. In the topological theory the $U(1)$ $R$--current $J(z)$ becomes anomalous. It has an anomalous background charge $\hat{c}$, which is related to the central charge of the untwisted theory by $\hat{c}=\frac{c}{3}$.\\
Given an $N=(2,2)$ superconformal algebra we can make two distinct choices for the topological twist. Performing the twist (\ref{eq-twist}) with the same sign for the left--moving and the right--moving sector it is called an $A$--twist. Choosing different signs in the two sectors is called a $B$--twist. The corresponding topological field theories are the A--model and the B--model. \\
We observe that the twisted $N=(2,2)$ superconformal algebra remains invariant under the following $\mathbbm{Z}_2$ transformation:
\begin{align}
J(z)\rightarrow J(z)&\qquad \bar{J}(\bar{z})\rightarrow -\bar{J}(\bar{z})
\end{align}
This exchanges the A--model and the B--model. This is the definition of mirror symmetry at the level of the algebra.

\subsubsection{The A--model}
We now discuss the $A$--twist and its consequences for the non--linear sigma model. In order to obtain the A--model we redefine the spins of the fields in the following way
\begin{align}
&\psi_+^i\equiv\chi^i\in\Gamma(\phi^{\ast}(T^{1,0}X))\qquad \:\quad\quad\psi_-^{\bar{i}}\equiv\chi^{\bar{i}}\in\Gamma(\phi^{\ast}(T^{0,1}X)) \nonumber\\
&\psi_+^{\bar{i}}\equiv\psi_z^{\bar{i}}\in\Gamma(\phi^{\ast}(K\otimes T^{0,1}X))\qquad\psi_-^{i}\equiv\psi_{\bar{z}}^{{i}}\in\Gamma(\phi^{\ast}(K\otimes T^{1,0}X))
\end{align}
The fermionic fields thus combine into a scalar $\chi\in\Phi^{\ast}(TX)$ and a holomorphic and an antiholomorphic one--form. The action written in terms of the newly defined quantities looks as follows:
\begin{align}
S&=2t\int_{\Sigma}d^2z \left(g_{i\bar{i}}(\partial_z\phi^i\partial_{\bar{z}}\phi^{\bar{i}}+\partial_{\bar{z}}\phi^i\partial_z\phi^{\bar{i}})+i\psi_z^{\bar{i}}\partial_{\bar{z}}\chi^ig_{\bar{i}i}+i\psi_{\bar{z}}^i\partial_z\chi^{\bar{i}}g_{\bar{i}i}-R_{i\bar{i}j\bar{j}}\psi_{\bar{z}}^i\psi_z^{\bar{i}}\chi^j\chi^{\bar{j}}\right)
\end{align}
Now we set $\epsilon_+=\bar{\epsilon}_-=0$ and $\epsilon_-=\bar{\epsilon}_+=-\epsilon$, where $\epsilon$ is a constant. Then we have $\delta=\epsilon(Q_++\bar{Q}_-)$. This A--type supersymmetry is now a scalar symmetry and can be defined on arbitrary worldsheets. We define the BRST operator
\begin{align}
\label{eq-asusy}
&Q_A=Q_++\bar{Q}_-,
\end{align}
which acts on fields via the anticommutator. The fields transform as follows under this symmetry:
\begin{align}
\begin{array}{lllllll}
\delta\phi^i&=&\epsilon\chi^i&\qquad&\delta\phi^{\bar{i}}&=&\epsilon\chi^{\bar{i}}\\
\delta\chi^{i}&=&0&\qquad&\delta\chi^{\bar{i}}&=&0\\
\delta \psi_z^{\bar{i}}&=&-i\epsilon\partial_z\phi^{\bar{i}}+\epsilon\chi^{\bar{j}}\Gamma^{\bar{i}}_{\phantom{i}\bar{j}\bar{m}}\psi_z^{\bar{m}}&\qquad&\delta\psi_{\bar{z}}^i&=&-i\epsilon\partial_{\bar{z}}\phi^i+\epsilon\chi^j\Gamma^i_{\phantom{i}jm}\psi_{\bar{z}}^m
\end{array}
\end{align}
The crucial observation is that the action can be written as follows\footnote{This holds modulo the equations of motion.}:
\begin{align}
S=it\int_{\Sigma}d^2z\{Q,V\}+t\int_{\Sigma}\phi^{\ast}(K),
\end{align}
where
\begin{align}
V&=g_{i\bar{j}}\left(\psi_z^{\bar{i}}\partial_{\bar{z}}\phi^j+\partial_z\phi^{\bar{i}}\psi_{\bar{z}}^j \right)
\end{align}
and 
\begin{align}
\int_{\Sigma}\phi^{\ast}(J)&=\int_{\Sigma}d^2z\left(\partial_z\phi^i\partial_{\bar{z}}\phi^{\bar{j}}g_{i\bar{j}}-\partial_{\bar{z}}\phi^i\partial_z\phi^{\bar{j}}g_{i\bar{j}} \right).
\end{align}
This is the integral of the pullback of the K\"ahler form $J=-ig_{i\bar{j}}dz^idz^{\bar{j}}$. It only depends on the cohomology class of $J$ and the homotopy class of the map $\phi$ and is thus a purely topological quantity. If $H^2(X,\mathbbm{Z})=\mathbbm{Z}$, then
\begin{align}
\label{eq-instanton}
\int_{\Sigma}\phi^{\ast}(J)=2\pi n,
\end{align}
given a proper normalization of the periods of $J$; $n$ will be identified as the instanton number. If we had a non--vanishing $B$--field we would have to replace $J$ by the complexified K\"ahler form $\omega=B+iJ$, where $B=\frac{1}{2}B_{i\bar{j}}d\phi^id\phi^{\bar{j}}$. \\
We now discuss the local physical operators of this theory, which are functions of $\phi$ and $\chi$ only. Considering an $n$--form $W=W_{i_1\ldots i_n}d\phi^{i_1}\ldots d\phi^{i_n}$ on $X$ we can define a local operator:
\begin{align}
\mathcal{O}_W(P)&=W_{i_1\ldots i_n}\chi^{i_1}\ldots\chi^{i_n}(P),
\end{align}
where $P$ is a point on the worldsheet.
One can then show that:
\begin{align}
\{Q,\mathcal{O}_W\}&=-\mathcal{O}_{dW}
\end{align}
Thus, the BRST cohomology of the A--model is isomorphic to the deRham cohomology on $X$. We can calculate correlators of these physical states, using the path integral:
\begin{align}
\langle\prod_a\mathcal{O}_a \rangle&=e^{-2\pi n t}\int D\phi D\chi D\psi e^{-it\{Q,\int V\}}\prod_a\mathcal{O}_a
\end{align}
Apart from the factor we pulled out, the path integral is independent of $t$ since differentiation of the correlator with respect to $t$ brings down irrelevant factors of the form $\{Q,\ldots\}$. Restricting to genus $0$, we can therefore compute the path integral in the limit of large $\mathrm{Re}\:t$, which corresponds to the weak coupling limit of the theory. The path integral thus localizes on the zero modes of the action and the action is minimized by holomorphic maps of $\Sigma$ of genus $0$ to $X$:
\begin{align}
&\partial_{\bar{z}}\phi^i=\partial_z\phi^{\bar{i}}=0
\end{align}
Since the action is topological there exist topologically non--trivial maps, called worldsheet instantons. They are classified by $n$ in (\ref{eq-instanton}). 
This implies that the path integral reduces to an integral over the moduli space $\mathcal{M}_{0,n}$ of holomorphic maps of degree $n$. Furthermore note that the path integral is independent of the complex structure of $\Sigma$ and $X$ and depends only on the cohomology classes of the K\"ahler form. This is because the complex structure dependence only appears in $Q$--exact terms in the path integral.

\subsubsection{The B--model}
We now describe the $B$--twisted non--linear sigma model. We have the following definitions:
\begin{align}
&\psi_{\pm}^{\bar{i}}\in\Gamma(\phi^{\ast}(T^{0,1}X))\nonumber\\
&\psi_+^i\in \Gamma(K\otimes\phi^{\ast}(T^{1,0}X))\nonumber\\
&\psi_-^i\in\Gamma(\bar{K}\otimes\phi^{\ast}(T^{1,0}X))
\end{align}
It is convenient to make the following field redefinitions:
\begin{align}
\eta^{\bar{i}}&=\psi_+^{\bar{i}}+\psi_-^{\bar{i}}\nonumber\\
\theta_i&=g_{i\bar{i}}(\psi_+^{\bar{i}}-\psi_-^{\bar{i}})
\end{align}
Furthermore, we combine $\psi_{\pm}^i$ into a one--form $\rho\in\Phi^{\ast}(T^{1,0}X)$, where $\psi_+^i\equiv\rho_z^i$ and $\psi_-^{i}\equiv \rho_{\bar{z}}^i$.
The action in terms of the new fields looks as follows:
\begin{align}
S=t\int_{\Sigma}d^2z\left(g_{i\bar{i}}(\partial_z\phi^i\partial_{\bar{z}}\phi^{\bar{i}}+\partial_{\bar{z}}\phi^i\partial_z\phi^{\bar{i}})+i\eta^{\bar{i}}\left(D_z\rho_{\bar{z}}^i+D_{\bar{z}}\rho_z^i\right)g_{i\bar{i}}+i\theta_i\left(D_{\bar{z}}\rho_z^i-D_z\rho_{\bar{z}}^i \right)+R_{i\bar{i}j\bar{j}}\rho_z^i\rho_{\bar{z}}^j\eta^{\bar{i}}\theta_kg^{k\bar{j}} \right)
\end{align}
Now we set $\epsilon_+=\epsilon_-=0$ and $\bar{\epsilon}_-=-\bar{\epsilon}_+=\epsilon$. Thus we get $\delta=\epsilon(\bar{Q}_++\bar{Q}_-)$. As in the A--model, this B--type supersymmetry can be globally defined on every worldsheet. The BRST operator for the B--model is:
\begin{align}
\label{eq-bsusy}
Q_B&=\bar{Q}_++\bar{Q}_-
\end{align}
The supersymmetry variations of the B--model look as follows:
\begin{align}
\delta\phi^i&=0\nonumber \\
\delta\bar{\phi}^{\bar{i}}&=-\epsilon\eta^{\bar{i}}\nonumber\\
\delta\eta^{\bar{i}}&=\delta\theta_i=0\nonumber\\
\delta\rho^i&=-i\epsilon\:d\phi^i
\end{align}
The action can be rewritten in the following way:
\begin{align}
S&=it\int\{Q,V\}+tW,
\end{align}
where
\begin{align}
V&=g_{i\bar{j}}\left(\rho_z^i\partial_{\bar{z}}\phi^{\bar{j}}+\rho_{\bar{z}}^i\partial_z\phi^{\bar{j}} \right)
\end{align}
and
\begin{align}
W&=\int_{\Sigma}\left(-\theta_iD\rho^i-\frac{i}{2}R_{i\bar{i}j\bar{j}}\rho^i\wedge\rho^j\eta^{\bar{i}}\theta_kg^{k\bar{j}} \right).
\end{align}
In the expression for $W$, $D$ is the exterior derivative on the worldsheet. The theory is a topological theory in the sense that it is independent of the complex structure of $\Sigma$ and the K\"ahler metric on $X$: $W$ is entirely independent of the complex structure, since it is written in terms of forms. Under a change of the K\"ahler metric of $X$ the Lagrangian is invariant up to terms of the form $\{Q,\ldots\}$ in the action \cite{Witten:1991zz}. \\
Next, we discuss the local observables of the B--model. We consider $(0,p)$--forms $V$ in $X$ with values in $\wedge^q T^{1,0}X$:
\begin{align}
V&=d\bar{z}^{\bar{i}_1}\ldots d\bar{z}^{\bar{i}_p}V_{\bar{i}_1\ldots\bar{i}_p}^{\phantom{\bar{i}_1\ldots\bar{i}_p}j_1\ldots j_q}\psi_{j_1}\ldots\psi_{j_q}
\end{align}
With that we can define the following operator:
\begin{align}
\mathcal{O}_V&=\eta^{\bar{i}_1}\ldots\eta^{\bar{i}_p}V_{\bar{i}_1\ldots \bar{i}_p}^{\phantom{\bar{i}_1\ldots\bar{i}_p }j_1\ldots j_q}\psi_{j_1}\ldots\psi_{j_q}
\end{align}
It can be shown that:
\begin{align}
\{Q,\mathcal{O}_V\}&=-\mathcal{O}_{\bar{\partial}V}
\end{align}
This implies that the BRST cohomology of the B--model can be related to Dolbeault cohomology on $X$.\\
We can now go on to compute correlation function with insertions of these operators. We observe that the path--integral is independent of the coupling $t$. Under a change of $t$ the term $\{Q,V\}$ changes by a $Q$--exact expression. The $t$--factor in front of $W$ can be removed by a field redefinition $\theta\rightarrow \theta/t$. This yields a polynomial dependence of $t$ in the correlation functions. Thus, there are no instanton contributions in the B--model; B--model calculations are classical.\\
As in the A--model we can compute the path integral in the weak coupling limit. One finds that the path integral localizes on constant maps $\Phi:
\Sigma\rightarrow X$:
\begin{align}
\partial_z\phi^i&=\partial_{\bar{z}}\phi^i=0
\end{align}
Quantizing the B--model one finds an anomaly. There is an anomaly cancellation condition which requires the vanishing of the first Chern class: $c_1(X)=0$. Thus, for the B--model to be consistent, $X$ has to be Calabi--Yau. The A--model, on the other hand, makes sense for arbitrary K\"ahler manifolds.\\\\
Mirror symmetry relates the A--model and the B--model in the following way: We have seen that the A--model only depends on the K\"ahler moduli whereas the B--model depends on the complex structure moduli of the Calabi--Yau target. On a Calabi--Yau manifold $X$ the dimension of the K\"ahler moduli space is encoded in the Hodge number $h^{1,1}$ of the manifold. The dimension of the complex structure moduli space is given by $h^{1,2}$. Since, in general, $h^{1,1}$ and $h^{2,1}$ are different for a Calabi--Yau, the Calabi--Yau in the A--model has to be different from the one in the B--model. The remarkable statement of mirror symmetry is that the A--model on a Calabi--Yau $X$ is equivalent to the B--model on the mirror Calabi--Yau $X^{\ast}$. As compared to $X$, the Hodge numbers of $X^{\ast}$ are exchanged. The mirror map gives an explicit isomorphism between the complex structure moduli on $X^{\ast}$ and the K\"ahler moduli on $X$. Mirror symmetry provides an important tool for explicit calculations. Remember that the A--model receives quantum corrections, while the B--model does not. We can use mirror symmetry to calculate quantum--corrected amplitudes in the A--model by making a classical calculation in the B--model and then applying the mirror map.\\\\So far, we have only discussed topological field theories. In order to obtain a string theory we have to couple the theory to gravity. This means that we have to gauge diffeomorphism invariance. This can be done by the Noether procedure, which yields the Beltrami differentials $\mu$. These couple, in the topological string, to the $G(z)$. At the level of correlation functions, coupling to gravity amounts to integrating the amplitudes over the moduli space of the Riemann surfaces. At genus $g$ with $g>1$  we can define a free energy $F_g$ as follows:
\begin{align}
\label{eq-fg}
F_g&=\int_{\mathcal{M}_g}\left\langle \prod_{k=1}^{3g-3}(\int G\mu_k)(\int \bar{G}\bar{\mu}_k) \right\rangle
\end{align}
Since we will only focus on tree level amplitudes in this thesis, we will not discuss higher genus amplitudes here.\\\\
Note that mirror symmetry extends to the full string theory. There, the statement is that the type IIA string theory compactified on a certain Calabi--Yau manifold is equivalent to the type IIB string compactified on the mirror Calabi--Yau.

\subsection{Phases of $N=2$ Theories}
We now briefly discuss the Landau--Ginzburg/Calabi--Yau correspondence \cite{Witten:1993yc}, which identifies the non--linear sigma model with Calabi--Yau target and a Landau--Ginzburg model with a certain superpotential as two 'phases' of a supersymmetric gauge theory, the {\em linear sigma model}. The order parameter which is relevant for the phase transition parameterizes the size of the Calabi--Yau.\\
The linear sigma model is a quite general $N=2$ supersymmetric abelian gauge theory in two dimensions. The action contains the following terms:
\begin{align}
S&=S_{kin}+S_W+S_{gauge}+S_{FI,\theta}
\end{align}
The first term is the kinetic energy of the chiral superfields, the second term is the superpotential interaction, the third term is the kinetic energy of the gauge fields and the last term includes the Fayet--Iliopoulos term and the term with the theta angle. We can write these expressions in terms of superfields. The kinetic term is:
\begin{align}
S_{kin}&=\int d^2xd^4\:\theta \bar{\Phi}e^V\Phi,
\end{align}
where $\Phi$ is the chiral superfield and $V$ is the vector superfield. The superpotential term is:
\begin{align}
S_W&=\int d^2xd^2\:\theta W+c.c
\end{align}
The gauge kinetic term looks as follows:
\begin{align}
S_{gauge}=-\frac{1}{4e^2}\int d^2xd^4\:\theta\bar{\Sigma}\Sigma,
\end{align}
where $e$ is the gauge coupling and $\Sigma$ is the twisted chiral superfield, satisfying $\bar{D}_+\Sigma=D_-\Sigma=0$. The last term in the action can be written as follows:
\begin{align}
S_{FI,\theta}&=\frac{it}{2\sqrt{2}}\int d^2xd\theta^+d\bar{\theta}^-\Sigma+c.c,
\end{align}
where
\begin{align}
\label{eq-tpar}
t=ir+\frac{\theta}{2\pi}.
\end{align}
The parameter $r$ is the Fayet--Iliopoulos parameter and $\theta$ is the theta angle. \\
We now consider a theory with $n$ chiral superfields $S_i$ of charge $1$ and one field $P$ of charge $-n$. We denote the bosonic components of $S_i$ and $P$ with $s_i$ and $p$, respectively. We choose the superpotential to be
\begin{align}
W&=P\cdot G(S_1,\ldots,S_n),
\end{align}
where $G$ is a homogeneous polynomial of degree $n$. $W$ is a quasihomogeneous polynomial. We demand that the equations
\begin{align}
\label{eq-transversality}
&\frac{\partial G}{\partial S_1}=\ldots=\frac{\partial G}{\partial S_n}
\end{align}
have no common roots except at $S_i=0$. This ensures that the hypersurface $X$ in $\mathbbm{CP}^{n-1}$ defined by the equation $G=0$ is smooth. In order to probe the low--energy physics of this theory we have to minimize the bosonic superpotential, which looks as follows:
\begin{align}
U&=|G(s_i)|^2+|p|^2\sum_i{\vline\frac{\partial G}{\partial s_i}\vline}^2+2|\sigma|^2\left(\sum_i|s_i|^2+n^2|p|^2\right),
\end{align} 
where
\begin{align}
D=-e^2\left(\sum_i \bar{s}_is_i-n\bar{p}p-r \right).
\end{align}
We first consider the case where $r\gg0$. Minimizing the $D$--term implies that not all $s_i$ can vanish, which entails that not all ${\vline\frac{\partial G}{\partial s_i}\vline}$ vanish. We are thus forced to set $p=0$. From this, it follows that $|\sigma|=0$ and $G=0$. The vanishing of the $D$--term then implies:
\begin{align}
\sum_i\bar{s}_is_i=r
\end{align}
In order to identify the ground state we also have to take into account the $U(1)$ gauge symmetry:
\begin{align}
(s_1,\ldots,s_n)&\sim (e^{i\varphi}s_1,\ldots,e^{i\varphi}s_n)
\end{align}
This implies that the $s_i$ live in $\mathbbm{CP}^{n-1}$. We conclude that the space of classical vacua is isomorphic to the hypersurface $X\subset\mathbbm{CP}^{n-1}$, defined by $G=0$. A smooth hypersurface of degree $n$ in $\mathbbm{CP}^{n-1}$ is Calabi--Yau. The size of the Calabi--Yau is governed by the parameter $r$, which can be identified with the K\"ahler parameter determining the radius of the Calabi--Yau. Thus, at large radius, the low energy theory is described by a $\sigma$--model with Calabi--Yau target. \\
Now we consider the case $r\ll0$. The vanishing of $D$ requires that $P\neq0$. Given the condition (\ref{eq-transversality}), this implies that all $s_i=0$. Then it follows that $|p|=\sqrt{\frac{-r}{n}}$. So, up to gauge transformation the theory has a unique classical vacuum. We can expand around this vacuum and find that the $s_i$ are massless for $n\geq 3$. Integrating out the massive field $p$ by setting it to its vacuum expectation value we get an effective superpotential for the $s_i$, which is $\tilde{W}=\sqrt{-r}G(s_i)$. The prefactor can be absorbed into the definition of the $s_i$. The theory is governed by the superpotential $\tilde{W}$, which has a degenerate critical point at the origin. We thus have obtained a Landau--Ginzburg theory in the limit $r\ll0$. In our case we obtain a Landau--Ginzburg orbifold: the vacuum expectation value of $p$ breaks the $U(1)$ gauge group down to a $\mathbbm{Z}_n$ subgroup, which acts as $s_i\rightarrow \omega s_i$, where $\omega$ is an $n$--th root of unity.\\\\
To conclude, we have found that the low energy limit of the linear sigma model is described by a Calabi--Yau sigma model at large radius, in the non-geometric regime ($r\ll 0$) it is realized as a Landau--Ginzburg model. The two regions are separated by a singularity at $r=0$. One can view these two models as phases of the linear sigma model, which undergoes a phase transition at $r=0$. One can smoothly interpolate between those phases by varying the parameter $t$ in (\ref{eq-tpar}) \cite{Witten:1993yc}.\\
The above discussion has an interesting implication on the topological string. We have argued before, that the topological B--model is independent of the K\"ahler moduli. It is thus insensitive to the variations of the size of the Calabi--Yau governed by $r$. This entails that the topological observables of the B--model are the same for the Landau--Ginzburg and the Calabi--Yau description. There are thus two equivalent ways to realize the B--model, which correspond to the large and small radius regimes in K\"ahler moduli space.\\
We can generalize this discussion in many ways, in particular it also holds for curves in weighted projective space, hypersurfaces in toric varieties and complete intersections of hypersurfaces. These cases may yield a more complicated phase structure than we found here.
\section{D--branes}
\label{sec-branes}
A D--brane is a boundary condition. D--branes are objects where an open string can end. Open strings live on worldsheets, which are Riemann surfaces with boundaries. As a toy example, consider a bosonic sigma model which maps from a cylinder $\Sigma=S^1\times\mathbbm{R}$ to $\mathbbm{R}$:
\begin{align}
S&=\int_{\Sigma}d^2x\partial_{\mu}\phi\partial^{\mu}\phi
\end{align}
Computing $\delta S=0$ one gets the following boundary contribution:
\begin{align}
\delta\phi\partial_{\perp}\phi|_{\partial\Sigma}&=0,
\end{align}
where $\partial_{\perp}$ is the derivative in the direction normal to the boundary.\\
There are two possibilities to satisfy this condition. The {\em Neumann} (N) boundary condition implies that the ends of the string move freely:
\begin{align}
\partial_{\perp}\phi\vert_{\partial\Sigma}&=0
\end{align}
The {\em Dirichlet} (D) boundary conditions entail that the ends of the string are fixed on a subspace of the target space:
\begin{align}
\delta\phi\vert_{\partial\Sigma}&=0
\end{align}
In general, a D$p$--brane is defined as a $p$--dimensional spatial subspace of the target space where the end of an open string is confined to. One thus has Neumann boundary conditions in $p$ directions and Dirichlet boundary conditions in the remaining spatial directions.\\
Dirichlet boundary conditions break Lorentz invariance. In supersymmetric theories this causes supersymmetry breaking. A generic D--brane configuration breaks supersymmetry completely. We will be interested in D--branes which only partly break supersymmetry, in particular we want to break $N=2$ supersymmetry down to $N=1$.\\
At the level of conformal field theory a boundary is introduced as follows. Given one boundary component, we can make a conformal transformation to map the worldsheet into the upper half--plane and its boundary to $z=\bar{z}$. The left-- and rightmoving currents have to match at the boundary. For $N=(2,2)$ superconformal theories, one has two possibilities to define boundary conditions which break half of the supersymmetry. These are called A--type and B--type boundary conditions \cite{Ooguri:1996ck}. A--branes are specified by the following conditions on the superconformal currents at $z=\bar{z}$:
\begin{align}
&T(z)=\bar{T}(\bar{z})\qquad G^+(z)=\pm\bar{G}^-(\bar{z})\qquad G^{-}(z)=\pm\bar{G}^+(\bar{z})\qquad J(z)=-\bar{J}(\bar{z})
\end{align}
The B--type boundary conditions are defined as follows:
\begin{align}
&T(z)=\bar{T}(\bar{z})\qquad G^+(z)=\pm\bar{G}^+(\bar{z})\qquad G^{-}(z)=\pm\bar{G}^-(\bar{z})\qquad J(z)=\bar{J}(\bar{z})
\end{align}
The mode expansions of these boundary conditions determine the {\em Ishibashi states} on the boundary. A general boundary state can be expanded in terms of Isibashi states. A--type and B--type boundary conditions are related via the mirror map. Boundary conformal field theory has been an active area of research and has lead to a lot of insight on D--branes. In this thesis we will look at D--branes from a different angle, so we refrain from giving a detailed review of boundary conformal field theory.\\
Although we will consider D--branes mostly from a worldsheet perspective, let us briefly mention some target space aspects \cite{polchinski2}. It can be shown that D--branes satisfying these conditions are BPS states of the supersymmetric theory. The world--volumes of $p$--branes couple to $(p+1)$--form Ramond--Ramond potentials $C_{p+1}$. D--branes carry conserved charges, the Ramond--Ramond charges. The potentials can be integrated over the world--volumes of the branes:
\begin{align}
\int C_{p+1}
\end{align}
These potentials are sources for gauge field strengths. In analogy with electromagnetism, a D$p$--brane and a $(6-p)$--brane are electric and magnetic sources for the same field strength\footnote{We consider superstring theory and therefore a 10--dimensional target space.}. For example, the free field equations and Bianchi identity for a two--form field strength, $d\ast F_2=dF_2=0$, are symmetric between $F_2$ and $(\ast F)_8$, and can be written in terms of a $1$--form and a $7$--form potential:
\begin{align}
F_2=dC_1&\qquad d\wedge\ast dC_1=0\nonumber\\
\ast F_2=(\ast F)_8=d C_7&\qquad d\wedge\ast dC_7=0
\end{align} 
An electric source is a D$0$--brane for $C_1$ or a D$6$--brane for $C_7$. A magnetic source is a D$6$--brane for $C_1$ or a D$0$--brane for $C_7$. The electric, respectively magnetic, charges are the Ramond--Ramond charges. 

\subsection{Topological D--branes}
We now turn to D--branes in topological string theory. We will discuss boundary conditions in the non--linear sigma model and thus restrict ourselves to the large radius limit. B--type D--branes in Landau--Ginzburg models, which are realized as matrix factorizations of the Landau--Ginzburg superpotential, will be discussed in the following chapters of this thesis. In line with standard conventions for the topological string, when speaking of a D$p$--brane we only take into account boundaries in the Calabi--Yau $X$, thus neglecting the three spatial directions of $\mathbbm{R}^{1,3}$. Topological D--branes on a Calabi--Yau threefold are thus D$0$--D$6$--branes in these conventions. Our discussion will follow \cite{Aspinwall:2004jr}.\\
Boundary conditions in topological string theory must be compatible with the topological twist. In particular, this implies that branes in the A--model must preserve A--type supersymmetry (\ref{eq-asusy}) and branes in the B--model must preserve B--type supersymmetry (\ref{eq-bsusy}). Which objects do these boundary conditions define?
Let us write the bosonic part of the action (\ref{eq-glsm}) in a more condensed notation:
\begin{align}
S_{bos}&=t\int_{\Sigma}d^2z g_{IJ}(\phi)\partial_z\phi^I\partial_{\bar{z}}\phi^J
\end{align}
Here $I,J$ denote real coordinates and $g_{IJ}$ incorporates the K\"ahler metric and the $B$--field. The boundary conditions connect the left-- and the rightmoving sector. We can write this as follows:
\begin{align}
\partial_z\phi^I&=R^I_{\phantom{I}J}(\phi)\partial_{\bar{z}}\phi^J\nonumber\\
\psi_+^I&=R^I_{\phantom{I}J}(\phi)\psi_-^J
\end{align}
The matrix $R$ is orthogonal with respect to the metric. Eigenvectors with eigenvalue $-1$ give Dirichlet boundary conditions. If we picture a D--brane as a submanifold $L$ of the Calabi--Yau $X$, these vectors span the directions normal to $L$. Eigenvectors associated to eigenvalues $+1$ of $R$ are associated to directions tangent to the D--brane.
\subsubsection{A--branes}
Now we turn to the A--model. The boundary conditions which are consistent with the A--twist are:
\begin{align}
\psi^i_+&=R^{i}_{\phantom{i}\bar{j}}\psi_-^{\bar{j}}\nonumber\\
\psi_+^{\bar{i}}&=R^{\bar{i}}_{\phantom{i}j}\psi_-^j\nonumber\\
R^i_{\phantom{i}j}&=R^{\bar{i}}_{\phantom{i}\bar{j}}=0
\end{align}
Now we choose a vector $v$ tangent to the D--brane, i.e. with eigenvalue $+1$. Now consider the complex structure $J$ with
\begin{align}
J_n^m=i\delta_n^m&\qquad J_{\bar{n}}^{\bar{m}}=-i\delta_{\bar{n}}^{\bar{m}}
\end{align}
The vector $Jv$ has eigenvalue $-1$ with respect to $R$, the vector $J^2v=-v$ is again in the tangent direction. Thus $J$ exchanges the directions tangent and normal to the D--brane $L$. In order for this to make sense, $L$ must be of middle dimension, which means of real dimension $3$, if $X$ is a Calabi--Yau threefold. Now consider two vectors $v$ and $w$, tangent to $L$. The vector $w$ is orthogonal to $Jv$ with respect to the metric. The K\"ahler form on $X$ can be written as $\omega=\frac{1}{2}g_{LM}J_N^Md\phi^Ld\phi^N$. The previous arguments imply that the K\"ahler form restricted to $L$ is zero. A manifold with such properties is called a {\em Lagrangian submanifold}. \\
Worldsheets with boundary lead to additional degrees of freedom. In the A--model we can include these extra gauge degrees of freedom into the action by defining a one--form $A$ on $X$. The additional term in the action looks as follows:
\begin{align}
S_{\partial\Sigma}&=t\int_{\partial\Sigma}\Phi^{\ast}(A)
\end{align}
$A$ is a gauge connection and we define $F=dA$. In order to maintain BRST invariance we must have $F=0$. Thus $A$ has to be a flat connection. Upon quantization one encounters an anomaly. The associated anomaly cancellation condition is related to the vanishing of the {\em Maslov class}:
\begin{align}
\xi_{\ast}:\pi_1(L)\rightarrow\pi_1(S^1)\cong\mathbbm{Z}
\end{align}
The Maslov class is $0$ if the fundamental group of $L$ vanishes: $\pi_1(L)=0$. We thus can make the following statement:
\begin{center}
{\em An A--type D--brane wraps a {\em Lagrangian submanifold} $L$, which has the properties:}
\end{center}
\begin{align}
&\omega|_L=0\nonumber\\
&F=0\nonumber\\
&\textrm{trivial Maslov class}
\end{align}

\subsubsection{B--branes}
We now investigate boundary conditions in the B--model. We consider the non--linear sigma model, the present discussion is valid in the large radius limit. \\
The following boundary conditions on the fermions are compatible with the B--twist:
\begin{align}
\psi_+^i&=R^i_{\phantom{i}j}\psi_-^j\nonumber\\
\psi_+^{\bar{i}}&=R^{\bar{i}}_{\phantom{i}\bar{j}}\psi_-^{\bar{j}}\nonumber\\
R^i_{\phantom{i}\bar{j}}&=R^{\bar{i}}_{\phantom{i}j}=0
\end{align}
In the B--model, the complex structure thus preserves the tangent and normal directions of the D--brane. Therefore we can make the statement:
\begin{center}
{\em A B--type D--brane wraps holomorphic cycles in $X$.}
\end{center}
A B--brane can thus be a D$0$--, D$2$--, D$4$-- or D$6$--brane. Having branes of different dimensions is troublesome. In order to see the problem we first discuss a D$6$--brane, i.e. a brane which fills $X$. Taking into account the additional boundary degrees of freedom we introduce a bundle $E\rightarrow X$ over the D--brane. Similar to the A--model case the condition of BRST invariance of the action implies that the curvature $F$ of the bundle is a two--form of type $(1,1)$. Thus $E\rightarrow X$ is a holomorphic bundle. What about the other branes, which correspond to true subspaces of $X$? The notion of a vector bundle over a submanifold is not well--defined. The correct mathematical framework to address this issue turns out to be sheaves. A B--brane at the large radius limit can be described as a {\em coherent sheaf}. We will say some more about this in the following section.\\
Finally, let us briefly mention that in order to define a physical D--brane, a topological D--brane has to satisfy additional stability conditions. In particular, A--branes have to wrap {\em special Lagrangian} submanifolds, which imposes the constraint that $\mathrm{Re}e^{i\theta}\Omega|_{L}=0$, where $\Omega$ is the holomorphic threeform (if we consider a Calabi--Yau threefold) and $\theta$ is an arbitrary phase. For B--branes the sheaves have to be {\em stable} in order to be physical. The information on stability is not encoded in the topological sector. Details on this issue can be found for example in \cite{Aspinwall:2004jr}.

\subsection{Mathematical Description}
D--branes naturally fit into the mathematical framework of categories. Categories are a very abstract concept and one may ask how this can be of use to a physicist. It turns out that categories are the suitable framework to describe systems of multiple D--branes and phenomena like tachyon condensation, which cannot be captured by other approaches, that are, in some sense, more 'physical'. It turns out that matrix factorizations realize all these concepts in a very explicit way. We will discuss the details in chapter \ref{chap-mf}, in this section we only intend to discuss the basic setup. We will closely follow \cite{Aspinwall:2004jr} and \cite{Sharpe:2003dr}.\\
Let us start with the definition of a category:
\begin{definition} A {\em category} $\mathcal{C}$ consists of a class $\mathrm{obj}(\mathcal{C})$ of {\em objects}, a set $\mathrm{Hom}_{\mathcal{C}}(A,B)$ of {\em morphisms} for every ordered pair $(A,B)$ of objects, an {\em identity morphism} $\mathrm{id}_A\in\mathrm{Hom}_{\mathcal{C}}(A,A)$, and a {\em composition function}
\begin{align}
\mathrm{Hom}_{\mathcal{C}}(A,B)\times\mathrm{Hom}_{\mathcal{C}}(B,C)&\rightarrow\mathrm{Hom}_{\mathcal{C}}(A,C),
\end{align}
for every ordered triplet $(A,B,C)$ of objects. If $f\in\mathrm{Hom}(A,B)$, $g\in\mathrm{Hom}(B,C)$, the composition is denoted $gf$. The above data is subject to two axioms:
\begin{enumerate}
\item {\em Associativity axiom}: $(hg)f=h(gf)$ for $f\in\mathrm{Hom}_{\mathcal{C}}(A,B)$, $g\in\mathrm{Hom}_{\mathcal{C}}(B,C)$ and $h\in\mathrm{Hom}_{\mathcal{C}}(C,D)$.
\item {\em Unit axiom}: $\mathrm{id}_Bf=f=f\mathrm{id}_A$ for $f\in\mathrm{Hom}_{\mathcal{C}}(A,B)$.
\end{enumerate}
\end{definition}
We can now make the following identification:
\begin{center}
{\em D--branes are objects in a category, open string states are morphisms.}
\end{center}
How this category is realized depends on the model under consideration.
\subsubsection{A--branes}
A--type D--branes on a Calabi--Yau are objects in the {\em Fukaya category} $\mathrm{Fuk}(X)$. \\
We will (unfortunately) not have to say much about this category since only a few examples are known. This is related to the fact that it is in general quite difficult to identify Lagrangian submanifolds. One simple example where this is possible is the torus $T^2$, where the D$1$--branes are just lines winding around the torus. The Fukaya category for the torus was defined in \cite{Polishchuk:1998db}. We will discuss the torus in detail in chapter \ref{chap-torus}.\\
The Fukaya category is defined in terms of symplectic geometry which reflects the independence of the complex structure. Fukaya's category is endowed with an $A_{\infty}$--structure. An $A_{\infty}$--algebra is a non--associative algebra. We will give more details on this in the following chapters.

\subsubsection{B--branes}
We have already mentioned in the previous section that B--branes at the large radius limit are related to sheaves. The precise statement is that B--type D--branes on a Calabi--Yau $X$ are objects in the {\em derived category of coherent sheaves} $D(X)$. This category is defined in terms of algebraic geometry and is independent of the K\"ahler structure. In contrast to the Fukaya category, the derived category of coherent sheaves is quite well--understood. We will now summarize the essential definitions related to this category, putting emphasis on the material we will need later on. We refer to \cite{Sharpe:2003dr,Aspinwall:2004jr} for the details.\\
We start by giving the definition of a sheaf. This is done in two steps:
\begin{definition}
A {\em presheaf} $\mathscr{F}$ on $X$ consists of the following data:
\begin{itemize}
\item For every open set $U\subset X$ we associate an abelian group $\mathscr{F}(U)$.
\item If $V\subset U$ are open sets there is a 'restriction' homomorphism $\rho_{UV}:\mathscr{F}(U)\rightarrow\mathscr{F}(V)$.
\end{itemize}
The following conditions have to be satisfied:
\begin{enumerate}
\item $\mathscr{F}(\emptyset)=0$.
\item $\rho_{UU}$ is the identity map.
\item If $W\subset V\subset U$ then $\rho_{UW}=\rho_{VW}\rho_{UV}$.
\end{enumerate}
\end{definition}
If $\sigma\in\mathscr{F}(U)$ we denote the restriction $\rho_{UV}(\sigma)$ by $\sigma|_{V}$. 
\begin{definition}
A {\em sheaf} $\mathscr{F}$ on $X$ is a presheaf satisfying the conditions:
\begin{enumerate}
\item If $U,V\subset X$ and $\sigma\in\mathscr{F}(U)$, $\tau\in\mathscr{F}(V)$ such that $\sigma_{U\cap V}=\tau_{U\cap V}$, then there exists $\nu\in\mathscr{F}(U\cup V)$ such that $\nu|_{U}=\sigma$ and $\nu|_{V}=\tau$.
\item If $\sigma\in\mathscr{F}(U\cap V)$ and $\sigma_U=\sigma_V=0$, then $\sigma=0$.
\end{enumerate}
\end{definition}
Our first goal is to explain the notion of a coherent sheaf. If $\mathscr{F}(U)$ is the group of holomorphic functions over $U$, we can define a sheaf of holomorphic functions $\mathscr{O}_X$, called the {\em structure sheaf}. Note that the holomorphic functions are not only a group but they also define a ring. This motivates the following definition:
\begin{definition}
Let $R$ be a ring with a multiplicative identity $1$. An $R$--module is an abelian group $M$ with an $R$--action given by a mapping $R\times M\rightarrow M$ such that
\begin{enumerate}
\item $r(x+y)=rx+ry$
\item $(r+s)x=rx+sx$
\item $(rs)x=r(sx)$
\item $1x=x$
\end{enumerate}
for any $r,s\in R$ and $x,y\in M$. 
\end{definition}
Roughly speaking, a module is the algebraic generalization of the notion of a vector space. This allows us to define a {\em sheaf of $\mathscr{O}_X$--modules}. This can be done because holomorphic functions over $U$ have a ring structure under multiplication. A sheaf $\mathscr{E}$ is a sheaf of $\mathscr{O}_X$--modules if $\mathscr{E}(U)$ is an $\mathscr{O}_X(U)$--module for any open $U\in X$. A {\em free $\mathscr{O}_X$--module of rank $n$} is given by:
\begin{align}
\mathscr{O}_X^{\oplus n}&=\underbrace{\mathscr{O}_X\oplus\ldots\oplus\mathscr{O}_X}_n
\end{align}
A sheaf $\mathscr{E}$ is called {\em locally free} of rank $n$ if there is an open covering $\{U_{\alpha}\}$ of $X$ such that $\mathscr{E}(U_{\alpha})\cong\mathscr{O}_X(U_{\alpha})^{\oplus n}$. There is a one--to--one correspondence between locally free sheaves of rank $n$ and sections of holomorphic vector bundles of rank $n$.\\
Locally free sheaves are thus the algebraic way of describing holomorphic vector bundles. However, a B--brane is more than a locally free sheaf. What we must consider instead are {\em coherent sheaves}. We refrain from giving the technical definition, which can be found in the reviews cited above. Coherent sheaves form a category. As compared to locally free sheaves, the category of coherent sheaves contains sheaves whose ranks may not be globally defined. In particular the category includes the {\em skyscraper sheaf} $\mathscr{O}_P$. $\mathscr{O}_P$ is defined such that $\mathscr{O}_P(U)$ is the trivial group if $U$ does not contain $P$. If it does we have $\mathscr{O}_P(U)=\mathbbm{C}$. It can be associated to a vector bundle which has fiber $\mathbbm{C}$ over the origin and trivial fiber elsewhere. Such objects describe D$0$--branes on the Calabi--Yau $X$. D--branes can be viewed as objects in the category of coherent sheaves. In the following we will continue working with locally free sheaves. The results given below can then be generalized to coherent sheaves.\\
Before we define the derived category, we turn to open string states. As we discussed earlier, the physical states in the $B$--model are related to the cohomology of the operator $Q_B$. It is possible to extend the notion of cohomology to categories of sheaves. We refer to the literature for the details and just state the result \cite{Aspinwall:2004jr}:\\\\
{\em An open string from the B--brane associated to the locally free sheaf $\mathscr{E}$ to another B--brane associated to the locally--free sheaf $\mathscr{F}$ is given by an element of the group $\mathrm{Ext}^q(\mathscr{E},\mathscr{F})$}.\\\\
The groups $\mathrm{Ext}$ can be related to the sheaf cohomology group $H^q$:
\begin{align}
H^q(X,\mathscr{H}om(\mathscr{E},\mathscr{F}))&=\mathrm{Ext}^q(\mathscr{E},\mathscr{F})
\end{align}
Here, $\mathscr{H}om(\mathscr{E},\mathscr{F})$ is a locally free sheaf, which can be obtained in the following way. The locally free sheaves $\mathscr{E},\mathscr{F}$ are related to holomorphic sections of bundles $E,F$. The sheaf $\mathscr{H}om(\mathscr{E},\mathscr{F})$ is then associated to $\mathrm{Hom}(E,F)$. The crucial point is now that one can relate the sheaf cohomology group $H^q$ to Dolbeault cohomology:
\begin{align}
H^{0,q}(X,\mathrm{Hom(E,F)})&=H^q(X,\mathscr{H}om(\mathscr{E},\mathscr{F}))
\end{align}
So we have succeeded in expressing the cohomology relevant for the B--model in terms of the mathematical language of categories and sheaves. But we are not quite done yet. It turns out that we have to refine the category in order to take into account all the B--branes. Given a locally free sheaf $\mathscr{E}$, we can define a collection of D--branes as follows:
\begin{align}
\mathscr{E}&=\bigoplus_{n\in\mathbbm{Z}}\mathscr{E}^n,
\end{align}
where $n$ is the 'ghost number' of the B--brane. For matrix factorizations this grading will reduce to a $\mathbbm{Z}_2$--grading. We can define morphisms $d_n:\mathscr{E}^n\rightarrow\mathscr{E}^{n+1}$, where:
\begin{align}
d_n\in\mathrm{Ext}^0(\mathscr{E}^n,\mathscr{E}^{n+1})=\mathrm{Hom}(\mathscr{E}^n,\mathscr{E}^{n+1})
\end{align}
These morphisms satisfy\footnote{We will meet this again in the next chapter, when we discuss Laundau--Ginzburg models, where this will be related to the matrix factorization condition.}:
\begin{align}
d_{n+1}d_n&=0
\end{align}
This defines a complex $\mathscr{E}^{\bullet}$:
\begin{diagram}
\ldots&\rTo^{d_{n-1}}&\mathscr{E}^n&\rTo^{d_n}&\mathscr{E}^{n+1}&\rTo^{d_{n+1}}&\ldots
\end{diagram}
Complexes are quite common in algebraic geometry, so let us give some details at this point. Complexes can be defined for modules, sheaves, groups, etc. Let us denote these objects by $\mathcal{A}_n$. If one has maps $\phi_n:\mathcal{A}_n\rightarrow\mathcal{A}_{n+1}$, which satisfy the property that $\phi_{n+1}\circ\phi_n=0$, this defines a complex:
\begin{diagram}
\ldots&\rTo^{\phi_{n-1}}&\mathcal{A}_n&\rTo^{\phi_n}&\mathcal{A}_{n+1}&\rTo^{\phi_{n+1}}&\ldots
\end{diagram}
The image of any map in this diagram is a subset of the kernel of the subsequent map. We obtain an {\em exact sequence} when the image of every map is the same as the kernel of the next map and not just a subset. A special case is a {\em short exact sequence}:
\begin{diagram}
0&\rTo&\mathcal{A}&\rTo^{\phi_1}&\mathcal{B}&\rTo^{\phi_2}&\mathcal{C}&\rTo&0
\end{diagram}
This means that $\phi_1$ is injective, $\phi_2$ is surjective and the image of $\phi_1$ is the kernel of $\phi_2$. Note that coherent sheaves have a nice description in terms of complexes. For a coherent sheaf $\mathscr{F}$ there exists a complex
\begin{align}
0\longrightarrow\mathscr{E}^n\longrightarrow\mathscr{E}^{n-1}\longrightarrow\ldots\longrightarrow\mathscr{E}^1\longrightarrow\mathscr{E}^0\longrightarrow\mathscr{F}\longrightarrow0,
\end{align}
called free resolution where the $\mathscr{E}_i$ are locally free sheaves.\\\\
Returning to B--branes, one can show that the spectrum of open string states related to B--branes defined by the complexes $\mathscr{E}^{\bullet},\mathscr{F}^{\bullet}$ is computed by $\mathrm{Ext}^n(\mathscr{E}^{\bullet},\mathscr{F}^{\bullet})$, where $n$ goes from $0$ to $\mathrm{dim}X$. It can furthermore be shown that all this structure can be cast into a category, the {\em derived category of locally free sheaves}. It turns out it is possible to extend this to coherent sheaves, so that we finally end up with the statement we made at the beginning that, at the large radius limit, the category of B--branes is the derived category of coherent sheaves. 

\subsubsection{Homological Mirror Symmetry}
The homological mirror symmetry conjecture is due to Kontsevich \cite{alg-geom/9411018}. It states the equivalence between the Fukaya category and the derived category of coherent sheaves:
\begin{align}
\mathrm{Fuk}(X)\cong D(X^{\ast}),
\end{align}
where $X$ and $X^{\ast}$ denote two Calabi--Yaus which are mirror to each other.\\
By now only a few explicit examples for homological mirror symmetry are known. Homological mirror symmetry on the torus will be discussed in chapter \ref{chap-torus}.

\chapter{Boundary Landau--Ginzburg Models and Matrix Factorizations}
\label{chap-mf}
\section{Introduction}
In this chapter we introduce D--branes in B--type topological Landau--Ginzburg models. Such D--branes have a description in terms of matrix factorizations of the Landau--Ginzburg superpotential. The idea goes back to an unpublished proposal by Kontsevich who suggested an equivalence of the category of coherent sheaves and the category of matrix factorizations. This equivalence was later proven by Orlov \cite{Orlov:2003yp} based upon previous work by Eisenbud \cite{eisenbud}. A physics derivation was first given in \cite{Kapustin:2002bi} and later rederived in \cite{Brunner:2003dc,Lazaroiu:2003zi,Herbst:2004ax}. Matrix factorizations provide a solution to the ``Warner problem'' \cite{Warner:1995ay}.\\
Matrix factorizations give a description of D--branes in the non--geometric regime of K\"ahler moduli space. In this regime, quantum effects become important and many examples imply that matrix factorizations give a powerful tool for understanding these effects. At the large radius point, B--type D--branes are described by coherent sheaves and Orlov's proof of Kontsevich's proposal makes contact between these two descriptions. In the physics literature the connection between matrix factorizations and coherent sheaves has been discussed in \cite{Aspinwall:2006ib}. In physics, the relation between the geometric description of Calabi--Yau compactifications at large radius and the non--geometric description in terms of Landau--Ginzburg models is realized as a phase transition in linear sigma models \cite{Witten:1993yc}. This should provide a suitable framework to prove Kontsevich's proposal in physics language and to establish an explicit mapping between matrix factorizations and coherent sheaves. First steps in this direction were made in \cite{Aspinwall:2007cs} and an exhaustive discussion is on the way \cite{hhp}. \\
The description of D--branes in terms of categories of matrix factorizations has several advantages as compared to other approaches, such as $K$--theory and boundary conformal field theory. The Ramond-Ramond charges of D--branes are encoded in $K$--theory \cite{Witten:1998cd}. Matrix factorization contain the information about RR--charges \cite{Walcher:2004tx} but, in contrast to $K$--theory, categories are also sensitive to the positions of D--branes. This is also reflected in the category of matrix--factorizations, which is defined in the non--geometric regime of K\"ahler moduli space. Such data is crucial if one wants to understand deformations of the D--brane moduli space and tachyon condensation. An alternative description of D--branes is in rational conformal field theories. These theories are well--understood and provide a reference point for comparing the properties of matrix factorizations with known results \cite{Brunner:2005fv,Brunner:2005pq,Enger:2005jk,Brunner:2006tc,Fredenhagen:2006qw,Keller:2006tf,Schmidt-Colinet:2007vi}. A description of D--branes in terms of boundary conformal field theory is only possible at the Gepner point, which is the point in moduli space where the Landau--Ginzburg model is a tensor product of A--type minimal models. 
Deformations away from the Gepner point are not captured by the CFT description, whereas this can be easily realized in terms of matrix factorizations. The most important benefit of matrix factorizations, at least from a physics point of view, is that they give an application--oriented realization of the abstract concepts of categories and sheaves which are essential for homological mirror symmetry but hard to access for explicit calculations. \\
There are, however, some drawbacks of the description of D--branes in terms of matrix factorizations. The most obvious technical problem is that there is no one--to--one correspondence between a matrix factorization and a certain D--brane. In general, many matrix factorizations describe the same D--brane configuration. This gives rise to a classification problem, which, so far, has only been solved for the simplest cases \cite{greuel-knoerrer}. The problem of the classification of matrix factorizations has two aspects: One problem to find all inequivalent matrix factorizations, where inequivalent means that the matrix factorizations cannot be related to each other by certain transformations. The second problem is to find a 'minimal set' of matrix factorizations from which one can obtain all the others by tachyon condensation processes. This amounts to finding a matrix factorization description of D--branes that generate the lattice of RR--charges. This problem was solved for the elliptic curve in \cite{Brunner:2004mt,Govindarajan:2005im,Knapp:2007kq}. Due to the redundancies that come with matrix factorizationsv, one can easily get lost if one is not aware of the mathematical background. The mathematical literature on this subject is vast and often hard to access for physicists. A further problem is that one cannot say much about the stability of the D--brane given by a matrix factorization. This is a conceptual problem due to the topological nature of these D--branes. B--type D--branes are insensitive to deformations in the K\"ahler moduli space but the stability of a D--brane depends crucially on the K\"ahler moduli in the sense that at different points in K\"ahler moduli space different D--brane configurations are stable. See e.g. \cite{Aspinwall:2004jr} for a review. Nevertheless, it is possible to extract some information on the stability of D--branes from matrix factorization. A proposal was made in \cite{Walcher:2004tx} where the notion of $R$--stability was introduced.\\\\
Over the years, research on matrix factorizations has brought a lot of insight into this subject and a considerable amount of literature has accumulated. Apart from the papers already mentioned here, further publications are \cite{Kapustin:2003ga,Kapustin:2003rc,Ashok:2004zb,Herbst:2004jp,Ashok:2004xq,Hori:2004ja,Herbst:2004zm,Herbst:2006kt,Knapp:2006rd,Hori:2006ic,Diaconescu:2006id,Jockers:2006sm}. There are also two reviews available: \cite{Hori:2004zd,Cargese}. In this chapter we collect the relevant facts about matrix factorizations, summarizing the most important results of the papers mentioned here. In section \ref{sec-bdrylg} we review the physics derivation of the matrix factorization condition. Section \ref{sec-mfphys} states the most important properties from a physics point of view and section \ref{sec-mfmath} makes contact with the mathematical side of the subject. In section \ref{sec-mfconstr} we give a short account of possible constructions of matrix factorizations and discuss some aspects of their relation to boundary conformal field theory. Finally, in section \ref{sec-mfother} we mention some further applications of matrix factorizations. 
\section{The Boundary Landau Ginzburg Model}
\label{sec-bdrylg}
In this section we summarize how matrix factorizations are related to boundary conditions in topological B--type Landau--Ginzburg models. We will mostly follow \cite{Brunner:2003dc}.\\
Consider an (untwisted) Landau--Ginzburg model:
\begin{align}
S_{\Sigma}&=\int_{\Sigma}d^2xd^4\theta K(\Phi,\bar{\Phi})+\int_{\Sigma}d^2xd^2\theta W(\Phi)+\:c.c.,
\end{align}
where $K$ is the K\"ahler potential and $W$ is the superpotential. $\Phi$ is a superfield with component expansion
\begin{align}
\Phi(y^{\pm},\theta^{\pm})&=\phi(y^{\pm})+\theta^+\psi_+(y^{\pm})+\theta^-\psi_-(y^{\pm})+\theta^{+}\theta^-F(y^{\pm}),
\end{align}
where $y^{\pm}=x^{\pm}-i\theta^{\pm}\bar{\theta}^{\pm}$ and $x^{\pm}=x^0\pm x^1$. We define the supercharges
\begin{align}
Q_{\pm}=\frac{\partial}{\partial\theta^{\pm}}+i\bar{\theta}^{\pm}\frac{\partial}{\partial x^{\pm}}&\qquad \bar{Q}_{\pm}=-\frac{\partial}{\partial\bar{\theta}^{\pm}}-i\theta^{\pm}\frac{\partial}{\partial x^{\pm}},
\end{align}
and the corresponding covariant derivatives
\begin{align}
D_{\pm}=\frac{\partial}{\partial\theta^{\pm}}-i\bar{\theta}^{\pm}\frac{\partial}{\partial x^{\pm}}&\qquad\bar{D}_{\pm}=-\frac{\partial}{\partial\bar{\theta}^{\pm}}+i\theta^{\pm}\frac{\partial}{\partial x^{\pm}}.
\end{align}
Then $\Phi$ is a chiral superfield: $\bar{D}_{\pm}\Phi=0$. Similarly, $\bar{\Phi}$ is antichiral. The supersymmetry variations of the fields with respect to $\delta=\epsilon_+Q_--\epsilon_-Q_+-\bar{\epsilon}_+\bar{Q}_-+\bar{\epsilon}_-\bar{Q}_+$ are:
\begin{align}
\begin{array}{lll}
\delta\phi_{\phantom{+}}=\phantom{+}\epsilon_+\psi_--\epsilon_-\psi_+&\qquad\qquad&\delta\bar{\phi}_{\phantom{+}}=-\bar{\epsilon}_+\bar{\psi}_-+\bar{\epsilon}_-\bar{\psi}_+\\
\delta\psi_+=\phantom{+}2i\bar{\epsilon}_-\partial_+\phi+\epsilon_+F&\qquad\qquad&\delta\bar{\psi}_+=-2i\epsilon_-\partial_+\bar{\phi}+\bar{\epsilon}_+\bar{F}\\\delta\psi_-=-2i\bar{\epsilon}_+\partial_-\phi+\epsilon_-F&\qquad\qquad&\delta\bar{\psi}_-=\phantom{+}2i\epsilon_+\partial_-\bar{\phi}+\bar{\epsilon}_-\bar{F}
\end{array}
\end{align}
$F$ is an auxiliary field which satisfies the algebraic equation of motion $F=-\frac{1}{2}\bar{W}'(\bar{\phi})$. Making use of this equation, the component form of the action looks as follows:
\begin{align}
S_{\Sigma}=&\int_{\Sigma}d^2x\left\{-\partial^{\mu}\bar{\phi}\partial_{\mu}\phi+\frac{i}{2}\bar{\psi}_-(\stackrel{\leftrightarrow}{\partial_0}+\stackrel{\leftrightarrow}{\partial_1})\psi_-+\frac{i}{2}\bar{\psi}_+(\stackrel{\leftrightarrow}{\partial_0}-\stackrel{\leftrightarrow}{\partial_1})\psi_+-\frac{1}{4}|W'|^2-\frac{1}{2}W''\psi_+\psi_--\frac{1}{2}\bar{W}''\bar{\psi}_-\bar{\psi}_+\right\}
\end{align}
Here we chose $K=\bar{\Phi}\Phi$.\\
We now introduce boundary conditions, that break the supersymmetry down to $N=1$. Since we are interested in the topological B--model we impose B--type boundary conditions which are compatible with the twist. B--type supersymmetry preserves the supercharge
\begin{align}
Q=\bar{Q}_++\bar{Q}_-&\qquad\mathrm{with}\qquad \bar{\epsilon}_-=-\bar{\epsilon}_+=\epsilon
\end{align}
The explicit form of B--model supercharges is:
\begin{align}
\bar{Q}=\frac{\partial}{\partial\theta^0}+i\bar{\theta}^0\frac{\partial}{\partial x^0}&\qquad Q=-\frac{\partial}{\partial\bar{\theta}^0}-i\theta^0\frac{\partial}{\partial x^0},
\end{align}
where $\theta^0=\frac{1}{2}(\theta^-+\theta^+)$ and $\bar{\theta}^0=\frac{1}{2}(\bar{\theta}^-+\bar{\theta}^+)$. 
Defining $\eta=\psi_-+\psi_+$ and $\theta=\psi_--\psi_+$, the B--type supersymmetry transformations read:
\begin{align}
\begin{array}{lll}
\delta\phi=\phantom{-}\epsilon\eta&\qquad\qquad&\delta\bar{\phi}=-\bar{\epsilon}\bar{\eta}\\
\delta\eta=-2i\bar{\epsilon}\partial_0\phi&\qquad\qquad&\delta\bar{\eta}=\phantom{-}2i\epsilon\partial_0\bar{\phi}\\
\delta\theta=\phantom{-}2i\bar{\epsilon}\partial_1\phi+\epsilon\bar{W}'(\bar{\phi})&\qquad\qquad&\delta\bar{\theta}=-2i\epsilon\partial_1\bar{\phi}+\bar{\epsilon}W'(\phi),
\end{array}
\end{align}
where $\delta=\epsilon\bar{Q}-\bar{\epsilon}Q$.\\
For $W=0$, one gets a supersymmetric Lagrangian by adding a boundary term
\begin{align}
S_{\partial\Sigma}&=\frac{i}{4}{\int_{\partial\Sigma}dx^0\left\{\bar{\theta}\eta-\bar{\eta}\theta\right\}\vline}_{0}^{\pi}
\end{align}
With $W\neq 0$ one gets a non--vanishing boundary contribution, the 'Warner term' \cite{Warner:1995ay}:
\begin{align}
\label{eq-warner}
\delta(S_{\sigma}+S_{\partial\Sigma})=\frac{i}{2}{\int_{\partial\Sigma}dx^0\left\{\epsilon\bar{\eta}\bar{W}'+\bar{\epsilon}\eta W'\right\}\vline}_{0}^{\pi}
\end{align}
So, without further modifications, supersymmetry cannot be preserved. This is known as the 'Warner problem'. In \cite{Kapustin:2002bi} this problem was solved by introducing matrix factorizations, as we will now show.\\
In order to cancel this term one has to introduce additional fermionic degrees of freedom on the boundary:
\begin{align}
\Pi(y^0,\theta^0,\bar{\theta}^0)&=\pi(y^0)+\theta^0l(y^0)-\bar{\theta}^0\left[E(\phi)+\theta^0\eta(y^0)E'(\phi)\right]
\end{align}
Note that this boundary fermion is not chiral but satisfies $D\Pi=E(\Phi')$, where $\Phi'=\phi(y^0)+\theta^0\eta(y^0)$. The boundary degrees of freedom account for the following additional terms for the action:
\begin{align}
\tilde{S}_{\partial\Sigma}&=-\frac{1}{2}{\int_{\partial\Sigma} dx^0d^2\theta\bar{\Pi}\,\Pi\vline}_0^{\pi}-\frac{i}{2}{\int_{\partial\Sigma}dx^0 d\theta \Pi J(\Phi)|_{\bar{\theta}=0}\vline}_0^{\pi}+c.c.
\end{align}
We can integrate out $l$ by substituting the algebraic equations of motion $l=-i\bar{J}(\bar{\phi})$. Then the component form of the boundary action reads:
\begin{align}
\tilde{S}_{\partial\Sigma}&={\int_{\partial\Sigma}dx^0\left\{i\bar{\pi}\partial_0\pi-\frac{1}{2}\bar{J}J+\frac{1}{2}\bar{E}E+\frac{i}{2}\pi\eta J'+\frac{i}{2}\bar{\pi}\bar{\eta}\bar{J}'-\frac{1}{2}\bar{\pi}\eta E'+\frac{1}{2}\pi\bar{\eta}\bar{E}'\right\}\vline}_0^{\pi}
\end{align}
The boundary fermion has the following supersymmetry transformations:
\begin{align}
\delta\pi&=-i\epsilon\bar{J}(\bar{\phi})-\bar{\epsilon}E(\phi)\nonumber\\
\delta\bar{\pi}&=\phantom{-}i\bar{\epsilon}J(\phi)-\epsilon\bar{E}(\bar{\phi})
\end{align}
Due to the non--chirality of $\Pi$ $\tilde{S}_{\partial\Sigma}$ is not supersymmetry invariant:
\begin{align}
\delta\tilde{S}_{\partial\Sigma}&=-\frac{i}{2}{\int_{\partial\Sigma}dx^0\left\{\epsilon\bar{\eta}(\bar{E}\bar{J})'+\bar{\epsilon}\eta(EJ)'\right\}\vline}_{0}^{\pi}
\end{align}
Combining this with (\ref{eq-warner}), we can get a supersymmetric action if we impose the condition \cite{Kapustin:2002bi}:
\begin{align}
\label{eq-mf}
W&=E\cdot J
\end{align}
Here one could add an integration constant, which we set to $0$. \\
We managed to construct a supersymmetric action, but this symmetry is not globally defined. 
Under the B--twist, we can replace the variation $\delta$ by a scalar fermionic BRST operator $Q$. The transformation rules become:
\begin{align}
\begin{array}{lll}
Q\phi=\phantom{-}0&\qquad\qquad&Q\bar{\phi}=\phantom{-}\bar{\eta}\\
Q\eta=\phantom{-}2i\partial_0\phi&\qquad\qquad&Q\bar{\eta}=\phantom{-}0\\
Q\theta=-2i\partial_1\phi&\qquad\qquad&Q\bar{\theta}=-W'(\phi)
\end{array}
\end{align}
We still have not imposed any boundary conditions on the system. B--type boundary conditions allow for D$0$-- and D$2$--branes. For D$0$--branes, i.e. Dirichlet boundary conditions, the boundary fermion $\pi$ decouples. For D$2$--branes, i.e. for Neumann boundary conditions, $\pi$ does not decouple and one gets:
\begin{align}
Q\pi=E(\phi)&\qquad\qquad Q\bar{\pi}=-iJ(\phi)
\end{align}
So, every factorization (\ref{eq-mf}) defines a boundary condition. It is not yet clear, why the boundary potentials $E$ and $J$ should be matrices. This arises if we canonically quantize the the boundary fermion $\pi$ by imposing the anticommutation relations:
\begin{align}
\{\pi,\bar{\pi}\}=1&\qquad\qquad\{\pi,\pi\}=\{\bar{\pi},\bar{\pi}\}=0
\end{align}
If $W=W(x_1,\ldots,x_N)$ then this Clifford algebra has a $2^N$--dimensional representation and we can write for the boundary BRST operator $Q$:
\begin{align}
Q&=E\bar{\pi}-iJ\pi
\end{align}
It satisfies the matrix factorization condition
\begin{align}
Q^2&=-iEJ=-iW
\end{align}
It turns out that one does not capture all the possible D--branes by restricting to matrix factorizations of rank $2^N$. The condition can be extended to matrices with arbitrary rank \cite{Lazaroiu:2003zi}. A D--brane is then characterized by the rank $2r$ matrix $Q$ and an involution $\sigma$, satisfying $\sigma^2=\mathbbm{1}_{2r\times 2r}$, implying a $\mathbbm{Z}_2$--grading:
\begin{align}
Q^2&=W\cdot\mathbbm{1}_{2r\times 2r}\qquad\qquad Q\sigma+\sigma Q=0
\end{align}
Choosing a diagonal representation
\begin{align}
\sigma=\left(\begin{array}{cc}
\mathbbm{1}_{r\times r}&0\\
0&-\mathbbm{1}_{r\times r}
\end{array}
\right),
\end{align}
we can write $Q$ as a block off--diagonal matrix:
\begin{align}
Q&=\left(\begin{array}{cc}
0&E\\
J&0
\end{array}
\right),
\end{align}
where the boundary potentials $E$ and $J$ are $r\times r$--matrices.
\section{Matrix Factorizations from a Physics Point of View}
\label{sec-mfphys}
This section contains a collection of facts about matrix factorizations. For derivations and proofs we refer to the original literature. Here we discuss the properties of matrix factorizations in physics terms. The mathematical aspects will be discussed in section \ref{sec-mfmath}. 
\subsection{Branes and Antibranes}
Given a D--brane with label $A$ and a matrix factorization $W=E^AJ^A$ with
\begin{align}
\label{eq-mf1}
Q^A&=\left(\begin{array}{cc}
0&E^A\\
J^A&0
\end{array}\right),
\end{align}
its antibrane is characterized by a matrix factorizations where $E$ and $J$ are exchanged:
\begin{align}
\bar{Q}^A&=-\left(\begin{array}{cc}
0&J^A\\
E^A&0
\end{array}
\right).
\end{align}
\subsection{The Open String Vacuum}
The open string vacuum is described by the trivial factorization:
\begin{align}
\label{eq-vac}
Q^{vac}=\left(\begin{array}{cc}
0&1\\
W&0
\end{array}
\right)
\end{align} 
This matrix factorization describes a situation where there are no D--branes at all and consequently no open string states. (\ref{eq-vac}) is not the only description of the open string vacuum. Tensor products and the ``antibrane'' of $Q^{vac}$ also describe the open string vacuum.\\
Constant entries in matrix factorizations usually imply that one can make transformations such that at least a block in the matrix has the form (\ref{eq-vac}), which entails that the matrix factorization contains a trivial piece. A matrix factorization which does not contain constant entries is called {\em reduced} \cite{Walcher:2004tx}.
\subsection{Equivalent Matrix Factorizations}
Matrix factorizations $(E^A,J^A)$ and  $(E^B,J^B)$ are called {\em equivalent} if they are related by a similarity transformation
\begin{align}
\label{eq-eqivmf}
E^B=U_1E^AU_2^{-1}&\qquad\qquad J^B=U_2J^AU_1^{-1},
\end{align}
where $U_1,U_2$ are invertible\footnote{This means that they are invertible over the polynomials.} matrices with polynomial entries. Thus, if one can transform one matrix factorization into another by elementary row-- and column manipulations, these two matrix factorizations describe the same D--brane. 

\subsection{R--Charge}
In \cite{Hori:2004ja,Walcher:2004tx} a prescription was given how to associate to a matrix factorization a matrix $R$, which encodes the $U(1)$ $R$--charges. Assuming that the superpotential $W(x_i)$, $i=1,\ldots,N$ is a quasi--homogeneous polynomial, we assign $R$--charges to the $x_i$ such that $W$ has $R$--charge 2:
\begin{align}
W(e^{i\lambda q_i}x_i)&=e^{2i\lambda}W(x_i)\qquad\qquad\forall\lambda\in\mathbbm{R}
\end{align}
This defines the $R$--charge of the bulk theory as the $U(1)$--action on the space of polynomials with respect to which $W$ is equivariant. The action closes for $\lambda=\pi H$, where the smallest integer such that $Hq_i\in2\mathbbm{Z}$ for all $i$.\\
The $U(1)$--action can be extended to the boundary. To every matrix factorization $Q^A$ we can associate an '$R$--matrix' $R^A$ such that
\begin{align}
\label{eq-rmat}
EQ^A+[R^A,Q^A]=Q^A,
\end{align}  
where 
\begin{align}
\label{eq-euler}
E=\sum_iq_ix_i\frac{\partial}{\partial x_i}
\end{align}
is the Euler derivative. The matrix $R$ is unambiguously defined if we demand that it is diagonal, traceless and has constant entries.

\subsection{Orbifolds}
In the context of matrix factorizations orbifolds were first discussed in \cite{Ashok:2004zb,Hori:2004ja}, an explicit construction of the representation of the orbifold group on matrix factorization was given in \cite{Walcher:2004tx}.\\
If $W(x_i)$ is homogeneous of degree $H$, i.e. if we can assign weights $w_i$ to the $x_i$ such that $\sum_{i=1}^Nw_i=H$ we can impose the action of the orbifold group $\Gamma=\mathbbm{Z}_H$. It is generated by the action $x_i\rightarrow e^{2i\pi w_i/H}x_i\equiv e^{i\pi q_i }x_i$. We can extend the orbifold action to the boundary by associating to every matrix factorization $Q^A$ a matrix $\gamma^A$ which is defined through the condition:
\begin{align}
\gamma^AQ^A(\omega_ix_i)(\gamma^A)^{-1}&=Q^A(x_i)
\end{align}
$\gamma^A$ has an explicit expression in terms of the involution $\sigma$ and the $R$--matrix:
\begin{align}
\label{eq-orbrep}
\gamma^A=\sigma e^{i\pi R^A}e^{-i\pi\varphi},
\end{align}
where the phase $\varphi$ is fixed by the condition $(\gamma^A)^H=\mathbbm{1}$. For each matrix factorization of $W$ there are thus $H$ $\mathbbm{Z}_H$--equivariant factorizations. 

\subsection{Orientifolds}
\label{sec-orientifold}
Orientifolds were discussed in \cite{Hori:2006ic,Diaconescu:2006id} in the context of matrix factorizations. A B--type orientifold projection is defined as:
\begin{align}
P=\tau\Omega,
\end{align}
where $\Omega$ is worldsheet parity and $\tau$ acts on target space variables. For Landau--Ginzburg models $\tau$ acts such that it reverses the sign of the superpotential:
\begin{align}
W(\tau x_i)\equiv\tau^{\ast}W&=-W(x_i)
\end{align}
Under orientation reversal on the worldsheet matrix factorizations transform into their {\em graded transpose}:
\begin{align}
Q^A(x_i)&\rightarrow -Q^A(\tau x_i)^{T}
\end{align}
For matrices the graded transpose $(\ldots)^{T}$ is defined as follows. Given a $\mathbbm{Z}_2$ graded matrix
\begin{align}
A&=\left(\begin{array}{cc}
a&b\\
c&d
\end{array}\right),
\end{align}
where the block matrix $a$ maps even states to even states, $d$ maps odd to odd and $b$, $c$ map between even and odd states, the graded transpose of $A$ is
\begin{align}
\label{eq-gradtrans}
A^T&=\left(\begin{array}{cc}
a^T&-c^T\\
b^T&d^T
\end{array}\right),
\end{align}
where $a^T,\ldots$ are the transposes of the matrices. It can be shown that the so defined action satisfies all the properties of an orientifold action and is compatible with $R$--charge, $\mathbbm{Z}_2$--grading and the orbifold projection.\\
In order to be able to define parity invariant configurations we define an operator $U$ with the following properties \cite{Hori:2006ic}:
\begin{align}
\label{eq-orientpro}
U\sigma^TU^{-1}&=\sigma \nonumber\\
U(-\tau^{\ast}Q^T)U^{-1}&=Q,
\end{align}
If we want the brane to be orbifold invariant in addition we introduce a character $\chi:\Gamma\rightarrow \mathbbm{C}^{\ast}$ and impose the following condition:
\begin{align}
\label{eq-orborient}
U(\chi(\gamma^{-1})^T)U^{-1}&=\gamma
\end{align}

\subsection{Open String States}
To every matrix factorization $Q^A$ we can assign a $\mathbbm{Z}_2$--graded differential $D_A$ which acts on an open string state $\Psi$ as the graded commutator with $Q^A$:
\begin{align}
\label{eq-diffa}
D_A\Psi&=Q^A\Psi-(-1)^{|\Psi|}\Psi Q^A.
\end{align}
where $|\Psi|$ is the $\mathbbm{Z}_2$--degree of $\Psi$.\\
Physical open string states which begin and end on the same brane $A$ are defined by the cohomology of $D_A$:
\begin{align}
H(D_A)&=\frac{\mathrm{Ker}(D_A)}{\mathrm{Im}(D_A)}=H^{even}(D_A)\oplus H^{odd}(D_A)
\end{align}
``Even'' states with $\mathbbm{Z}_2$--charge 0 will be referred to as bosons and are usually\footnote{We will stick to these conventions as long as there are no bulk fields which are by standard conventions also denoted by $\phi$.} denoted by $\phi$. If we choose the block matrix representation (\ref{eq-mf1}) for $Q^A$ the even states correspond to block--diagonal matrices. ``Odd'' states have $|\Psi|=1$ and are referred to as fermions and are denoted by $\psi$. They can be represented as block--offdiagonal matrices.\\
So far, we have only considered open string states which begin and end on the same brane. In order to describe boundary changing operators we pick two matrix factorizations $Q^A$ and $Q^B$ which define a graded differential $D_{AB}$ which acts on boundary changing states $\Psi^{AB}$ as follows:
\begin{align}
\label{eq-diffab}
D_{AB}\Psi^{AB}=Q^A\Psi^{AB}-(-1)^{|\Psi^{AB}|}\Psi^{AB}Q^B
\end{align}
The boundary changing spectrum is then defined by the cohomology of $D_{AB}$. Setting $A=B$, we recover the definitions for the boundary preserving sector.\\
The $R$--charge $q_{\Psi^{AB}}$ of an open string state $\Psi^{AB}$ is defined by
\begin{align}
E\Psi^{AB}+R^A\Psi^{AB}-\Psi^{AB}R^B=q_{\Psi^{AB}}\Psi^{AB}.
\end{align} 
An open string state is orbifold invariant if
\begin{align}
\label{eq-orbmorph}
\gamma^A\Psi^{AB}(\omega_ix_i)(\gamma^B)^{-1}=\Psi^{AB}(x_i)
\end{align}
Under the orientifold action, open string states transform into their graded transpose:
\begin{align}
\Psi^{AB}(x_i)\rightarrow\Psi^{AB}(\tau x_i)^{T}.
\end{align}
For parity invariant D--branes, which are characterized by the action of the operator $U$ of (\ref{eq-orientpro}), open string states transform as:
\begin{align}
\label{eq-orientmorph}
\Psi^{AB}(x_i)\rightarrow U^A\Psi^{AB}(\tau x_i)^{T}(U^B)^{-1}.
\end{align}
Note that the states are {\em not} invariant under the orientifold action. The action also extends to orbifold invariant states \cite{Hori:2006ic}.

\subsection{Tachyon Condensation}
Open string states induce tachyon condensation. In the context of matrix factorizations tachyon condensation is described as follows. Consider two matrix factorizations $Q^A$ and $Q^B$ and an open string state $T$ stretching between the branes $A$ and $B$. Then we can define a matrix 
\begin{align}
\label{eq-boundstate}
Q^C&=\left(\begin{array}{cc}
Q^A&T\\
0&Q^B
\end{array}
\right),
\end{align}
which defines a {\em bound state} of $Q^A$ and $Q^B$. Using the physical state condition (\ref{eq-diffab}) for $T$, one finds that $Q^C$ satisfies the matrix factorization condition $(Q^C)^2=W\cdot\mathbbm{1}$. \\
Note that this is not a dynamical process, which would require a string field theory description. In the context of the topological string we have no dynamical means to turn on the tachyon $T$: without $T$ we have two branes $Q^A$ and $Q^B$, for non--zero $T$ the D--brane configuration is given by the bound state.

\subsection{Residue Formula for Correlators}
For the bulk theory there exists a residue formula which computes correlators in topological Landau--Ginzburg theories \cite{Vafa:1990mu}. Since we are mostly interested in tree--level correlators, we only give the three--point function on the sphere which is defined as follows:
\begin{align}
\label{eq-bulk3pt}
\langle\phi_i\phi_j\phi_k\rangle&=\frac{1}{(2\pi i)^N}\oint d^Nx\frac{\phi_i\phi_j\phi_k}{\partial_1W\ldots\partial_N W},
\end{align}
where the $\phi_i$ are bulk fields  and the contour goes around $W=0$.\\
This formula can be used to calculate the three--point correlators explicitly. On the sphere, we cannot calculate correlators with a higher number of insertions this way because these insertions have to be integrated descendants\footnote{See section \ref{sec-constr-effsupo} for a discussion.}.\\
In \cite{Kapustin:2003ga}, a generalization of (\ref{eq-bulk3pt}) for topological Landau--Ginzburg models with boundary was proposed. A derivation using localization techniques was given in \cite{Herbst:2004ax}. It computes disk correlators with three boundary insertions or one boundary and one bulk insertion. The bulk--boundary correlator is given by
\begin{align}
\label{eq-kapustin1}
\langle\phi_i\Psi_a^{AA}\rangle&=\frac{1}{(2\pi i)^N}\oint d^Nx\frac{\phi_i\mathrm{STr}\left((\partial Q^A)^{\wedge N}\Psi_a^{AA}\right)}{\partial_1W\ldots\partial_N W},
\end{align}
and similarly, for the case of three boundary insertions, we have:
\begin{align}
\label{eq-kapustin2}
\langle\Psi_a^{AB}\Psi_b^{BC}\Psi_c^{CA}\rangle&=\frac{1}{(2\pi i)^N}\oint d^Nx\frac{\mathrm{STr}\left((\partial Q^A)^{\wedge N}\Psi_a^{AB}\Psi_b^{BC}\Psi_c^{CA}\right) }{\partial_1W\ldots\partial_N W},
\end{align}
where $\mathrm{STr}$ denotes the supertrace and $(\partial Q^A)^{\wedge N}$ is the antisymmetrized derivative of $Q^A$:
\begin{align}
(\partial Q^A)^{\wedge N}&=\frac{1}{N!}\sum_{\sigma\in S_n}(-1)^{|\sigma|}\partial_{\sigma(1)}Q^A\cdot\ldots\cdot\partial_{\sigma(N)}Q^A,
\end{align}
with $\sigma$ being an element of the permutation group $S_n$. The residues defined above are non--zero if and only if the integrand is proportional to the Hessian $H=\mathrm{det}\partial_i\partial_jW$ of the superpotential.\\
We can use (\ref{eq-kapustin2}) to define a topological metric on the boundary fields:
\begin{align}
\label{eq-bdrymet1}
\omega_{ab}&=\langle\Psi_a^{AB}\Psi_b^{BA}\rangle\equiv\langle\mathbbm{1}\Psi_a^{AB}\Psi_b^{BA}\rangle
\end{align}

\subsection{RR--Charges}
A prescription to compute $RR$--charges was given in \cite{Walcher:2004tx}. The $RR$--charge is defined as the correlation function on the disk with the $RR$--groundstate inserted in the bulk. The bulk $RR$--ground states were computed in \cite{Vafa:1989xc}. We consider models with integer central charge $\hat{c}=\sum_i(1-q_i)$ and orbifold group $\mathbbm{Z}_H$. $B$--branes couple to Ramond--Ramond ground states with opposite left-- and right--moving R--charge: $q_L=-q_R$. In order to find these states, we consider the $l$--th twisted sector with respect to the orbifold action. We can divide the fields $x_i$ into fields $\{x_i^u\}$ which have $R$--charges $lq_i\in 2\mathbbm{Z}$ and fields $\{x_i^t\}$ with $lq_i\neq 2\mathbbm{Z}$. The important observation is that the ground states with $q_L=q_L^u+q_L^t=-q_R=q_R^u+q_R^t$ from the $l$--th twisted sector are the neutral ground states of the effective potential $W_l(x_i^u)$ obtained by setting the fields with $lq_i\neq 2\mathbbm{Z}$ to $0$. A basis of these ground states is given by:
\begin{align}
|l;\alpha\rangle :&\qquad \phi_l^{\alpha}=\prod_{i=1}^{r_l}(x_i^u)^{\alpha_i},
\end{align} 
where $r_l$ is the number of fields with $lq_i\in 2\mathbbm{Z}$ and $\alpha$ ranges over a basis of the subspace of the chiral ring $\mathbbm{C}[x_i^u]/\partial W_l$ with $R$--charge $q_L^u=q_R^u=\sum_{lq_i\in 2\mathbbm{Z}}\alpha_i\frac{q_i}{2}=\frac{\hat{c}^u}{2}$, where $\hat{c}^u=\sum_{lq_i\in 2\mathbbm{Z}}(1-q_i)$. With these definitions, the $RR$--charge for a matrix factorization $Q^A$ is determined by the following disk correlator:
\begin{align}
\mathrm{ch}(Q^A)|l;\alpha\rangle&=\langle l;\alpha|Q\rangle\nonumber\\
&=\frac{1}{r_l!}\oint\frac{\phi_l^{\alpha}\mathrm{Str}\left(\gamma^l(\partial Q_l)^{\wedge r_l}\right)}{\partial_1W_l\ldots\partial_{r_l}W_l},
\end{align}
where $\gamma^l$ is the generator of the orbifold group for the $l$--th twisted sector.

\subsection{GSO Projections}
\label{sec-gso}
In the bulk theory, a Landau--Ginzburg model with superpotential $W(x_i)$ is equivalent to a Landau--Ginzburg model with a superpotential $W(x_i,z)=W(x_i)-z^2$, where a quadratic term in a new variable is added. Here, ``equivalent'' means, that adding a quadratic term does not change the chiral ring and consequently the physical spectrum remains unchanged\footnote{Of course, adding a variable does change the geometric interpretation which relates the Landau--Ginzburg superpotential $W$ via the LG/CY--correspondence to a hypersurface in projective space defined by $W$=0.}.  This situation changes when one adds a boundary. Given a matrix factorization $W(x_i)\mathbbm{1}=E(x_i)J(x_i)$ of $W(x_i)$ the most obvious matrix factorization of $W(x_i,z)$ is:
\begin{align}
W(x_i,z)\mathbbm{1}&=\left(\begin{array}{cc}
E(x_i)&z\\
-z&J(x_i)
\end{array}\right)
\left(\begin{array}{cc}
J(x_i)&-z\\
z&E(x_i)
\end{array}\right)
\end{align}
These two models have different boundary spectra. It was shown in \cite{Brunner:2005pq} that these two models correspond to two different GSO projections (type 0A/B) 
in the conformal field theory description. The result generalizes to superpotentials with more than one variable.
\section{Matrix Factorizations from a Mathematics Point of View}
\label{sec-mfmath}
B--type D--branes are objects in the derived category of coherent sheaves, which is equivalent to the category of matrix factorizations. Open string states stretching between two D--branes correspond to morphisms between these objects. Open strings which begin and end on the same brane correspond to endomorphisms.\\
In this section we give the basic properties of these categories and relate them to the physics terminology introduced in the previous section.

\subsection{The Category of Matrix Factorizations}
The construction of the category of matrix factorizations of $W(x_i)$, which we will denote by $MF(W)$ goes back to Kontsevich \cite{klo}. The objects of these categories are triples $(M,\sigma, Q)$, where $M$ is a free $\mathcal{R}=\mathbbm{C}[x_1,\ldots,x_N]$--module with $\mathbbm{Z}_2$--grading $\sigma$ and an odd endomorphism $Q$ satisfying $Q^2=W\mathbbm{1}$. Decomposing $M$ into homogeneous components $M=M_-\oplus M_+$ of rank $r$, one can write $Q$ in the usual block off--diagonal form with the boundary potentials $E$ and $J$:
\begin{align}
Q=\left(\begin{array}{cc}
0&E\\
J&0
\end{array}\right)
\end{align}
Here $E\in\mathrm{Hom}(M_-,M_+)$ and $J\in\mathrm{Hom}(M_+,M_-)$ have an interpretation as morphisms. This can be pictured by the following diagram:
\begin{align}
\begin{diagram}
M&=&\big(M_-&\pile{\rTo^E\\\lTo_J}&M_+\big)
\end{diagram}
\end{align}
This can be given a physical interpretation \cite{Kapustin:2002bi,Hori:2004ja,Herbst:2004ax,Cargese}. $M_+$ and $M_-$ describe a brane anti--brane pair and $J$ and $E$ can be interpreted a tachyon field configuration, which forces the two branes to condense into the brane $M$.\\
The transformations (\ref{eq-eqivmf}) are then automorphisms of $M$. $MF(W)$ is a differential graded category with differential $D$ defined in (\ref{eq-diffab}), which acts on the morphism space $\mathrm{Hom}_{\mathcal{R}}(M^A,M^B)$, defined by the cohomology of this differential.

\subsection{Triangulated Categories}
The category of matrix factorization is triangulated. Triangulated categories have additional structure which reflect important physical properties of D--branes. See for instance \cite{Aspinwall:2004jr} for a review.\\ 
One important property is the existence of a shift functor $[1]$, which reverses the $\mathbbm{Z}_2$--grading: $[1]:\sigma\rightarrow -\sigma$. In the physical description the shift functor relates branes and antibranes:
\begin{align}
Q[1]&=\left(\begin{array}{cc}
0&E\\
J&0
\end{array}
\right)[1]=
\left(\begin{array}{cc}
0&-J\\
-E&0
\end{array}
\right)
\end{align}
In particular, the existence of the shift functor implies that a fermionic/bosonic state stretching between a brane $A$ and a brane $B$ is equivalent to a bosonic/fermionic state stretching between $\bar{A}$ and $B$ or $A$ and $\bar{B}$, where $\bar{A},\bar{B}$ denote the respective antibranes.\\
The second characteristic of a triangulated category is the existence of {\em distinguished triangles}. One particular triangle is:
\begin{align}
\label{diag-cone}
\begin{diagram}
&&Q^C&&\\
&\ldTo&&\luTo&\\
Q^A&&\rDashto_T&&Q^B\\
\end{diagram}
\end{align}
The statement is, that for every morphism $T$ going from the object $Q^A$ to the object $Q^B$ there exist $Q^C$ and the morphisms from $Q^B$ to $Q^C$ and from $Q^C$ to $Q^A$.\\
The dashed arrow in (\ref{diag-cone}) corresponds to a map to $Q^B[1]$. The triangle can also be written as the following exact sequence:
\begin{align}
\begin{diagram}
Q^B&\rTo&Q^C&\rTo&Q^A&\rTo^T&Q^B[1]\\
\end{diagram}
\end{align}
We can associate a matrix factorization to the triangle (\ref{diag-cone}):
\begin{align}
Q^C=\left(\begin{array}{cc}
Q^A&T\\
0&Q^B
\end{array}\right)
\end{align}
This construction, which is known as the {\em cone construction}, provides the mathematical setup for tachyon condensation. $Q^C$ is called the {\em mapping cone}. In physics language the diagram (\ref{diag-cone}) means the following: Given an open string state $T$ between branes $A$ and $B$, there exists a bound state $C$ of the branes $A$ and $B$, which is formed by tachyon condensation of $T$.

\subsection{Serre Duality}
Serre duality gives an equivalence relation between cohomologies:
\begin{align}
H^*(D_{AB})=H^{*+N}(D_{BA}),
\end{align}
where $D_{AB}$ is the differential (\ref{eq-diffab}) and $N$ is the number of variables in the superpotential $W$. Thus, to every open string state going from brane $A$ to brane $B$, we can associate a Serre dual state going from $B$ to $A$. If $W$ has an odd number of variables bosons pair up with fermions and vice versa. If $W$ has an even number of variables bosons pair up with bosons and fermions with fermions.\\
Serre duality ensures that the boundary metric (\ref{eq-bdrymet1}) is non--degenerate.

\subsection{Kn\"orrer Periodicity}
In section \ref{sec-gso} we saw that adding a quadratic term to the superpotential $W(x_i)$ we get a different theory on the boundary. If one adds two more variables to $W(x_i)$ with $\bar{W}(x_i,z_1,z_2)=W(x_i)+z_1^2+z_2^2$ or $\tilde{W}(x_i,z_1,z_2)=W(x)+z_1z_2$ one obtains theories which are equivalent to the theory with $W(x)$. The mathematical structure behind this is an equivalence of categories, called Kn\"orrer periodicity \cite{Knorrer}. The statement is that the category $MF(W)$ is equivalent to $MF(\tilde{W})$ or $MF(\bar{W})$. Given a matrix factorization $(E(x_i),J(x_i))$ of $W(x_i)$, one can easily construct a matrix factorization of $W(x_i,z_1,z_2)$:
\begin{align}
W(x_i,z_1,z_2)&=\left(\begin{array}{cc}
E(x_i)&z_1\\
-z_2&J(x_i)
\end{array}\right)
\left(\begin{array}{cc}
J(x_i)&-z_1\\
z_2&E(x_i)
\end{array}\right)
\end{align}
In \cite{Hori:2006ic} it was shown that in the context of orientifolds Kn\"orrer periodicity doubles, i.e. that one has to add quadratic terms in four variables in order to get an equivalence of categories. 

\subsection{Refining the Category of Matrix Factorizations}
In order to incorporate the additional information of $R$--charges, orbifolds and orientifolds into the category theoretic description, the category of matrix factorizations has to be extended. \\\\
Including $R$--charges promotes the $\mathbbm{Z}_2$--graded category $MF(W)$ to a $\mathbbm{Z}$--graded category \cite{Walcher:2004tx}, with $\mathbbm{Z}$--graded matrix factorizations, which were denoted $MF_{R}(W)$ in \cite{Kajiura:2005yu}, where an exhaustive discussion of the category from the mathematics point of view is given. The $\mathbbm{Z}$--grading is generated by the Euler vector field (\ref{eq-euler}).\\
In order to include the orbifold action, we introduce a triangulated differential graded category $MF_{\Gamma}(W)$, where $\Gamma$ is the orbifold group. Its objects are defined by the data
\begin{align}
(M,\sigma,Q,\gamma),
\end{align}
where $\gamma$ is the representation (\ref{eq-orbrep}) of the orbifold group. Its morphisms satisfy (\ref{eq-orbmorph}). \\
Including the orientifold action, we can define a category $MF^{\epsilon}_{\mathscr{P}}$ \cite{Hori:2006ic} (see also \cite{Diaconescu:2006id}) with objects:
\begin{align}
(M,\sigma,Q,U),
\end{align}
where $U$ was defined in (\ref{eq-orientpro}). It is a map $U:M^{\ast}\rightarrow M$, where $M^{\ast}$ is the dual of $M$. $(\ldots)^T$ is the graded transpose defined in (\ref{eq-gradtrans}) and $\tau$ as defined in section \ref{sec-orientifold}. $\mathscr{P}$ defines a parity functor $\mathscr{P}:MF(W)\rightarrow MF(W)$ with action:
\begin{align}
(M,\sigma,Q)&\rightarrow(M^{\ast},\sigma^T,-\tau^{\ast}Q^T)\nonumber\\
\Psi^{AB}\in\mathrm{Hom}(M^A,M^B)&\rightarrow \tau^{\ast}(\Psi^{AB})^T\in\mathrm{Hom}(M_B^{\ast},M_A^{\ast}),
\end{align}
where $\tau^{\ast}Q(x_i)\equiv Q(\tau x_i)$.\\
Given two objects labeled by $A$ and $B$ in $MF^{\epsilon}_{\mathscr{P}}$, the parity transformation of open string states is given by a map $P:\mathrm{Hom}(M^A,M^B)\rightarrow \mathrm{Hom}(M^B,M^A)$ defined by (\ref{eq-orientmorph}). The category is parameterized by a sign $\epsilon$ which arises from the condition that $P$ acts as a projector. \\
Finally, we can define a category $MF_{\mathscr{P}_{\chi}}^{\pm c}$ which incorporates orientifolds and orbifolds. Its objects are:
\begin{align}
(M,\sigma,Q,\gamma,U),
\end{align}
subject to the conditions (\ref{eq-orientpro}), (\ref{eq-orborient}). The function $c$ is a phase which parameterizes the category \cite{Hori:2006ic}. 
The functor $\mathscr{P}_{\chi}$ acts as
\begin{align}
\mathscr{P}_{\chi}:& (M,\sigma,Q,\gamma)\rightarrow (M^{\ast},\sigma^T,-\tau^{\ast}Q^T,\chi(\rho^{-1})^T).
\end{align}

\subsection{Maximal Cohen--Macaulay Modules}
\label{sec-mcm}
A different construction of matrix factorization is given in the context of maximal Cohen--Macaulay modules over local rings of hypersurface singularities. This description will become important in the deformation theory construction of the effective superpotential which we will discuss in section \ref{sec-def-effsupo}. The discussion here is mostly taken from \cite{Walcher:2004tx}. The ring which is relevant for us is given by $\tilde{\mathcal{R}}_m=\mathcal{R}_m/W$ where $\mathcal{R}_m=\mathbbm{C}[[x_1,\ldots,x_N]]$ is the ring of formal power series and $m=(x_1,\ldots,x_N)$ is the maximal ideal.\\
We now consider a matrix factorization $(E,J)$ of $W$ and the $\mathcal{R}_m$--module $M=\mathrm{Coker}(E)$. The $\mathcal{R}_m$--free resolution of $M$ is given by the short exact sequence\footnote{We write the sequence from right to left  in accord with the conventions of \cite{Siqveland1}, which will be our main reference in section \ref{sec-def-effsupo} }:
\begin{align}
\begin{diagram}
0&\lTo&M&\lTo&F&\lTo^E&G&\lTo&0,\\
\end{diagram}
\end{align}  
where $F\cong G\cong (\mathcal{R}_m)^r$ are rank $r$ free modules. $M$ descends to an $\tilde{R}_m$--module with infinite free resolution
\begin{align}
\begin{diagram}
0&\lTo&M&\lTo&\tilde{F}&\lTo^E&\tilde{G}&\lTo^J&\tilde{F}&\lTo^E&\tilde{G}&\lTo&\cdots,\\
\end{diagram}
\end{align}
where $\tilde{F}\cong\tilde{G}\cong(\tilde{\mathcal{R}}_m)^r$. The characteristic property of this sequence is that it becomes periodic after the first step. $M$ is called a maximal Cohen--Macaulay module over $\tilde{\mathcal{R}}_m$. The maximal Cohen--Macaulay modules form a category $MCM(W)$ and a theorem of Eisenbud \cite{eisenbud} states that the objects of this category are given by matrix factorizations.\\
In the mathematics literature matrix factorizations are usually seen as free resolutions of maximal Cohen--Macaulay modules. A standard reference is \cite{yoshino}. This description turns out to be particularly useful if one wants to extract bundle data from matrix factorizations \cite{Laza,Govindarajan:2005im}.
\section{Construction of Matrix Factorizations}
\label{sec-mfconstr}
In this section we present some basic constructions of matrix factorizations. 

\subsection{Tensor Product Branes}
\label{sec-tensor}
The simplest matrix factorizations are given by the factorization of the minimal models of type $A_{k-2}$, $k\geq 3$ which have the superpotential $W=x^k$. This has an obvious factorization:
\begin{align}
\label{eq-amodfact}
W=EJ&\qquad E=x^r&J=x^{k-r}\qquad 
Q=\left(\begin{array}{cc}
0&x^r\\
x^{k-r}&0
\end{array}\right) \quad r=1,\ldots k-1
\end{align}
These matrix factorizations were discussed in \cite{Brunner:2003dc,Herbst:2004zm}. They were identified with certain boundary states in conformal field theory. It was also shown that all the matrix factorizations of the $A$--minimal models are given by direct sums of these one--dimensional factorizations.\\\\
In \cite{Ashok:2004zb}, a construction was given to calculate matrix factorizations for tensor products of $A$--type minimal models. Let us consider the simplest example of the tensor product of two minimal models with superpotential 
\begin{align}
W(x_1,x_2)&=W(x_1)+W(x_2)=E_1(x_1)J_1(x_1)+E_2(x_2)J_2(x_2).
\end{align}
We can now construct a matrix factorization $W(x_1,x_2)=E(x_1,x_2)J(x_1,x_2)$ with:
\begin{align}
E(x_1,x_2)=\left(\begin{array}{cc}
E_1(x_1)&E_2(x_2)\\
J_2(x_2)&-J_1(x_1)
\end{array}\right)&\qquad
J(x_1,x_2)=\left(\begin{array}{cc}
J_1(x_1)&E_2(x_2)\\
J_2(x_2)&-E_1(x_1),
\end{array}\right)
\end{align}
where $E_i(x_i),J_i(x_i)$ are of type (\ref{eq-amodfact}). This is easily generalized to tensor products of $n$ minimal models. Consider a superpotential
\begin{align}
W(x_1,\ldots,x_n)=W_1(x_1,\ldots,x_m)+W_2(x_{m+1},\ldots,x_n),
\end{align}
and let $(E_1(x_1,\ldots,x_m),J_1(x_1,\ldots,x_m))$ and $(E_2(x_{m+1},\ldots,x_n),J_2(x_{m+1},\ldots,x_n))$ be tensor product matrix factorizations of $W_1$ and $W_2$, respectively. Then we can construct a factorization $(E,J)$ of $W$, where:
\begin{align}
E=\left(\begin{array}{cc}
E_1\otimes\mathbbm{1}&\mathbbm{1}\otimes E_2 \\
\mathbbm{1}\otimes J_2 &-J_1\otimes\mathbbm{1}
\end{array}\right)&\qquad
J=\left(\begin{array}{cc}
J_1\otimes\mathbbm{1}&\mathbbm{1}\otimes E_2\\
\mathbbm{1}\otimes J_2&-E_1\otimes\mathbbm{1},
\end{array}\right)
\end{align}
This type of matrix factorizations could be identified with the Recknagel--Schomerus branes in the corresponding Gepner model \cite{Ashok:2004zb,Brunner:2005fv,Enger:2005jk}. The Recknagel--Schomerus branes are those branes in boundary conformal field theory, which preserve the different $N=2$ superconformal algebras of the minimal model components separately \cite{Recknagel:1997sb}.\\
Given a superpotential $W(x_1,\ldots,x_n)$, which is a homogeneous polynomial of degree $H$ there is an obvious way to factorize $W$ which yields a tensor product brane:
\begin{align}
\label{eq-permsupo}
W=\sum_{i=1}^n \frac{w_i}{H}x_i\frac{\partial W}{\partial x_i},
\end{align}
where the $w_i$ are the homogeneous weights of the $x_i$. This factorization also exists if one turns on deformations of $W$.

\subsection{Permutation Branes}
A further standard construction of matrix factorizations was discussed in \cite{Brunner:2005fv,Enger:2005jk}. They arise for superpotentials of the form
\begin{align}
W&=x_1^k+x_2^k
\end{align}
Apart from factorizing each minimal model component individually, we can also factorize $W$ as follows:
\begin{align}
W&=\prod_{m=0}^{k-1}(x_1-\eta_m x_2),
\end{align}
where
\begin{align}
\eta_m&=e^{-\pi i\frac{2m+1}{k}}\qquad m=0,\ldots,k-1
\end{align}
are the $k$--th roots of $-1$. We can then define a rank one matrix factorization $W=E J$ of the superpotential (\ref{eq-permsupo}) with
\begin{align}
E=\prod_{m\in I}(x_1-\eta_m x_2)&\qquad J=\prod_{n\in D\backslash I}(x_1-\eta_nx_2),
\end{align}
where $D=\{0,\ldots,k-1\}$ and $I$ is a subset of $D$. In \cite{Brunner:2005fv,Enger:2005jk} a subset of these branes was identified with the permutation branes \cite{Recknagel:2002qq}. In Gepner models, these branes preserve the different $N=2$ superconformal algebras up to permutation.

\subsection{Generalized Permutation Branes}
In \cite{Caviezel:2005th,Fredenhagen:2006qw} a further type of rank one matrix factorizations was introduced. Consider the following superpotential:
\begin{align}
W&=x_1^{k_1}+x_2^{k_2}=x_1^{dr_1}+x_2^{dr_2},
\end{align}
where $d=\mathrm{gcd}(k_1,k_2)$ and $d\geq 2$. The superpotential can be factorized in the following way:
\begin{align}
W=\prod_{m=0}^{d-1}(x_1^{r_1}-\eta_mx_2^{r_2}),
\end{align}
where $\eta_m$ are the $d$--th roots of $-1$. This gives rise to new matrix factorizations. A subset of these which correspond to generalized permutation branes whose conformal field theory description was given in \cite{Fredenhagen:2006qw}. \\\\
In \cite{Caviezel:2005th} it was shown that the three constructions given above account for all the $RR$--charges in Gepner models which are tensor products of $A$--type minimal models. One can construct more general D--branes as tensor products of (generalized) permutation branes with tensor product branes. Such branes are called {\em transposition branes}.\\
Note that the mapping between matrix factorizations and boundary states does not lead to a complete classification of matrix factorizations because there are matrix factorizations which can not be given an interpretation in terms of boundary states.

\subsection{Cone Construction}
As we have already mentioned before, it is possible to obtain a new matrix factorization from two given ones via tachyon condensation as was shown in (\ref{eq-boundstate}).

\section{Further Applications of Matrix Factorizations}
\label{sec-mfother}
In this section we want to account for some further interesting applications of matrix factorizations in physics which will not be discussed in detail in this thesis.\\
One interesting application is concerned with the resolution of singularities. Such problems were discussed in \cite{curtoPHD,Curto:2006}. A hypersurface singularity is defined as a polynomial of the form
\begin{align}
\label{eq-singdef}
f=x_1^2+g(x_2,\ldots,x_n)=0
\end{align}
Via the McKay correspondence \cite{McKay} one can associate to each resolution of the singularity a maximal Cohen--Macaulay module, which can be described in terms of matrix factorizations of (\ref{eq-singdef}). In \cite{Curto:2006} this formulation was used to describe flop transitions. \cite{curtoPHD} discusses resolutions of singularities using matrix factorizations in the context of the correspondence between matrix models to Calabi--Yau compactifications.\\
Another application is in the context of knot theory. In \cite{khovanov1,khovanov2} a relation between knot homology and matrix factorizations was established, which was discussed from a physics point of view in \cite{Gukov:2005qp}, where Landau--Ginzburg superpotentials associated to knot homology were investigated.\\
Recently, matrix factorizations proved to be useful in connection with moduli stabilization with fluxes in non--geometric backgrounds \cite{Becker:2006ks}. 
\chapter{The Effective Superpotential}
\label{chap-effsupo}
This chapter is devoted to the effective superpotential $\mathcal{W}_{eff}$, which plays a central role in $N=1$ string compactifications. The effective superpotential can be interpreted as an effective four--dimensional space time superpotential for a string theory compactified on a Calabi--Yau manifold. 
This is one of the examples where the topological string computes important physical quantities. At present however, the formalism we are working with is not sufficiently well--understood to be applied to a full fledged Calabi--Yau compactification. In this chapter we will thus focus on two interpretations of the effective superpotential, which also make sense for the toy models we are considering here.\\
We will start the discussion with an introductory section which deals with the various interpretations of $W_{eff}$. In section \ref{sec-def-effsupo} we explain how the effective superpotential can be calculated by means of deformation theory. In section \ref{sec-constr-effsupo} we show how to compute $\mathcal{W}_{eff}$ by solving consistency conditions for open topological strings. In this chapter we only introduce the relevant techniques, their applications in concrete examples will be postponed until chapter \ref{chap-minmod}.
\section{Interpretation of the Effective Superpotential}
\label{sec-prop-effsupo}

\subsection{Obstructions to Deformations}
\label{sec-obstr}
\begin{figure}
\begin{center}
\includegraphics{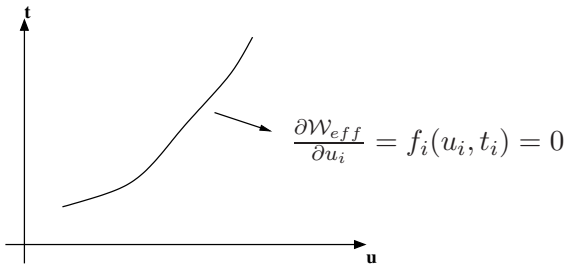}
\caption{The critical locus in the moduli space of a single D--brane.}
\label{fig-critloc}
\end{center}
\end{figure}

A very interesting aspect of the effective superpotential is that it encodes the obstructions to deformations of D--branes. This can be derived in a string field theory context \cite{Lazaroiu:2001nm}. In general, deformations of D--branes will be obstructed and these obstructions are encoded in the critical locus of the effective superpotential. The effective superpotential will in general depend on boundary deformation parameters $u_i$ and bulk deformation parameters $t_i$, which are the moduli of the theory. The critical locus is defined as follows:
\begin{align}
\frac{\partial \mathcal{W}_{eff}(u_i,t_i)}{\partial u_i}&=f_i(u_i,t_i)=0
\end{align}
Note that we only take derivatives of $\mathcal{W}_{eff}$ with respect to the boundary moduli $u_i$ and treat the bulk deformations as background fields \cite{Brunner:2003dc,Herbst:2004zm}.\\
In \cite{Herbst:2004zm} it was shown that the critical locus of the effective superpotential coincides with the factorization locus of a matrix factorization. In general, one cannot expect the matrix factorization $Q(x_i)=E(x_i)J(x_i)$ to be satisfied in the presence of bulk and boundary deformations. But along the critical locus one indeed has:
\begin{align}
W(x_i;t_i)\cdot\mathbbm{1}&=E(x_i;u_i,t_i)J(x_i;u_i,t_i),
\end{align} 
with deformed matrices $E(x_i;u_i,t_i)$, $J(x_i;u_i,t_i)$. The above relation also implies that $u_i$ and $t_i$ cannot be independent.\\
Since the matrix factorization condition is directly related to $N=1$ supersymmetry (see section \ref{sec-bdrylg}), one can deduce that, along the factorization locus and hence along the critical locus, $N=1$ supersymmetry is preserved, whereas it is broken at every other point in parameter space. For a single D--brane one can illustrate this as shown in fig. \ref{fig-critloc}.
Only along the line where $f_i(u_i,t_i)=0$ a deformed matrix factorization exists and $N=1$ supersymmetry is preserved. If one considers a system of multiple D--branes the critical line can have various branches corresponding to different tachyon condensates \cite{Herbst:2004zm}.\\\\
Now the question arises, whether it is possible to calculate the effective superpotential by computing the most general, and generically non--linear, deformations of a matrix factorization. The answer is yes, but it turns out that one has to employ the powerful, rather abstract, mathematical concept of deformation theory in order to solve this problem. The idea is to extract the critical locus by iteratively calculating the deformations of a matrix factorization and their obstructions, which can (under certain conditions) be integrated to give $W_{eff}$. The reason why one has to use this complicated mathematical framework instead of methods which are more physically motivated is twofold: Firstly, when computing deformations of D--branes, one moves away from the critical point. This is why CFT methods will not be of use in this approach to the problem. A second complication is that applying deformation theory methods forces one to use operators which map on the full off--shell Hilbert space. In $N=1$ compactificaltion the unphysical Hilbert space does not decouple, which leads to significant complications as compared to $N=2$ theories. In particular this implies that we can no longer consider on--shell techniques like in the closed string case. This problem has been addressed in \cite{Ashok:2004xq,Cargese} in the context of matrix factorizations. We will now briefly summarize these results.\\\\
The dynamics of off--shell open string modes in B--type topological theories is governed by the holomorphic Chern--Simons action, which defines a cubic string field theory \cite{Witten:1992fb}. On a Calabi--Yau threefold $X$, the action looks as follows:
\begin{align}
S_{CS}&=\int_X \Omega_X \mathrm{Tr}\left(\frac{1}{2}a Q a+\frac{1}{3}a^3\right),
\end{align} 
with fields $a$, BRST operator Q and the holomorphic three--form $\Omega_X$. The off--shell Hilbert space $\mathcal{H}$ has a Hodge decomposition with respect to $Q$:
\begin{align}
\mathcal{H}&=\mathcal{H}_{phys}\oplus\mathcal{H}_{exact}\oplus\mathcal{H}_{unphys}\nonumber\\
&=H\oplus\mathrm{Im}(Q)\oplus\mathrm{Im}(Q^{\dag}),
\end{align}
where $H$ is the physical Hilbert space defined by the cohomology of $Q$ and $Q^{\dag}$ is the hermitian conjugate of $Q$. We can now define an expansion of the off--shell fields $a$:
\begin{align}
a&=\Psi+\tilde{\Psi},
\end{align}
where $\Psi\in H$ and $\tilde{\Psi}\in\mathrm{Im}(Q)\oplus\mathrm{Im}(Q^{\dag})$. We furthermore impose a Siegel--type gauge fixing condition: $Q^{\dag}\tilde{\Psi}=0$. It has been shown in \cite{Lazaroiu:2001nm} that the effective superpotential is given by the following expression:
\begin{align}
\label{eq-sfteffsupo}
\mathcal{W}_{eff}&=\sum_{n=2}^{\infty}\frac{(-1)^{n(n+1)/2}}{n+1}\langle \Psi,m_n(\Psi^{\otimes n})\rangle,
\end{align}
where the inner product is defined as $\langle\Psi\Phi\rangle=\int_X \mathrm{Tr}(\Psi\Phi)$. The 'higher products' $m_n:H^{\otimes n}\rightarrow H$ are subject to the constraints 
\begin{align}
\label{eq-sftcritloc}
\sum_{n=1}^{\infty}(-1)^{n(n+1)/2}m_n(\Psi^{\otimes n})&=0.
\end{align}
The $m_n$ define a cyclic $A_{\infty}$ structure with cyclicity condition
\begin{align}
\langle \Psi_{a_{n+1}},m_n(\Psi_{a_1},\ldots,\Psi_{a_n})\rangle&=(-1)^{n(\tilde{a}_1+1)}\langle \Psi_{a_1},m_n(\Psi_{a_2},\ldots,\Psi_{a_{n+1}})\rangle,
\end{align}
where $\tilde{a}$ denotes the 'suspended grade' of $\Psi_{a}$, see section \ref{sec-constr-effsupo}. The $m_n$ satisfy an $A_{\infty}$--algebra:
\begin{align}
\label{eq-algainf}
\sum^m_{\stackrel{k+l=m+1}{j=0,\ldots,k-1}}(-1)^{\tilde{a_1}+\ldots\tilde{a_j}}m_k(\Psi_{a_1}\ldots\Psi_{a_j},m_l(\Psi_{a_{j+1}\ldots\Psi_{a_{j+l}}}),\Psi_{a_{j+l+1}}\ldots\Psi_{a_m})&=0
\end{align}
Equation (\ref{eq-sftcritloc}) describes the critical locus of the effective superpotential. In deformation theory, these equations are generalized Maurer--Cartan equations. 
In the string field theory context these are the equations of motion. Note that the sums in (\ref{eq-sfteffsupo}) and (\ref{eq-sftcritloc}) generically contain infinitely many terms, which makes them hard to compute. In some simple cases, namely the minimal models to be discussed in chapter \ref{chap-minmod}, these expressions are actually polynomial.\\
The calculations of the effective superpotential thus boils down to determining the higher products $m_n$. Formally, this can be done as follows (see for instance \cite{Merkulov,Kajiura:2005sn}): Define a restriction $Q|_{\mathrm{Im}Q^{\dag}}:\mathrm{Im}(Q^{\dag})\rightarrow\mathrm{Im}(Q)$. It is invertible and its inverse is denoted by $Q^{-1}$. Furthermore we define a projector $\pi$ from the full Hilbert space to the exact states: $\pi:\mathcal{H}\rightarrow\mathrm{Im}(Q)$. Using these expressions one can also define a projector $U$ into the unphysical Hilbert space: $U=Q^{-1}\pi:\mathcal{H}\rightarrow\mathrm{Im}(Q^{\dag})$. Then one defines multilinear maps $\lambda_n:\mathcal{H}^{\otimes n}\rightarrow\mathcal{H}$ which act on the fields $a_i$ that live in the full Hilbert space:
\begin{align}
\label{eq-algampl}
\lambda_2(a_1,a_2)=&a_1\cdot a_2\nonumber\\
\vdots& \nonumber\\
\lambda_n(a_1,\ldots,a_n)=&(-1)^{n-1}\left[U\lambda_{n-1}(a_1,\ldots,a_{n-1})\right]\cdot a_n-(-1)^{n\tilde{a}_1}a_1\cdot\left[U\lambda_{n-1}(a_2,\ldots,a_n)\right]\nonumber \\
&-\sum_{\stackrel{k,l\geq 2}{k+l=n}}(-1)^{k+(l-1)(\tilde{a}_1+\ldots+\tilde{a}_n)}\left[U\lambda_k(a_1,\ldots,a_k)\right]\cdot\left[U\lambda_l(a_{k+1}\ldots,a_n)\right]
\end{align}
The $\lambda_n$ can be related to the $m_n$ via a projector which maps to the physical Hilbert space: $\Pi:\mathcal{H}\rightarrow H$, $\Pi=\mathbbm{1}-[Q,U]$, where $[\cdot,\cdot]$ is a graded commutator. The $m_n$ are then defined as follows:
\begin{align}
m_1=0\qquad m_n=\Pi\lambda_n
\end{align}
How can this construction be related to the deformation of matrix factorizations? A general deformation of a matrix factorization $Q$ is given by:
\begin{align}
\label{eq-qdefformal}
Q_{def}=Q+\sum_{n=1}^{\infty}Q_n.
\end{align}
At linear order the deformations are given by the fermionic open string states: $Q_1=u_a\Psi_a$. In general this will spoil the factorization condition:
\begin{align}
Q_{def}^2-W=u^2[\Psi,\Psi]\equiv\lambda_2(\Psi,\Psi)
\end{align}
As long as $\lambda_2(\Psi,\Psi)$ is exact, one can cancel the term at the next order by adding
\begin{align}
Q_2=-U\lambda_2(\Psi,\Psi)
\end{align}
If $\lambda_2(\Psi,\Psi)$ gives a physical state, the contribution cannot be cancelled by adding a term to $Q$ and thus the deformation is obstructed. These obstructions are encoded in the equations 
\begin{align}
\label{eq-obstrformal}
\sum m_n(\Psi^{\otimes n})=\sum\Pi\lambda_n(\Psi^{\otimes n})=0. 
\end{align}
In general the $Q_n$ are given by \cite{Cargese}:
\begin{align}
\label{eq-masseyformal}
Q_n=-U\lambda_n(\Psi^{\otimes n}).
\end{align}
This construction is rather formal because the explicit form of the operators $U$ and $\Pi$ is in general not known. Luckily, one does not need the explicit form of these operators in order to calculate the effective superpotential. All one needs to know is how they act on certain states. In \cite{Ashok:2004xq,Aspinwall:2007cs} the above constructions were used to calculate the effective superpotential for a certain D--brane configuration on the quintic.\\
In section \ref{sec-def-effsupo} we will describe an algorithm which recursively computes the $Q_n$ and $m_n$ \cite{siqvelandPHD,Siqveland1} and is equivalent to the methods used in \cite{Ashok:2004xq,Aspinwall:2007cs}. 
\subsection{Generating Function for Disk Amplitudes}
The effective superpotential is defined as the generating functional of open string disk amplitudes. In \cite{Lazaroiu:2001nm} this was derived in a string field theory context as already mentioned in section \ref{sec-obstr}. It was shown that the higher products $m_n$ appearing in (\ref{eq-sfteffsupo}) are related to the tree--level diagrams and the relations (\ref{eq-algampl}) encode how an amplitude decomposes into amplitudes with a lower number of insertions. \\
The idea is now to calculate $\mathcal{W}_{eff}$ by explicitly calculating the values of the open string disk amplitudes via consistency constraints. This was discussed in \cite{Hofman:2000ce,Herbst:2004jp}. \\
In closed topological string theories, the analogue of the effective superpotential is the WDVV potential $\mathcal{F}(t)$ which is the generating function of the genus zero amplitudes. All the correlators can be obtained as derivatives with respect to the bulk parameters $t_i$. $\mathcal{F}(t)$ is constrained by the WDVV equations, which arise form the crossing--symmetry of the four--point correlator \cite{Dijkgraaf:1990dj}:
\begin{align}
\label{eq-WDVV}
\partial_i\partial_k\partial_m\mathcal{F}\eta^{mn}\partial_n\partial_j\partial_l\mathcal{F}=\partial_i\partial_j\partial_m\mathcal{F}\eta^{mn}\partial_n\partial_k\partial_l\mathcal{F},
\end{align}
where $\eta_{mn}=C_{0mn}$ is the topological metric, $C_{ijk}$ are the three--point functions and $\partial_i=\frac{\partial}{\partial t_i}$.\\
The integrability of the closed string amplitudes is linked to the fact that the correlators are completely symmetric with respect to their insertions, and to the existence of flat coordinates which are determined by $N=2$ special geometry.
For open strings the situation is much more complicated. In particular, it is not  possible to integrate the correlators with respect to the boundary deformation parameters because the disk correlators are only cyclically symmetric with respect to boundary insertions, but they are still integrable with respect to the bulk parameters. In general, one also lacks flat coordinates on the boundary moduli space.  \\
Consider an amplitude with $m$ boundary insertions and $n$ bulk insertions. We can write this as
\begin{align}
B_{a_0\ldots a_m;i_0\ldots i_n}=\partial_{i_0}\ldots\partial_{i_n}\mathcal{F}_{a_0,\ldots,a_m}|_{t=0},
\end{align}
where $\mathcal{F}_{a_1\ldots a_m}$ are the disk amplitudes. One can define a formal generating function of these amplitudes \cite{Herbst:2004jp}:
\begin{align}
\hat{\mathcal{W}}&=\sum_{m\geq 1}\frac{1}{m}\hat{s}_{a_m}\ldots\hat{s}_{a_0}\mathcal{F}_{a_0,\ldots,a_m}(\hat{t}),
\end{align}
where $\mathcal{F}_{a_1,\ldots,a_m}(\hat{t})$ is viewed as a formal power series and the $\hat{s}_a$ are non--commuting parameters. We have also introduced formal parameters $\hat{t}$, associated to the bulk parameters. This object does not have a physical interpretation, but it can be evaluated over the super--commuting variables $s_a=\{u_a,v_a\}$, where the $u_a$ are commuting and the $v_a$ are anti--commuting. We define:
\begin{align}
\mathcal{W}_{eff}(s;t)&:=\sum_{m\geq 1}\frac{1}{m}s_{a_m}\ldots s_{a_0}\mathcal{F}_{a_0\ldots a_m}(t)
\end{align}
The parameters $s_a$ can then be given a physical interpretation as deformations parameters related to odd/even boundary insertions\footnote{The parameters $u_i$ are related to the fermionic boundary insertions, the anti--commuting $v_i$ are related to the even boundary insertions.}. Now define a supersymmetrized combination of the amplitudes:
\begin{align}
\label{eq-symcorr}
\mathcal{A}_{a_0\ldots a_m}&:=(m-1)!\mathcal{F}_{(a_0,\ldots,a_m)}:=\frac{1}{m}\sum_{\sigma\in S_m}\eta(\sigma;a_0,\ldots,a_m)\mathcal{F}_{a_{\sigma(0)}\ldots a_{\sigma(m)}},
\end{align}
where $\sigma$ is a permutation and $\eta$ is the sign one obtains from permuting the variables $s_a$. Then one can write:
\begin{align}
\label{eq-effsupo-corr}
\mathcal{W}_{eff}(s;t)&=\sum_{m\geq 1}\frac{1}{m}s_{a_m}\ldots s_{a_1}\mathcal{A}_{a_1\ldots a_m}(t)
\end{align}
One can now get the $\mathcal{A}_{a_1\ldots a_m}$ as partial derivatives $\partial_a\equiv\frac{\partial}{\partial s_a}$ of the effective superpotential:
\begin{align}
\mathcal{A}_{a_0\ldots a_m}=\partial_{a_0}\ldots\partial_{a_m}\mathcal{W}_{eff}(s;t)|_{s=0}
\end{align}
Thus, in contrast to the closed string case, the derivative of the effective superpotential does {\em not} yield the amplitudes but only their symmetrized combinations. The amplitudes therefore contain more information than the effective superpotential.\\
In order to calculate $\mathcal{W}_{eff}$ in this approach, one has to find a way to calculate the values of all the correlators by solving generalized consistency constraints analogous to those in the closed string case. Such consistency constraints were derived in \cite{Herbst:2004jp} and will be discussed in section \ref{sec-constr-effsupo}. The constraints are more complicated than (\ref{eq-WDVV}) due to the cyclic structure of the boundary insertions and because a larger number of sewing constraints has to be satisfied by correlators with bulk and boundary insertions. In particular, \cite{Herbst:2004jp} gives a field theory derivation of the $A_{\infty}$--relations (\ref{eq-algainf}). There are two further sets of constraints given in \cite{Herbst:2004jp}, the bulk--boundary crossing constraint and the Cardy--condition, where they were successfully tested for the $A$--minimal models. In \cite{Herbst:2006nn,Knapp:2006rd,Keller:2006tf}  it was pointed out that they will have to be modified for more complicated models.\\
Note that such consistency constraints can also be derived for higher genus amplitudes \cite{Herbst:2006kt}. They are called 'quantum--$A_{\infty}$ relations'. 
\section{Construction of the Effective Superpotential via Deformation Theory}
\label{sec-def-effsupo}
In this section we present what is probably the most sophisticated, but from a physics point of view also the most abstract way to calculate the effective superpotential. The algorithm is called the 'method of computing formal moduli' and was introduced in \cite{ siqvelandPHD,Siqveland1,Siqveland2} based upon \cite{laudal1,laudal2}. It was extended in \cite{Siqveland3}. The method uses general concepts of deformation theories with the following specifications:
\begin{itemize}
\item It computes the versal deformations of {\em modules} and is thus tailored for matrix factorizations.
\item The method considers deformations of D--branes with fermionic open string states, which are related to commuting deformation parameters $u_i$. In order to include deformations with bosonic states, one would have to find a non--commutative extension of the formalism. For the case of minimal models, which will be our focus, one does not loose information by neglecting bosonic deformations because $\mathcal{W}_{eff}$ is independent of the associated non--commuting deformation parameters \cite{Herbst:2004jp}.
\item Bulk deformations are not included in the formalism. For a single D--brane, this means in particular that the equations for the critical locus $f_i(u_i)=\frac{\partial\mathcal{W}_{eff}}{\partial u_i}=0$ only have the trivial solution $u_i=0$. This means this is only a formal deformation. This changes if one considers systems of multiple D--branes \cite{Herbst:2004zm} or if one turns on bulk moduli. It was shown in \cite{Knapp:2006rd} and will be reviewed in section \ref{chap-minmod} that bulk--deformations can be incorporated into the formalism. 
\end{itemize}
In the following subsection we attempt to give some intuition for the abstract concepts of deformation theory. Then we move on to describe the mathematical construction underlying the algorithm. In section \ref{subsec-sumalg} we summarize the results of the mathematical construction which are relevant for the application of the algorithm.

\subsection{An Intuitive Picture}
\label{sec-darkroom}
\begin{figure}
\begin{center}
\includegraphics{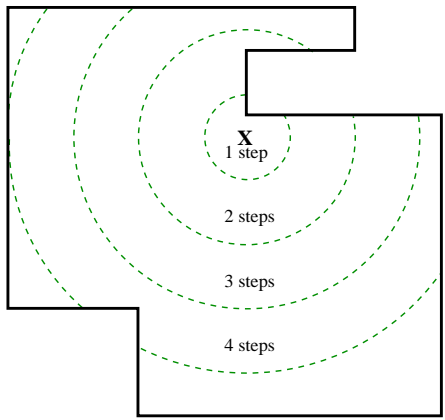}
\caption{Exploring the shape of a dark room as an analogy for deformation theory.}
\label{fig-darkroom}
\end{center}
\end{figure}

A nice way to think about deformation theory is the analogy of trying to explore the shape of a dark room as depicted in fig. \ref{fig-darkroom}.\\
The procedure is as follows: In order to investigate the room in the most efficient way, one first takes one step in every possible direction, say, left, right, forward, backwards. In the mathematical formalism, the directions are encoded, at linear level, in terms of the fermionic cohomology elements. At higher orders, which is related to taking more steps in the room, this is governed by higher products in the cohomology, the so--called Massey products. The allowed steps correspond to the possible deformations and are thus encoded in the $Q_n$ in (\ref{eq-qdefformal}).\\
Hitting a wall of the room corresponds to encountering an obstruction of the deformation. One has to make sure to exclude this obstructed direction when increasing the number of steps. In the mathematical formalism these obstructions are encoded in the relations (\ref{eq-masseyformal}). There may however be some subtleties: the room to explore could have some hidden areas, like the top right corner in fig. \ref{fig-darkroom}. In order to find this hidden corner one has to find a way to navigate around the obstacle one finds after the first step. Also this is encoded in the formalism as information about obstructions at lower order enters in higher order computations. \\
Note that fig. \ref{fig-darkroom} implies that after a certain number of steps one cannot move any further because one has reached the end of the room in all possible directions. In the deformation theory formalism this would mean that the sums in (\ref{eq-qdefformal}) and (\ref{eq-masseyformal}) only contain finitely many terms. This is true for the minimal models, which have only massive deformations, where the algorithm really terminates after a finite number of steps. For theories with marginal deformations, there are in general infinitely many terms (unless one makes a clever choice for the input of the algorithm \cite{Aspinwall:2007cs}). One should thus regard the room as infinitely large.

\subsection{Mathematical Construction}
In this section we summarize the mathematical construction of the 'method of computing formal moduli', which allows us to calculate deformations of matrix factorizations. We closely follow \cite{siqvelandPHD,Siqveland1}, further literature on the deformation theory of modules and related subjects can be found in \cite{Siqveland2,Siqveland3} and \cite{eriksen,math.AG/0303166,eriksen2,eriksen3}. 
We will summarize the essential steps of the algorithm in section \ref{subsec-sumalg}. \\\\
We start by defining the deformation of a module. We denote by $A$ a $k$--algebra, where $k$ is a field. Furthermore, $E$ is an $A$--module\footnote{In this section we adopt the notation of \cite{Siqveland1}, which differs slightly from the notation we used in section \ref{sec-mfmath}.}.
We denote by $\ell$ the category of local Artinian\footnote{An Artinian ring is a ring which satisfies a 'descending chain relation on ideals'.  These are rings whose underlying sets are finite or finite--dimensional vector spaces over fields \cite{wiki:artin}. The ring $\tilde{\mathcal{R}}_m$ defined  in section \ref{sec-mcm} is precisely what we are working with here.} $k$--algebras with residue field $k$. For an object $S$ in $\ell$ we can then define an infinitesimal deformation functor $\mathrm{Def_E(S)}$ taking values in the category of sets:
\begin{align}
\mathrm{Def}_E(S)&=\left\{(E,\theta)|E\mathrm{~is~an~}A\otimes_k S\mathrm{~module,~flat~over~}S\mathrm{,~and~}E\otimes_Sk \stackrel{\theta}{\cong}E\right\}/\cong
\end{align}
Then there is a theorem due to Schlessinger \cite{Schlessinger}, which states that there exists a hull $\hat{H}_E$ of $\mathrm{Def}_E$, termed local formal moduli of $E$, which is a complete local $k$--algebra together with a smooth morphism $\mathrm{Mor}(\hat{H}_E,-)\rightarrow\mathrm{Def}_E$ such that $\mathrm{Mor}(\hat{H}_E,k[x]/x^2)\rightarrow \mathrm{Def}(k[x]/x^2)$ is a bijection.\\
We now have to construct the hull $\hat{H}_E$. We start by fixing a free resolution $L^{\bullet}$ of $E$ as an $A$--module. In our case this amounts to choosing a matrix factorization. One can define a differential
\begin{align}
d_i&:L_i\rightarrow L_{i-1}.
\end{align}
We set
\begin{align}
\mathrm{Hom}_A^p(L^{\bullet},L^{\bullet})&=\prod_{m\geq p}\mathrm{Hom}(L_m,L_{m-p}),
\end{align}
and then $d^p:\mathrm{Hom}_A^p(L^{\bullet},L^{\bullet})\rightarrow\mathrm{Hom}_A^{p+1}(L^{\bullet},L^{\bullet})$ is defined by:
\begin{align}
\label{eq-homdiff}
d^p\left(\left\{\alpha_i^p\right\}_{i\geq p}\right)&=\left\{d_i\circ\alpha_{i-1}^p-(-1)^p\alpha_i^p\circ d_{i-p} \right\}
\end{align}
$\mathrm{Hom}^{\bullet}_A$ is a graded differential associative $A$--algebra, where the multiplication is the composition of $\mathrm{Hom}_A^{\bullet}(L^{\bullet},L^{\bullet})$. For matrix factorizations, (\ref{eq-homdiff}) is just the (component version of the) physical state condition (\ref{eq-diffab}).\\
An important result states that that there is a natural isomorphism between $\mathrm{Ext}$--groups and cohomology:
\begin{align}
\label{eq-extcoh}
\mathrm{Ext}^i_A(E,E)&\cong H^i(\mathrm{Hom}^{\bullet}_A(L^{\bullet},L^{\bullet})),\qquad i\geq 0
\end{align}
In \cite{Siqveland1} the following theorem is proven:
\begin{thm}
Given an object $S$ in $\ell$ assume there exists a lifting $\left\{L^{\bullet}\otimes_kS,d_i(S) \right\}$ of the complex $\{L^{\bullet},d_i\}$ of the free resolution $L^{\bullet}$ of $E$. This means that there exists a commutative diagram of the form
\begin{align}
\begin{diagram}
0&\lTo&E_S&\lTo&L_0\otimes S&\lTo^{d_1(S)}&L_1\otimes S&\lTo^{d_2(S)}&L_2\otimes S&\lTo^{d_3(S)}&\cdots\\
&&&&\dTo&&\dTo&&\dTo\\
0&\lTo&E&\lTo&L_0&\lTo^{d_1}&L_1&\lTo^{d_2}&L_2&\lTo^{d_3}&\cdots\\
\end{diagram}
\end{align}
where for every $i$ the composition $d_{i+1}(S)\circ d_i(S)=0$ and $d_{i}(S)$ is $A\otimes_k S$--linear. Then $\left\{L^{\bullet}\otimes_kS,d_i(S) \right\}$ is an $A\otimes_k S$--free resolution of $E_S$ and $E_S$ is a lifting of $E$ to $S$.
\end{thm}
In the context of matrix factorizations, which corresponds to the special case where the complexes above are periodic, i.e. if $E$ is a maximal Cohen--Macaulay module. This means that, if a deformation is possible, one can construct a deformed matrix factorization for which the factorization condition holds. The rest of this section will be devoted to the explicit construction of this lifting.\\
At first we have to define what an obstruction is. The following theorem relates obstructions to elements of $\mathrm{Ext}^2(E,E)$, which is for matrix factorizations related to the even cohomology elements via (\ref{eq-extcoh}).
\begin{thm}
Let $\pi:R\rightarrow S$ be a small morphism in $\ell$, where a small morphism satisfies $m_R\cdot\mathrm{ker}\pi=0$, $m_R$ being the maximal ideal. Then let $E_S\in\mathrm{Def}_E(S)$ correspond to the lifting $\left\{L^{\bullet}\otimes S,d_i(S) \right\}$ of $L^{\bullet}$. Then there is a uniquely defined obstruction
\begin{align}
o(E_S,\pi)&\in\mathrm{Ext}_A^2(E,E)\otimes_k I
\end{align}
given in terms of the 2--cocycle $o\in\mathrm{Hom}_A^{\bullet}\otimes_k I$ such that $o(E_S,\pi)=0$ if and only if $E_S$ may be lifted to $R$. Here $I=\mathrm{ker}\pi$. Moreover, if $o(E_S,\pi)=0$, then the set of liftings of $E_S$ to $R$ is a principal homogeneous space\footnote{A principal homogeneous space for a group $G$ is a set $X$ for which $G$ acts freely and transitively \cite{wiki:phs}.} over $Ext_A^1(E,E)$.
\end{thm}
The linear, infinitesimal deformations are given by the odd cohomology elements, related to $\mathrm{Ext}^1$:
\begin{align}
\mathrm{Ext}^1_A(E,E)&\cong\mathrm{Def}_E(k[x]/x^2)\cong\mathrm{Def}_E(k[\varepsilon])
\end{align}
The next step is the proof of a structure theorem due to Laudal \cite{laudal1,laudal2}. This proof is constructive and yields an algorithm to calculate explicitly the deformations of modules. The theorem states the following:
\begin{thm}
There exists a morphism of complete local $k$--algebras
\begin{align}
o:T^2&=\mathrm{Sym}_k\left(\mathrm{Ext}^2(E,E)^*\right){\hat{}}\rightarrow T^1=\mathrm{Sym}_k\left(\mathrm{Ext}^1(E,E)^*\right){\hat{}}
\end{align}
(where $\mathrm{Sym}$ is the symmetric power and the $\ast$ denotes the dual of the group)
such that $\hat{H}_E\cong T^1\otimes_{T^2}k$. The $\hat{}$ in the above equation indicates that we consider the completion of the algebra\footnote{This means that we go from ordinary power series to formal power series.}. Furthermore, for any small morphism $\pi:R\rightarrow S$ in the diagram
\begin{align}
\begin{diagram}
\label{diag-laudal}
\mathrm{Ext}^2(E,E)^*\subseteq m_{T^2}\subseteq T^2&\rTo^{o}&T^1&\rTo^{\bar{\phi}}&R\\
&&\dTo&&\dTo^{\pi}\\
&&\hat{H}&\rTo^{\phi}&S\\
\end{diagram}
\end{align}
the obstruction for lifting $E_{\phi}\in\mathrm{Def}_E(S)$ to $R$ is given by the restriction of $o\circ\bar{\phi}$ to $\mathrm{Ext}^2(E,E)^{\ast}$.
\end{thm}
The conclusion to draw from this rather abstract theorem is that the hull $\hat{H}_E$ is isomorphic to the quotient ring
\begin{align}
\hat{H}_E&\cong k[u_1,\ldots,u_d]/(f_1,\ldots,f_r),
\end{align}
where $d=\mathrm{dim}(\mathrm{Ext}^1(E,E))$ and $r\leq\mathrm{dim}(\mathrm{Ext}^2(E,E))$. The vanishing relations $f_i$ actually determine the critical locus of $\mathcal{W}_{eff}$ and their coefficients can be constructed from {\em matric Massey products}\footnote{A Massey product is a higher operation in cohomology, generalizing the cup product between cohomology elements.}. \\\\
In order to explicitly construct this quotient ring we choose bases $\{x_1,\ldots,x_d\}$ for $\mathrm{Ext}^1_A(E,E)^{\ast}$, and $\{y_1,\ldots,y_r\}$ for $\mathrm{Ext}^2_{A}(E,E)^{\ast}$ and the corresponding dual bases $\{x_1^{\ast},\ldots,x_d^{\ast}\}$, $\{y_1^{\ast},\ldots,y_r^{\ast}\}$, which are the fermionic and bosonic open string states in the physical description.\\
A lemma due to Nakayama states that\footnote{In the following we will skip the subscript $E$ in $\hat{H}_E$.} 
\begin{align}
\hat{H}/m^2&\cong k[u_1,\ldots,u_d]/m^2,
\end{align}
where $m=(u_1,\ldots,u_d)$. We now start a stepwise construction by setting
\begin{align}
S_1=k[u_1,\ldots,u_d]/m^2\equiv k[u]/m^2&\qquad R_2=k[u]/m^3
\end{align}
We let $E_{\phi_1}\in\mathrm{Def}_E(S_1)$ correspond to $\phi_1:\hat{H}\rightarrow \hat{H}/m^2$, $\phi_1(x_i)=u_i$ -- thus, this construction associates a deformation parameter $u_i$ to every odd cohomology element. Then, if $a\subseteq R_2$ is a least ideal such that $E_{\phi_1}$ can be lifted to $R_2/a=S_2$ then $R_2/a\cong \hat{H}/m^3$.\\
Now consider the following diagram 
\begin{align}
\begin{diagram}
&&R_2=k[u]/m^3\\
&&\dTo_{\pi_2'}\\
\hat{H}&\rTo^{\phi_1}&S_1=k[u]/m^2
\end{diagram}
\end{align}
Note that these diagrams are intermediate steps in the diagram (\ref{diag-laudal}). These constructions implicitly implement the actions of the operators $U$ and $\Pi$ of section \ref{sec-obstr}.\\
Now pick a monomial basis for $S_1$ of the form $\{u^{\vec{n}}\}_{\vec{n}\in\bar{B}_1}$, $\bar{B}_1=\{\vec{n}\in \mathbbm{N}^d:|\vec{n}|\leq 1\}$, where $|\vec{n}|=\sum_{i=1}^d n_i$ and $u^{\vec{n}}=\prod_{i=1}^du_i^{n_i}$. We furthermore choose a basis $\{u^{\vec{n}}\}_{\vec{n}\in B_2'}$, $B_2'=\{\vec{n}\in\mathbbm{N}^d:|\vec{n}|=2\}$ for $\mathrm{ker}\:\pi_2'$. Then the obstructions are:
\begin{align}
o(E_{\phi_1},\pi_2')\in\mathrm{Ext}_A^2(E,E)\otimes_k\mathrm{ker}\pi_2'&=\sum_{\vn\in B_2'}a_{\vn}\otimes u^{\vn}=\sum_{i=1}^r y_i^{\ast}\otimes f_i^2(u).
\end{align}
Here we have the following definitions: $\phi_1$ is called a defining system for the Massey products
\begin{align}
\langle x^{\ast};\vn\rangle&=a_{\vn},\qquad \vn\in B'_2.
\end{align}
How these Massey products can be calculated explicitly will be shown at the end of this section. The $f_i^2$ (the $2$ indicates the order in u) are defined as follows:
\begin{align}
f_i^2(u)=\sum_{\vn\in B_2'}y_i(\langle x^{\ast};\vn\rangle)u^{\vn}
\end{align}
Thus, whenever a Massey product is non--zero in the cohomology, i.e. if it is proportional to an even cohomology element, the proportionality constant determines the coefficient in the vanishing relations $f_i$ whereas the powers in the $u_i$ are defined by the basis vectors $\vn$.\\
This implies that
\begin{align}
\hat{H}/m^3\cong R_2/(f_1^2,\ldots,f_r^2)=k[u]/(m^3+(f_1^2,\ldots,f_r^2))=S_2.
\end{align} 
In order to lift $E_{\phi_2}$ to $S_2$ we consider the following diagram:
\begin{align}
\begin{diagram}
&&R_3=k[u]/(m^4+m(f_1^2,\ldots,f_r^2))\\
&&\dTo_{\pi_3'}\\
\hat{H}&\rTo^{\phi_2}&S_2\\
&\rdTo&\dTo_{\pi_2}\\
&&S_1\\
\end{diagram}
\end{align}
The lift of $E_{\phi_1}$ to $S_2$ is done by the composition $\kappa$, which does nothing but impose the relation defining the obstruction to vanish at the given order: $o(E_{\phi_1})=\kappa o(E_{\phi_2},\pi_2')=\sum y_i^{\ast}\otimes f_i^2(u)=0$. This implies that $\phi_2:\hat{H}\rightarrow S_2$ can be constructed.\\
Now pick a monomial basis $\{u^{\vn}\}_{\vn\in B_2}$ for $\mathrm{ker}\pi_2=m^2/(m^3+(f_1^2,\ldots,f_2^2))$ and set $\bar{B}_2=\bar{B}_1\cup B_2$. Then $\bar{B}_2$ is a basis for $S_2$, which implies that for every $\vn$ with $|\vn|\leq 2$ there is a unique relation\footnote{The vector $\vm$ in the following equations is not to be confused with the maximal ideal $m$!} in $S_2$:
\begin{align}
u^{\vn}&=\sum_{\vm \in \bar{B}_2}\beta_{\vn,\vm}u^{\vm}
\end{align}
Furthermore, $o(E_{\phi_1},\pi_2)=0$ implies for $\vm\in\bar{B_2}$ that
\begin{align}
\beta_{\vm}=\sum_{\vn\in B_2'}\beta_{\vn,\vm}\langle x^{\ast};\vn\rangle=0.
\end{align}
This finishes the discussion at order 2. Since at this order non--trivial relations $f_i$ appear, we also have to consider the next order before performing the induction step to prove Laudal's structure theorem.\\\\
At order 3, one has
\begin{align}
\mathrm{ker}\pi_3'&=(f_1^2,\ldots,f_r^2)/m(f_1^2,\ldots,f_r^2)\oplus m^3/(m^4+m^3\cap m(f_1^2,\ldots,f_r^2))=a\oplus I_3
\end{align}
Now we pick a basis $\{u^{\vn}\}_{\vn\in B_3'}$ for $I_3$, where on may choose $\vn\in B_3'$ such that $u^{\vn}=u_ku^{\vm}$ for some $\vm\in B_2$ and one of the $u$--parameters\footnote{Thus, we just add unit vectors to the vectors defining $B_2$.}, and we set $\bar{B}_3'=\bar{B}_2\cup B_3'$. Then for every $\vn$ with $|\vn|\leq 3$ there is a unique relation in $R_3$:
\begin{align}
u^{\vn}&=\sum_{\vm\in\bar{B}_3'}\beta'_{\vn,\vm}u^{\vm}+\sum_j\beta'_{\vn,j}f_j^2
\end{align}
The second term means that one gets additional contributions from the relations $f_i$, where the index $j$ indicates something which is not in $\bar{B}'_3$. In the analogy of the dark room, given in the previous section, this amounts to going around the barrier in the upper right corner of fig. \ref{fig-darkroom}.\\
The obstruction is then:
\begin{align}
o(E_{\phi_2},\pi_3')&=\sum y_j^{\ast}\otimes f_j^2+\sum_{\vn\in B_3'}a_n\otimes u^{\vn}=\sum y_j^{\ast}\otimes f_j^3,
\end{align}
where we call any map $\phi_2$ a defining system for the Massey products
\begin{align}
\langle x^{\ast};\vn\rangle&=a_{\vn},\qquad \vn\in B_3'.
\end{align}
We can then write:
\begin{align}
f_j^3&=\sum_{\vn\in B_2'}y_j\langle x^{\ast};\vn\rangle u^{\vn}+\sum_{\vn\in B_3'}y_j\langle x^{\ast};\vn\rangle u^{\vn}.
\end{align}
Now we have:
\begin{align}
\hat{H}/m^4&=S_3=R_3/(f_1^3,\ldots,f_r^3)=k[u]/(m^4+(f_1^3,\ldots,f_r^3))
\end{align}
Next, we pick a monomial basis $\{u^{\vn}\}_{\vn\in B_3}$ for $\ker\pi_3$, where $\pi_3:S_3\rightarrow S_2$, such that $B_3\subseteq B_3'$ and we set $\bar{B}_3=\bar{B}_2\cup B_3$. Then, there exists a unique relation in $S_3$ for every $\vn$ with $|\vn|\leq 3$:
\begin{align}
u^{\vn}=\sum_{\vm\in\bar{B}_3}\beta_{\vn,\vm}u^{\vm}
\end{align}
As before $o(E_{\phi_2},\pi_3)=0$ implies the relations
\begin{align}
\beta_{\vm}&=\sum_{l=2}^3\sum_{\vn\in B_l'}\beta_{\vn,\vm}\langle x^{\ast};\vn\rangle=0
\end{align}
Now that we have captured all the essential structure, we can continue by induction. For any $k\geq 1$, assume that we have found $S_{2+k}=k[u]/(m^{2+k+1}+(f_1^{2+k},\ldots,f_r^{2+k}))$ such that $\hat{H}/m^{2+k+1}\cong S_{2+k}$. Consider the diagram
\begin{align}
\begin{diagram}
&&R_{2+k+1}=k[u]/(m^{2+k+2}+m(f_1^{2+k},\ldots,f_r^{2+k}))\\
&&\dTo_{\pi_{2+k+1}'}\\
\hat{H}&\rTo^{\phi_{2+k}}&S_{2+k}=k[u]/(m^{2+k+1}+(f_1^{2+k},\ldots,f_r^{2+k}))\\
&\rdTo_{\ldots}&\dTo_{\ldots}\\
&&S_1\\
\end{diagram}
\end{align}
Here the $[\ldots]$ stand for the intermediate steps. Now write
\begin{align}
\mathrm{ker}\pi_{2+k+1}'&=(f_1^{2+k},\ldots,f_r^{2+k})/m(f_1^{2+k},\ldots,f_r^{2+k})\oplus m^{2+k+1}/(m^{2+k+2}+m^{2+k+1}\cap m(f_1^{2+k},\ldots,f_r^{2+k}))\nonumber \\
&=a\oplus I_{2+k+1}
\end{align}
Pick a monomial basis for $I_{2+k+1}$ of the form $\{u^{\vn}\}_{\vn\in B'_{2+k+1}}$, where one may assume that for $\vn\in B'_{2+k+1}$, one has $u^{\vn}=u_ku^{\vm}$ for some $\vm\in B_{2+k}$. Then put $\bar{B}_{2+k+1}=\bar{B}_{2+k}\cup B'_{2+k+1}$. For every $\vn$ with $|\vn|\leq 2+k+1$ there is a unique relation
\begin{align}
u^{\vn}&=\sum_{\vm\in\bar{B}'_{2+k+1}}\beta'_{\vn,\vm}u^{\vm}+\sum_j\beta'_{\vn,j}f_j^{2+k}
\end{align}
The obstructions are:
\begin{align}
o(E_{\phi_{2+k}},\pi'_{2+k+1})&=\sum_j y_j^{\ast}\otimes f_j^{2+k}+\sum_{\vn\in B'_{2+k+1}}a_{\vn}\otimes u^{\vn}=\sum_j y_j^{\ast}\otimes f_j^{2+k+1}
\end{align}
Any map $\phi_{2+k}$ is called a defining system for the Massey products
\begin{align}
\langle x^{\ast};\vn\rangle&=a_n,\qquad \vn\in B'_{2+k+1}
\end{align}
Then we have
\begin{align}
f_{j}^{2+k+1}&=\sum_{l=0}^{2+k+1}\sum_{\vn\in B'_{2+l}}y_j\langle x^{\ast};\vn\rangle u^{\vn}
\end{align}
One has
\begin{align}
\hat{H}/m^{2+k+2}&\cong S_{2+k+1}=k[u]/(m^{2+k+2}+(f_1^{2+k+2},\ldots,f_r^{2+k+1})).
\end{align}
Set $\pi_{2+k+1}:S_{2+k+1}\rightarrow S_{2+k}$ and pick a monomial basis $\{u^{\vn}\}_{\vn\in B_{2+k+1}}$ for $\mathrm{ker}\pi_{2+k+1}$ such that $B_{2+k+1}\subseteq B'_{2+k+1}$ and furthermore set $\bar{B}_{2+k+1}=\bar{B}_{2+k}\cup B_{2+k+1}$. Then there is a unique relation for all $\vn$ with $|\vn|\leq 2+k+1$:
\begin{align}
u^{\vn}&=\sum_{\vm\in\bar{B}_{2+k+1}}\beta_{\vn,\vm}u^{\vm},
\end{align}
and due to $o(E_{\phi_{2+k}},\pi_{2+k+1})=0$ we have:
\begin{align}
\label{eq-urel}
\beta_{\vm}&=\sum_{l=2}^{2+k+1}\sum_{\vn\in B_l'}\beta_{\vn,\vm}\langle x^{\ast};\vn\rangle=0.
\end{align}
By induction, this proves
\begin{align}
\label{eq-betadef}
\hat{H}\cong\mathrm{lim}_k S_{2+k}=k[[u]]/(\bar{f}_1,\ldots,\bar{f}_r),
\end{align}
where $\bar{f}_j=\lim_k f_j^{2+k}$. This proves Laudal's theorem, when one sets $o(y_j)=\bar{f}_j$ in (\ref{diag-laudal}). Note that this is a formal expression and $k$ will in general become infinite. However, for the special case of the minimal models the $\bar{f}_j$ are polynomial expressions, see chapter \ref{chap-minmod}.\\\\
We still miss one piece in order to be able to calculate $\hat{H}$ explicitly. We have yet to find out how one can calculate the Massey products. These can be obtained from the lifting of the operator $d$ to $d(S)$ and the conditions (\ref{eq-urel}) and (\ref{eq-betadef}). At linear order one has $S_1=k[u_1,\ldots,u_d]/m^2$ and one can pick a monomial basis $\{u^{\vm}\}_{\vm\in\bar{B}_1}$. Then let $E_1\in\mathrm{Def}_E(S_1)$ and set:
\begin{align}
\alpha_{\vec{0}}=d_i&\qquad \alpha_{\vec{e}_i}={x^{\ast}_{j,i}}\in\mathrm{Hom}_A^1(L^{\bullet},L^{\bullet}),
\end{align}
 where the $\vec{e}_i$ are the canonical basis vectors of $\mathbbm{R}^d$. From the construction of $\hat{H}$ it follows that $E_1$ is represented by the lifting $\{L^{\bullet}\otimes_k S_1,d(S_1)\}$ of $\{L^{\bullet},d\}$, where
\begin{align}
d(S_1)_i|_{L_1\otimes 1}=\sum_{\vm\in\bar{B}_1}\alpha_{\vm}\otimes u^{\vm}
\end{align}
This means that we have deformed our matrix factorization with the odd cohomology elements at linear order. Up to order $u^1$ the new differential satisfies the matrix factorization condition but there are terms remaining at order $u^2$, which may be cancelled by adding further terms to the differential, unless they are in the even cohomology, i.e. they correspond to obstructions. What we have to do in this case has been explained above. The $\alpha_{\vm}$ are identified as the defining system of Massey products. Taking into account the obstructions we end up with the following proposition \cite{siqvelandPHD,Siqveland1}:
\begin{prop}
Given a defining system $\{\alpha_{\vm}\}$ for the $k$--th Massey products $\langle x^{\ast};\vn\rangle$ with $\vn\in B'_{2+k}$ then $\langle x^{\ast}; \vn\rangle$ is represented by
\begin{align}
y(\vn)&=\sum_{|\vm|\leq 2+k}\sum_{\stackrel{\vm_1+\vm_2=\vm}{\vm_i\in B_{2+k-1}}}\beta'_{\vm,\vn}\alpha_{\vm_1}\circ\alpha_{\vm_2}
\end{align}
One can then compute explicitly the polynomials
\begin{align}
f_j^{2+k}&=\sum_{l=0}^{k}\sum_{\vn\in B'_{2+k-1}}y_j\left(\langle x^{\ast};\vn\rangle\right)u^{\vn}, \qquad j=1,\ldots,r,
\end{align}
which induce the identities (\ref{eq-urel}) and (\ref{eq-betadef}), such that for every $\vm\in B_{2+k}$ we pick an $\alpha_{\vm}\in\mathrm{Hom}_A^1(L^{\bullet},L^{\bullet})$ such that
\begin{align}
\label{eq-newalpha}
d\:\alpha_{\vm}&=-\beta_{\vm}=-\sum_{\vn\in B'_{2+k}}\beta_{\vn,\vm}y(\vn).
\end{align}
Then the family $\{\alpha_{\vm}\}_{\vm\in\bar{B}_{2+k}}$ is a defining system for the Massey products $\langle x^{\ast};\vn\rangle$ with $\vn\in B'_{2+k+1}$. 
\end{prop}
The deformed differential is then:
\begin{align}
d(S)&=\sum_{m\in \bar{B}}\alpha_{\vm}u^{\vm},\qquad \bar{B}=\bigcup_{i}{B}_i.
\end{align}

Equation (\ref{eq-newalpha}) calculates the deformations of the matrix factorization at higher order. It follows directly from requiring that (\ref{eq-betadef}) is satisfied which in turn is linked to the condition that $o=0$. Note that the choice of $\alpha_{\vm}$ is not unique, which is related to the fact that $\mathcal{W}_{eff}$ and the underlying $A_{\infty}$--structure are defined only up to field redefinitions.\\
We now have all the ingredients to calculate the $\hat{H}\cong k[[u]]/(\bar{f}_1,\ldots,\bar{f}_r)$ order by order for matrix factorizations. However, the formalism does not guarantee that (homogeneous linear combinations of) the $\bar{f}_i$ can be integrated to give $\mathcal{W}_{eff}$. The arguments of \cite{Herbst:2004zm} imply that this is possible and we give explicit examples for minimal models in chapter \ref{chap-minmod}.

\subsection{Summary and Algorithm}
\label{subsec-sumalg}
Let us now summarize the general idea and the essential results of the mathematical construction of \cite{siqvelandPHD,Siqveland1}, discussed in the previous section. Here, we will focus on matrix factorizations. \\
Consider a matrix factorization $Q$ with $Q^2=W$ and calculate the open string spectrum:
\begin{align}
\psi_i\in H^{odd}\qquad \phi_i\in H^{even}&\qquad \mathrm{dim}H^{even}=\mathrm{dim}H^{odd}=N
\end{align}
We now want to calculate the most general non--linear deformation of this matrix factorization, taking into account only deformations with odd states. We make the following ansatz:
\begin{align}
\label{eq-qdefans}
Q_{def}&=Q+\sum_{\vec{m}\in\bar{B}}\alpha_{\vec{m}}u^{\vec{m}}
\end{align}
Here, $\vec{m}$ is a multi index: $u^{\vec{m}}=u_1^{m_1}u_2^{m_2}\ldots u_N^{m_N}$ and we define $|\vec{m}|=\sum_{i=1}^N m_i$. $\bar{B}$ describes the allowed set of vectors $\vm$.
$u_1,\ldots, u_N$ are deformation parameters associated to $\psi_1,\ldots,\psi_n$ and  $\alpha_{\vec{m}}$ are matrices to be determined. At the order $|\vec{m}|=1$ (linear deformations) they are given by the odd cohomology elements:
\begin{align}
\alpha_{(1,0,\ldots,0)}=\psi_1\quad \alpha_{(0,1,\ldots,0)}=\psi_2\quad\ldots\quad \alpha_{(0,\ldots,0,1)}=\psi_{N} 
\end{align}
Now we impose the matrix factorization condition on $Q_{def}$:
\begin{align}
\label{eq-qdeffact1}
Q_{def}^2&\stackrel{!}{=}W\cdot\mathbbm{1}+\sum_{i=1}^N{f'}_i(u)\phi_i\\
\label{eq-qdeffact2}
&\sim Q^2+\sum_{\vec{m}}[Q,\alpha_{\vec{m}}]u^{\vec{m}}+\sum_{\vec{m}_1+\vec{m}_2=\vec{m}}\underbrace{\alpha_{\vec{m}_1}\alpha_{\vec{m}_2}}_{y(\vec{m})}u^{\vec{m}}
\end{align}
Note that imposing $Q_{def}^2=W\cdot\mathbbm{1}$ does not work and we must employ the more general condition in (\ref{eq-qdeffact1}). Obviously, the matrix factorization condition only holds if we demand that ${f'}_i(u)=0$. The relations ${f'}_i(u)$ are equivalent to the vanishing relations $f_i(u)$ of the polynomial ring of deformations $k[u_i]/(f_i(u_i))$. At the same time these relations determine the critical locus of the effective superpotential.\\
In the second line of the above equation we naively inserted the ansatz (\ref{eq-qdefans}). $y(\vm)$ is called 'matric Massey product' \cite{siqvelandPHD,Siqveland1}. Equation (\ref{eq-qdeffact2}) is actually only correct up to order $|\vm|=2$. At higher orders the definition of the Massey products gets modified due to the presence of the $f_i(u)$, as we will show below. The 'method of computing formal moduli' of \cite{siqvelandPHD,Siqveland1} provides an algorithm to calculate the $f_i(u)$ and the $\alpha_{\vm}$ explicitly for all orders in $\vm$.\\
Let us first look at the lowest orders, where (\ref{eq-qdeffact2}) is correct. At linear order $|\vm|=1$ in the deformation parameters, the second term in (\ref{eq-qdeffact2}) is zero, since the $\alpha_{\vm}$ are the fermionic cohomology elements and the second term becomes the physical state condition. The first Massey product $y(\vm)$ appears at order $|\vm|=2$. We can calculate this product explicitly, since all the $\alpha_{\vm}$ at order $|\vm|=1$ are known. In the analogy introduced in section \ref{sec-darkroom}, calculating these products amounts to making steps in the dark room. $y(\vm)$ can take the following values:
\begin{itemize}
\item $y(\vm)\notin H^{even}$. In this case we can find an $\alpha_{m}$ with $|\vm|=2$ such that
\begin{align}
[Q,\alpha_{\vec{m}}]\equiv\beta_{\vm}&=-y(\vec{m}).
\end{align}
Thus, the second and the third term in (\ref{eq-qdeffact2}) cancel at order $|\vm|=2$ and we produced new $\alpha_{\vm}$'s and thus can calculate Massey products at higher order. In the dark room analogy the corresponding situation is that we have made a step in a certain direction and have not hit an obstacle, which implies that we can make further steps in this direction.
\item $y(\vm)\in H^{even}$, i.e. $y(\vm)=c\:\phi_k$, where $c$ is some number. Clearly, this cannot be cancelled by a term $[Q,\alpha_{\vec{m}}]$ since the $\phi_i$ are by definition not exact. Thus, we have encountered an {\em obstruction}. This amounts to hitting a wall in the dark--room analogy and we have to make sure that we do not take further steps in this direction. The obstructions are encoded in the polynomial $f_k(u)$ associated to $\phi_k$ in the following way:
\begin{align}
f_k=c u^{\vm}
\end{align}
\end{itemize}
We can now continue this algorithm at higher orders in $|\vm|$. There, however, some subtleties arise due to the presence of the $f_i(u)$. They impose relations among the $\vm$, which have to be incorporated into the algorithm. One has to introduce various bases for allowed vectors $\vm$. Furthermore the definition of the higher order Massey products has to be modified as compared to the naive definition of (\ref{eq-qdeffact2}). The deformation theory construction of \cite{siqvelandPHD,Siqveland1} yields the following results:\\
For a vector $\vn\in B'_{i+1}$, $i>0$ ($B'_{i+1}$ to be defined momentarily) the Massey product $y(\vn)$ is given by:
\begin{align}
\label{eq-massey}
y(\vn)\equiv\langle\psi;\vn\rangle&=\sum_{|\vm|\leq i+1}\sum_{\stackrel{\vm_1+\vm_2=\vm}{\vm_i\in B_{i}}}\beta'_{\vm,\vn}\alpha_{\vm_1}\circ\alpha_{\vm_2},
\end{align}
where the composition '$\circ$' is the matrix product. The coefficients $\beta'_{\vm,\vn}$ can be determined from the unique relation
\begin{align}
\label{eq-masseycoeff1}
u^{\vn}&=\sum_{\vm\in\bar{B}'_{i+1}}\beta'_{\vn,\vm}u^{\vm}+\sum_j\beta'_{\vn,j}f_j
\end{align}
for each $\vn\in\mathbbm{N}^N$ with $|\vn|\leq i+1$. If the Massey product is $y(\vn)=c\cdot\phi_k$ then we get a contribution to the $k$--th polynomial $f_k(u)$:
\begin{align}
\label{eq-masseyobs}
f^{i+1}_j=f_j^{i}+\sum_{\vn\in B'_{i+1}}c\cdot u^{\vn},
\end{align}
where the exponent gives the order in $u$.\\
The $\alpha_{\vm}$ are defined as follows. For each vector $\vm$ in a basis $B_{i+1}$ we can find a matrix $\alpha_{\vm}$  such that:
\begin{align}
\label{eq-masseydef}
[Q,\alpha_{\vm}]&=-\beta_{\vm}=-\sum_{l=0}^{i+1-2}\sum_{\vn\in B'_{2+l}}\beta_{\vn,\vm}y(\vn),
\end{align}
where the coefficients $\beta_{\vn,\vm}$ are given by the unique relation
\begin{align}
\label{eq-masseycoeff2}
u^{\vn}&=\sum_{\vm\in\bar{B}_{i+1}}\beta_{\vn,\vm}u^{\vm}
\end{align}
The various bases $B$, $\bar{B}$, $B'$, $\bar{B}'$ are defined recursively. One starts by setting $\bar{B}_1=\{\vn\in\mathbbm{N}^{N}|\:|\vn|\leq1\}$ and $B_1=\{\vn\in\mathbbm{N}^N|\:|\vn|=1\}$. For $i>1$ $B'_{i+1}$ is then defined as a basis for $m^{i+1}/(m^{i+2}+m^{i+1}\cap m(f_1,\ldots,f_N))$, where $m=(u_1,\ldots,u_N)$ defines the maximal ideal. The elements $\{u^{\vn}\}_{\vn\in B'_{i+1}}$ can be chosen such that $u^{\vn}=u_k\cdot u^{\vm}$ for some $\vm\in B_{\bar{B}_i}$ and $u_k\in\{u_1,\ldots ,u_N\}$. One defines $\bar{B}'_{i+1}=\bar{B}_i\cup B'_{i+1}$. Finally, $B_{i+1}$ is a basis for $(m^{i+1}+(f_1^i,\ldots,f_N^i))/(m^{i+2}+(f_1^{i+1},\ldots,f_N^{i+1}))$ such that $B_{i+1}\subseteq B'_{i+1}$. We set $\bar{B}_{i+1}=\bar{B}_i\cup B_{i+1}$.\\\\
With these definitions it is now possible to calculate the critical locus $f_i(u)$ of the effective superpotential along with the deformed matrix factorization $Q^{def}$. The algorithm, which we will refer to as the 'Massey product algorithm', looks as follows \cite{Siqveland1}:
\begin{itemize}
\item Choose a matrix factorization $Q$ and calculate the open string spectrum, where $\psi_i\in H^{odd}$ and $\phi_{i}\in H^{even}$, where $i=1,\ldots,N$.
\item Set $\alpha_{\vec{e}_i}=\psi_i$, where $\vec{e}_i$ are the canonical basis vectors of $\mathbbm{R}^N$. Furthermore associate a deformation parameter $u_k$ to every $\psi_k$.
\item For each $i\geq 0$ perform the following steps:
\begin{itemize}
\item Calculate the bases $B'_{i+1}$ and $\bar{B}'_{i+1}$.
\item Determine the coefficients $\beta'_{\vm,\vn}$ from the relations (\ref{eq-masseycoeff1}).
\item Calculate the Massey products $y(\vn)$ defined in (\ref{eq-massey}).
\item Determine $f_j^{i+1}$ using (\ref{eq-masseyobs}).
\item Pick bases $B_{i+1}$ and $\bar{B}_{i+1}$.
\item Calculate the coefficients $\beta_{\vm,\vn}$ from the relations (\ref{eq-masseycoeff2}).
\item Pick suitable $\alpha_{\vm}$ according to (\ref{eq-masseydef}).
\end{itemize}
\item If the algorithm terminates at a given order, integrate (homogeneous linear combinations of) the $f_i$ in order to obtain $\mathcal{W}_{eff}$.
\item Calculate the deformed matrix factorization:
\begin{align}
\label{eq-masseyqdef}
Q_{def}&=Q+\sum_{\vm\in B}\alpha_{\vm}u^{\vm}, \qquad \vm\in\bigcup_i B_i
\end{align}
\end{itemize}
We conclude this section with a few comments. First, note that $f_i=0$ corresponds to (\ref{eq-obstrformal}) and (\ref{eq-masseyqdef}) is (\ref{eq-qdefformal}). Thus, we have indeed succeeded in finding a way to explicitly calculate the formal expressions introduced in section \ref{sec-obstr}, without explicitly knowing the operators $U$ and $\Pi$.\\
Furthermore note that the choice of $\alpha_{\vm}$ is ambiguous. Taking different $\alpha_{\vm}$ also results in different $f_i$. The effective superpotentials obtained from this different choices are related via field redefinitions of the $u_i$, where field redefinition means in this case that every $u_k$ can be replaced by a homogeneous polynomial in terms of the $u_i$. This freedom reflects the reparameterization freedom one has in the underlying $A_{\infty}$--structure.\\
For matrix factorizations in general Landau--Ginzburg models we cannot expect that the algorithm terminates due to the presence of marginal deformations. For the minimal models the algorithm terminates at a given order and the $f_i(u)$ are homogeneous polynomials which can be integrated to give $\mathcal{W}_{eff}$. \\
Furthermore note that some of the $f_i(u)$ may remain zero throughout the calculation. The deformed polynomial ring of deformations is then defined as $k[u_1,\ldots,u_N]/(f_1,\ldots,f_r)$, where $r\leq N$. \\
A detailed discussion of examples can be found in \cite{siqvelandPHD,Siqveland1,Knapp:2006rd} and in chapter \ref{chap-minmod}.

\section{The Effective Superpotential from Consistency Constraints}
\label{sec-constr-effsupo}
In this section we discuss the consistency constraints for open topological strings which were derived in \cite{Herbst:2004jp}. 
\subsection{Topological Amplitudes and Selection Rules}
Let us first define a disk amplitude with an arbitrary number of bulk and boundary insertions\footnote{For the sake of readability we denote here bosonic and fermionic operators by $\psi$ and bulk operators by $\phi$ in contrast to previous conventions.}:
\begin{align}
\label{eq-bdef}
B_{a_0\ldots a_m;i_1\ldots i_n}:&=(-1)^{\tilde{a}_1+\ldots\tilde{a}_{m-1}}\left\langle\psi_{a_0}\psi_{a_1}P\int\psi^{(1)}_{a_2}\ldots\int\psi^{(1)}_{a_{m-1}}\psi_{a_m}\int\phi^{(2)}_{i_1}\ldots \int\phi^{(2)}_{i_n} \right\rangle \nonumber \\
&=-\left\langle\phi_{i_1}\psi_{a_0}P\int\psi^{(1)}_{a_1}\ldots\int\psi^{(1)}_{a_m}\int\phi_{i_2}^{(2)}\ldots\int\phi_{i_n}^{(2)}\right\rangle,
\end{align}
where 
\begin{align}
\int\phi_i^{(2)}&\equiv\int\phi_i^{(1,1)}=\int_{D_2}[G,[\bar{G},\phi_i]]dz\:d\bar{z}
\end{align}
are the bulk descendants, with $D_2$ the disk and $G$ the twisted fermionic current, and
\begin{align}
\int\psi_a^{(1)}&=\int_{\tau_l}^{\tau_r}[G,\psi_a]d\tau
\end{align}
are the boundary descendants, where the integral runs, from a suitably chosen position $\tau_l$ to the left of the operator to a position $\tau_r$ to its right, along the boundary of the disk. The boundary integrals in (\ref{eq-bdef}) have to be path ordered, and $P$ denotes the path ordering operator. Furthermore one can fix the positions of three boundary insertions or one bulk and one boundary insertion due to the $PSL(2,\mathbbm{R})$--invariance of the disk. We also introduced a {\em suspended grade} $\tilde{a}$ of the boundary fields $\psi_a$:
\begin{align}
\tilde{a}:=|\psi_a|+1,
\end{align}
where $|\psi_a|$ is the $\mathbbm{Z}_2$--degree of the boundary field. The grading of the correlators is thus determined by the grades of the descendants rather than the fields themselves.\\
The equality of the two correlators given in (\ref{eq-bdef}) can be derived from Ward identities \cite{Herbst:2004jp}.\\
Similar to the closed string case, we can introduce a topological metric on the boundary:
\begin{align}
\label{eq-bdrymet}
\omega_{ab}&=\langle\psi_a\psi_b\rangle=(-1)^{\tilde{a}}B_{0ab}=(-1)^{\tilde{\omega}}(-1)^{\tilde{a}\tilde{b}}\omega_{ba},
\end{align}
where the ``0'' stands for the insertion of the unit operator.
It can be used to raise and lower indices:
\begin{align}
B^a_{\phantom{a}a_1\ldots a_m}:=\omega^{ab}B_{ba_1\ldots a_m}.
\end{align}
The correlators (\ref{eq-bdef}) are cyclic in the boundary insertions:
\begin{align}
\label{eq-bdrycyc}
B_{a_0\ldots a_m;i_1\ldots i_m}&=(-1)^{\tilde{a}_m(\tilde{a}_0+\ldots+\tilde{a}_{m-1})}B_{a_ma_0\ldots a_{m-1};i_1\ldots i_m}
\end{align}
Furthermore they are symmetric under permutations of the bulk indices.\\
It turns out to be convenient to define
\begin{align}
B_{a_0a_1}&=B_{a_0}=B_i=0.
\end{align}
A correlator (\ref{eq-bdef}) satisfies the following selection rules, generalizing the selection rules for correlators on the sphere \cite{Dijkgraaf:1990dj}:
\begin{itemize}
\item R--Charge selection rule. \\
Due to the anomalous $U(1)$--current, we have a non--vanishing background charge, which has to be saturated by the correlator. The $R$--charges of the integrated insertions are:
\begin{align}
\label{eq-rcharge}
q^I_{\psi}&=q_{\psi}-1 \nonumber\\
q^I_{\phi}&=q_{\phi}-2
\end{align}
\item The correlators must have the same suspended degree as the boundary metric:
\begin{align}
\label{eq-z2sel}
B_{a_0a_1\ldots a_m;i_1\ldots i_n}&=0\qquad \mathrm{unless}\;\;\tilde{a}_0+\ldots\tilde{a}_m=\tilde{\omega}
\end{align}
\item Insertions of the unit operator are only allowed if there are no integrated insertions.
\begin{align}
\label{eq-unitsel}
B_{0a_1\ldots a_m;i_1\ldots i_n}&=0\qquad \mathrm{for}\;\;m\geq3\;\;\mathrm{or}\;\;n\geq 1
\end{align}
\end{itemize}
Since the amplitudes are completely symmetric with respect to the bulk insertions we can introduce generating functions for the bulk perturbations which satisfy the following property:
\begin{align}
B_{a_0\ldots a_m;i_1\ldots i_n}&=\partial_{i_1}\ldots\partial_{i_n}\mathcal{F}_{a_0\ldots a_m}(t)|_{t=0}
\end{align}
For $m\geq 2$ the generating functions are given by:
\begin{align}
\label{eq-fdef}
\mathcal{F}_{a_0\ldots a_m}&=(-1)^{\tilde{a}_1+\ldots+\tilde{a}_{m-1}}\langle\psi_{a_0}\psi_{a_1}P\int\psi^{(1)}_{a_2}\ldots\int\psi^{(1)}_{a_{m-1}}\psi_{a_m}e^{\sum_pt_p\int\phi^{(2)}_p}\rangle\nonumber\\
&=(-1)^{\tilde{a}_1+\ldots+\tilde{a}_{m-1}}\sum_{N_0\ldots N_{h_c-1}=0}^{\infty}\prod_{p=0}^{h_c-1}\frac{t_p^{N_p}}{N_p!}\langle\psi_{a_0}\psi_{a_1}P\int\psi^{(1)}_{a_2}\ldots\int\psi^{(1)}_{a_{m-1}}\psi_{a_m}\left[\int\phi^{(2)}_p \right]^{N_p}\rangle\nonumber\\
\end{align}
For $m=0$ and $m=1$ we define $\mathcal{F}_a(t)$ and $\mathcal{F}_{ab}(t)$ through:
\begin{align}
\partial_i\mathcal{F}_a(t)&=-\langle\phi_i\psi_ae^{\sum_pt_p\int\phi^{(2)}_p}\rangle\\
\partial_i\mathcal{F}_{ab}(t)&=-\langle\phi_i\psi_aP\int\psi^{(1)}_be^{\sum_pt_p\int\phi^{(2)}_p} \rangle
\end{align}
\subsection{$A_{\infty}$--relations}
\begin{figure}
\begin{center}
\includegraphics{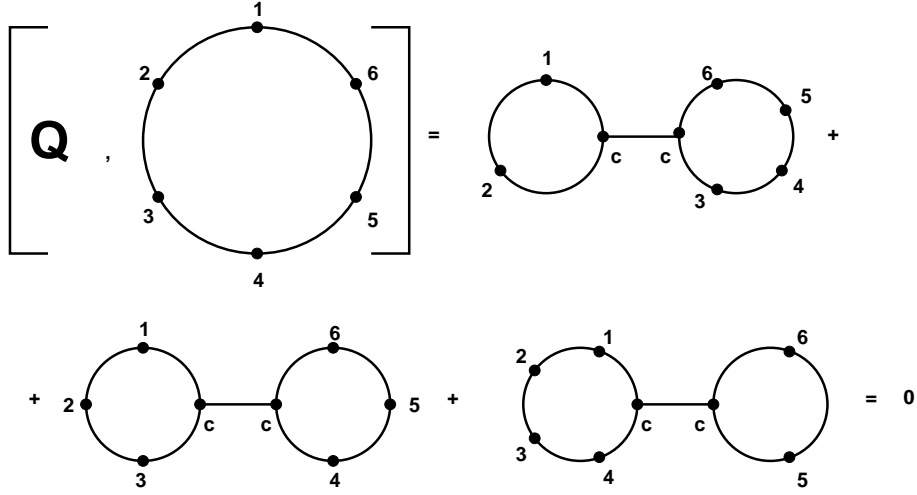}
\end{center}
\caption{$A_{\infty}$--relations arising from the Ward identity of the BRST operator $Q$.}\label{fig-ainf}
\end{figure}

The $A_{\infty}$--relations can be derived from the Ward identity of the BRST operator $Q$:
\begin{align}
\left\langle [Q, B_{a_0\ldots a_m;i_1\ldots i_m}]\right\rangle&=0
\end{align}
The condition infers a series of algebraic constraints which encode how an amplitude with a certain number of insertions factorizes into amplitudes with less insertions. This factorization is due to contact terms which arise when boundary insertions approach each other. A contact term arises whenever $Q$ hits an operator $G$ of an integrated insertion. Fig. \ref{fig-ainf} gives a schematic picture of what happens.
In terms of the generating functions $\mathcal{F}_{a_0\ldots a_m}$ the $A_{\infty}$--constraints look as follows \cite{Herbst:2004jp}:
\begin{align}
\label{eq-ainf1}
\sum_{\stackrel{k,j=0}{k\leq j}}^m(-1)^{\tilde{a}_1+\ldots\tilde{a}_k}\mathcal{F}^{a_0}_{\phantom{a_0}a_1\ldots a_kca_{j+1}\ldots a_m}(t)\mathcal{F}^{c}_{\phantom{c}a_{k+1}\ldots a_j}(t)&=0,
\end{align}
where the indices are raised and lowered with the topological metric (\ref{eq-bdrymet}).
In terms of the correlators $B_{a_0\ldots a_m;i_0\ldots i_n}$ this is:
\begin{align}
\label{eq-ainf2}
\sum_{I\subseteq I_{0,n}}\sum_{\stackrel{k,j=0}{k\leq j}}^m(-1)^{\tilde{a}_1+\ldots\tilde{a}_k}B^{a_0}_{\phantom{a_0}a_1\ldots a_kca_{j+1}\ldots a_m;I_{0;n}\backslash I}B^c_{\phantom{c}a_{k+1}\ldots a_j;I}&=0,
\end{align}
where $I_{p,q}=\{i_p,i_{p+1},\ldots,i_q\}$. This is called a {\em weak} $A_{\infty}$ algebra. Without bulk insertions the equations reduce to:
\begin{align}
\label{eq-ainfmin}
\sum_{\stackrel{k,l=2}{k-m+2<l\leq k}}(-1)^{\tilde{a}_1+\ldots+\tilde{a}_{l-2}}B^{a_0}_{\phantom{a_0}\ldots a_{l-2}ca_{k+1}a_m}B^c_{\phantom{c}a_{l-1}\ldots a_k}&=0\quad m\geq 2
\end{align}
This structure is called a {\em minimal} $A_{\infty}$--algebra. \\
There is a direct relation between (\ref{eq-ainfmin}) and the relations (\ref{eq-algainf}) in terms of the higher products $m_n$ defined in section \ref{sec-obstr}. These products are related to the amplitudes via \cite{Herbst:2004jp}:
\begin{align}
m_n(\psi_{a_1}\ldots\psi_{a_m})&=B^{a_0}_{\phantom{a_0}a_1\ldots a_m}\psi_{a_0}
\end{align}
A generalization for amplitudes with bulk insertions is also possible \cite{Herbst:2004jp}.\\
Note that these relations are valid for both boundary preserving and boundary changing insertions. In the boundary changing sector one introduces additional indices for every boundary insertion, which label the boundary: $\psi_a^{AB}$. Only 'cyclically closed' correlators $\langle \psi_{a_1}^{A_1 A_2}\psi_{a_2}^{A_2 A_3} \ldots \psi_{a_m}^{A_mA_1}\rangle$ are well--defined. 
\subsection{Bulk--Boundary Crossing Constraint and the Cardy Condition}
The $A_{\infty}$--relations alone do not uniquely determine the values of the correlators. There are two more constraints which can be derived for disk amplitudes. \\
The bulk--boundary crossing constraint encodes the factorization of an amplitude when bulk operators approach each other or a bulk operator approaches the boundary. It is a generalization of the bulk--boundary crossing constraint of topological field theory \cite{Lazaroiu:2000rk}. It looks as follows:
\begin{align}
\label{eq-bbcr}
\partial_i\partial_j\partial_k\mathcal{F}\eta^{kl}\partial_l\mathcal{F}_{a_0\ldots a_m}&=
\end{align}
\begin{align}
&=\sum_{0\leq m_1\leq\ldots m_4\leq m}(-1)^{\tilde{a}_{m_1}+\ldots+\tilde{a}_{m_3}}\mathcal{F}_{a_0\ldots a_{m_1}ba_{m_2+1}\ldots a_{m_3}ca_{m_4+1}\ldots a_m}\partial_i\mathcal{F}^b_{\phantom{b}a_{m_1+1}\ldots a_{m_2}}\partial_j\mathcal{F}^c_{\phantom{c}a_{m_3+1}\ldots a_{m_4}}\nonumber
\end{align}
Here, $\mathcal{F}(t)$ is the bulk WDVV potential and $\eta^{kl}$ is the inverse of the topological bulk metric $\eta_{kl}=\langle \phi_0\phi_k\phi_l\rangle$. \footnote{Note that only those equations are consistent where the operator product defining $c_{ijk}=\partial_i\partial_j\partial_k\mathcal{F}(t)$, $\phi_i(t)\phi_j(t)=c_{ij}^k\phi_k(t)\:\mathrm{mod}\:dW$, does not contain any exact terms.}\\
In \cite{Knapp:2006rd}, we found a discrepancy between the effective superpotential calculated using the bulk--boundary crossing constraint and the results obtained from deformation theory methods. We will discuss this in detail in chapter \ref{chap-minmod}. In \cite{Keller:2006tf} it was shown explicitly for the $E_6$ minimal model that this constraint yields wrong values for certain amplitudes. This implies that the constraint as given in (\ref{eq-bbcr}) only works for the $A$--minimal models.\\\\
The minimal models of the $A$--series satisfy an additional constraint, the Cardy Condition \cite{Herbst:2004jp}:
\begin{align}
\label{eq-cardy}
\partial_i\mathcal{F}_{a_0\ldots a_n}\eta^{ij}\partial_j\mathcal{F}_{b_0\ldots b_m}&=
\end{align}
\begin{align}
=\sum_{\stackrel{{0\leq n_1\leq n_2\leq n}}{0\leq m_1\leq m_2\leq m}}(-1)^{(\tilde{c}_1+\tilde{a}_0)(\tilde{c}_2+\tilde{b}_0)+\tilde{c_1}+\tilde{c_2}}\omega^{c_1d_1}\omega^{c_2d_2}\mathcal{F}_{a_0\ldots a_{n_1}d_1b_{m_1+1}\ldots m_2c_2a_{n_2+1}\ldots a_n}\mathcal{F}_{b_0\ldots b_{m_1}c_1a_{n_1+1}\ldots a_{n_2}d_2b_{m_2+1}\ldots b_m}&\nonumber
\end{align}
This is the generalization of the Cardy condition of topological field theory, which is derived from the factorization of the cylinder amplitude into an open or a closed string channel. This constraint is not topological in the sense that it is not independent of the metric. This is why one cannot expect it to hold for topological correlators. However, it turns out \cite{Herbst:2004jp,Knapp:2006rd} that it can be imposed on the $A$--minimal models. In combination with the other constraints (\ref{eq-ainf1}) and (\ref{eq-bbcr}) it uniquely determines the values of all the correlators. The fact that this is possible also reflects the non--topological nature of the Cardy condition: in the topological theory we expect that the amplitudes are fixed only up to field redefinitions or ``$A_{\infty}$--morphisms'' and that, in contrast to $N=2$ theories, we need to specify the K\"ahler potential in order to completely fix the values of the amplitudes.

\subsection{A recipe for calculating $\mathcal{W}_{eff}$ using Consistency Constraints}
To conclude our discussion of consistency constraints we list the necessary steps to calculate the effective superpotential by solving these constraints, focusing on matrix factorizations and B--type topological Landau--Ginzburg models. Concrete examples will be given in chapter \ref{chap-minmod} and appendix \ref{app-amod}. Furthermore, in appendix \ref{app-mathematica} we will describe how the following recipe can be implemented in Mathematica. To calculate $\mathcal{W}_{eff}$ one proceeds as follows:
\begin{itemize}
\item Pick a matrix factorization and calculate the open string spectrum. Furthermore determine the $R$--charges of the states. 
\item Calculate the topological metric, the boundary three--point function and the bulk--boundary two--point function using the Kapustin--Li formula (\ref{eq-kapustin1}),(\ref{eq-kapustin2}).
\item Determine which are the allowed correlators by applying the R--charge selection rule and (\ref{eq-z2sel}), (\ref{eq-unitsel}).
\item If bulk deformations are turned on, calculate the topological bulk metric and the WDVV potential $\mathcal{F}(t)$.
\item Set up the constraint equations (\ref{eq-ainf1}), (\ref{eq-bbcr}) and (\ref{eq-cardy}), using the correlators without integrated insertions as input, and solve for the remaining correlators.
\item Sum up the correlators as described in (\ref{eq-effsupo-corr}) and (\ref{eq-symcorr}). For the $A$--minimal models the result does not contain any free parameters. 
\end{itemize}

\chapter{D--Branes in Topological Minimal Models}
\label{chap-minmod}
This chapter is devoted to the calculation of the effective superpotential by applying the methods described in chapter \ref{chap-effsupo} for explicit examples. We will focus on the Landau--Ginzburg formulation of topological minimal models. These models do not describe realistic Calabi--Yau compactifications and are thus not of direct phenomenological interest. Nevertheless the minimal models are valuable as toy models, since they have many features of full--fledged Calabi--Yau compactifications but avoid many technical difficulties. In particular, minimal models do not have any marginal deformations, which entails the we can work in a purely algebraic setup, thus avoiding the difficulties arising due to moduli. One can build more interesting models for string compactifications out of minimal models by tensoring and orbifolding them, which gives rise to the well--known, and well--understood, Gepner models. Studying matrix factorizations in minimal models may thus provide useful information on Calabi--Yau compactifications and may help to get a better understanding of various aspects of matrix factorizations. \\
In section \ref{sec-minmod} we discuss the Landau--Ginzburg description of topological minimal models. Section \ref{sec-a3} deals with the simplest non--trivial case, the $A_3$ minimal model. Although minimal models of type $A$ have been discussed already in the matrix factorization literature \cite{Brunner:2003dc,Herbst:2004jp,Herbst:2004zm} we will go through this example in great detail since it will enable us to describe the methods to calculate the effective superpotential in a most explicit way. In section \ref{sec-e6} we discuss the $E_6$ minimal model which is just slightly more complicated than the models of type $A$. The aim is to test if the methods described in chapter \ref{chap-effsupo} also work for more general cases. A surprising result is that the consistency constraints discussed in section \ref{sec-constr-effsupo} only work for minimal models of type $A$. In particular, the Cardy constraint (\ref{eq-cardy}) yields inconsistent equations. We find that the deformation theory methods we discussed in section \ref{sec-def-effsupo} can be applied for all types of minimal models. Using these methods we are able to compute $\mathcal{W}_{eff}$ for many examples. We state these results in section \ref{sec-results}. These results actually pass a non--trivial check because it turns out that the {\em effective} superpotentials of the minimal models are in fact the {\em Landau--Ginzburg} superpotentials of certain Kazama--Suzuki type coset models, which were discussed in \cite{Lerche:1991re,Eguchi:1996nh,Lerche:1996an,Eguchi:2001fm}.\\\\
This chapter summarizes the results of \cite{Knapp:2006rd}.
\section{Minimal Models}
\label{sec-minmod}
Minimal models are exactly solvable conformal field theories. They are particularly simple since they only have massive, i.e. relevant, deformations. This means that these models do not have ``true'' moduli. Topological minimal models have a Landau--Ginzburg description. The corresponding superpotential can be classified in terms of the simply--laced Lie groups. This is known as the ADE classification, see e.g. \cite{Zuber:2000ia} for some background. The superpotentials are given by the following polynomials:
\begin{align}
\label{eq-adesupo}
A_k:\qquad & W=x^{k+2}\nonumber \\
D_k:\qquad & W=x^{k-1}+xy^2-z^2\nonumber\\
E_6:\qquad & W=x^3+y^4-z^2 \nonumber \\
E_7:\qquad & W=x^3+xy^3-z^2\nonumber\\
E_8:\qquad & W=x^3+y^5-z^2
\end{align}
The background charge for these models can be expressed in terms of the dual Coxeter number $k$:
\begin{align}
\label{eq-bgcharge}
\hat{c}&=\frac{k-2}{k}
\end{align}
Obviously, this is always smaller than $1$. The bulk theory of these models was discussed in \cite{Dijkgraaf:1990dj}. Computing the chiral rings of these models, immediately shows that there are no marginal deformations, i.e. there are no ring elements which have the same $R$--charge as the superpotential itself. In particular, this means that, given the superpotential has charge $2$,  the $R$--charges of the bulk fields are smaller than 2 and the $R$--charges of the boundary fields are smaller than $1$. By (\ref{eq-rcharge}) this implies that all integrated insertions have negative $R$--charge. As a consequence there is a maximal number of insertions a correlator can have. From this it follows that the effective superpotential is a polynomial. \\
We can actually relate the degree of $\mathcal{W}_{eff}$ to the dual Coxeter number of the Lie group associated to the minimal model. This is done as follows.  By the selection rules, the allowed correlators must have an $R$--charge which is equal to the background charge $\hat{c}$. We can now determine the correlator with the maximal number of insertions. This disk correlator will have three unintegrated insertions of the boundary field with the highest charge $\hat{c}$, we will call it $\Psi_{\hat{c}}$, and a certain number of integrated operators, which have negative charge. To get the maximum number of insertions one must use only insertions of $\int\Psi_{\hat{c}}$, which has charge ${\hat{c}}-1$, which is the least negative. From the charge selection rule we can now calculate the the number $x$ of integrated insertions of this top element:
\begin{align}
3\cdot \hat{c}+x\cdot(\hat{c}-1)&\stackrel{!}{=}\hat{c}
\end{align}
This yields $x=\frac{2\hat{c}}{1-\hat{c}}$. Now take into account that for the minimal models the background charge is related to the Coxeter number $k$ via (\ref{eq-bgcharge}). Inserting this, we find that the number of integrated insertions of the top--element is $k-2$. Adding the three unintegrated operators, one finds that the top--correlator is a $k+1$--point function. Looking more closely, one also finds that the selection rule for the $\mathbb{Z}_2$--charge is satisfied and that this correlator will not vanish and contribute to the superpotential. The deformation parameter $u$ associated to $\Psi_{\hat{c}}$ has weight one and we will get a term $u_1^{k+1}$ in the superpotential. Since the effective superpotential is a homogeneous polynomial, we conclude:\\\\
{\it The effective superpotentials for the ADE minimal models always have degree $k+1$, where $k$ is the Coxeter number.}\\\\
The first step in calculation $\mathcal{W}_{eff}$ is to find the matrix factorizations of the Landau--Ginzburg superpotential. The classification of matrix factorizations is in general an unsolved problem. For the minimal models, however, this problem has been solved some time ago. In \cite{greuel-knoerrer} the authors give an algorithm to explicitly calculate all the inequivalent matrix factorizations of the superpotentials (\ref{eq-adesupo}). It turns out that there is one matrix factorization for every node in the Dynkin diagram of the associated Lie group. A list of these matrix factorizations can be found for instance in \cite{yoshino,Kajiura:2005yu}.
\section{The $A_3$ Minimal Model -- A Toy Example}
\label{sec-a3}
A simple non--trivial minimal model is of type $A$ at level $k=3$. The superpotential is\footnote{The factor $\frac{1}{5}$ is convention.}:
\begin{align}
W=\frac{x^5}{5}
\end{align}
The associated Lie group has Coxeter number $5$, the $U(1)$ background charge is $\hat{c}=\frac{3}{5}$.\\
The field $x$ has $R$--charge $\frac{2}{5}$. The closed string spectrum is determined by the chiral ring $\frac{\mathbbm{C}[x]}{\partial W}$, which is generated by the monomials $\{1,x,x^2,x^3\}$. 

\subsection{Matrix Factorizations and Open String Spectrum}
As we already mentioned in section \ref{sec-tensor} all the matrix factorizations of a superpotential $W=x^n$ are generated by rank one factorizations $W=EJ=x^rx^{n-r}$ for $r=1,\ldots,n-1$. In our case this yields four matrix factorizations which split into two brane--antibrane pairs. We choose, for definiteness, the following matrix factorization:
\begin{align}
Q&=\left(\begin{array}{cc}
0&x^2\\
\frac{x^3}{5}&0
\end{array}\right)
\end{align}
We can calculate the $R$--matrix by solving (\ref{eq-rmat}), which yields:
\begin{align}
R&=\left(\begin{array}{cc}
\frac{1}{10}&0\\
0&-\frac{1}{10}
\end{array}\right)
\end{align}
The open string states are determined by the cohomology of $Q$. There are two bosons $\phi_i$ and two fermions $\psi_i$:
\begin{align}
\phi_1^{(0)}=\left(\begin{array}{cc}
1&0\\
0&1
\end{array}\right)&\qquad
\phi_2^{(\frac{2}{5})}=\left(\begin{array}{cc}
x&0\\
0&x
\end{array}\right)
\end{align}
\begin{align}
\psi_1^{(\frac{1}{5})}=\left(\begin{array}{cc}
0&1\\
-\frac{x}{5}&0
\end{array}\right)&\qquad
\psi_2^{(\frac{3}{5})}=\left(\begin{array}{cc}
0&x\\
-\frac{x^2}{5}&0
\end{array}\right)
\end{align}
The superscripts are the $R$--charges of the states. 

\subsection{The Effective Superpotential via Deformation Theory}
We now compute the effective superpotential for this model by applying the Massey product algorithm we have discussed in section \ref{sec-def-effsupo}. At the linear level, we assign matrices $\alpha_{\vec{e}_i}$ to every fermionic cohomology element:
\begin{align}
\alpha_{(1,0)}=\psi_1\qquad \alpha_{(0,1)}=\psi_2,
\end{align}
where we have defined a basis $B_1$ with elements in $\vm\in\mathbbm{N}^2$:
\begin{align}
|\vec{m}|=1:\qquad& B_1=\{(1,0),(0,1)\}
\end{align}
The first Massey products arise at order $|\vm|=2$, where we define the basis $B'_2$:
\begin{align}
|\vec{m}|=2:\qquad& B'_2=\{(2,0),(1,1),(0,2)\}
\end{align}
Thus, there are three Massey products at order $2$:
\begin{align}
\label{eq-massey2a3}
y(2,0)&=\alpha_{(1,0)}\cdot\alpha_{(1,0)}=-\frac{1}{5}\phi_2\qquad \nonumber\\
y(0,2)&=\alpha_{(0,1)}\cdot\alpha_{(0,1)}=-\frac{1}{5}\left(\begin{array}{cc}x^3&0\\0&x^3\end{array}\right)\nonumber\\
y(1,1)&=\alpha_{(1,0)}\cdot\alpha_{(0,1)}+\alpha_{(0,1)}\cdot\alpha_{(1,0)}=-\frac{2}{5}\left(\begin{array}{cc}x^2&0\\0&x^2\end{array}\right)
\end{align}
The first product, $y(2,0)$ is proportional to a bosonic open string state. Since this cannot be cancelled at higher order, this is an obstruction and contributes to the vanishing relations $f_i$:
\begin{align}
\label{eq-crit2a3}
 f_i:\qquad f_1^{(2)}=0\qquad& f_2^{(2)}=-\frac{1}{5}u_2^2
\end{align}
Here we have associated deformation parameters $u_2$ and $u_1$ to $\psi_1$ and $\psi_2$, respectively. The indices of the parameters $u_i$ correspond to their homogeneous weights, which is related as follows to the $R$--charge of the states:
\begin{align}
w_{u_i}&=\frac{1}{2}k(1-q_{\psi_i}),
\end{align} 
where the Coxeter number $k=5$, in the case at hand. Note that with this labelling we have\footnote{This may seem unnatural at the moment but regarding the effective superpotential it is the natural choice.} $u^{\vm}=u_2^{m_1}u_1^{m_2}$. The last two Massey products in (\ref{eq-massey2a3}) are not proportional to bosonic cohomology elements and can thus be canceled by introducing new $\alpha_{\vm}$ at order $2$, which satisfy $\{Q,\alpha\}=-y$:
One possible choice is:
\begin{align}
\alpha_{(1,1)}=\frac{2}{5}\left(\begin{array}{cc}0&0\\1&0\end{array}\right)\qquad &
\alpha_{(0,2)}=\frac{1}{5}\left(\begin{array}{cc}0&0\\x&0\end{array}\right)
\end{align}
For the basis $B_2$ we thus get:
\begin{align}
B_2=\{\cancel{(2,0)},(1,1),(0,2)\}
\end{align}
At order $3$, we introduce a basis $B'_3$:
\begin{align}
|\vec{m}|=3:\qquad& B'_3=\{(1,2),(0,3),\cancel{(2,1)},\cancel{(3,0)}\}
\end{align}
We must not use the vectors $(2,1)$ and $(3,0)$ in $B'_3$ because we have to take into account the relation $f_2^{(2)}$ at this order, which reduces the basis. So, there are only two Massey products to compute:
\begin{align}
y(1,2)&=\alpha_{(1,0)}\cdot\alpha_{(0,2)}+\alpha_{(0,2)}\cdot\alpha_{(1,0)}+\alpha_{(1,1)}\cdot\alpha_{(0,1)}+\alpha_{(0,1)}\cdot\alpha_{(1,1)}=\frac{3}{5}\phi_2\nonumber \\
y(0,3)&=\alpha_{(0,1)}\cdot\alpha_{(0,2)}+\alpha_{(0,2)}\cdot\alpha_{(0,1)}=\frac{1}{5}\left(\begin{array}{cc}x^2&0\\0&x^2\end{array}\right)
\end{align}
The first product contributes to the vanishing relations, which now read:
\begin{align}
\label{eq-crit3a3}
 f_i:\qquad f_1^{(3)}=0\qquad &f_2^{(3)}=-\frac{1}{5}u_2^2+\frac{3}{5}u_2u_1^2
\end{align}
We get one new matrix $\alpha_{(0,3)}$:
\begin{align}
\alpha_{(0,3)}&=\frac{1}{5}\left(\begin{array}{cc}0&0\\1&0\end{array}\right)
\end{align}
Since there are no new vanishing relations at this order, we have $B_3=B'_3$.\\
At order 4, the basis is given by:
\begin{align}
 |\vec{m}|=4:\qquad& B'_4=\{(1,3),(0,4)\}
\end{align}
The two Massey products are:
\begin{align}
y(1,3)=&\alpha_{(1,1)}\cdot\alpha_{(0,2)}+\alpha_{(0,2)}\cdot\alpha_{(1,1)}+\overbrace{3(\alpha_{(1,1)}\alpha_{(1,0)}+\alpha_{(1,0)}\cdot\alpha_{(1,1)})}^{\textrm{from }f_2:\:u_1^2u_2=3u_1u_2^3}\nonumber\\
&+\alpha_{(1,0)}\cdot\alpha_{(0,3)}+\alpha_{(0,3)}\cdot\alpha_{(1,0)} =\phi_1\nonumber\\
y(0,4)&=\alpha_{(0,3)}\cdot\alpha_{(0,1)}+\alpha_{(0,1)}\cdot\alpha_{(0,3)}+\alpha_{(0,2)}\cdot\alpha_{(0,2)}=-\frac{1}{5}\phi_2
\end{align}
In $y(1,3)$ we also get a contribution from the vector $(2,1)$, which is not an element of the basis $\bar{B}'=\bigcup_i B'_i$ due to the relation $f_2^{(3)}$ in (\ref{eq-crit3a3}). Both of these products contribute to the vanishing relations, there are no new $\alpha$--matrices.
\begin{align}
f_i:\qquad f_1^{(4)}=u_2u_1^3\qquad& f_2^{(4)}=-\frac{1}{5}u_2^2+\frac{3}{5}u_2u_1^2-\frac{1}{5}u_1^4
\end{align}
Having now a non--zero relation $f_1$, the basis $B_4$ reads:
\begin{align}
B_4=\{\cancel{(1,3)},(0,4)\}
\end{align}
Thus, at order $5$ the basis $B'_5$ only contains one element:
\begin{align}
|\vec{m}|=5\qquad B'_5=\{\cancel{(1,4)},(0,5)\}
\end{align}
The only Massey product at this order is:
\begin{align}
y(0,5)=\alpha_{(0,2)}\cdot\alpha_{(0,3)}+\alpha_{(0,3)}\cdot\alpha_{(0,2)}-(\alpha_{(1,1)}\cdot\alpha_{(1,0)}+\alpha_{(1,0)}\cdot\alpha_{(1,1)})=-\frac{2}{5}\phi_1
\end{align}
The term in parentheses again arises due to the vanishing relations. These look as follows at the order 5:
\begin{align}
\label{eq-crit5a3}
f_i:\qquad f_1^{(5)}=u_2u_1^3-\frac{2}{5}u_1^5 \qquad &f_2^{(5)}=-\frac{1}{5}u_2^2+\frac{3}{5}u_2u_1^2-\frac{1}{5}u_1^4 
\end{align}
At order $6$, we have:
\begin{align}
|\vec{m}|=6:\qquad& B'_6=\{(0,6)\}
\end{align}
The Massey product is:
\begin{align}
y(0,6)&=\alpha_{(0,3)}\cdot\alpha_{(0,3)}+\frac{2}{5}\left(\alpha_{(1,1)}\cdot\alpha_{(0,3)}+\alpha_{(0,3)}\cdot\alpha_{(1,1)}\right)=0
\end{align}
Thus, there are no non--vanishing Massey products at order $6$. Given the matrices $\alpha_i$ at our disposal it is not possible to define any Massey products at higher orders. Thus, the algorithm terminates at order $6$.\\
We can now write down the deformed matrix factorization:
\begin{align}
\label{eq-qdefa3}
Q_{def}&=Q+\sum_{\vec{m}}\alpha_{\vec{m}}u^{\vec{m}}=Q+\alpha_{(1,0)}u_2+\alpha_{(0,1)}u_1+\alpha_{(1,1)}u_2u_1+\alpha_{(0,2)}u_1^2+\alpha_{(0,3)}u_1^3\nonumber\\
&=\left(\begin{array}{cc}
0&x^2+u_2+u_1x\\
\frac{x^3}{5}-\frac{1}{5}u_2x+\frac{2}{5}u_1x^2+\frac{2}{5}u_2u_1+\frac{1}{5}u_1^2x-\frac{u_2^3}{5}&0
\end{array}\right)
\end{align}
One easily checks that this satisfies the matrix factorization condition (\ref{eq-qdeffact1}). The critical locus of $\mathcal{W}_{eff}$ is defined by the vanishing relations $f_i\equiv f_i^{(5)}$ in (\ref{eq-crit5a3}). One can read off that $f_1$ is a homogeneous polynomial of degree $5$ and that $f_2$ is homogeneous of degree $4$. From the arguments presented in section \ref{sec-minmod} we know that $\mathcal{W}_{eff}$ is homogeneous of degree $6$. This would imply that we could obtain the effective superpotential by integrating $f_1$ with respect to $u_1$ and $f_2$ with respect to $u_2$. It is, however, not that simple. What one has to do is to integrate homogeneous linear combinations of the $f_i$. We therefore define:
\begin{align}
r_1=\int \mathrm{d}u_2\:c_1\:f_2\qquad &r_2=\int\mathrm{d}u_1\:c_2\:f_1+c_3\:u_1\:f_2,
\end{align}
where $c_1,\ldots,c_3$ are parameters, two of which can be determined by solving the integrability condition $\partial_{u_1}\partial_{u_2}r_1=\partial_{u_1}\partial_{u_2} r_2$. This determines $\mathcal{W}_{eff}$ up to an overall numerical constant. Giving this constant a particular value, the result is:
\begin{align}
\label{eq-weff1a3}
\mathcal{W}_{eff}&=\frac{1}{30}u_1^6+\frac{3}{10}u_1^2u_2^2-\frac{1}{5}u_1^4u_2-\frac{1}{15}u_2^3
\end{align}
Integrating the polynomials $f_i'$ which arise in $Q_{def}^2=W\mathbbm{1}+f_i'\phi_i$ yields an equivalent superpotential. In this example we have: 
\begin{align}
f_1'&=\frac{2}{5}u_1u_2^2-\frac{1}{5}u_1^3u_2 \qquad f_2'=-\frac{1}{5}u_2^2+\frac{3}{5}u_2u_1^2-\frac{1}{5}u_1^4
\end{align}
Note that the choice of $\alpha$--matrices is not unique. A different choice entails a different deformed $Q$--operator and different vanishing relations, which also lead to a different expression for the effective superpotential. These superpotentials are equivalent up to field redefinitions. In general, one has a lot of freedom in choosing the $\alpha$--matrices, but our example is so simple that there is actually only one more choice. The second set of possible $\alpha_{\vm}$ is:
\begin{align}
\tilde{\alpha}_{(1,0)}=\psi_1\qquad &\tilde{\alpha}_{(0,1)}=\psi_2
\end{align}
\begin{align}
\tilde{\alpha}_{(1,1)}=\frac{2}{5}\left(\begin{array}{cc}
0&0\\
1&0
\end{array}\right)\qquad
\tilde{\alpha}_{(0,2)}=\left(\begin{array}{cc}
0&1\\
0&0
\end{array}\right)\qquad
\tilde{\alpha}_{(0,3)}=\frac{1}{5}\left(\begin{array}{cc}
0&0\\
1&0
\end{array}\right)
\end{align}
The vanishing relations are:
\begin{align}
\tilde{f}_1=u_2u_1^3+\frac{3}{5}u_1^5\qquad &\tilde{f}_2=-\frac{1}{5}u_2^2+\frac{1}{5}u_2u_1^2+\frac{1}{5}u_1^5
\end{align}
This can be integrated to give the following effective superpotential:
\begin{align}
\label{eq-weff2a3}
\tilde{\mathcal{W}}_{eff}&=\frac{1}{15}u_1^6+\frac{1}{10}u_1^2u_2^2+\frac{1}{5}u_1^4u_2-\frac{1}{15}u_2^3
\end{align}
One gets from (\ref{eq-weff1a3}) to (\ref{eq-weff2a3}) by the following non--linear field redefinition:
\begin{align}
u_1\longrightarrow u_1\qquad & u_2\longrightarrow u_2+u_1^2
\end{align}

\subsection{Massey Products and Bulk Deformations}
We now show how to calculate the effective superpotential with bulk deformations by extending the Massey product algorithm. \\
The most generic deformed Landau--Ginzburg superpotential for the $A_3$--model looks as follows:
\begin{align}
\label{eq-lgbulka3}
W(x;s)&=\frac{x^5}{5}+s_5+s_4x+s_3x^2+s_2x^3,
\end{align}
where $s_2,\ldots,s_5$ are dimensionful deformation parameters, whose indices indicate their homogeneous weights.\\
Given a deformed matrix factorization $Q_{def}(x;u)$ we can now include bulk deformations by making the following ansatz:
\begin{align}
Q_{def}(x;u;s)&=Q_{def}(x;u)+\sum_{\vm}\bar{\alpha}_{\vm}s^{\vm}
\end{align}
We then demand that this squares to the deformed Landau--Ginzburg potential:
\begin{align}
Q_{def}(x;u;s)^2&\stackrel{!}{=}W(x;s)+\bar{f}_1\phi_1+\bar{f}_2\phi_2
\end{align}
Starting with $Q^2_{def}(x;u)=W(x)+f_1'\phi_1+f_2'\phi_2$, one sees that those terms in the bulk deformed Landau--Ginzburg potential (\ref{eq-lgbulka3}), that are proportional to $\phi_i\mathbbm{1}$ can be produced by adding terms to $f'_i$, which yields the $\bar{f}_i$ at leading order in the $s_i$. All the other terms of $W(x;s)$ can only be built by constructing matrices $\bar{\alpha}_{\vm}$ corresponding to the deformations of $Q_{def}(x;u)$ at linear order in $s$. This is the starting point of an algorithm which is analogous to the Massey product algorithm for the boundary deformations.\\
We now consider the $A_3$--example. Our starting point is (\ref{eq-qdefa3}). In (\ref{eq-lgbulka3}) the terms proportional to $s_5$ and $s_4$ lie in the even cohomology of the boundary and enter into the definition of the vanishing relations $\bar{f}_i$. The other terms must be produced by the $\bar{\alpha}_{\vm}$ at order $|\vm|=1$. We thus have to find $\alpha$--matrices such that:
\begin{align}
\{Q_{def}(x;u),\bar{\alpha}_{(1,0)}\}&=x^2\cdot\mathbbm{1}\qquad\mathrm{mod}\:H^{even}\nonumber\\
\{Q_{def}(x;u),\bar{\alpha}_{(0,1)}\}&=x^3\cdot\mathbbm{1}\qquad\mathrm{mod}\:H^{even}
\end{align}
A convenient choice is:
\begin{align}
\bar{\alpha}_{(1,0)}=\left(\begin{array}{cc}
0&0\\
1&0
\end{array}\right)\qquad &
\bar{\alpha}_{(1,0)}=\left(\begin{array}{cc}
0&0\\
x-u_1&0
\end{array}\right)
\end{align}
Now we got to order $|\vm|=2$. It turns out that all the Massey products at this order are $0$. So, the algorithm terminates after the first step. The bulk--deformed matrix factorization is:
\begin{align}
Q_{def}(x;s;u)&=Q_{def}(x;u)+s_3\bar{\alpha}_{(1,0)}+s_2\bar{\alpha}_{(0,1)}\nonumber\\
&=\left(\begin{array}{cc}
0&x^2+u_2+u_1x\\
\frac{x^3}{5}-\frac{1}{5}u_2x+\frac{2}{5}u_1x^2+\frac{2}{5}u_2u_1+\frac{1}{5}u_1^2x-\frac{u_2^3}{5}+s_3+s_2(x-u_2)&0
\end{array}\right)
\end{align}
Squaring this matrix and subtracting $W(x;s)$ one can read off the vanishing relations $\bar{f}_i$:
\begin{align}
\bar{f}_1&=\phantom{-}\frac{2}{5}u_1u_2^2-\frac{1}{5}u_1^3u_2-s_5+s_3u_2-s_2u_1u_2 \nonumber \\
\bar{f}_2&=-\frac{1}{5}u_2^2+\frac{3}{5}u_2u_1^2-\frac{1}{5}u_1^4-s_4+s_2u_2+s_3u_1-s_2u_1^2
\end{align}
Integrating these polynomials, one gets:
\begin{align}
\label{eq-weffa3bulk}
\mathcal{W}_{eff}(u;s)=&\frac{1}{30}u_1^6+\frac{3}{10}u_1^2u_2^2-\frac{1}{5}u_1^4u_2-\frac{1}{15}u_2^3+s_2\left(\frac{u_2^2}{2}-u_1^2u_2+\frac{u_1^4}{4}\right)\nonumber\\
&+s_3\left(u_1u_2-\frac{u_1^3}{3}\right)+s_4\left(\frac{u_1^2}{2}-u_2\right)-s_5u_1
\end{align}
Setting the bulk parameters $s_i$ to 0 we recover (\ref{eq-weff1a3}).

\subsection{The Effective Superpotential via Consistency Constraints}
In \cite{Herbst:2004jp} the effective superpotential was calculated for the $A_3$ minimal model by solving the consistency constraints discussed in section \ref{sec-constr-effsupo}. The constraint equations have a unique solution, which reads:
\begin{align}
\label{eq-weffa3constr}
\mathcal{W}_{eff}(u;t)=&-\frac{1}{30}u_1^6-\frac{3}{10}u_1^2u_2^2-\frac{1}{5}u_1^4u_2-\frac{1}{15}u_2^3+t_2\left(\frac{u_2^2}{2}+u_1^2u_2+\frac{u_1^4}{4}\right)\nonumber \\
&+t_3\left(u_1u_2\frac{u_1^3}{3}\right)+(t_4-t_2^2)\left(\frac{u_1^2}{2}+u_2\right)+(t_5-t_2t_3)u_1
\end{align}
The result is given in terms of the 'flat' coordinates $t_i$. These define the special coordinate system where the bulk correlators are given as partial derivatives of the WDVV potential $\mathcal{F}(t)$ with respect to the $t_i$. They are related to the generic deformation parameters $s_i$ by the following transformations \cite{Dijkgraaf:1990dj}:
\begin{align}
s_2\longrightarrow -t_2\quad s_3\longrightarrow -t_3\quad s_4\longrightarrow -t_4+t_2^2\quad s_5\longrightarrow -t_5+t_2t_3
\end{align}
Upon a field redefinition $u_1\rightarrow -u_1,u_2\rightarrow -u_2$ (\ref{eq-weffa3constr}) and (\ref{eq-weffa3bulk}) are identified.\\\\
We give further results of effective superpotentials for the minimal models of type $A$ in appendix \ref{app-amod}

\section{The $E_6$ Minimal Model}
\label{sec-e6}
In this section we discuss the calculation of the effective superpotential for the $E_6$ minimal model. The Landau--Ginzburg superpotential is:
\begin{align}
\label{eq-we6lg}
W&=x^3+y^4+\varepsilon z^2,
\end{align}
where $\varepsilon=\pm 1$. Introducing this parameter is just for convenience since we can always choose the matrix factorization to be real. The choice of sign has no influence on the dimensions and charges of the spectrum or the form of the superpotential.\\
The Coxeter number $k$ of the $E_6$--model is 12, the background charge is $\hat{c}=\frac{5}{6}$.\\
This model is only a slight complication as compared to the minimal models of type $A$. We now have three variables instead of one. This brings us closer to Gepner models. Since the $E_6$ model is still a minimal model one could expect that all the methods to calculate the effective superpotential which work for the minimal models of type $A$ also apply for the $E_6$ case. However, it turns out to be worthwhile to consider 'yet another example': The surprising result we get is that the consistency constraints for the correlators fail for this model. 

\subsection{Matrix Factorizations and Open String Spectrum}
Matrix factorizations and cohomology for the $E_6$--model have already been discussed in \cite{Kajiura:2005yu}. There are six matrix factorizations, one for each node in the Dynkin diagram of $E_6$. The following matrices satisfy $W\mathbbm{1}=E\cdot J$:
\begin{align}
E_1=J_2=\left(
\begin{array}{cc}
-y^2-\sqrt{-\varepsilon}z&x\\
x^2&y^2-\sqrt{-\varepsilon}z
\end{array}
\right)&\quad
E_2=J_1=\left(
\begin{array}{cc}
-y^2+\sqrt{-\varepsilon}z&x\\
x^2&y^2+\sqrt{-\varepsilon}z
\end{array}
\right)
\end{align}
\begin{align}
\label{eq-m3}
E_3=J_3&=\left(
\begin{array}{cccc}
-\sqrt{\varepsilon}z&0&x^2&y^3\\
0&-\sqrt{\varepsilon}z&y&-x\\
x&y^3&\sqrt{\varepsilon}z&0\\
y&-x^2&0&-\sqrt{\varepsilon}z
\end{array}
\right)
\end{align}
\begin{align}
\label{eq-m4}
E_4=J_5=\left(
\begin{array}{cccc}
-y^2+\sqrt{-\varepsilon}&0&xy&x\\
-xy&y^2+\sqrt{-\varepsilon}&x^2&0\\
0&x&\sqrt{-\varepsilon}z&y\\
x^2&-xy&y^3&\sqrt{-\varepsilon}z
\end{array}
\right)
\end{align}
\begin{align}
E_5=J_4&=\left(
\begin{array}{cccc}
-y^2-\sqrt{-\varepsilon}z&0&xy&x\\
-xy&y^2-\sqrt{-\varepsilon}z&x^2&0\\
0&x&-\sqrt{-\varepsilon}z&y\\
x^2&-xy&y^3&-\sqrt{-\varepsilon}z
\end{array}
\right)
\end{align}
\begin{align}
E_6=\left(
\begin{array}{cccccc}
-\sqrt{-\varepsilon}z&-y^2&xy&0&x^2&0\\
-y^2&-\sqrt{-\varepsilon}z&0&0&0&x\\
0&0&-\sqrt{-\varepsilon}z&-x&0&y\\
0&xy&-x^2&-\sqrt{-\varepsilon}z&y^3&0\\
x&0&0&y&-\sqrt{-\varepsilon}z&0\\
0&x^2&y^3&0&xy^2&-\sqrt{-\varepsilon}z
\end{array}
\right)
\end{align}
\begin{align}
J_6=\left(
\begin{array}{cccccc}
\sqrt{-\varepsilon}z&-y^2&xy&0&x^2&0\\
-y^2&\sqrt{-\varepsilon}z&0&0&0&x\\
0&0&\sqrt{-\varepsilon}z&-x&0&y\\
0&xy&-x^2&\sqrt{-\varepsilon}z&y^3&0\\
x&0&0&y&\sqrt{-\varepsilon}z&0\\
0&x^2&y^3&0&xy^2&\sqrt{-\varepsilon}z
\end{array}
\right)
\end{align}
 We summarize the data of the boundary preserving spectrum in table \ref{tab-threevarcoh}.\\
\begin{table}[h]
\begin{center}
\begin{tabular}{|c|c|c|c|}
\hline
\vrule width 0pt height 12pt depth 6ptFactorization& Rank & Spectrum bosonic & Spectrum fermionic\\ \hline
\vrule width 0pt height 12pt depth 6pt $M_1$ & $2$ & $0\:6$ & $4\:10$\\ \hline
\vrule width 0pt height 12pt depth 6pt $M_2$ & $2$ & $0\:6$ & $4\:10$\\ \hline
\vrule width 0pt height 12pt depth 6pt $M_3$ & $4$ & $0\:4\:6\:10$ & $0\:4\:6\:10$\\ \hline
\vrule width 0pt height 12pt depth 6pt $M_4$ & $4$ & $0\:2\:4\:6^2\:8$ & $2\:4^2\:6\:8\:10$\\ \hline
\vrule width 0pt height 12pt depth 6pt $M_5$ & $4$ & $0\:2\:4\:6^2\:8$ & $2\:4^2\:6\:8\:10$\\ \hline 
\vrule width 0pt height 12pt depth 6pt $M_6$ & $6$ & $0\:2^2\:4^3\:6^3\:8^2\:10$ & $0\:2^2\:4^3\:6^3\:8^2\:10$ \\
\hline
\end{tabular}
\end{center}
\caption{Boundary preserving spectrum of the $E_6$ model.}\label{tab-threevarcoh}
\end{table}
We labelled the matrix factorizations by $M_i$, the second column indicates the ranks of the matrices. The last two columns give the even and the odd spectrum. The numbers correspond to the $R$--charges, multiplied by the Coxeter number $k=12$ of $E_6$, and the exponents give the multiplicities. Note that there are six possible values of the charges, $q_{\psi}\in\{0,2,4,6,8,10\}$. To fermionic states with these charges we will associate fermionic deformation parameters $u_i$ with weights $w_{u_i}=\frac{1}{2}(12-q_{\psi_i})\in\{6,5,4,3,2,1\}$.  We observe that, concerning the spectrum, there are two types of matrix factorizations. The factorizations $M_1,M_2$ and $M_4,M_5$ have the same spectra, respectively. These factorizations form pairs of branes and antibranes. $M_3$ and $M_6$ belong to a second class of D--branes. The even spectrum is identical to the odd spectrum, these branes are ``self--dual'' ---  the brane is its own antibrane \cite{Kapustin:2003rc}. \\
We note that the highest charge, which is equal to the background charge, is always in the fermionic sector, whereas the charge $0$ state is always in the bosonic sector. Serre duality is realized in the following way: For every boson $\phi$ there is a fermion $\psi$ such that $q_{\phi}+q_{\psi}=\hat{c}$.

\subsection{The Effective Superpotential via Deformation Theory}
We now go on to calculate $\mathcal{W}_{eff}$ for the simplest $E_6$ matrix factorization, $M_1$. Choosing the parameter $\varepsilon=1$ in (\ref{eq-we6lg}), we have:
\begin{align}
E_1=\left(
\begin{array}{cc}
-y^2-z&x\\
x^2&y^2-z
\end{array}
\right)\qquad&
J_1=\left(
\begin{array}{cc}
-y^2+z&x\\
x^2&y^2+z
\end{array}
\right)
\end{align}
The open string spectrum defined by the operator $Q=\left(\begin{array}{cc}0&E\\J&0\end{array}\right)$ is easily found:
\begin{align}
\label{eq-e6bos}
\phi_1^{(0)}=\left(\begin{array}{cccc}
1&0&0&0\\
0&1&0&0\\
0&0&1&0\\
0&0&0&1
\end{array}
\right)\qquad&
\phi_2^{(\frac{1}{2})}=\left(\begin{array}{cccc}
y&0&0&0\\
0&y&0&0\\
0&0&y&0\\
0&0&0&y
\end{array}
\right)
\end{align}
\begin{align}
\label{eq-e6fer}
\psi_3^{(\frac{1}{3})}=\left(\begin{array}{cccc}
0&0&0&1\\
0&0&-x&0\\
0&1&0&0\\
-x&0&0&0
\end{array}
\right)\qquad&
\psi_4^{(\frac{5}{6})}=\left(\begin{array}{cccc}
0&0&0&y\\
0&0&-xy&0\\
0&y&0&0\\
-xy&0&0&0
\end{array}
\right)
\end{align}
We have two fermionic operators which we can use to deform the matrix factorization. We associate a parameter $u_1$ to $\psi_4$ and $u_4$ to $\psi_3$. Then we set:
\begin{align}
\alpha_{(1,0)}=\psi_3\qquad& \alpha_{(0,1)}=\psi_4\qquad B_1=\{(1,0),(0,1)\}
\end{align} 
The Massey products for order $2$ with $B'_2=\{(2,0),(1,1),(0,2)\}$ are:
\begin{align}
y_{(2,0)}&=\alpha_{(1,0)}\alpha_{(1,0)}=-x\mathbbm{1}\nonumber \\
y_{(0,2)}&=\alpha_{(0,1)}\alpha_{(0,1)}=-{xy^2}{\mathbbm{1}}\nonumber \\
y_{(1,1)}&=\alpha_{(1,0)}\alpha_{(0,1)}+\alpha_{(0,1)}\alpha_{(1,0)}=-2xy{\mathbbm{1}}
\end{align}
None of these products is in the even cohomology and thus we do not get a contribution to the vanishing relations $f_1,f_2$ of the ring of deformations $k[u_1,u_4]/(f_1,f_2)$. Therefore we have $B_2=B'_2$. The matrix factorization condition is satisfied at order $2$ if we choose the following $\alpha_{\vm}$ for $|\vm|=2$:
\begin{align}
\alpha_{(2,0)}=\left(\begin{array}{cccc}
0&0&0&0\\
0&0&1&0\\
0&0&0&0\\
1&0&0&0
\end{array}
\right)\quad
\alpha_{(1,1)}=\left(\begin{array}{cccc}
0&0&0&0\\
0&0&2y&0\\
0&0&0&0\\
2y&0&0&0
\end{array}
\right)\quad
\alpha_{(0,2)}=\left(\begin{array}{cccc}
0&0&0&0\\
0&0&y^2&0\\
0&0&0&0\\
y^2&0&0&0
\end{array}
\right)
\end{align}
At order $|\vm|=3$, the basis is $B'_3=\{(3,0),(2,1),(1,2),(0,3)\}$. For the four Massey products we find:
\begin{align}
y_{(3,0)}&=\alpha_{(2,0)}\alpha_{(1,0)}+\alpha_{(1,0)}\alpha_{(2,0)}=\phi_1 \nonumber \\
y_{(2,1)}&=\alpha_{(2,0)}\alpha_{(0,1)}+\alpha_{(0,1)}\alpha_{(2,0)}+\alpha_{(1,1)}\alpha_{(1,0)}+\alpha_{(1,0)}\alpha_{(1,1)}=3\phi_2 \nonumber \\
y_{(1,2)}&=\alpha_{(1,1)}\alpha_{(0,1)}+\alpha_{(0,1)}\alpha_{(1,1)}+\alpha_{(0,2)}\alpha_{(1,0)}+\alpha_{(1,0)}\alpha_{(0,2)}=3y^2\mathbbm{1} \nonumber \\
y_{(0,3)}&=\alpha_{(0,2)}\alpha_{(0,1)}+\alpha_{(0,1)}\alpha_{(0,2)}=y^3\mathbbm{1}
\end{align}
$y_{(3,0)}$ and $y_{(2,1)}$ are in the even cohomology and thus yield the following contributions to $f_1,f_2$:
\begin{align}
f_1^{(3)}=u_4^3\qquad& f_2^{(3)}=3u_1u_4^2
\end{align}
The basis $B_3$ is:
\begin{align}
B_3=\{\cancel{(3,0)},\cancel{(2,1)},(1,2),(0,3)\}
\end{align}
The remaining two Massey products give rise to two new alpha matrices:
\begin{align}
\alpha_{(1,2)}=\left(\begin{array}{cccc}
0&0&\frac{3}{2}&0\\
0&0&0&-\frac{3}{2}\\
\frac{3}{2}&0&0&0\\
0&-\frac{3}{2}&0&0
\end{array}
\right)\qquad&
\alpha_{(0,3)}=\left(\begin{array}{cccc}
0&0&\frac{y}{2}&0\\
0&0&0&-\frac{y}{2}\\
\frac{y}{2}&0&0&0\\
0&-\frac{y}{2}&0&0
\end{array}
\right)
\end{align}
The non--zero values of $f_1,f_2$ reduce the basis of vectors $\vm$ at order 4:
\begin{align}
B'_{4}&=\{\cancel{(4,0)},\cancel{(3,1)},\cancel{(2,2)},(1,3),(0,4)\}
\end{align}
The procedure continues completely analogous as compared to the $A_3$--case. The algorithm terminates after 13 steps. We get only one more $\alpha$--matrix at order $|\vm|=6$:
\begin{align}
\alpha_{(0,6)}=\left(\begin{array}{cccc}
0&0&\frac{1}{8}&0\\
0&0&0&-\frac{1}{8}\\
\frac{1}{8}&0&0&0\\
0&-\frac{1}{8}&0&0
\end{array}
\right)
\end{align}
The vanishing relations defining the critical locus are:
\begin{align}
\label{eq-crit1e6}
f_1&=u_4^3-\frac{3}{4}u_4u_1^8-\frac{5}{64}u_1^{12}\nonumber \\
f_2&=3u_4^2u_1+\frac{3}{2}u_4u_1^5+\frac{1}{8}u_1^9
\end{align}
The deformed matrix factorization is:
\begin{align}
\label{eq-e6qdef}
Q_{def}(x;u)=Q+u_4\alpha_{(1,0)}+u_1\alpha_{(0,1)}+u_4^2\alpha_{(2,0}+u_1^2\alpha_{(0,2)}+u_1u_4\alpha_{(1,1)}+u_4u_1^2\alpha_{(1,2)}+u_1^3\alpha_{(0,3)}+u_1^6\alpha_{(0,6)}
\end{align}
Explicitly, we have:
\begin{align}
E_{def}(x;u)&=\left(\begin{array}{cc}
-y^2-z+\frac{1}{2}yu_1^3+\frac{3}{2}u_4u_1^2+\frac{1}{8}u_1^6 &
x+u_4+yu_1\\
x^2- xu_4-xyu_1+2yu_4u_1+y^2u_1^2+u_4^2& y^2-z-\frac{1}{2}yu_1^3-\frac{3}{2}u_4u_1^2-\frac{1}{8}u_1^6 
\end{array}
\right)
\end{align}
\begin{align}
J_{def}(x;u)&=\left(\begin{array}{cc}
-y^2+z+\frac{1}{2}yu_1^3+\frac{3}{2}u_4u_1^2+\frac{1}{8}u_1^6 & x+\frac{1}{c_1}u_4+yu_1\\
 x^2-xu_4-xyu_1+yu_4u_1+y^2u_1^2+u_4^2&y^2+z-\frac{1}{2}yu_1^3-\frac{3}{2}u_4u_1^2-\frac{1}{8}u_1^6
\end{array}
\right)
\end{align}
The deformed BRST operator satisfies $Q_{def}^2=W\mathbbm{1}+f_1'\phi_1+f_2'\phi_2$, where the polynomials
\begin{align}
f_1'&=u_4^3+\frac{9}{4}u_1^4u_4^2+\frac{3}{8}u_1^8u_2+\frac{1}{64}u_1^{12}\nonumber \\
f_2'&=3u_4^2u_1+\frac{3}{2}u_4u_1^5+\frac{1}{8}u_1^9
\end{align}
are equivalent to (\ref{eq-crit1e6}). Integration of (\ref{eq-crit1e6}) yields the following expression for the effective superpotential:
\begin{align}
\label{eq-weff1e6}
\mathcal{W}_{eff}(u)&=u_4^3u_1+\frac{3}{4}u_4^2u_1^5+\frac{1}{8}u_4u_1^9+\frac{5}{832}u_1^{13}
\end{align}
The choice of matrices $\alpha_{\vm}$ is highly ambiguous. One example for an equivalent deformation of $Q$ is:
\begin{align}
\tilde{E}_{def}=\left(\begin{array}{cc}
-y^2-z-\frac{1}{2}xu_1^2+u_4u_1^2-\frac{1}{8}u_1^6 &x+u_4+yu_1-\frac{1}{4}u_1^4\\
x^2-xu_4-xyu_1+2yu_4u_1-\frac{1}{4}yu_1^5+u_4^2+\frac{3}{4}u_4u_1^4+\frac{1}{8}u_1^8 & y^2-z+xu_1^2-u_4u_1^2+\frac{1}{8}u_1^6
\end{array}
\right)
\end{align}
\begin{align}
\tilde{J}_{def}=\left(\begin{array}{cc}
-y^2+z-\frac{1}{2}xu_1^2+u_4u_1^2-\frac{1}{8}u_1^6 & x+yu_1+u_4-\frac{1}{4}u_1^4\\
x^2-xu_4-xyu_1+2yu_4u_1-\frac{1}{4}yu_1^5+u_4^2+\frac{3}{4}u_4u_1^3+\frac{1}{8}u_4^8 & y^2+z+\frac{1}{2}xu_1^2-u_4u_1^2+\frac{1}{8}u_1^6
\end{array}
\right)
\end{align}
Furthermore one finds:
\begin{align}
\tilde{f}_1&=u_4^3-\frac{9}{16}u_4u_1^8+\frac{5}{64}u_1^{12}\nonumber \\
\tilde{f}_2&=u_4^2u_1-\frac{1}{16}u_1^9
\end{align}
These polynomials can be integrated to give the following effective superpotential:
\begin{align}
\label{eq-weff2e6}
\tilde{\mathcal{W}}_{eff}(u)&=u_4^3u_1-\frac{1}{16}u_4u_1^9+\frac{5}{832}u_1^{13}
\end{align}
(\ref{eq-weff1e6}) and (\ref{eq-weff2e6}) are related by the field redefinition
\begin{align}
u_1\longrightarrow u_1 &\qquad u_4\longrightarrow u_4+u_1^4
\end{align}

\subsection{Bulk Deformations and Deformation Theory}
We now calculate the effective superpotential with bulk deformations using the extended Massey product algorithm. The bulk--deformed Landau--Ginzburg superpotential is:
\begin{align}
\label{eq-lgbulke6}
W(x;s)&=x^3+y^4-z^2+s_{12}+s_9u+s_8x+s_6y^2+s_5xy+s_2xy^2
\end{align}
As in the $A_3$--model we make an ansatz for the bulk--deformed $Q$--operator, such that:
\begin{align}
Q_{def}^2(x;u;s)&=\left(Q_{def}(x;u)+\sum_{\vm}\bar{\alpha}_{\vm}s^{\vm} \right)^2\stackrel{!}{=}W(x;s)+ \bar{f}_1\phi_1+\bar{f}_2\phi_2,
\end{align}
where we choose (\ref{eq-e6qdef}) for the boundary deformed $Q$--operator.\\
The terms in (\ref{eq-lgbulke6}) which are proportional to $s_9$ and $s_{12}$ lie in the even cohomology and can be absorbed into $\bar{f}_1$ and $\bar{f}_2$. Thus, for $|\vm|=1$, we look for block--offdiagonal matrices $\bar{\alpha}_{\vm}$ which satisfy:
\begin{align}
{[}Q_{def}(x;u),\bar{\alpha}_{(1,0,0,0)}{]}&=x\phantom{y^2}\qquad\mathrm{mod}\:H^{even}\nonumber  \\
{[}Q_{def}(x;u),\bar{\alpha}_{(0,1,0,0)}{]}&=y^2\phantom{x} \qquad\mathrm{mod}\:H^{even}\nonumber\\
{[}Q_{def}(x;u),\bar{\alpha}_{(0,0,1,0)}{]}&=xy\phantom{{}^2} \qquad\mathrm{mod}\:H^{even}\nonumber\\
{[}Q_{def}(x;u),\bar{\alpha}_{(0,0,0,1)}{]}&=xy^2 \qquad\mathrm{mod}\:H^{even}
\end{align}
One possible choice is:
\begin{align}
\bar{\alpha}_{(1,0,0,0)}=\left(\begin{array}{cccc}
0&0&0&0\\
0&0&1&0\\
0&0&0&0\\
1&0&0&0
\end{array}
\right)\quad
\bar{\alpha}_{(0,1,0,0)}=\left(\begin{array}{cccc}
0&0&-\frac{1}{2}&0\\
0&0&0&\frac{1}{2}\\
-\frac{1}{2}&0&0&0\\
0&\frac{1}{2}&0&0
\end{array}
\right)\quad
\bar{\alpha}_{(0,0,1,0)}=\left(\begin{array}{cccc}
0&0&\frac{u_1}{2}&0\\
0&0&y&-\frac{u_1}{2}\\
\frac{u_1}{2}&0&0&0\\
y&-\frac{u_1}{2}&0&0
\end{array}
\right)
\end{align}
\begin{align}
\bar{\alpha}_{(0,0,0,1)}&=\left(\begin{array}{cccc}
0&0&-\frac{x}{2}-\frac{u_1^4}{8}&-\frac{u_1^2}{4}\\
\frac{u_1}{2}&0&0&0\\
0&0&\frac{1}{4}xu_1^2+u_1^2u_4+\frac{u_1^6}{8}&\frac{x}{2}+\frac{u_1^4}{8}\\
-\frac{x}{2}-\frac{u_1^4}{8}&-\frac{u_1^2}{4}&0&0
\end{array}\right)
\end{align}
Computing the Massey products at order 2, one finds that there are three products which are neither in the even cohomology nor $0$. This leads to three more $\bar{\alpha}_{\vm}$ at order $|\vm|=2$:
\begin{align}
\bar{\alpha}_{(0,1,0,1)}=\left(\begin{array}{cccc}
0&0&0&0\\
0&0&-\frac{1}{2}&0\\
0&0&0&0\\
-\frac{1}{2}&0&0&0
\end{array}
\right)\qquad&
\bar{\alpha}_{(0,0,1,1)}=\left(\begin{array}{cccc}
0&0&0&0\\
0&0&\frac{u_1}{2}&0\\
0&0&0&0\\
\frac{u_1}{2}&0&0&0
\end{array}
\right)
\end{align}
\begin{align}
\bar{\alpha}_{(0,0,0,2)}=\left(\begin{array}{cc}
0&a\\
a&0
\end{array}\right)\qquad&
a=\left(\begin{array}{cc}
\frac{u_1^2}{4}&\frac{1}{12}\\
-\frac{x}{3}+\frac{5}{12}(yu_1+u_4)-\frac{u_1^4}{16}&-\frac{u_1^2}{4}
\end{array}\right)
\end{align}
We get two more $\bar{\alpha}_{\vm}$ at orders 3 and 4:
\begin{align}
\bar{\alpha}_{(0,0,0,3)}=\left(\begin{array}{cccc}
0&0&0&0\\
0&0&\frac{7u_1^2}{48}&0\\
0&0&0&0\\
\frac{7u_1^2}{48}&0&0&0
\end{array}\right)\qquad&
\bar{\alpha}_{(0,0,0,4)}=\left(\begin{array}{cccc}
0&0&0&0\\
0&0&\frac{1}{36}&0\\
0&0&0&0\\
\frac{1}{36}&0&0&0
\end{array}\right)
\end{align}
The algorithm terminates at order $8$. The deformed matrix factorization is:
\begin{align}
Q_{def}(x;u;s)=&Q_{def}(x;u)+s_8\bar{\alpha}_{{1,0,0,0}}+s_6\bar{\alpha}_{(0,1,0,0)}+s_5\bar{\alpha}_{(0,0,1,0)}+s_2\bar{\alpha}_{(0,0,0,1)}\nonumber \\
&+s_6s_2\bar{\alpha}_{(0,1,0,1)}+s_5s_2\bar{\alpha}_{(0,0,1,1)}+s_2^2\bar{\alpha}_{(0,0,0,2)}+s_2^3\bar{\alpha}_{(0,0,0,3)}+s_2^4\bar{\alpha}_{(0,0,0,4)}
\end{align}
Squaring this matrix, we can read off $\bar{f}_1,\bar{f}_2$, which can be integrated to the following effective superpotential:
\begin{align}
\label{eq-weffe6bulk}
\mathcal{W}_{eff}(u;s)=&u_4^3u_1+\frac{3}{4}u_4^2u_1^5+\frac{1}{8}u_4u_1^9+\frac{5}{832}u_1^{13} \nonumber \\
&+u_1\left(-s_{12}+\frac{1}{4}s_6^2+\frac{1}{12}s_2^2s_8-\frac{1}{24}s_2^3s_6+\frac{1}{432}s_2^6\right)+u_1^2\left(-\frac{1}{4}s_5s_6+\frac{1}{48}s_2^3s_5\right)\nonumber \\
&+u_1^3\left(\frac{1}{12}s_5^2-\frac{1}{12}s_2s_8+\frac{1}{24}s_2^2s_6+\frac{1}{576}s_2^5\right)+u_1^4\left(-\frac{1}{8}s_9+\frac{1}{24}s_2^2s_5 \right)\nonumber\\
&+u_4\left(-s_9+\frac{1}{12}s_2^2s_5 \right)+u_1^5\left(\frac{1}{10}s_8-\frac{1}{40}s_2s_6+\frac{1}{96}s_2^4\right)\nonumber\\
&+u_1u_4\left(s_4-\frac{1}{2}s_2s+\frac{1}{16}s_2^4\right)+\frac{1}{4}u_1^2u_4s_2s_5+u_1^7\left(-\frac{3}{56}s_6-\frac{1}{448}s_2^3 \right)\nonumber\\
&+u_1^3u_4\left(-\frac{1}{2}s_6+\frac{1}{24}s_2^3\right)+\frac{3}{64}u_1^8s_5+\frac{1}{2}u_1^4u_4s_5+\frac{1}{2}u_4^2s_5+\frac{1}{64}u_1^9s_2^2+\frac{1}{2}u_1u_4^2s_2^2\nonumber\\
&+\frac{3}{16}u_1^5u_4s_2^2-\frac{1}{352}u_1^{11}s_2+\frac{1}{4}u_1^3u_4^2s_2
\end{align}


\subsection{Combined Deformation Theory and CFT Constraints}
In order to check the correctness of the result (\ref{eq-weffe6bulk}) it would be useful to reproduce it by different methods. For the $A_3$ minimal model we could show that the results for the effective superpotential derived by deformation theory methods on the one hand and by solving consistency constraints on the other hand, agreed. \\
The relevant steps to set up and solve the consistency constraints for open topological strings were discussed in section \ref{sec-constr-effsupo}. A convenient approach to set up and solve the equations is to start with the $A_{\infty}$--relations (\ref{eq-ainf2}) and solve them for the correlators. Since these equations are not enough to uniquely determine the correlators, one inserts the result into the bulk--boundary constraint equations (\ref{eq-bbcr}), solves those and inserts the result into the Cardy condition (\ref{eq-cardy}), whose solution uniquely determines the correlators.\\
Performing this procedure for the $E_6$ minimal model it turns out that the Cardy condition is in contradiction with the other two constraints. By contradiction we mean the following: The bulk--boundary crossing constraint gives definite values to some correlators, whereas it sets others to $0$. The Cardy--condition now produces equations where a sum of terms is supposed to be zero and all the values are already known from the other constraints such that the Cardy equation cannot be satisfied. In some cases the other constraints set all summands but one to zero and thus the equation cannot be satisfied. In other cases the Cardy equations are contradictory among themselves, one equation giving a certain value to a correlator and another one assigning to it a different value. See also appendix \ref{app-amod} for an explicit example of the failure of the Cardy condition in the boundary--changing sector of the $A$--type minimal models. The $A_{\infty}$--relations and the bulk--boundary crossing constraints do not yield any contradictory equations. So one may assume that they are correct, although we will have to qualify this assumption. \\\\
In order to check the correctness of (\ref{eq-weffe6bulk}), the idea is to combine deformation theory methods and the consistency constraints and check if the result agrees with the results obtained from the deformation theory techniques alone. In order to do so, we proceed as follows:
\begin{itemize}
\item If $t=0$, respectively $s_i=0$, i.e. without bulk insertions, the bulk--boundary crossing constraint does not contain any information. This is why we may assume that in this case the superpotential coming from the $A_{\infty}$--relations and the one coming from the versal deformation of the $Q$--operator will agree. We thus start by computing the superpotentials for $t=0$ with either method.
\item The superpotential obtained from solving the $A_{\infty}$--relations will contain as undetermined parameters non--linear expressions in the unknown correlators. Comparing with the result of the Massey product algorithm one obtains an overdetermined system of non--linear equations for the correlators which has, at least for the examples we checked, a unique solution.
\item Next we set up the  $A_{\infty}$--constraints and the bulk--boundary crossing relations with $t\neq 0$ and use the boundary correlators whose values we found by comparison of the two superpotentials as input for solving the equations. It turns out that this is enough to uniquely determine all the values of the remaining correlators and the full superpotential is fixed. 
\end{itemize}
Going through this procedure for examples of the minimal models of the $A$--series and comparing with the results that come from the deformation of the matrix factorization with $t\neq 0$ we find agreement up to field redefinitions.\\
Turning to the $E_6$--example we set up and solve the $A_{\infty}$--relations and sums up the correlators. Explicitly, one finds this effective superpotential:
\begin{align}
\label{eq-ainftysupo}
\mathcal{W}_{eff}^{A_{\infty}}(u)=&B_{3334}u_4^3u_1+\left(-6B_{3334}B_{243244}+12B_{3334}^2B_{22224}\right)u_4^2u_1^5\nonumber \\
&+\left(12B_{3334}B_{243244}^2-48B_{3334}^2B_{22224}B_{243244}+32B_{3334}^3B_{22224}^2 \right)u_4u_1^9\nonumber\\
&+\big(8B_{3334}B_{243244}^3+48B_{3334}^2B_{22224}B_{243244}^2-64B_{3334}^3B_{22224}^2B_{243244}\nonumber\\
&+\frac{320}{13}B_{3334}^4B_{22224}^3\big)u_1^{13}
\end{align}
The indices of the correlators $B_{a_0\ldots a_m}$ refer to the indices of the inserted boundary fields (\ref{eq-e6bos}), (\ref{eq-e6fer}). We have three undetermined correlators $B_{3334},B_{22224},B_{243244}$. The form of this expression of course depends on how one solves the $A_{\infty}$ relations. Since, without bulk insertions, there are only the $A_{\infty}$ constraints that determine the superpotential this result must be equal to (\ref{eq-weff1e6}). From this, we get an overdetermined non--linear system of equations for the unknown correlators, which has the following {\em unique} solution:
\begin{align}
B_{3334}=1\qquad B_{22224}=\frac{1}{16}\qquad B_{243244}=0
\end{align}
Now one turns on bulk deformations and sets up the full $A_{\infty}$ and bulk--boundary crossing constraints. Using the data above as input it turns out that one can now uniquely fix all the correlators. Summing up the results to a superpotential one finds:
\begin{align}
\label{eq-weffe6bbcr}
\overline{\mathcal{W}}_{eff}(u;t)=&u_4^3u_1+\frac{3}{4}u_4^2u_1^5+\frac{1}{8}u_4u_1^9+\frac{5}{832}u_1^{13} \nonumber\\
&+\frac{1}{2}u_1\left(t_{12}-\frac{1}{2}t_6^2-\frac{1}{2}t_2t_5^2-\frac{1}{2}t_2^2t_8+\frac{1}{6}t_2^3t_6+\frac{1}{8}\left(t_6^2-t_2^3t_6+\frac{1}{4}t_2^6 \right) \right)\nonumber \\
&+\frac{1}{2}  \left(u_4+\frac{1}{8}u_1^4\right)\left(t_9-t_2^2t_5\right)\nonumber\\
&+\frac{1}{4}  \left(u_4u_1^3+\frac{3}{28}u_1^7\right)\left(t_6-\frac{1}{2}t_2^3\right)
\end{align}
This result is given in terms of the flat coordinates $t_i$, which are related to the parameters $s_i$ as follows \cite{Dijkgraaf:1990dj,Klemm:1991vw}:
\begin{align}
&s_2\longrightarrow -t_2\qquad s_2\longrightarrow -t_5\qquad s_6\longrightarrow -t_6+\frac{1}{2}t_2^3\qquad s_8\longrightarrow-t_8+t_2t_6+\frac{1}{12}t_2^4\nonumber\\
&s_9\longrightarrow-t_9\qquad s_{12}\longrightarrow -t_{12}+\frac{1}{2}t_6^2-\frac{1}{6}t_2^3t_6+\frac{1}{2}t_2^2t_8+\frac{1}{2}t_2t_5^2
\end{align}
Comparing (\ref{eq-weffe6bulk}) and (\ref{eq-weffe6bbcr}), one finds that $\overline{\mathcal{W}}_{eff}$ has a much simpler form. It looks like a reduced version of (\ref{eq-weffe6bulk}). Indeed, when relating the two results by a field redefinition one has to make a truncation of the bulk fields in addition. The reason for this result is the bulk--boundary crossing constraint. The $A_{\infty}$--relations alone only give relations between the correlators, in the present example only two of the amplitudes are set to zero explicitly. By just looking at the $A_{\infty}$ constraint we would obtain precisely the structure of (\ref{eq-weffe6bulk}). The bulk--boundary crossing constraint sets many of the correlators, which would be allowed by the selection rules, to zero and thus truncates many of the terms in the superpotential (\ref{eq-weffe6bulk}).\\
Since the bulk--boundary--crossing equations are consistent with the $A_{\infty}$--relations this result suggests that the bulk--boundary crossing equations impose an additional constraint on the superpotential\footnote{The Mathematica programs discussed in appendix \ref{app-mathematica} produce $536$ $A_{\infty}$--equations and $195$ equations from the bulk--boundary crossing constraints, for altogether $280$ correlators. That this large system of non--linear equations is actually consistent may justify such a claim.}. It might be that if one deforms the matrix factorization one only captures the $A_{\infty}$--structure and that one gets additional constraints from the interaction between bulk and boundary. This is incorporated in a mathematical structure termed Open Closed Homotopy Algebra (OCHA) which has recently been introduced in the literature \cite{Kajiura:2005sn}.\\
However, in \cite{Keller:2006tf} it was shown explicitly that the result (\ref{eq-weffe6bbcr}) for $\mathcal{W}_{eff}$ which was determined by combining deformation theory methods with the constraint equations implies that a certain correlator is $0$ but can be proven to be non--vanishing by CFT methods. This implies that the bulk--boundary crossing constraint needs to be extended in a way which modifies the equations such that one can get agreement with the result obtained from the Massey product method. This may also be possible for the Cardy constraint \cite{getzlerpriv}.
\section{Relation to Kazama--Suzuki Coset Models}
\label{sec-kazuk}
In this section we show that the {\em effective} superpotentials associated with rank two matrix
factorizations can be related to the {\em Landau--Ginzburg} potentials of
simply--laced, level one, hermitian symmetric space (SLOHSS)
models~\cite{Lerche:1991re}\footnote{Quite recently an interesting relation between Kazama--Suzuki superpotentials, matrix factorizations and knot homology was discovered \cite{khovanov1,khovanov2,Gukov:2005qp}. See also section \ref{sec-mfother}.}. These Kazama--Suzuki type models are represented by
cosets $G/H$, where the group $G$ is divided by its maximal subgroup $H$. We can associate SLOHSS models to the following cosets:
\begin{align}
&\frac{SU(n+m)}{SU(n)\times SU(m)\times U(1)}\qquad\frac{SO(2n)}{SO(2n-2)\times U(1)}\qquad \frac{SO(2n)}{SU(n)\times U(1)}\nonumber \\
&\frac{E_6}{SO(10)\times U(1)} \qquad\frac{E_7}{E_6\times U(1)}
\end{align}
Landau--Ginzburg potentials for the deformed SLOHSS models were
derived in \cite{Eguchi:2001fm}. They are obtained by 
first expressing the Casimirs $V_i$ of the group $G$, in terms of the Casimirs
of $H$, which are called $x_i$. Next, we set $V(x_i)=v_i$, identifying
the Casimirs with deformation parameters $v_i$ of the superpotential.
This yields a system of equations, where the $x_i$ that appear linearly
can be eliminated. The remaining equations can then be integrated to a superpotential.\\
The explicit form for the $E_6$--coset was given in \cite{Eguchi:2001fm}:
\begin{align}
\label{eq-weffslohss}
W(x,z,w)=&x^{13}-\frac{25}{169}x\,z^3+\frac{5}{26}x^2\,w_2\nonumber \\
&+z\left(x^9+x^7\,w_1+\frac{1}{3}x^5\,w_1^2-x^4\,w_2-\frac{1}{3}x^2\,w_1\,w_2+\frac{1}{12}x^3\,w_3-\frac{1}{6}x\,w_4+\frac{1}{3}w_5 \right)\nonumber \\
&+\frac{247}{165}x^{11}\,w_1+\frac{13}{15}x^9\,w_1^2-\frac{39}{20}x^8\,w_2+\frac{169}{945}x^7\,w_1^3+\frac{13}{105}x^7\,w_3-\frac{26}{15}x^6\,w_1\,w_2\nonumber \\
&+\frac{13}{225}x^5\,w_1\,w_3-\frac{13}{50}x^5\,w_4-\frac{91}{180}x^4\,w_1^2\,w_2+\frac{13}{30}x^4\,w_5+\frac{13}{15}x^3w_2^2-\frac{13}{90}x^3\,w_1\,w_4\nonumber \\
& -\frac{13}{120}x^2\,w_2\,w_3+\frac{13}{90}x^2\,w_1\,w_5-\frac{13}{270}x\,w_6-\frac{13}{360}w_1^4\,w_2+\frac{13}{90}w_1^2\,w_5
\end{align}
This superpotential is, up to a (quite complicated) field redefinition precisely the effective superpotential associated
to the $2\times 2$--factorization of the minimal model in our results.
The ansatz for such a field redefinition looks as follows:
\begin{align}
&u_1\longrightarrow \alpha_1 x\qquad u_4\longrightarrow \alpha_2z+\alpha_3x^4+\alpha_4w_1^2+\alpha_5w_1x^2\nonumber \\
&s_2\longrightarrow\beta_{15}w_1\qquad s_5\longrightarrow\beta_{14}w_2\qquad s_6\longrightarrow\beta_{12}w_3+\beta_{13}w_1^3\nonumber \\
&s_8\longrightarrow\beta_9w_4+\beta_{10}w_3w_1+\beta_{11}w_1^4\qquad s_9\longrightarrow\beta_7w_5+\beta_8w_2w_1^2 \nonumber \\
&s_{12}\longrightarrow\beta_1w_6+\beta_2w_4w_1^2+\beta_3w_3^2+\beta_4w_3w_1^3+\beta_5w_2^2w_1+\beta_6w_1^6
\end{align}
Inserting this into (\ref{eq-weffe6bbcr}) we can fix all the free parameters $\alpha_i$ and $\beta_i$ to match with (\ref{eq-weffslohss}), proving that the two expressions are equivalent.\\\\
It seems that only the effective superpotentials associated to $2\times 2$ matrix factorizations have a direct connection to the superpotentials coming from SLOHSS models. It is interesting to ask why and how the matrix factorizations are encoded in these coset models. We now give a qualitative explanation of how this happens. One way to calculate such superpotentials (at least in principle) is to eliminate as many variables as possible \cite{Eguchi:2001fm}. Take for instance the $E_6$--example:
\begin{align}
W&=x^3+y^4-z^2\;\mbox{(+ bulk deformations)}.
\end{align}
The equation $W=0$ describes an ALE space, a two complex dimensional surface in $\mathbb{C}^3$. We now look for lines and quadrics on this surface.
Using the ansatz
\begin{align}
x&=\lambda y+\alpha(\lambda),
\end{align}
where $\lambda$ is a variable of weight one and $\alpha$ is a homogeneous polynomial of weight four. The variable $z$ can be eliminated from $W$:
\begin{align}
z&=\sqrt{x^3+y^4\;\mbox{(+ bulk deformations)}}
\end{align}
In this relation the ansatz for $x$ can be inserted. The equation on $z$ describes a quadric if the expression under the square root is a perfect square. In this case one gets:
\begin{align}
z&=y^2+\gamma_1(\lambda) y+\gamma_2(\lambda)
\end{align}
Lines and quadrics on the surface $W=0$ are then given by
\begin{align}
A_1&=\lambda y+\alpha(\lambda)\:=\:0\nonumber \\
A_2&=y^2+\gamma_1(\lambda) y+\gamma_2(\lambda)\:=\:0.
\end{align}
The Nullstellensatz tells us that this is consistent with $W=0$ if $W$ has the form
\begin{equation}
W=A_1B_1+A_2B_2,
\end{equation}
for some $B_1,B_2$. This is precisely a $2\times 2$ matrix factorization of the superpotential! Our results for the ADE minimal models thus suggest that there is a direct relation between the coset model Landau--Ginzburg superpotentials and $\mathcal{W}_{eff}$ for $2\times 2$ matrix factorizations of the ADE minimal models. The results are therefore consistent with the argument presented above.\\
Note in addition that there is no SLOHSS model associated with $E_8$. However, there is also no rank $2$ matrix factorization of the $E_8$ Landau--Ginzburg potential. 
\section{More Results}
\label{sec-results}
In this section we list some more results for the superpotentials of the minimal models of types $E_6,E_7,E_8$. Due to the technical complexity of these calculations we set all the bulk parameters to 0. Still, it is not possible to do the deformation theory calculations by hand for any but the simplest example presented here. To produce the results given in this section, the computer algebra program Singular \cite{GPS05} was used. Singular provides the library \texttt{deform.lib} \cite{GPdeform} which can handle such calculations. \\
The input for our $E_6$--example is the following:
\begin{center}
\begin{algorithmic}
\STATE  $\mathtt{ LIB }\mathtt{ "deform.lib";}$
\STATE $\mathtt{int}\; \mathtt{p = printlevel;}$
\STATE $\mathtt{printlevel = 1;}$
\STATE $\mathtt{ring}$ $\mathtt{Ro = 0,(x,y,z),dp;}$
\STATE $\mathtt{ideal}\; \mathtt{Io = x3+y4-z2;}$
\STATE $\mathtt{matrix}\; \mathtt{Mo[2][2] = -y2-z,x,x2,y2-z;}$
\STATE $\mathtt{list}\; \mathtt{L = mod\_ versal(Mo,Io);}$
\end{algorithmic}
\end{center}
This computes the deformations of the module characterized by the matrix factorization. The result first gives the linear deformations, then indicates the steps in the iteration in the algorithm of the Massey products. The output is stored in lists and can be accessed as follows:
\begin{center}
\begin{algorithmic}
\STATE $\mathtt{def}\; \mathtt{Qx=L[2];}$
\STATE $\mathtt{setring}\; \mathtt{Qx;}$
\STATE $\mathtt{print(Ms);}$
\STATE $\mathtt{print(Js);}$
\end{algorithmic}
\end{center}
{\tt print(Ms)} yields the deformed matrix factorization, {\tt print(Js)} gives the $f$--polynomials. All that is left to do in order to obtain the effective superpotential is to integrate these polynomials.\\\\
We start with two more examples for $E_6$. First we consider the the self--dual matrix factorization $M_3$ given in (\ref{eq-m3}). We choose $\varepsilon=-1$ in order to have only real entries in the matrix factorization. The fermionic spectrum can be read off from table (\ref{tab-threevarcoh}). We associate to the four fermionic states deformation parameters $u_i$ with charges $\{1,3,4,6\}$. The Massey product algorithm yields the ring $k[u_i]/(f_i)$ with four polynomials $f_i$ of degrees $\{12,10,9,7\}$:
\begin{align}
f_1&=-u_6^2+u_4^3-u_3^4-18u_4u_3^2u_1^2+7u_6u_3u_1^3-6u_4^2u_1^4-26u_3^2u_1^6-9u_4u_1^8-3u_1^{12}\nonumber\\
f_2&=4u_3^3u_1+12u_4u_3u_1^3+8u_3u_1^7\nonumber\\
f_3&=3u_4^2u_1-6u_3^2u_1^3-u_1^9\nonumber\\
f_4&=-2u_6u_1+4u_3u_1^4
\end{align}
These can be integrated to the following effective superpotential:
\begin{align}
\label{eq-supom3}
\mathcal{W}_{eff}(u)&=-u_6^2u_1+u_4^3u_1-u_3^4u_1-6u_4u_3^2u_1^3+4u_6u_3u_1^4-8u_3^2u_1^7-u_4u_1^9-\frac{5}{13}u_1^{13}
\end{align}
Our last example for the $E_6$ model is the factorization $M_4$ (\ref{eq-m4}), where we set $\varepsilon=-1$. There are six deformation parameters with charges $\{1,2,3,4,4,6\}$. We find the following expression for the superpotential:
\begin{align}
\label{eq-supom4}
\mathcal{W}_{eff}&=\frac{5}{1664}u_1^{13}-\frac{1}{4}u_1^8u_5+\frac{3}{4}u_1^6u_2u_5-2u_2^4u_5+2u_2u_3^2u_5-u_4^2u_5-2u_3u_5^2\nonumber\\
&+u_1^3\left(-u_2u_3u_5-\frac{1}{2}u_5^2\right)+u_1^9\left(-\frac{1}{16}u_2^2+\frac{1}{16}u_4+\frac{1}{16}v_4\right)-u_4u_5v_4+u_5v_4^2\nonumber\\
&+u_1^4\left(-\frac{3}{2}u_2^2u_5+u_4u_5+\frac{1}{2}u_5v_4\right)
+u_2^2\left(3u_4u_5+3v_4u_5\right)\nonumber\\
&+u_1^5\left(-\frac{3}{8}u_2^4-\frac{3}{8}u_4^2-\frac{1}{2}u_3u_5+u_2^2\left(\frac{3}{4}u_4+\frac{3}{4}v_4\right)-\frac{3}{4}u_4v_4-\frac{3}{8}v_4^2\right)\nonumber\\
&+u_1^2\left(2u_2^3u_5-u_3^2u_5+u_2\left(-2u_4u_5-u_5v_4\right)\right)+u_1\bigg(-\frac{1}{2}u_2^6+\frac{1}{2}u_4^3+u_2u_5^2+\frac{3}{2}u_4^2v_4\nonumber\\
&+\frac{3}{2}u_4v_4^2+\frac{1}{2}v_4^3+u_2^4\left(\frac{3}{2}u_4+\frac{3}{2}v_4\right)+u_3\left(u_4u_5+2u_5v_4\right)+u_2^2\left(-\frac{3}{2}u_4^2+u_3u_5-3u_4v_4-\frac{3}{2}v_4^2\right)\bigg)
\end{align}
With (\ref{eq-weff1e6}), (\ref{eq-supom3}) and (\ref{eq-supom4}) we have actually given the superpotentials of five of the six branes since a brane yields the same superpotential as its anti--brane. Unfortunately, we were not able to calculate the effective superpotential for the rank four matrix factorization of the $E_6$--model, because the calculation exceeds the capabilities of an ordinary PC.\\
The relation of the superpotentials coming from matrix factorizations of rank greater than $2$ to coset models is not quite clear. One can, for instance, try to eliminate the parameters $\{u_2,u_3,v_4,u_5\}$ by solving $\frac{\partial\mathcal{W}}{\partial u_2}=\ldots=\frac{\partial\mathcal{W}}{\partial u_5}=0$ but this yields non--rational expressions for some of these variables.\\\\
Next, we will give an example for the superpotential of the simplest matrix factorization for the $E_7$--model. The superpotential for $E_7$ is:
\begin{align}
W&=x^3+xy^3+z^2
\end{align}
The dual Coxeter number of $E_7$ is $k=18$.\\
The simplest matrix factorization is the following self--dual rank $2$ factorization:
\begin{align}
E=J&=\left(\begin{array}{cc}
z&x\\
x^2+y^3&-z
\end{array}\right)
\end{align}
The odd spectrum consists of three states with charges $\{0,8,16\}$ to which we associate parameters $u_i$ with charges $\{1,5,9\}$. The deformed matrix factorizations are given by:
\begin{align}
E_{def}=&\bigg(\begin{array}{c}
z + u_1y^2 + u_5y + u_9\\
y^3 + x^2 - u_1^2x y - 2u_1u_5x + u_1^4y^2 + 4u_1^3u_5y - 8u_1^3u_9 + 
  u_1^6x - 2u_1^8y + 20u_1^7u_5 - 11u_1^{12}
\end{array}\nonumber\\
&\begin{array}{c}
x + u_1^2y + 2u_1u_5 - u_1^6\\
-z + u_1y^2 + u_5y + u_9
\end{array}
\bigg)
\end{align}
The polynomials defining the critical locus of the effective superpotential are:
\begin{align}
f_1&=-u_9^2 - 16u_1^4u_5u_9 + 40u_1^8u_5^2 + 8u_1^9u_9 - 42u_1^{13}u_5 + 
    11u_1^{18}\nonumber\\
f_2&=-2u_5u_9 + 8u_1^4u_5^2 - 8u_1^5u_9 + 12u_1^9u_5 - 9u_1^{14}\nonumber\\
f_3&=-u_5^2 - 2u_1u_9 + 6u_1^5u_5 - 3u_1^{10}
\end{align}
These polynomials of degrees 18, 14, 10 are easily integrated to the following effective superpotential of degree 19:
\begin{align}
\label{eq-e7supo}
\mathcal{W}_{eff}(u)&=-\frac{55}{19}u_1^{19}+12u_1^{14}u_5-15u_1^9u_5^2+5u_1^4u_5^3-3u_1^{10}u_9+6u_1^5u_5u_9-u_5^2u_9-u_1u_9^2
\end{align}
This can be related the SLOHSS model associated with the following coset:
\begin{align}
\frac{E_7}{E_6\times U(1)}
\end{align}
\cite{Eguchi:2001fm} gives the following expression for the Landau--Ginzburg potential of this deformed coset model (we set all the bulk parameters to 0):
\begin{align}
W(x,y,z)&=\frac{1016644}{817887699}x^{19}+\frac{33326}{177147}x^{14}y+\frac{266}{6561}x^{10}z+\frac{16850}{2187}x^9y^2+\frac{80}{27}x^5yz+\frac{124}{9}x^4y^3+xz^2+\frac{3}{2}y^2z
\end{align}
One obtains this expression from (\ref{eq-e7supo}) by a simple field redefinition.\\\\
Finally we also give an example for the $E_8$ model. The Landau--Ginzburg potential for this model is:
\begin{align}
W&=x^3+y^5+z^2
\end{align}
The dual Coxeter number is $k=30$. The simplest matrix factorization has rank $4$:
\begin{align}
E=J&=\left(\begin{array}{cccc}
z& 0& x& y\\
0& z& y^4& -x^2\\
x^2& y& -z& 0\\
y^4& -x& 0& -z
\end{array}
\right)
\end{align}
We have four bosonic and four fermionic states. We associate to the fermionic states deformation parameters $u_i$ with charges $\{1,6,10,15\}$. We get four polynomials with degrees $\{30,25,21,16\}$:
\begin{align}
f_1=&u_{15}^2 + u_{10}^3 + u_{6}^5 - 30u_{10}u_{6}^3u_1^2 + 4u_{15}u_{6}^2u_1^3 + 
  21u_{10}^2u_{6}u_1^4 + 2u_{15}u_{10}u_1^5 + 140u_{6}^4u_1^6 - 
  155u_{10}u_{6}^2u_1^8\nonumber \\
&- 16u_{15}u_{6}u_1^9 - 3u_{10}^2u_1^{10} + 40u_{6}^3u_1^{12} +185u_{10}u_{6}u_1^{14} + 6u_{15}u_1^{15} - 359u_{6}^2u_1^{18} - 45u_{10}u_1^{20}\nonumber\\
& + 
    215u_{6}u_1^{24} - 35u_1^{30}\nonumber\\
f_2=&5u_{6}^4u_1 - 30u_{10}u_{6}^2u_1^3 + u_{15}u_{6}u_1^4 + 125u_{6}^3u_1^7 - 
    30u_{10}u_{6}u_1^9 + 10u_{6}^2u_1^{13} + 15u_{10}u_1^{15}\nonumber\\
& - \frac{117}{2}u_{6}u_1^{19} + 
    14u_1^{25}\nonumber \\
f_3=&-3u_{10}^2u_1 + 10u_{6}^3u_1^3 + 2u_{15}u_1^6 + 25u_{6}^2u_1^9 - 
    25u_{6}u_1^{15} + 6u_1^{21}\nonumber\\
f_4=&-2u_{15}u_1 - 10u_{6}^2u_1^4 + 10u_{6}u_1^{10} - 3u_1^{16}
\end{align}
Integration yields the following superpotential:
\begin{align}
\mathcal{W}_{eff}=&-\frac{11}{31}u_1^{31}+u_1^{25}u_6-10u_1^{19}u_6^2+45u_1^{13}u_6^3-55u_1^7u_6^4-u_1u_6^5+3u_1^{21}u_{10}-15u_1^{15}u_6u_{10}\nonumber\\
&+15u_1^9u_6^2u_{10}+10u_1^3u_6^3u_{10}-u_1u_{10}^3-3u_1^{16}u_{15}+10u_1^{10}u_6u_{15}-10u_1^4u_6^2u_{15}-u_1u_{15}^2\nonumber\\
&+4u_{1}^{21}u_{10}-u_1u_{10}^3-3u_1^{16}u_{15}
\end{align}
We cannot relate this result to a coset model, since such models do not exist for $E_8$. This supports the conjecture that matrix factorizations of rank greater than $2$ can not be related to such coset models and provide a more general structure.

\chapter{The Torus and Homological Mirror Symmetry}
\label{chap-torus}
\section{Introduction}
This chapter is concerned with the simplest example of homological mirror symmetry, realized on the two--dimensional torus. The torus has the usual bulk modulus and also a brane modulus. \\
In the A--model, the D$1$ branes wrapping the torus are just straight lines in the fundamental domain. The brane modulus is a combination of position shifts of the lines with respect to the origin and Wilson loops around the circumferences determined by intersecting branes. The instanton--corrected correlators have a simple geometric interpretation as the areas enclosed by intersecting branes. The torus is thus an example where the correlators on the A--model side can actually be calculated by computing these areas and summing them up. In the $B$--Model the D$1$--branes map into D$0$ and D$2$ branes. The brane modulus encodes the positions of the $D0$--branes. The torus has a description in terms of Landau--Ginzburg orbifolds. There are three such models. They describe special points of enhanced symmetry in the K\"ahler moduli space. The $\mathbbm{Z}_3$--orbifold is described by a cubic curve, the $\mathbbm{Z}_4$--orbifold is realized in terms of a quartic curve. Furthermore there exists a $\mathbbm{Z}_6$--orbifold.  Explicitly, the
associated Landau--Ginzburg superpotentials take the following form (see e.g. \cite{Lerche:1989cs}):
\begin{align}
\label{eq-cubic}
W_{\mathbbm{Z}_3}&=x_1^3+x_2^3\:+x_3^3-a\:x_1x_2x_3\\
\label{eq-quartic}
W_{\mathbbm{Z}_4}&=x_1^4+x_2^4\:-x_3^2-a\:x_1^2x_2^2\\
W_{\mathbbm{Z}_6}&=x_1^6+x_2^3\:- x_3^2-a\:x_1x_2^4.
\end{align}
All these curves are special cases of the elliptic curve. The parameter $a$ is related to the complex structure modulus $\tau$.\\
A classification of D--branes on the torus is possible. In \cite{atiyah}, Atiyah showed that vector bundles $\mathcal{E}$ on the elliptic curve can be classified in terms of three numbers:
\begin{align}
(r(\mathcal{E}),c_1(\mathcal{E}),\zeta)\equiv(N_2,N_0,\zeta)
\end{align} 
Here, $r(\mathcal{E}),c_1(\mathcal{E})$ are the rank and the first Chern class of the vector bundle, which are related to the number of D$2$-- and D$0$--branes on the torus. $\zeta$ is a number which encodes the location of the D$0$--brane. In the physics literature this was discussed in \cite{Govindarajan:2005im}.\\
Matrix factorizations on the cubic curve have been classified in \cite{Laza}, This classification was done at the Gepner point for $a=0$. In \cite{Hori:2004ja}, matrix factorizations on the cubic curve were computed away from the Gepner point. In particular, it was shown how to include bulk {\em and} boundary moduli into the matrix factorization formalism.\\
There has been an exhaustive discussion of homological mirror symmetry on the torus in the math literature \cite{Polishchuk:1998db,PolishchukFukaya,PolishchukAppell,PolishchukMirror,Polishchuk:2000kx,PolishchukTheta,math.AG/0012018,Kajiura:2007zz}. These papers do not use the matrix factorization language but describe homological mirror symmetry as the equivalence between the Fukaya category and the derived category of coherent sheaves. Homological mirror symmetry and matrix factorizations on the cubic curve from the physics point of view was discussed in \cite{Brunner:2004mt,Govindarajan:2005im,Herbst:2006nn,Jockers:2006sm}. In \cite{Brunner:2004mt} 'long' and 'short' branes, which wrap the long and short diagonals of the fundamental domain of the torus in the A--model were identified with certain matrix factorizations in the B--model with Landau--Ginzbug potential (\ref{eq-cubic}). Furthermore the three--point correlators were calculated in the A--model and the B--model and they were shown to match. In \cite{Govindarajan:2005im} the authors discuss how one can extract bundle data out of matrix factorizations and it is shown how to construct higher rank matrix factorizations out of a minimal set via tachyon condensation. The description of branes in terms of matrix factorizations at different points in moduli space was also discussed in this paper. The paper \cite{Herbst:2006nn} discusses how to calculate the effective superpotential for the cubic torus. This calculation is done by computing correlators in the A--model, since it is at present not known how to calculate correlators with more than three insertions in the B--model. The correctness of these correlators is tested by various consistency checks: it was shown that the correlators satisfy the $A_{\infty}$--relations and the quantum $A_{\infty}$--equations \cite{Herbst:2006kt}. Their analytical properties, i.e. where and how the amplitudes degenerate, were also investigated. See also \cite{Cremades:2003qj} for the calculation of three--point functions in the A--model. In this paper the Yukawa couplings of the MSSM were obtained from toroidal compactifications with intersecting D--branes.  The effective superpotential for $B$--branes was discussed in \cite{Govindarajan:2006uy}. In \cite{Jockers:2006sm} monodromies in K\"ahler moduli space were discussed in the context of matrix factorizations, using the torus and the quintic as examples.\\\\
In this chapter we discuss the quartic curve (\ref{eq-quartic}) in detail. Mirror symmetry for $T^2$ at the $\mathbbm{Z}_4$--symmetric point has been discussed in \cite{Dell'Aquila:2005jg}. The difference to what we will discuss here is that this paper deals with the torus at the $\mathbbm{Z}_{4}$ symmetric point, where a $\mathbbm{Z}_4$--orbifold is taken. The mirror is a Landau--Ginzburg model with superpotential $W=x^4+y^4$. We will discuss mirror symmetry of $T^2$, where in the A--model the complex structure modulus is set to $e^{\frac{2\pi i}{4}}$. In the B--model this corresponds to fixing the K\"ahler modulus to $e^{\frac{2\pi i}{4}}$. At this point in moduli space the B--model has a description in terms of a Landau--Ginzburg orbifold with superpotential (\ref{eq-quartic}). The CFT aspects of the quartic torus have been investigated in \cite{Schmidt-Colinet:2007vi}, see also \cite{Brunner:2006tc} for a discussion in the context of $K3$ surfaces.\\
As compared to the cubic curve, which was treated thoroughly in the literature, the quartic curve has some novel features. The most obvious one is that the variables of the quartic superpotential do not have equal weights. This has an influence
on the uniformization procedure which relates the moduli dependent parameters
of the B--model to those natural in the A--model.
Another difference between the quartic and the cubic torus is that the superpotential
contains only terms with even exponents. This implies that there will be a
self-dual matrix factorization, i.e. a brane which is its own antibrane.
The existence of self-dual matrix factorizations then entails that we must include
antibranes into the description if we want to compute correlators on the quartic torus.
Furthermore, selection rules tell us that the three--point functions for the 
quartic torus always have insertions of two different kinds of factorizations, corresponding to the 'long' and 'short' branes  whereas for the cubic curve all calculations were done just for the 'long' branes. \\\\
In section \ref{sec-quarticmf} we discuss the matrix factorizations of the quartic superpotential (\ref{eq-quartic}) and we identify the 'long' and 'short' branes. Section \ref{sec-quarticcoh} deals with the open string spectrum on the torus. In section \ref{sec-quarticcorr} we calculate the three--point functions in the B--model. In \ref{sec-quarticd2} we discuss an exceptional matrix factorization, which describes a pure D$2$--brane wrapping the torus. In section \ref{sec-quartica} we calculate three--point functions in the A--model and verify homological mirror symmetry by comparison with the B--model calculation.\\\\
This chapter summarizes my contribution to \cite{Knapp:2007kq}.
\section{Matrix Factorizations}
\label{sec-quarticmf}
We consider the following three--variable Landau--Ginzburg superpotential for the quartic torus:
\begin{align}
\label{eq-supoquartic}
W&=x_1^4+x_2^4-x_3^2-ax_1^2x_2^2
\end{align}
In order to incorporate moduli in a natural manner we introduce
parameters $\alpha_1^i\equiv\alpha_1(u_i,\tau)$, $\alpha_2^i\equiv\alpha_2(u_i,\tau)$ and $\alpha_3^i\equiv\alpha_3(u_i,\tau) $,
which depend on the boundary modulus $u$ of the brane (we label the brane by an index $i$) and
the complex structure modulus $\tau$ of the torus \cite{Hori:2004ja,Brunner:2004mt}. The matrix factorization condition constrains the $\alpha_i$ to lie on the Jacobian of the torus. The parameters therefore have to satisfy the following relation:
\begin{align}
\label{eq-alpharel}
(\alpha_1^i)^4+(\alpha_2^i)^4-(\alpha_3^i)^2-a(\alpha_1^i)^2(\alpha_2^i)^2&=0
\end{align}
We now give the matrix factorizations which correspond to the long and short branes in the A--model. It is a priori not obvious to see from the structure of the factorization whether it is a long brane or a short brane. One way to find out is by computing the RR--charges \cite{Walcher:2004tx,Govindarajan:2005im}. We choose a different  approach, identifying the branes by their spectra. From the A--model picture we know that long branes intersect twice in the fundamental domain of the torus, which implies that we have two states stretching between two long branes. Short branes intersect only once and we expect one open string state between two short branes. We will show explicitly in the following section and in the appendix that the matrix factorizations given below satisfy these properties.\\
For the brane anti--brane pair of the short branes we find: $Q^{\mathcal{S}}_i=\left(\begin{array}{cc}0&E_i^{\mathcal{S}}\\J_i^{\mathcal{S}}&0\end{array}\right)$, where $i=\{1,\ldots,4\}$:
\begin{align}
\label{eq-short1}
E_i^{\mathcal{S}}&=\left(\begin{array}{cc}
\alpha_1^ix_1+\alpha_2^ix_2&\alpha_3^ix_3+\frac{(\alpha_3^i)^2}{\alpha_1^i\alpha_2^i}x_1x_2\\
\frac{1}{\alpha_1^i\alpha_2^i}x_1x_2-\frac{1}{\alpha_3^i}x_3&-\frac{1}{\alpha_1^i}x_1^3+\frac{\alpha_1^i}{(\alpha_2^i)^2}x_1x_2^2-\frac{1}{\alpha_2^i}x_2^3+\frac{\alpha_2^i}{(\alpha_1^i)^2}x_1^2x_2
\end{array}\right)
\end{align} 
\begin{align}
\label{eq-short2}
J_i^{\mathcal{S}}=\left(\begin{array}{cc}
\frac{1}{\alpha_1^i}x_1^3-\frac{\alpha_1^i}{(\alpha_2^i)^2}x_1x_2^2+\frac{1}{\alpha_2^i}x_2^3-\frac{\alpha_2^i}{(\alpha_1^i)^2}x_1^2x_2&\alpha_3^ix_3+\frac{(\alpha_3^i)^2}{\alpha_1^i\alpha_2^i}x_1x_2\\
\frac{1}{\alpha_1^i\alpha_2^i}x_1x_2-\frac{1}{\alpha_3^i}x_3&-\alpha_1^ix_1-\alpha_2^ix_2
\end{array}\right)
\end{align}
This has the structure of a linear permutation brane. This is in accordance with the case of the cubic curve \cite{Brunner:2004mt}, where the short branes were also identified with linear permutation branes.\\
For the long branes we take the following expression: 
$Q^{\mathcal{L}}_i=\left(\begin{array}{cc}0&E_i^{\mathcal{L}}\\J_i^{\mathcal{L}}&0\end{array}\right)$, where:
\begin{align}
\label{eq-long1}
E_i^{\mathcal{L}}=\left(\begin{array}{cc}
\frac{\alpha_1^i}{\alpha_2^i}x_1^2-\frac{\alpha_2^i}{\alpha_1^i}x_2^2&\frac{1}{\alpha_3^i}x_3-\frac{1}{\alpha_1^i\alpha_2^i}x_1x_2\\
\alpha_3^ix_3+\frac{(\alpha_3^i)^2}{\alpha_1^i\alpha_2^i}x_1x_2&\frac{\alpha_2^i}{\alpha_1^i}x_1^2-\frac{\alpha_1^i}{\alpha_2^i}x_2^2
\end{array}\right)
\end{align}
\begin{align}
\label{eq-long2}
J_i^{\mathcal{L}}=\left(\begin{array}{cc}
\frac{\alpha_2^i}{\alpha_1^i}x_1^2-\frac{\alpha_1^i}{\alpha_2^i}x_2^2&-\frac{1}{\alpha_3^i}x_3+\frac{1}{\alpha_1^i\alpha_2^i}x_1x_2\\
-\alpha_3^ix_3-\frac{(\alpha_3^i)^2}{\alpha_1^i\alpha_2^i}x_1x_2&\frac{\alpha_1^i}{\alpha_2^i}x_1^2-\frac{\alpha_2^i}{\alpha_1^i}x_2^2
\end{array}\right)
\end{align}
These matrix factorizations correspond to Recknagel--Schomerus branes. \\
Note that, apart from the permutation branes, there is a standard construction for a matrix factorization. One can factorize the superpotential as follows: $W=\sum_i w_i x_i\frac{\partial W}{\partial x_i}$, where $w_i$ are the homogeneous weights of the $x_i$. In our case this would yield a $4\times 4$ matrix factorization. Comparing with the cubic curve, one might expect that a factorization of this kind would give the long branes. We will argue in section \ref{sec-quarticd2} why (\ref{eq-long1}), (\ref{eq-long2}) is the simplest choice for the long branes.\\
Since we have a $\mathbb{Z}_4$--orbifold action the index $i$ can take the values $i\in\{1,2,3,4\}$. The R--matrices are given by:
\begin{align}
R_1&=\mathrm{diag}\left(\frac{1}{4},-\frac{1}{4},-\frac{1}{4},\frac{1}{4}\right)\\
R_2&=\mathrm{diag}(0,0,0,0)
\end{align}
The orbifold matrices associated to the matrix factorizations above are \cite{Walcher:2004tx}:
\begin{align}
\gamma_{1,2}^i&=\sigma e^{i\pi R_{1,2}}e^{-i\pi\varphi_i},
\end{align}
where $\sigma=\left(\begin{array}{cc}\mathbbm{1}_2&0\\0&-\mathbbm{1}_2\end{array}\right)$ and the phase $\varphi$ is determined by the condition $\gamma^4=\mathbbm{1}$.
\section{Cohomology}
\label{sec-quarticcoh}
\begin{figure}
\begin{center}
\includegraphics{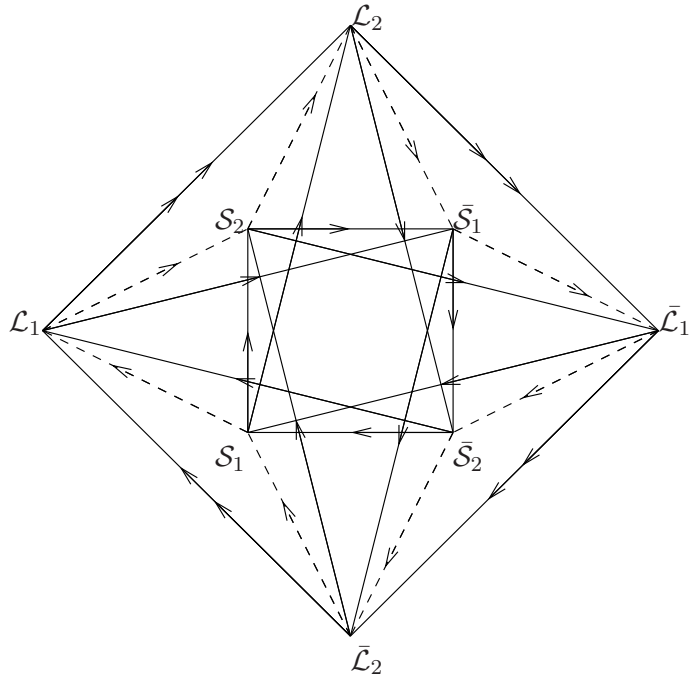}
\end{center}
\caption{The quiver diagram for the quartic torus.}\label{fig-bquartic1}
\end{figure}

\begin{figure}
\begin{center}
\includegraphics{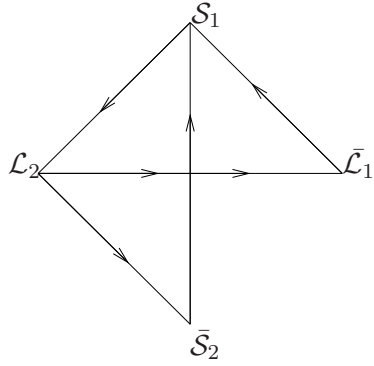}
\end{center}
\caption{Three--point functions on the B--side.}\label{fig-bquartic2}
\end{figure}

\begin{figure}
\begin{center}
\includegraphics{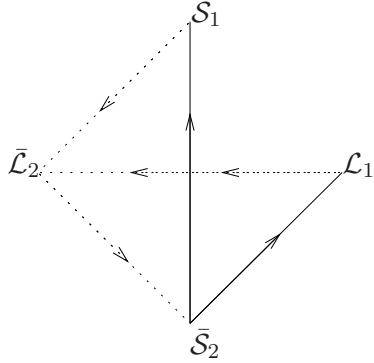}
\end{center}
\caption{Three--point functions on the B--side, including bosons.}\label{fig-bquartic3}
\end{figure}

We now calculate the open string spectrum. In order to do so, we make ans\"atze for the open string states, which are constrained by the condition of orbifold invariance (\ref{eq-orbmorph}). This confines the $R$--charges of the states which determines the $x_i$--dependence of the matrix representing the open string state. The moduli dependent factors have to be evaluated by solving the physical state condition modulo (\ref{eq-alpharel}).\\
The boundary changing spectrum is depicted in figure \ref{fig-bquartic1}.
Here, only the fermionic states have been drawn. By Serre duality, the bosonic states run in the opposite direction. The torus has background charge $\hat{c}=1$ so Serre duality implies that $q_{\psi}+q_{\phi}=1$. The states mapping between branes of the same type have charge $1/2$. This tells us that at a generic point in moduli space there will be no non--vanishing three--point functions if one considers only short branes or only long branes. For the open string states between long and short branes the solid lines represent fermions of charge $1/4$ and the dashed lines are fermions of charge $3/4$.\\
For every brane, there is also a fermionic boundary preserving operator of charge 1, which corresponds to a marginal boundary deformation. It is given by $\Omega=\partial_u Q$, where $u$ is the boundary modulus. This state will yield a contribution $\langle\mathbbm{1}\mathbbm{1}\Omega\rangle$ to the boundary metric for each of the long and short branes.\\
Furthermore, there exist additional states between a brane and its antibrane if the branes lie on top of each other. In particular, there will be a boson of charge 1 and a fermion of charge 0. The existence of these states comes from the fact that a boson, resp. fermion, beginning and ending on the same brane implies the existence of a fermion, resp. boson, stretching between the brane and its antibrane. In this sense, these states are related to $\mathbbm{1}$ and $\Omega$. We will not pursue this degenerate case any further in this work.\\\\
In the following, we will be interested in computing the non--vanishing three--point functions for the quartic torus. Clearly, only the fermions with charge $1/4$ and $1/2$ can contribute to the three--point functions. The possible three--point correlators correspond to oriented triangles in figure \ref{fig-bquartic1}.  Note that we did not draw the bosons in this graph. In order to identify all the three-point functions one has to keep in mind that there is also a bosonic arrow going in the opposite direction. The quiver in fig. 
\ref{fig-bquartic1} has an obvious $\mathbb{Z}_4$ symmetry. So, each type of correlator appears four times. \\
Let us first show some examples. We cut out a patch of figure \ref{fig-bquartic1} which contains all the information about correlators with only fermions. This is shown in figure \ref{fig-bquartic2}.
The possible correlators are thus:
\begin{align}
&\langle\psi_{\bar{\mathcal{S}}_2\mathcal{S}_1}\psi_{\mathcal{S}_1\mathcal{L}_2}\psi_{\mathcal{L}_2\bar{\mathcal{S}}_2}\rangle, \quad \langle\psi_{\bar{\mathcal{L}}_1\mathcal{S}_1}\psi_{\mathcal{S}_1\mathcal{L}_2}\psi_{\mathcal{L}_2\bar{\mathcal{L}}_1}\rangle,\quad \langle\psi_{\bar{\mathcal{L}}_1\mathcal{S}_1}\psi_{\mathcal{S}_1\mathcal{L}_2}\bar{\psi}_{\mathcal{L}_2\bar{\mathcal{L}}_1}\rangle.
\end{align}
To find correlators with bosonic insertions, we have to swap the directions of some arrows in the quiver. One has, for instance, a configuration depicted in figure \ref{fig-bquartic3} where bosons are represented by dotted lines.
Since the labeling of the branes is just convention and since we cannot tell the difference whether a state goes from brane to brane or from antibrane to antibrane in the B--model we will label the states in the correlators just with $\mathcal{L}$ and $\mathcal{S}$ and not with $\mathcal{L}_1,\bar{\mathcal{L}}_2,\bar{\mathcal{S}}_2$, etc.\\
There are two different types of correlators, those with two long branes and one short brane and those with two short branes and one long brane. In this thesis we will mostly be concerned with the first type.\\
There are eight different correlators of type long--long--short:
\begin{align}
\label{eq-fercorr}
&\langle\psi_{\mathcal{L}\mathcal{L}}\psi_{\mathcal{L}\mathcal{S}}\psi_{\mathcal{S}\mathcal{L}} \rangle\qquad \langle\bar{\psi}_{\mathcal{L}\mathcal{L}}\psi_{\mathcal{L}\mathcal{S}}\psi_{\mathcal{S}\mathcal{L}} \rangle\\
&\langle\psi_{\mathcal{L}\mathcal{L}}\phi_{\mathcal{L}\mathcal{S}}\phi_{\mathcal{S}\mathcal{L}} \rangle\qquad \langle\bar{\psi}_{\mathcal{L}\mathcal{L}}\phi_{\mathcal{L}\mathcal{S}}\phi_{\mathcal{S}\mathcal{L}} \rangle\\
&\langle\phi_{\mathcal{L}\mathcal{L}}\psi_{\mathcal{L}\mathcal{S}}\phi_{\mathcal{S}\mathcal{L}} \rangle\qquad \langle\bar{\phi}_{\mathcal{L}\mathcal{L}}\psi_{\mathcal{L}\mathcal{S}}\phi_{\mathcal{S}\mathcal{L}} \rangle\\
\label{eq-boscorr}
&\langle\phi_{\mathcal{L}\mathcal{L}}\phi_{\mathcal{L}\mathcal{S}}\psi_{\mathcal{S}\mathcal{L}} \rangle\qquad \langle\bar{\phi}_{\mathcal{L}\mathcal{L}}\phi_{\mathcal{L}\mathcal{S}}\psi_{\mathcal{S}\mathcal{L}} \rangle
\end{align}
Furthermore, there are four correlators of type short--short--long:
\begin{align}
&\langle\psi_{\mathcal{S}\mathcal{S}}\psi_{\mathcal{L}\mathcal{S}}\psi_{\mathcal{S}\mathcal{L}}\rangle \qquad \langle\psi_{\mathcal{S}\mathcal{S}}\phi_{\mathcal{L}\mathcal{S}}\phi_{\mathcal{S}\mathcal{L}}\rangle\\
&\langle\phi_{\mathcal{S}\mathcal{S}}\psi_{\mathcal{L}\mathcal{S}}\phi_{\mathcal{S}\mathcal{L}}\rangle \qquad \langle\phi_{\mathcal{S}\mathcal{S}}\phi_{\mathcal{L}\mathcal{S}}\psi_{\mathcal{S}\mathcal{L}}\rangle
\end{align}
Here we used a bar to distinguish between the two states stretching between two long branes.\\
We collect the explicit results for all the open string states in appendix \ref{app-torus}.
\section{Correlators in the B--model}
\label{sec-quarticcorr}
This section is concerned with the calculation of three--point correlators in the B--model. These can be determined by the residue formula (\ref{eq-kapustin2}).\\
The key difficulty in this computation is to find the correct normalization for the cohomology elements such that the correlators can be identified with the instanton sums in the A--model. This is related to finding flat coordinates on the moduli space. For the boundary preserving operator $\Omega=\partial_u Q$ the correct normalization can be deduced from the normalization of the superpotential which has to be scaled by a flattening normalization factor \cite{Lerche:1991wm,Brunner:2004mt}. The normalization of the boundary changing operators is more difficult to calculate. In \cite{Brunner:2004mt} it was argued that the correctly normalized three--point functions have to satisfy the heat equation:
\begin{align}
\left(\frac{\partial^2}{\partial u^2}-8\pi i \frac{\partial}{\partial \tau}\right)C_{ijk}(u,\tau)&=0
\end{align}
This implies that the correlators are theta functions, because these are the solutions to this equation.\\
In the following subsections we will calculate the normalization for the boundary preserving operator $\Omega$ and compute the correlators $\langle\Omega\rangle$. In order to be able to compute three--point functions, we first uniformize the moduli--dependent parameters $\alpha_i$, i.e. we express them in terms of the boundary modulus $u$ and the complex structure modulus $\tau$. Then we proceed to calculating the three--point functions in the B--model, making extensive use of theta function identities. 
\subsection{The flattening normalization factor}
The normalization of the boundary preserving operator $\Omega=\partial_u Q$ is related to the normalization of the superpotential $W$ via the matrix factorization condition. In \cite{Lerche:1991wm} it was shown that it is necessary to change the normalization of the superpotential by a modulus dependent prefactor, $W\rightarrow \frac{1}{q(\tau)}W$, in order to have vanishing connection terms in the differential equations satisfied by the periods. The matrix factorization condition then implies that we have to redefine $Q\rightarrow q(\tau)^{-\frac{1}{2}}Q$. This additional factor is then inherited by $\Omega$. We will now calculate the flattening normalization factor for the quartic torus, following the steps in \cite{Lerche:1991wm}.\\
We start by defining the following integrals, which are related to periods of differential forms, given a superpotential with $n$ variables:

\begin{align}
&u_0=(-1)^{\lambda}\Gamma(\lambda)\int \frac{q(s)}{W^{\lambda}}\mathrm{d}x_1\wedge\ldots\wedge\mathrm{d}x_n
\end{align} 
\begin{align}
&u_{\alpha}^{(\lambda)}=(-1)^{\lambda+1}\Gamma(\lambda+1)\int_{\gamma}\frac{\phi_{\alpha}(x_i,s)}{W^{\lambda+1}}\mathrm{d}x_1\wedge\ldots\wedge\mathrm{d}x_n
\end{align}
Here, $q(s)$ is a function of the moduli and the flattening factor we are looking for. It is given in generic, non--flat coordinates $s$. $\phi_{\alpha}(x_i,s)$ is a bulk cohomology element, $\gamma$ is a homology cycle and $\Gamma(\lambda)$ is the gamma function. It can then be shown \cite{Lerche:1991wm} that $u_0$ satisfies the following differential equation:
\begin{align}
\frac{\partial^2}{\partial s_i\partial s_j}u_0&=C_{ij}^{\alpha}u_{\alpha}^{(\lambda+1)}+\Gamma_{ij}^k\frac{\partial}{\partial s_k}u_0,
\end{align} 
where $C_{ij}^{\alpha}$ are the structure constants of the bulk chiral ring and $\Gamma_{ij}^k$ is the Gauss--Manin connection. Going from the generic coordinates $s_i$ to flat coordinates $t_i$ amounts to requiring that the connection vanishes: $\Gamma\equiv 0$. In particular, this condition leads to a differential equation which determines the flattening factor $q(t)$.\\
We will now specialize to the quartic torus. Thus, $W$ is given by (\ref{eq-supoquartic}) and $n=3$. Furthermore, we set for the modulus of the torus $t\equiv\tau$, as usual. Then we have $u_0=(-1)^{\lambda}\Gamma(\lambda)\int \frac{q(\tau)}{W^{\lambda}}\mathrm{d}x_1\wedge\mathrm{d}x_2\wedge\mathrm{d}x_3$. In the following we will drop the integral measure. Computing the second derivative of $u_0$ with respect to the modulus, one gets:
\begin{align}
\frac{\partial^2 u_0}{\partial\tau^2}&=\frac{q''}{q}u_0+(-1)^{\lambda+1}\Gamma(\lambda+1)\int\frac{1}{W^{\lambda+1}}\left(\frac{2q'}{q}+\frac{a''}{a'}\right)(-q\,a'\,x_1^2x_2^2)+(-1)^{\lambda+2}\Gamma(\lambda+2)\int\frac{q}{W^{\lambda+2}}(a')^2x_1^4x_2^4
\end{align}
Next, we partially integrate the third term, applying the following identities:
\begin{align}
(4-a^2)x_1^4x_2^4&=x_1x_2^3\left(x_2\partial_{x_1}W+\frac{1}{2}a\,x_1\partial_{x_2}W\right)\nonumber\\
x_2\partial_{x_2}W&=x_2\left(4x_2^3-2a\,x_1^2x_2\right)
\end{align}
The vanishing of the connection corresponds to the vanishing of the terms proportional to $\frac{1}{W^{\lambda+1}}$. This leads to a differential equation for $q(\tau)$ in terms of $a(\tau)$:
\begin{align}
&(-1)^{\lambda+1}\Gamma(\lambda+1)\int\frac{1}{W^{\lambda+1}}\left(\frac{2q'}{q}+\frac{a''}{a'}+\frac{2aa'}{4-a^2}\right)(-q\,a'\,x_1^2x_2^2)\stackrel{!}{=}0
\end{align}
This is easily integrated to give the following result for $q(\tau)$:
\begin{align}
q(\tau)&=\left(\frac{4-a(\tau)^2}{a'(\tau)}\right)^{\frac{1}{2}}
\end{align}
From this, one also gets another useful identity:
\begin{align}
\eta^6(\tau)&=\frac{(-1)^{\frac{7}{4}}}{2\pi^{\frac{3}{2}}}\frac{1}{q^2(\tau)}=\frac{(-1)^{\frac{7}{4}}}{2\pi^{\frac{3}{2}}}\frac{a'(\tau)}{4-a^2(\tau)}
\end{align}
\subsection{The correlators $\langle\Omega_{\mathcal{S}}\rangle$ and $\langle\Omega_{\mathcal{L}}\rangle$}
As argued above, the correctly normalized correlators look as follows \cite{Brunner:2004mt},
\begin{align}
&\langle\Omega_{\mathcal{S}/\mathcal{L}}\rangle=q(\tau)\int\frac{\mathrm{str}\left(\frac{1}{3!}(\mathrm{d}Q^{\mathcal{S}/\mathcal{L}})^{\wedge 3}\partial_u Q^{\mathcal{S}/\mathcal{L}}\right)}{\partial_1W\partial_2W\partial_3W}=\int\frac{f(u,\tau)H(x)}{\partial_1W\partial_2W\partial_3W}=f(u,\tau),
\end{align}
where $H(x)=\frac{1}{12}\mathrm{det}\partial_i\partial_jW$ is the Hessian. Going to the patch $\alpha_2(u,\tau)=1$ and using the relation coming from the vanishing of the $u$--derivative of (\ref{eq-alpharel}) in the selected patch,
\begin{align}
0&=\partial_u\:\left(\alpha_1(u,\tau)^4+1-\alpha_3(u,\tau)^2-a\:\alpha_1(u,\tau)^2\right),
\end{align}
one finds that the two correlators take the same value:
\begin{align}
\langle\Omega_{\mathcal{S}}\rangle&=\langle\Omega_{\mathcal{L}}\rangle=f(u,\tau)=q(\tau)\frac{1}{2}\frac{\partial_u\alpha_3(u,\tau)}{2\alpha_1(u,\tau)^3-a(\tau)\alpha_1(u,\tau)}
\end{align}
One can show that $\langle\Omega_{\mathcal{S}/\mathcal{L}}\rangle=1$ is satisfied if $u$ is a flat coordinate on the Jacobian \cite{Brunner:2004mt}. On the torus, the holomorphic one--form looks as follows:
\begin{align}
\label{eq-hol1form}
&\eta=q(\tau)\int_{\mathcal{C}}\frac{\omega}{W}, \qquad\textrm{where}\quad\omega=\sum_{i=1}^{3}(-1)^ix_i\mathrm{d}x_1\wedge\widehat{\mathrm{d}x_i}\wedge\mathrm{d}x_3,
\end{align}
and $\mathcal{C}$ is a contour winding around the hypersurface $W=0$. In our local patch, where $W=W(\alpha_1,1,
\alpha_3)$, we have $\omega=-\mathrm{d}\alpha_1\wedge\mathrm{d}\alpha_3$. We can solve this contour integral using the residue theorem: $\int_{\mathcal{C}}\frac{\mathrm{d}W}{W}=1$. This is due to the fact that $W$ is zero along the torus, which implies that $\frac{1}{W}$ has a first order pole on the hypersurface. From the relation $\mathrm{d}W=\sum_{i=1}^{3}\frac{\partial W}{\partial \alpha_i}\mathrm{d}\alpha_i\vert_{\alpha_2=1}$ we obtain:
\begin{align}
\mathrm{d}\alpha_1&=\frac{\mathrm{d}W-\frac{\partial W}{\partial\alpha_3}\mathrm{d}\alpha_3}{\frac{\partial W}{\partial \alpha_1}}
\end{align}
Inserting this into (\ref{eq-hol1form}) and using the residue formula, we get:
\begin{align}
&\eta=q(\tau)\frac{\mathrm{d}\alpha_3}{\partial_{\alpha_1}W(\alpha_1,1,\alpha_3)}=\frac{q(\tau)}{2}\frac{\mathrm{d}\alpha_3}{2\alpha_1(u,\tau)^3-a\alpha_1(u,\tau)}
\end{align}
The solution of $f(u,\tau)=1$ is thus given by:
\begin{align}
u&=\int_{\infty}^{\alpha_3}\eta,
\end{align}
which shows that this is a flat coordinate on the Jacobian.
\subsection{Uniformization of the $\alpha_i$}
We now give the explicit expression for the functions $\alpha_i(u,\tau)$ in terms of the boundary modulus $u$ and the complex structure modulus $\tau$. For the torus this amounts to computing the mirror map, i.e. we express the coordinates of the B--model in terms of flat coordinates which are the natural variables in the A--model. In the case of the torus these are the complex structure parameter $\tau$, which will be identified with the K\"ahler parameter on the mirror, and the brane positions $u_i$ which will be identified with shift-- and Wilson line moduli in the A--model. \\
The $\alpha_i$ will be expressed by theta functions which give a basis of global sections of line bundles on the torus. The theta functions with characteristics are defined as follows\footnote{We have collected the relevant definitions and identities in appendix \ref{app-theta}.}:
\begin{align}
\label{eq-thetadef}
&\Theta\left[\begin{array}{c}c_1\\c_2\end{array}\right](u,\tau)=\sum_{m\in\mathbb{Z}}q^{(m+c_1)^2/2}e^{2\pi i(u+c_2)(m+c_1)},
\end{align}
where $q=e^{2\pi i \tau}$. According to \cite{Polishchuk:1998db}, the $n$ functions $\theta[\frac{a}{n},0](nu,n\tau)$ with $a\in \mathbb{Z}/n\mathbb{Z}$ are the global sections of degree $n$ line bundles $L^n$. For $n=2$, the following relation holds: 
\begin{align}
\label{eq-2var}
\Theta^4\left[\begin{array}{c}0\\0\end{array}\right](0,2\tau)+\Theta^4\left[\begin{array}{c}\frac{1}{2}\\0\end{array}\right](0,2\tau)-a\:\Theta^2\left[\begin{array}{c}0\\0\end{array}\right](0,2\tau)\Theta^2\left[\begin{array}{c}\frac{1}{2}\\0\end{array}\right](0,2\tau)=0,
\end{align}
where $a$ can be found, for instance, in \cite{Harvey:1988ur,Lerche:1989cs,Giveon:1990ay}. The $a$--parameter defines a map $a:H^+/\Gamma(2)\rightarrow \mathbb{CP}^1$ from the fundamental region of the modular group $\Gamma(2)$ ($H^+$ denotes the upper half plane) to the Riemann sphere, given by $\mathbb{CP}^1$. In terms of the modular invariant $j(\tau)=\frac{1}{q}+744+\ldots$ it is given by the following expression \cite{Giveon:1990ay}: 
\begin{align}
\label{eq-jrel}
j(\tau)=\frac{16(a^2+12)^3}{(a^2-4)^2}
\end{align}
Here, one has to be careful to choose the correct branch of the solution for $a$. \\
Clearly, the relation (\ref{eq-2var}) is not what we are looking for, because if we identify $\alpha_1=\Theta[0,0](0,2\tau)$ and $\alpha_2=\Theta[1/2,0](0,2\tau)$, the relation is $\alpha_1^4+\alpha_2^4-a\alpha_1^2\alpha_2^2=0$ instead of (\ref{eq-alpharel}). This is the relation for the two--variable version of the quartic superpotential. One sees that (\ref{eq-2var}) is only satisfied at a single point, $u=0$, in the brane moduli space.\\\\
For the three--variable quartic torus we need to uniformize the $\alpha_i$ to satisfy (\ref{eq-alpharel}). The correct basis of theta functions is given by the Jacobi theta functions:
\begin{align}
\label{eq-jacobidef}
&\Theta_1(u,\tau)\equiv\Theta\left[\begin{array}{c}\frac{1}{2}\\\frac{1}{2}\end{array}\right](u,\tau)\quad \Theta_2(u,\tau)\equiv\Theta\left[\begin{array}{c}\frac{1}{2}\\0\end{array}\right](u,\tau) \nonumber\\
&\Theta_3(u,\tau)\equiv\Theta\left[\begin{array}{c}0\\0\end{array}\right](u,\tau)\quad\Theta_4(u,\tau)\equiv\Theta\left[\begin{array}{c}0\\\frac{1}{2}\end{array}\right](u,\tau)
\end{align}
It turns out that the correct solution looks as follows:
\begin{align}
\label{eq-uniformization}
&\alpha_1(u,\tau)=\Theta_1(2u,2\tau)\qquad \alpha_2(u,\tau)=\Theta_4(2u,2\tau)
\nonumber \\
&\alpha_3(u,\tau)=\frac{\Theta_4^2(2\tau)}{\Theta_2(2\tau)\Theta_3(2\tau)}\Theta_2(2u,2\tau)\Theta_3(2u,2\tau),
\end{align}
where we define $\Theta_i(\tau)\equiv\Theta_i(0,\tau)$.\\
Furthermore we write the parameter $a$ as:
\begin{align}
a&=\frac{\Theta_2^4(2\tau)+\Theta_3^4(2\tau)}{\Theta_2^2(2\tau)\Theta_3^2(2\tau)}
\end{align}
One can check that this expression also satisfies (\ref{eq-jrel}).\\
We can now show that with these definitions (\ref{eq-alpharel}) is satisfied. The most elegant way is to prove this analytically, using the following quadratic identities for the theta functions (see for example \cite{FarkasKra}):
\begin{align}
\Theta_3^2(u,\tau)\Theta_4^2(\tau)&=\Theta_4^2(u,\tau)\Theta_3^2(\tau)-\Theta_1^2(u,\tau)\Theta_2^2(\tau)\nonumber\\
\Theta_2^2(u,\tau)\Theta_4^2(\tau)&=\Theta_4^2(u,\tau)\Theta_2^2(\tau)-\Theta_1^2(u,\tau)\Theta_3^2(\tau)
\end{align}
Note that for the quartic curve the uniformization is slightly more complicated than for the cubic curve. This is due to the fact that not all the variables have the same weight. As a consequence, $\alpha_3(u,\tau)$, which has twice the weight of $\alpha_1(u,\tau)$ and $\alpha_2(u,\tau)$, has to be expressed as a composite of theta functions. It is to be expected that this is always the case whenever a quadratic term is added to the superpotential.
\subsection{Three--point correlators}
We now calculate the correlators (\ref{eq-fercorr})-(\ref{eq-boscorr}) using the Kapustin formula (\ref{eq-kapustin2}). 
Plugging the states into the residue formula, the actual value of the correlator will be multiplied by a rational function of the $\alpha_i$ which can be absorbed into the normalization of the states. Since this task is quite complicated we will simplify the problem stepwise:
\begin{itemize}
\item Simplify the open string states using theta function identities.
\item Insert these simplified states into (\ref{eq-kapustin2}) and make use of more identities for theta functions to identify the correlators.
\end{itemize}
Performing these steps, we will be able to extract the values of the correlators without knowing the exact normalization. In general, one needs to know the precise normalization in order to make the results comparable to the A--model. Without further input it cannot be determined. Our approach is to pull out factors $\alpha_{1/2}(u_i+u_j+u_k,\tau)$ from every contribution to the Hessian, which are expected to be the correct values for the correlators. We then verify that the results are correct by comparing with the results coming from the mirror calculation.
We will now give a more detailed description of the steps mentioned above.

\subsubsection*{Simplification of the open string states}
The open string states given above are quite complicated matrices whose entries contain sums of quotients of the $\alpha_i$. We can use the uniformization in terms of theta functions to simplify these expressions. In particular, we can apply theta function identities such that the $\alpha_i$--expressions give only one quotient instead of sums and we can pull out common factors which we may throw away, since the correlators are only defined up to factors in the $\alpha_i$. For the correlators it is thus possible to reduce the number of terms by a factor of 8. For this we made use of the addition formulas for the theta functions (see for instance \cite{Mumford1}). We collected the most important ones in the appendix. For the simplification of the states we used (\ref{eq-simp1}), (\ref{eq-simp2}).

\subsubsection*{The Correlators}
By plugging the simplified open string states into (\ref{eq-kapustin2}) and making further manipulations with theta function identities, we can pull out a factor in all terms in the supertrace which contribute to the Hessian. In that case we need to apply the most general identities for our theta functions. In particular, we make extensive use of (\ref{eq-mytheta}).\\\\
Here we compute the correlators which involve two long branes and one short brane. Applying the theta function identities, one sees that one term on the rhs of (\ref{eq-mytheta}) always vanishes\footnote{This comes from $\Theta_1(0,\tau)=0$.}, leaving us with a term of the form $\Theta_{(1/4)}(u_1+u_2+u_3,\tau)$ and some factors coming from the bad normalization of the states. Thus, up to the normalization factor, we find the following results for the correlators:
\begin{align}
\label{eq-corrfff}
\langle\psi_{\mathcal{L}\mathcal{L}}(u_1,u_2)\psi_{\mathcal{L}\mathcal{S}}(u_2,u_3)\psi_{\mathcal{S}\mathcal{L}}(u_3,u_1)\rangle&\sim\Theta_1(2(u_1+u_2-u_3),2\tau)\nonumber \\
\langle\bar{\psi}_{\mathcal{L}\mathcal{L}}(u_1,u_2)\psi_{\mathcal{L}\mathcal{S}}(u_2,u_3)\psi_{\mathcal{S}\mathcal{L}}(u_3,u_1)\rangle&\sim\Theta_4(2(u_1+u_2-u_3),2\tau)
\end{align}
\begin{align}
\label{eq-corrfbb}
\langle \psi_{\mathcal{L}\mathcal{L}}(u_1,u_2)\phi_{\mathcal{L}\mathcal{S}}(u_2,u_3)\phi_{\mathcal{S}\mathcal{L}}(u_3,u_1)\rangle&\sim\Theta_1(2(u_1+u_2+u_3),2\tau)\nonumber\\
\langle \bar{\psi}_{\mathcal{L}\mathcal{L}}(u_1,u_2)\phi_{\mathcal{L}\mathcal{S}}(u_2,u_3)\phi_{\mathcal{S}\mathcal{L}}(u_3,u_1)\rangle&\sim\Theta_4(2(u_1+u_2+u_3),2\tau)
\end{align}
\begin{align}
\langle\phi_{\mathcal{L}\mathcal{L}}(u_1,u_2)\psi_{\mathcal{L}\mathcal{S}}(u_2,u_3)\phi_{\mathcal{S}\mathcal{L}}(u_3,u_1)\rangle&\sim\Theta_1(2(u_1-u_2+u_3),2\tau)\nonumber\\
\langle\bar{\phi}_{\mathcal{L}\mathcal{L}}(u_1,u_2)\psi_{\mathcal{L}\mathcal{S}}(u_2,u_3)\phi_{\mathcal{S}\mathcal{L}}(u_3,u_1)\rangle&\sim\Theta_4(2(u_1-u_2+u_3),2\tau)
\end{align}
\begin{align}
\label{eq-corrbbf}
\langle\phi_{\mathcal{L}\mathcal{L}}(u_1,u_2)\phi_{\mathcal{L}\mathcal{S}}(u_2,u_3)\psi_{\mathcal{S}\mathcal{L}}(u_3,u_1)\rangle&\sim\Theta_1(2(u_1-u_2-u_3),2\tau)\nonumber \\
\langle\bar{\phi}_{\mathcal{L}\mathcal{L}}(u_1,u_2)\phi_{\mathcal{L}\mathcal{S}}(u_2,u_3)\psi_{\mathcal{S}\mathcal{L}}(u_3,u_1)\rangle&\sim\Theta_4(2(u_1-u_2-u_3),2\tau)
\end{align}
The correlators always come in pairs. One correlator vanishes for $u_i=0$, the other one does not. In the A--model these correlators correspond to the two triangles that can be enclosed by two long branes and one short brane in the fundamental domain of the torus. Furthermore note that the pairs of correlators differ only by the relative signs of the $u_i$. The reason is that a fermion stretching between brane $\mathcal{A}$ and brane $\mathcal{B}$ corresponds to a boson stretching between $\mathcal{A}$ and the antibrane $\bar{\mathcal{B}}$. Branes have opposite orientation as compared to antibranes in the A--model picture which amounts to a relative sign change in the $u_i$.\\
The calculation of the correlators involving two short branes and one long brane is more involved and seems to require the knowledge of the exact normalization of the states. We thus refrain from computing these here. 

\section{The ``exceptional'' D2--brane}
\label{sec-quarticd2}
As promised, we will now give an explanation why the matrix factorization (\ref{eq-long1}), (\ref{eq-long2}) gives a more convenient description for the long branes than the ``canonical'' factorization $W=\sum_i w_i x_i\frac{\partial W}{\partial x_i}$. This construction yields a $4\times 4$ matrix factorization $Q_i=\left(\begin{array}{cc}0&E_i\\J_i&0\end{array}\right)$ with:\footnote{Note that this factorization does not come exactly from $W=\sum_i q_i x_i\frac{\partial W}{\partial x_i}$. There is an additional term proportional to $x_1x_2$ in the entries with $x_3$. Altering the structure in this way does not change the properties of this factorization but it has the effect that the $\alpha_i$--dependent prefactors are just quotients of the $\alpha_i$ and not rational functions. This simplifies the calculations tremendously.}
\begin{align}
E_i&=\left(\begin{array}{cccc}
\alpha_1^ix_1&\alpha_2^ix_2&\alpha_3^ix_3+\frac{(\alpha_3^i)^2}{\alpha_1^i\alpha_2^i}x_1x_2&0\\
\frac{1}{\alpha_2^i}x_2^3-\frac{\alpha_2^i}{(\alpha_1^i)^2}x_1^2x_2&-\frac{1}{\alpha_1^i}x_1^3+\frac{\alpha_1^i}{(\alpha_2^i)^2}x_1x_2^2&0&\alpha_3^ix_3+\frac{(\alpha_3^i)^2}{\alpha_1^i\alpha_2^i}x_1x_2\\
\frac{1}{\alpha_3^i}x_3-\frac{1}{\alpha_1^i\alpha_2^i}x_1x_2&0&\frac{1}{\alpha_1^i}x_1^3-\frac{\alpha_1^i}{(\alpha_2^i)^2}x_1x_2^2&\alpha_2^ix_2\\
0&\frac{1}{\alpha_3^i}x_3-\frac{1}{\alpha_1^i\alpha_2^i}x_1x_2 &\frac{1}{\alpha_2^i}x_2^3-\frac{\alpha_2^i}{(\alpha_1^i)^2}x_1^2x_2&-\alpha_1^i x_1
\end{array}\right)
\\
J_i&=\left(\begin{array}{cccc}
\frac{1}{\alpha_1^i}x_1^3-\frac{\alpha_1^i}{(\alpha_2^i)^2}x_1x_2^2&\alpha_2^ix_2&-\alpha_3^ix_3-\frac{(\alpha_3^i)^2}{\alpha_1^i\alpha_2^i}x_1x_2&0\\
\frac{1}{\alpha_2^i}x_2^3-\frac{\alpha_2^i}{(\alpha_1^i)^2}x_1^2x_2&-\alpha_1^ix_1&0&-\alpha_3^ix_3-\frac{(\alpha_3^i)^2}{\alpha_1^i\alpha_2^i}x_1x_2\\
-\frac{1}{\alpha_3^i}x_3+\frac{1}{\alpha_1^i\alpha_2^i}x_1x_2&0&\alpha_1^ix_1&\alpha_2^ix_2\\
0&-\frac{1}{\alpha_3^i}x_3+\frac{1}{\alpha_1^i\alpha_2^i}x_1x_2&\frac{1}{\alpha_2^i}x_2^3-\frac{\alpha_2^i}{(\alpha_1^i)^2}x_1^2x_2&-\frac{1}{\alpha_1^i}x_1^3+\frac{\alpha_1^i}{(\alpha_2^i)^2}x_1x_2^2
\end{array}\right)
\end{align}
A straightforward calculation shows that this matrix factorization, together with the factorization (\ref{eq-short1}), (\ref{eq-short2}) for the short branes, yields the same spectrum as depicted in figure \ref{fig-bquartic1}. Thus, this factorization also represents the long branes. However, there is a catch: Let us compute the correlator $\langle\Omega\rangle$ of the marginal boundary preserving operator $\Omega=\partial_uQ$. In order to do this, we insert it into the residue formula for the three--point function (\ref{eq-kapustin2}), which for this case looks as follows:
\begin{align}
\langle\Omega\rangle&=\int\frac{\mathrm{str}\left(\frac{1}{3!}(\mathrm{d}Q)^{\wedge 3}\partial_u Q\right)}{\partial_1W\partial_2W\partial_3W}
\end{align} 
In order to give something non--vanishing, the supertrace should be proportional to the Hessian. Inserting into this formula, one finds that the supertrace is identically 0. \\
We interpret this as follows: Since $\Omega$ is the derivative of $Q$ with respect to the boundary modulus the vanishing of this correlator implies that the boundary modulus for this matrix factorization has a fixed value. Such matrix factorizations have already been discussed in \cite{Hori:2004ja,Govindarajan:2005im}. They are interpreted as a single rigid D$2$ brane wrapping the torus.\\
Although these special points in moduli space are an interesting issue, we do not want to restrict ourselves to a specific value of the boundary modulus but rather find the most general expression for the long branes. The discussion in \cite{Hori:2004ja,Govindarajan:2005im} implies that we have to add a pair of $D0\overline{D0}$ branes to the system. Then one of the branes can move freely on the torus, while its antibrane remains fixed and thus we have restored the boundary modulus as the relative distance between the D$0$--brane and the $\overline{D0}$--brane. The results of \cite{Hori:2004ja,Govindarajan:2005im} tell us that this can be done perturbing this matrix with the marginal boundary fermion, which will yield a reducible matrix factorization. \\
In order to find an expression for this operator (which is not $\partial_uQ$ but an equivalent description) we go to the Gepner point. There, the Landau--Ginzburg description of the torus is a tensor product of two $A_2$ minimal models and one (trivial) $A_0$ piece: $A_2(x_1)\otimes A_2(x_2)\otimes A_0(x_3)$. Our matrix factorization has the following form at the Gepner point (see also: \cite{Dell'Aquila:2005jg}): $Q_{gep}=\left(\begin{array}{cc}0&E_{gep}\\J_{gep}&0\end{array}\right)$:
\begin{align}
E_{gep}=\left(\begin{array}{cccc}
\label{eq-long-gep}
x_1&x_2&x_3&0\\
x_2^3&-x_1^3&0&x_3\\
x_3&0&x_1^3&x_2\\
0&x_3&x_2^3&-x_1
\end{array}\right)\qquad
J_{gep}=\left(\begin{array}{cccc}
x_1^3&x_2&-x_3&0\\
x_2^3&-x_1&0&-x_3\\
-x_3&0&x_1&x_2\\
0&-x_3&x_2^3&-x_3^3
\end{array}\right)
\end{align}
There is a unique fermionic state of weight $1$, which is the tensor product of the highest weight fermions of the two $A_2$ minimal models. We identify this state with the state $\Omega$: $\Omega_{gep}=\left(\begin{array}{cc}0&\Omega^{(0)}_{gep}\\\Omega^{(1)}_{gep}&0\end{array}\right)$, where
\begin{align}
\Omega^{(0)}_{gep}=\left(\begin{array}{cccc}
0&0&0&1\\
0&0&-x_1^2x_2^2&0\\
0&x_1^2&0&0\\
-x_2^2&0&0&0
\end{array}\right)\qquad
\Omega^{(1)}_{gep}\left(\begin{array}{cccc}
0&0&0&x_1^2\\
0&0&-x_2^2&0\\
0&1&0&0\\
-x_1^2x_2^2&0&0&0
\end{array}\right);
\end{align}
Perturbing (\ref{eq-long-gep}) with this state and turning the moduli back on we find the following reducible matrix factorization: $Q_i^{red}=\left(\begin{array}{cc}0&E_i^{red}\\J_i^{red}&0\end{array}\right)$, where
\begin{align}
E_i^{red}&=\left(\begin{array}{cccc}
x_1&(\alpha_2^i)^2x_2&\frac{1}{\alpha_3^1}x_3+\left(-\frac{1}{\alpha_1^i\alpha_2^i}+\frac{(\alpha_2^i)^3}{\alpha_1^i(\alpha_3^i)^2}\right)x_1x_2&\frac{\alpha_1^i}{\alpha_2^i}\\
(\alpha_2^i)^2x_2^3+\left(-\frac{(\alpha_2^i)^4}{(\alpha_1^i)^2}+\frac{(\alpha_3^i)^2}{(\alpha_1^i)^2}\right)x_1^2x_2&-(\alpha_3^i)^2x_1^3&-\frac{\alpha_1^i}{\alpha_2^i}x_1^2x_2^2&\alpha_3^ix_3\\
\frac{1}{\alpha_3^i}x_3&\frac{\alpha_1^i}{\alpha_2^i}x_1^2&\frac{1}{(\alpha_3^i)^2}x_1^3&\frac{1}{(\alpha_2^i)^2}x_2\\
-\frac{\alpha_1^i}{\alpha_2^i}x_2^2&\mu_1&\mu_2&-x_1
\end{array}\right),
\end{align}
where $\mu_1=\alpha_3^ix_3+\left(-\frac{(\alpha_2^i)^3}{\alpha_1^i}+
\frac{(\alpha_3^i)^2}{\alpha_1^i\alpha_2^i}\right)x_1x_2$ and $\mu_2=\frac{1}{(\alpha_2^i)^2}x_2^3+\left(\frac{1}{(\alpha_1^i)^2}-\frac{(\alpha_2^i)^4}{(\alpha_1^i)^2(\alpha_3^i)^2}\right)x_1^2x_2$.
\begin{align}
J_i^{red}&=\left(\begin{array}{cccc}
x_1^3&\frac{1}{(\alpha_2^i)^2}&-\alpha_3^1x_3&\frac{\alpha_1^i}{\alpha_2^i}x_1^2\\
\mu_2&-\frac{1}{(\alpha_3^1)^3}x_1&-\frac{\alpha_1^i}{\alpha_2^1}&-\frac{1}{\alpha_3^1}x_3-\left(-\frac{1}{\alpha_1^i\alpha_2^i}+\frac{(\alpha_2^i)^3}{\alpha_1^i(\alpha_3^i)^2}\right)x_1x_2\\
\mu_1&\frac{\alpha_1^i}{\alpha_2^i}&(\alpha_3^i)^2x_1&(\alpha_2^i)^2x_2\\
-\frac{\alpha_1^i}{\alpha_2^i}x_1^2x_2^2&-\frac{1}{\alpha_3^i}x_3&(\alpha_2^i)^2x_2^3+\left(-\frac{(\alpha_2^i)^4}{(\alpha_1^i)^2}+\frac{(\alpha_3^i)^2}{(\alpha_1^i)^2}\right)x_1^2x_2&-x_1^3
\end{array}\right)
\end{align}
These manipulations leave the spectrum unchanged, except for the condition $\langle\Omega\rangle\neq 0$. This matrix factorization is clearly reducible, since it contains two terms which are independent of $x_1,x_2$. Thus, we can make row-- and column manipulations to transform this factorization into a lower--dimensional one. A few steps of elementary operations yield the following simple result for $E_i^{red}$ (analogous steps lead to a corresponding expression for $J_i^{red}$): $E_i^{red}=\left(\begin{array}{cc}0&A_i\\B_i&0\end{array}\right)$, where
\begin{align}
A_i&=\left(\begin{array}{cc}
0&\frac{\alpha_1^i}{\alpha_2^i}\\
\frac{\alpha_2^i}{\alpha_1^i}\left(x_1^4+x_2^4-x_3^2+x_1^2x_2^2\left(-\frac{(\alpha_1^i)^2}{(\alpha_2^i)^2}-\frac{(\alpha_2^i)^2}{(\alpha_1^i)^2}+\frac{(\alpha_3^i)^2}{(\alpha_1^i)^2(\alpha_2^i)^2}\right)\right)&0
\end{array}\right)
\end{align}
\begin{align}
B_i=\left(\begin{array}{cc}\frac{1}{\alpha_3^1}x_3-\frac{1}{\alpha_1^i\alpha_2^i}x_1x_2&\frac{\alpha_1^i}{\alpha_2^i}x_1^2-\frac{\alpha_2^i}{\alpha_1^i}x_2^2\\
\frac{\alpha_2^i}{\alpha_1^i}x_1^2-\frac{\alpha_1^i}{\alpha_2^i}x_2^2&\alpha_3^ix_3+\frac{(\alpha_3^i)^2}{\alpha_1^i\alpha_2^i}x_1x_2\end{array}\right)
\end{align}
Thus, the canonical $4\times 4$ matrix factorization for the long branes at a generic point in moduli space is isomorphic to the given $2\times 2$ factorization (\ref{eq-long1}), (\ref{eq-long2}).
\section{The A--Model and Homological Mirror Symmetry}
\label{sec-quartica}
\begin{figure}
\begin{center}
\includegraphics{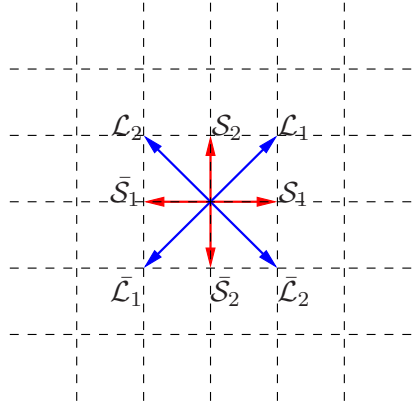}
\end{center}
\caption{'Long' and 'short' branes in the A--model.}\label{fig-abranes}
\end{figure}

\begin{figure}
\begin{center}
\includegraphics{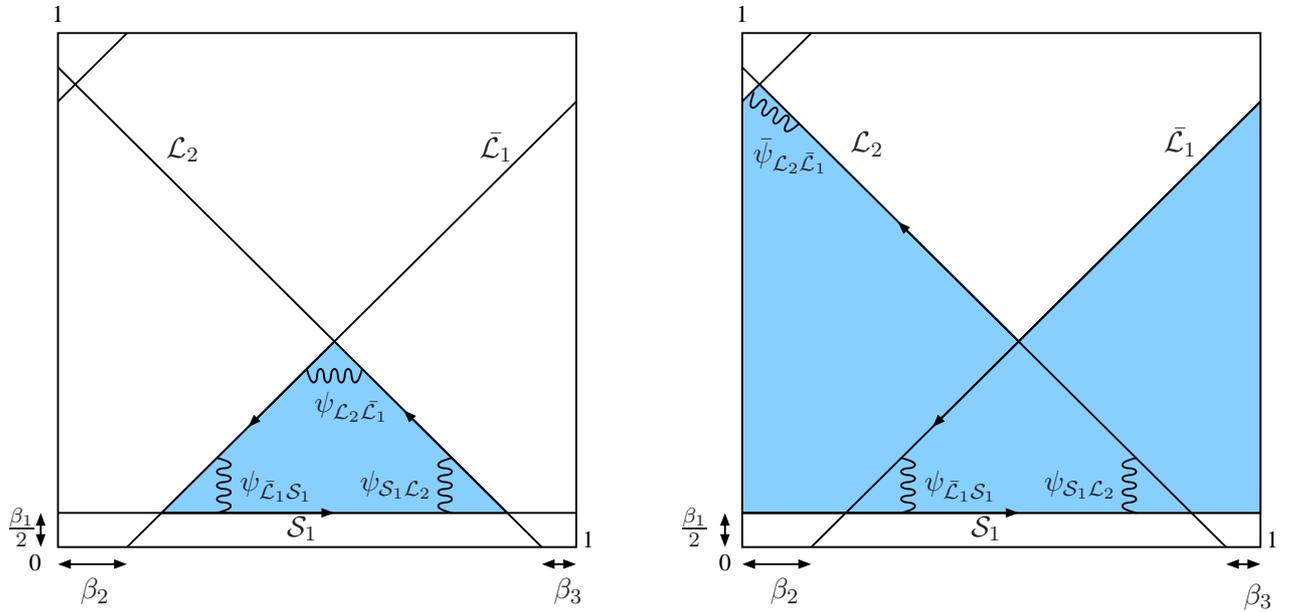}
\end{center}
\caption{The correlators $\langle\psi_{\mathcal{L}_2\bar{\mathcal{L}}_1}\psi_{\bar{\mathcal{L}}_1\mathcal{S}_1}\psi_{\mathcal{S}_1\mathcal{L}_2}\rangle$ and $\langle\bar{\psi}_{\mathcal{L}_2\bar{\mathcal{L}}_1}\psi_{\bar{\mathcal{L}}_1\mathcal{S}_1}\psi_{\mathcal{S}_1\mathcal{L}_2}\rangle$ in the A--model.}\label{fig-corrfff}
\end{figure}

\begin{figure}
\begin{center}
\includegraphics{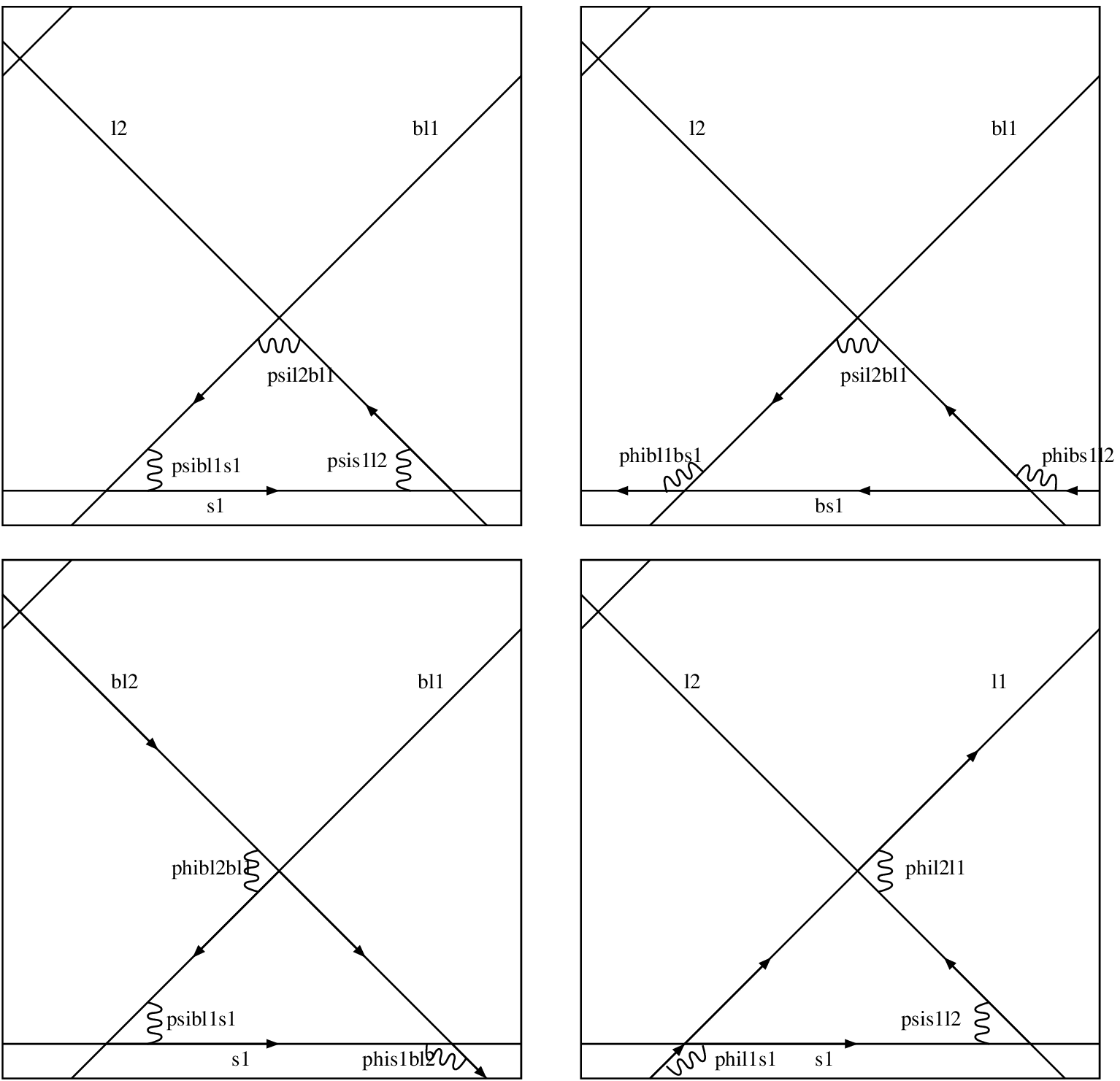}
\end{center}
\caption{The correlators $\langle\psi_{\mathcal{L}_2\bar{\mathcal{L}}_1}\psi_{\bar{\mathcal{L}}_1\mathcal{S}_1}\psi_{\mathcal{S}_1\mathcal{L}_2}\rangle$, $\langle \psi_{\mathcal{L}_2\bar{\mathcal{L}}_1}\phi_{\bar{\mathcal{L}}_1\bar{\mathcal{S}}_1}\phi_{\bar{\mathcal{S}}_1\mathcal{L}_2}\rangle$, $\langle\phi_{\bar{\mathcal{L}}_2\bar{\mathcal{L}}_11}\psi_{\bar{\mathcal{L}}_1\mathcal{S}_1}\phi_{\mathcal{S}_1\bar{\mathcal{L}}_2}\rangle$ and $\langle \phi_{\mathcal{L}_2\mathcal{L}_1}\phi_{\mathcal{L}_1\mathcal{S}_1}\psi_{\mathcal{S}_1\mathcal{L}_2}\rangle$ in the A--model.}\label{fig-acorrs}
\end{figure}

In this section we identify the correlators (\ref{eq-corrfff})--(\ref{eq-corrbbf}) of the B--model with instanton sums in the A--model. On $T^2$ the instantons have a simple geometrical interpretation: they are the areas enclosed by the D$1$--branes winding around the torus. The open string states are at the intersection points of the branes. Even and odd cohomology elements are identified with the angles enclosed by two intersecting branes. We choose the convention that the state is fermionic if the angle is oriented in the counter--clockwise direction. For bosonic states the angle is enclosed in clockwise direction.\\
In the A--model a three--point correlator on the disk is defined as follows (see for example \cite{Govindarajan:2006uy}):
\begin{align}
\label{eq-acorrdef}
C_{ijk}(\tau,\alpha_a,\beta_a)&=\sum_{n=-\infty}^{\infty}e^{2\pi i\tau A_{ijk}^{(n)}(\beta_a)}e^{2\pi i W_{ijk}^{(n)}(\alpha_a)},
\end{align}
where the sum is over all lattice shifts.
Here $\tau$ is the K\"ahler modulus and corresponds to the complex structure modulus in the B--model. The complex structure modulus is fixed to the value $e^{\frac{2\pi i}{4}}$, which makes the fundamental domain of the torus a square with side length~$1$. $A_{ijk}^{(n)}(\beta_a)$ in (\ref{eq-acorrdef}) denotes the area of the instanton depending on shift moduli $\beta_a$ of the branes. The areas are weighted by the total area of the fundamental domain, so that the exponent in (\ref{eq-acorrdef}) is dimensionless. The phases $W_{ijk}^{(n)}(\alpha_a)$ are Wilson line contributions. They are obtained by integrating the flat connection along the circumference of the instanton \cite{Govindarajan:2006uy}. The connection is parameterized by the modulus $\alpha_a$ associated to the brane with index $a$. The lengths are weighted by the length of the brane in the fundamental domain.\\
We start with calculating the correlators (\ref{eq-corrfff}) which have three fermionic insertions. We label the 'long' and 'short' D$1$--branes as depicted in figure \ref{fig-abranes}.\\
In order to calculate the three--point functions of interest we have to pick two 'long' branes and one 'short' brane. For the three--fermion correlator we choose $\bar{\mathcal{L}}_1$, $\mathcal{L}_2$ and $\mathcal{S}_1$. In the B--model this corresponds to the triangles depicted in figure \ref{fig-bquartic2}. With these conventions the correlators $\langle\psi_{\mathcal{L}_2\bar{\mathcal{L}}_1}\psi_{\bar{\mathcal{L}}_1\mathcal{S}_1}\psi_{\mathcal{S}_1\mathcal{L}_2}\rangle$ and $\langle\bar{\psi}_{\mathcal{L}_2\bar{\mathcal{L}}_1}\psi_{\bar{\mathcal{L}}_1\mathcal{S}_1}\psi_{\mathcal{S}_1\mathcal{L}_2}\rangle$ look as shown in figure \ref{fig-corrfff}.
In order to obtain the areas and circumferences of these triangles we have to calculate their side lengths in terms of the shift parameters $\beta_a$. We start by calculating the correlator $\langle\psi_{\mathcal{L}_2\bar{\mathcal{L}}_1}\psi_{\bar{\mathcal{L}}_1\mathcal{S}_1}\psi_{\mathcal{S}_1\mathcal{L}_2}\rangle$ on the lefthand side in figure \ref{fig-corrfff}.  Choosing the origin as indicated in the picture, one gets:
\begin{align}
a^{short}_{\mathcal{L}_2\bar{\mathcal{L}}_1\mathcal{S}_1}&=\frac{1}{\sqrt{2}}(-1+\beta)\nonumber\\
a^{long}_{\mathcal{L}_2\bar{\mathcal{L}}_1\mathcal{S}_1}&=-1+\beta,
\end{align}
where $\beta=-\beta_1+\beta_2+\beta_3$. Taking into account lattice shifts of the long branes, whose distance is $\sqrt{2}$, we get:
\begin{align}
a^{(n),short}_{\mathcal{L}_2\bar{\mathcal{L}}_1\mathcal{S}_1}&=\sqrt{2}\left(n+\frac{1}{2}+\frac{\beta}{2}\right)\nonumber\\
a^{(n),long}_{\mathcal{L}_2\bar{\mathcal{L}}_1\mathcal{S}_1}&=(n+\beta),
\end{align}
The area of the instanton, normalized by the area $1$ of the fundamental domain, is just $A=\frac{1}{2}(a^{short})^2$:
\begin{align}
A^{(n)}_{\mathcal{L}_2\bar{\mathcal{L}}_1\mathcal{S}_1}&=\left(n+\frac{1}{2}+\frac{\beta}{2}\right)^2
\end{align}
In order to calculate the Wilson line contribution, we assign moduli $\alpha_1$, $\alpha_2$ and $\alpha_3$ to the branes $\mathcal{L}_1$, $\mathcal{L}_2$ and $\mathcal{S}_1$, respectively. We furthermore assign negative values of these parameters to the antibranes. Scaling down the side lengths by $\sqrt{2}$ along the diagonal branes and by $1$ along the 'short' brane, we get the following contribution for the phase:
\begin{align}
W^{(n)}_{\mathcal{L}_2\bar{\mathcal{L}}_1\mathcal{S}_1}&=\left(n+\frac{1}{2}+\frac{\beta}{2}\right)\alpha+\left(\frac{\beta}{2}+\frac{1}{2}\right)\alpha_3,
\end{align}
where $\alpha=-\alpha_1+\alpha_2+\alpha_3$. We are now ready to compute the three--point function $C_{\mathcal{L}_2\bar{\mathcal{L}}_1\mathcal{S}_1}$:
\begin{align}
C_{\mathcal{L}_2\bar{\mathcal{L}}_1\mathcal{S}_1}&=\sum_{n=-\infty}^{\infty}e^{2\pi i\tau\left(n+\frac{1}{2}+\frac{\beta}{2}\right)^2}e^{2\pi i \left(n+\frac{1}{2}+\frac{\beta}{2}\right)}e^{2\pi i \left(\frac{\beta}{2}+\frac{1}{2}\right)\alpha_3}\nonumber \\
&=e^{2\pi i\tau \frac{\beta^2}{4}}e^{2\pi i\frac{\beta\alpha}{2}}e^{2\pi i \left(\frac{\beta}{2}+\frac{1}{2}\right)\alpha_3}\sum_{n=-\infty}^{\infty}e^{2\pi i\tau \left(n+\frac{1}{2}\right)^2}e^{2\pi i \left(\beta\tau+\alpha\right)\left(n+\frac{1}{2} \right)}\nonumber\\
&=\mathcal{C}\cdot\Delta_{\mathcal{L}_2\bar{\mathcal{L}}_1\mathcal{S}_1}
\end{align}
Here $\mathcal{C}$ is a non--holomorphic factor which is due to a holomorphic anomaly \cite{Govindarajan:2006uy}. This factor has to be dropped when computing the effective superpotential, which is a purely holomorphic quantity. In the following we will omit this factor, since it is not necessary for the comparison with the B--model quantities.\\
Recently there has been considerable progress in understanding the holomorphic anomaly for the open string case. As was shown in \cite{Bershadsky:1993cx} for the closed string, the holomorphic anomaly can be used to determine higher genus amplitudes because the holomorphic anomaly equations are recursion relations for these amplitudes. In \cite{Eynard:2007hf} such equations were found for open topological strings on non--compact Calabi--Yaus. In \cite{Walcher:2007tp} holomorphic anomaly equations for compact Calabi--Yaus have been found.\\
Now we perform the mirror map in order to make contact with the B--model results. We define:
\begin{align}
\tau\equiv\tau &\qquad 2u\equiv (\beta\tau+\alpha)
\end{align}
The first identity tells us that the K\"ahler modulus of the A--model becomes the complex structure modulus in the B--model. The second identity states that the brane modulus $u$ in the B--model corresponds to a combination of shift and Wilson line moduli in the A--model. Using these definitions and setting $q=e^{2\pi i\tau}$ we get:
\begin{align}
\label{eq-corrfff1}
\Delta_{\mathcal{L}_2\bar{\mathcal{L}}_1\mathcal{S}_1}&=\sum_{n=-\infty}^{\infty}q^{\left(n+\frac{1}{2}\right)^2}e^{2\pi i\left(n+\frac{1}{2}\right)2u}\nonumber\\
&=\Theta_2(2u,2\tau)
\end{align}
In the last line we used the definitions (\ref{eq-thetadef}), (\ref{eq-jacobidef}) for the theta functions. \\
Before comparing to the B--model we also calculate the second correlator with three fermionic insertions, $\langle\bar{\psi}_{\mathcal{L}_2\bar{\mathcal{L}}_1}\psi_{\bar{\mathcal{L}}_1\mathcal{S}_1}\psi_{\mathcal{S}_1\mathcal{L}_2}\rangle$ on the right hand side of figure \ref{fig-corrfff}. In order to see that the area drawn there is a triangle, one simply draws the picture on the covering space of the torus. Calculating the side lengths, one finds:
\begin{align}
\bar{a}^{(n),short}_{\mathcal{L}_2\bar{\mathcal{L}}_1\mathcal{S}_1}&=\sqrt{2}\left(n+\frac{\beta}{2}\right)\nonumber\\
\bar{a}^{(n),long}_{\mathcal{L}_2\bar{\mathcal{L}}_1\mathcal{S}_1}&=(n+\beta)
\end{align}
Summing up the instanton contributions and dropping the non--holomorphic factor, one finds:
\begin{align}
\label{eq-corrfff2}
\bar{\Delta}_{\mathcal{L}_2\bar{\mathcal{L}}_1\mathcal{S}_1}&=\sum_{n=-\infty}^{\infty}q^{n^2}e^{2\pi i\, 2un}=\Theta_3(2u,2\tau)
\end{align}
In order to find agreement with the B--model results we use the following identity for theta functions:
\begin{align}
\label{eq-thetashift}
\Theta_2(u,\tau)&=-\Theta_1(u+\frac{1}{2},\tau)\nonumber\\
\Theta_3(u,\tau)&=\phantom{-}\Theta_4(u+\frac{1}{2},\tau)
\end{align}
In the A--model, such a shift corresponds to a shift of the origin. Implementing these shifts and setting $u\equiv u_1+u_2-u_3$, the correlators (\ref{eq-corrfff1}), (\ref{eq-corrfff2}) are precisely the B--model correlators (\ref{eq-corrfff})\footnote{We cannot determine the overall sign since we do not know the precise normalization of the B--model correlators.}. Actually, the parameters $u_i$ in the B--model are related as follows to the Wilson line moduli $\alpha_i$ in the A--model:
\begin{align}
\label{eq-abshift}
&u_1\longleftrightarrow\alpha_2\qquad u_2\longleftrightarrow\alpha_1\qquad u_3\longleftrightarrow-\alpha_3
\end{align}
Now we discuss the remaining correlators we calculated in the B--model. These have one fermionic and two bosonic insertions. Since a fermion stretching between branes $\mathcal{A}$ and $\mathcal{B}$ is equivalent to a boson going from $\mathcal{A}$ to the antibrane $\bar{\mathcal{B}}$ we may obtain the correlators involving bosons from those with only fermionic insertions in the following way. If we start with a correlator with three fermionic insertions, we get a correlator with two bosons and one fermion, if we replace one of the three branes by its antibrane. Starting with the correlator $\langle\psi_{\mathcal{L}_2\bar{\mathcal{L}}_1}\psi_{\bar{\mathcal{L}}_1\mathcal{S}_1}\psi_{\mathcal{S}_1\mathcal{L}_2}\rangle$, which corresponds to the smaller triangle in figure \ref{fig-corrfff}, we can obtain from this the correlators 
\begin{align}
\langle \psi_{\mathcal{L}_2\bar{\mathcal{L}}_1}\phi_{\bar{\mathcal{L}}_1\bar{\mathcal{S}}_1}\phi_{\bar{\mathcal{S}}_1\mathcal{L}_2}\rangle\qquad \langle\phi_{\bar{\mathcal{L}}_2\bar{\mathcal{L}}_11}\psi_{\bar{\mathcal{L}}_1\mathcal{S}_1}\phi_{\mathcal{S}_1\bar{\mathcal{L}}_2}\rangle\qquad
\langle \phi_{\mathcal{L}_2\mathcal{L}_1}\phi_{\mathcal{L}_1\mathcal{S}_1}\psi_{\mathcal{S}_1\mathcal{L}_2}\rangle 
\end{align}
by exchanging branes with antibranes.
The corresponding configurations in the A--model are shown in figure \ref{fig-acorrs}.
Similar pictures can be drawn for the correlators with the second open string state $\bar{\psi}_{\mathcal{L}\mathcal{L}}$, which stretches between two long branes.\\
The correlators including bosons seem to correspond to unoriented triangles enclosed by the three branes. On the torus, however these correspond to triangles with a definite orientation. In order to see this, remember that the open string states correspond to angles between two branes and have a certain orientation. The states are indicated by wavy lines in the picture. In order to  see the correlators we have to follow the lines and take a turn to another line if there is an open string state between two branes. The angle corresponding to the state must be enclosed by the path. For the correlators including bosons one sees that the path leaves the fundamental domain at one point, and enters on the other side, by the periodicity on the lattice. Thus, for these correlators, one brane winds once around the torus before coming back to close the triangle, which is then oriented. The area of such a triangle is the area of the triangle corresponding to ther fermionic correlator plus the area of the full torus. Since we sum over all lattice shifts this this does not change the area part of the instanton sum. What does change is the contribution to the Wilson lines since we associate negative connection parameters to the antibranes. Computing the correlators we find:
\begin{align}
\Delta_{\mathcal{L}_2\bar{\mathcal{L}}_1\bar{\mathcal{S}}_1}&=\sum_{n=-\infty}^{\infty}q^{\left(n+\frac{1}{2}\right)^2}e^{2\pi i\left(n+\frac{1}{2}\right)(\beta\tau+(-\alpha_1+\alpha_2-\alpha_3)}=\Theta_2(2u_a,2\tau)\nonumber \\
\bar{\Delta}_{\mathcal{L}_2\bar{\mathcal{L}}_1\bar{\mathcal{S}}_1}&=\sum_{n=-\infty}^{\infty}q^{n^2}e^{2\pi i n(\beta\tau+(-\alpha_1+\alpha_2-\alpha_3))}=\Theta_3(2u_a,2\tau)\nonumber \\
\Delta_{\bar{\mathcal{L}}_2\bar{\mathcal{L}}_1\mathcal{S}_1}&=\sum_{n=-\infty}^{\infty}q^{\left(n+\frac{1}{2}\right)^2}e^{2\pi i\left(n+\frac{1}{2}\right)(\beta\tau+(-\alpha_1-\alpha_2+\alpha_3))}=\Theta_2(2u_b,2\tau)\nonumber \\
\bar{\Delta}_{\bar{\mathcal{L}}_2\bar{\mathcal{L}}_1\mathcal{S}_1}&=\sum_{n=-\infty}^{\infty}q^{n^2}e^{2\pi i n(\beta\tau+(-\alpha_1-\alpha_2+\alpha_3))}=\Theta_3(2u_b,2\tau)\nonumber\\
\Delta_{\mathcal{L}_2\mathcal{L}_1\mathcal{S}_1}&=\sum_{n=-\infty}^{\infty}q^{\left(n+\frac{1}{2}\right)^2}e^{2\pi i\left(n+\frac{1}{2}\right)(\beta\tau+(\alpha_1+\alpha_2+\alpha_3))}=\Theta_2(2u_c,2\tau)\nonumber \\
\bar{\Delta}_{\mathcal{L}_2\mathcal{L}_1\mathcal{S}_1}&=\sum_{n=-\infty}^{\infty}q^{n^2}e^{2\pi i n(\beta\tau+(\alpha_1+\alpha_2+\alpha_3))}=\Theta_3(2u_c,2\tau)
\end{align}
The definitions of $u_a,u_b,u_c$ can be read off from the sums. Using the correspondence (\ref{eq-abshift}) and (\ref{eq-thetashift}) we find agreement with (\ref{eq-corrfbb})--(\ref{eq-corrbbf}), thus verifying homological mirror symmetry.

\chapter{Outlook}
\label{chap-outlook}
In this thesis we discussed B--type D--branes in Landau--Ginzburg models which are realized in terms matrix factorizations of the Landau--Ginzburg superpotential. Since this subject is still relatively new, there are many open issues which call for a better understanding. \\
One central point is to find methods to compute the effective superpotential for examples which are more complicated than minimal models. One possibility would be to generalize the Massey product algorithm we discussed in chapter \ref{chap-effsupo}. The formalism would have to be extended in various ways. One issue is to generalize the formalism to handle systems of multiple D--branes. Such a generalization seems possible \cite{math.AG/0303166}. In the B--model we have do deal with Landau--Ginzburg orbifolds, so one has to find out if the Massey product algorithm is compatible with the orbifold action. Results in the mathematics literature imply that the odds are good concerning the generalization to orbifolds \cite{Siqveland3}. A probably more challenging problem is to include deformations of bosonic states into the Massey product formalism. This would lead to a noncommutative structure in the space of deformations. Finally one has to find a way to include moduli into the algorithm. \\
A different approach to obtain the effective superpotential would be to compute all the disk amplitudes. This may be done by extending the consistency constraints for open topological strings. Although this task is an interesting challenge, it is questionable if this approach is really applicable for complicated examples since the number of equations increases rapidly with the number of possible insertions of a correlator. \\
In \cite{Herbst:2006nn} the effective superpotential on the torus was computed by calculating all the $n$--point functions on the disk in the A--model. On the torus the A--model calculation is easy due to the simple geometric structure of the A--branes and the instantons on the torus. For more complicated Calabi--Yau manifolds is seems unlikely that one can calculate these amplitudes in the A--model. The way out would by to calculate these amplitudes in the B--model and then use homological mirror symmetry to obtain the instanton--corrected amplitudes in the A--model. This leads us to a large field of open problems in connection with homological mirror symmetry. Open string mirror symmetry is quite well understood for non--compact Calabi--Yaus but almost nothing is known for the compact case. See \cite{Walcher:2006rs} for recent results on the quintic.\\
A rather urgent problem in the context of homological mirror symmetry is that there are by now no methods known how to calculate disk amplitudes with more than three operator insertions in the B--model, not to mention higher genus amplitudes for compact Calabi--Yaus. Recently, open string amplitudes in the B--model have been computed for non--compact Calabi--Yaus using matrix model techniques \cite{Marino:2006hs}. In the closed string theory these amplitudes can be obtained by solving certain differential equations, but in the open string case no such differential equations are known.\\
A further issue is to find the mirror map for the open string case. It is not known how to construct this map in general. In the closed string case this is related to finding flat coordinates on the moduli space which is in turn linked to $N=2$ special geometry and the Picard--Fuchs differential equations. There exists a notion of $N=1$ special geometry (see e.g. \cite{Mayr:2001xk,Lerche:2001cw}), but the construction has only been explicitly realized for non--compact Calabi--Yau spaces with 'toric' D--branes.\\
Many of the open problems we have discussed so far are not directly related to matrix factorizations, although the beauty of this formalism makes it hard to believe that matrix factorizations will not play a role in understanding at least some of these problems. There are also many things to do in order to get a better understanding of D--branes characterized by matrix factorizations. For all but the simplest cases the classification of matrix factorizations is an unresolved issue. A related problem is how to make contact between the non--geometric B--branes at the Landau--Ginzburg point and the B--branes at the large radius point \cite{Aspinwall:2006ib,Aspinwall:2007cs,hhp}.\\
Most of the discussions in this thesis were concerned with B--branes. These are relatively well understood due to their description in terms of matrix factorizations or coherent sheaves. Compared to the well--founded, although not complete, knowledge on B--branes, almost nothing is known about the A--branes. Finding out more about A--branes is a challenging task. What is even more interesting is mirror symmetry in this context, that is, given a B--brane, how does one obtain the mirror A--brane? So far, this problem has been solved only for a few examples like the torus.

\appendix
\chapter{Implementing the Consistency Conditions in Mathematica}
\label{app-mathematica}
We now describe how to implement the consistency constraints for open topological strings which were discussed in section \ref{sec-constr-effsupo}. We refrain from giving the full code but give a short description of the various routines and discuss how they can be implemented. We will label the most important programs with a $({}^{\ast})$.
\section{Input Data and Bookkeeping}
We need to know the following data as input for the programs.
\begin{itemize}
\item One or more matrix factorizations of a Landau--Ginzburg superpotential and the open string spectrum.
\item $R$--charges and suspended $\mathbbm{Z}_{2}$ degree of all the open string states and of the integrated insertions.
\item If bulk insertions are included, we need the chiral ring and the $R$--charges of its elements as well as the three--point correlators in the bulk.
\item The parameters which specify the model: dimensions of bulk and boundary cohomology, background charge, the dual Coxeter number, the suspended grade of the boundary metric.
\end{itemize}
The basic input data is stored conveniently in three lists:
\begin{align}
\mathtt{states=}&\mathtt{\big\{\{1,q_{bos_1},sbrane,ebrane,bos_1\},\ldots,\{0,q_{fer_1},sbrane,ebrane,fer_1\},\ldots} \nonumber \\
&\mathtt{\ldots,\{0,q_{bulk_1},0,0,bulk_1\},\ldots\big\}}\nonumber\\
\mathtt{statesint=}&\mathtt{\big\{\{1,q_{bos_1}-1,sbrane,ebrane\},\ldots,\{0,q_{fer_1}-1,sbrane,ebrane\},\ldots} \nonumber \\
&\mathtt{\ldots,\{0,q_{bulk_1}-2,0,0\},\ldots\big\}}\nonumber \\
\mathtt{brst=}&\mathtt{\{Q_1,\ldots,Q_n\}}
\end{align}
We thus have two lists containing the data of the states and the integrated insertions. We choose the conventions that we list first the even boundary states, then the odd ones and in the end the bulk states. Each state is characterized by its suspended grade, its $R$--charge $q$, the brane {\tt sbrane} where, the open string starts and the brane {\tt ebrane}, where it ends, and, for unintegrated insertions, the value. \\
We now describe some useful routines for bookkeeping:
\subsection{{\tt degree}}
{\bf Purpose:} Return the $\mathbbm{Z}_2$--grade of a state.\\
{\bf Input:} An (integrated) state, i.e. {\tt states(int)[[1]]}.\\
{\bf Output:} The value of the $\mathbbm{Z}_2$--grade.
\subsection{{\tt charge}}
{\bf Purpose:} Return the $R$--charge of a state.\\
{\bf Input:} An (integrated) state, i.e. {\tt states(int)[[2]]}.\\
{\bf Output:} The value of the $R$--charge.
\subsection{{\tt sbrane}}
{\bf Purpose:} Return the starting brane of a state.\\
{\bf Input:} An (integrated) state, i.e. {\tt states(int)[[3]]}.\\
{\bf Output:} The label of the brane where the open string state begins.
\subsection{{\tt ebrane}}
{\bf Purpose:} Return the end brane of a state.\\
{\bf Input:} An (integrated) state, i.e. {\tt states(int)[[4]]}.\\
{\bf Output:} The label of the brane where the open string state ends.
\subsection{{\tt val}}
{\bf Purpose:} Return the value of a state.\\
{\bf Input:} An unintegrated state, i.e. {\tt states[[5]]}.\\
{\bf Output:} The cohomology element.
\subsection{{\tt pickall}${}^{\ast}$}
{\bf Purpose:} Return the value {\em or} the cyclic representative of a correlators $B_{a_0,\ldots,a_m;i_0,\ldots,i_n}$.\\
{\bf Input:} {\tt pickall[states,target,}$\mathtt{\{a_0,\ldots,a_m\},\{i_0,\ldots,i_n\}]}$, where {\tt target} is the list containing the information about the correlator.\\
{\bf Output:} The value of the correlator or the cyclic representative formatted as $\mathtt{b[\{a_0,\ldots,a_m\},\{i_0,\ldots,i_n\}]}$.\\
{\bf Implementation:} We have various cases depending in the input:
\begin{itemize}
\item Two boundary insertions and no bulk insertion. This is the boundary metric, the value is known. Go through the list on metrics, if nothing is found, permute the indices and include the appropriate sign (\ref{eq-bdrymet}). Return the value, or $0$ if there is no match.
\item Three boundary insertions and no bulk insertion. This is the boundary three--point function, whose value is known. Try out all cyclic permutations permutations and put the correct sign (\ref{eq-bdrycyc}). Return the value, or $0$ if there is no match.
\item One boundary and one bulk insertion. This is the bulk--boundary two--point function whose value is known. Return the value is the indices match, otherwise return $0$.
\item All other cases. First check if the bulk indices match. Then go through the cyclic permutations of the boundary indices, taking into account the sign (\ref{eq-bdrycyc}). If the indices match, return the variable for the cyclic representative multiplied with the appropriate sign which was picked up by the cyclic permutations. Return $0$ if there is no match.
\end{itemize}
\subsection{\tt pickF}
{\bf Purpose:} Return the value {\em or} the cyclic representative of a correlators $\mathcal{F}_{a_0,\ldots,a_m}$.\\
{\bf Input:} {\tt pickF[states,target,}$\mathtt{\{a_0,\ldots,a_m\}]}$, where {\tt target} is the list containing the information on the correlator.\\
{\bf Output:} The value of the correlator stored {\tt target} in terms of its cyclic representative.\\
{\bf Implementation:} Like {\tt pickall} but simpler because we need not take care of the bulk indices. This function may be included into {\tt pickall}.

\subsection{\tt pickbulk}
{\bf Purpose:} Return the value of a bulk correlator.\\
{\bf Input:} A list of bulk correlators whose values are known and a set of indices specifying the correlator.\\
{\bf Output:} The value of the correlator. 

\section{Selection Rules and Correlators}
Given a matrix factorization and a set of open string states we can calculate all the correlators without integrated insertions, using the Kapustin--Li residue formula. Furthermore we have to find out which correlators with integrated insertions are allowed by the selection rules.

\subsection{{\tt str}}
{\bf Purpose:} Compute the supertrace of a matrix.\\
{\bf Input:} A quadratic matrix of even rank.\\
{\bf Output:} The value of the supertrace.

\subsection{{\tt samecharge}}
{\bf Purpose:} Format the $R$--charges of the states.\\
{\bf Input:} The list of unintegrated states.\\
{\bf Output:} A list whose length is the Coxeter number $k$, where at th $i$--th position there is a list of labels of the states with $R$--charge $i/k$. This is minimal--model specific.

\subsection{{\tt corr2pt}}
{\bf Purpose:} Compute the boundary metric $\omega_{ab}$ using the Kapustin--Li formula (\ref{eq-kapustin2}).\\
{\bf Input:} The list of unintegrated states and the matrix factorizations.\\
{\bf Output:} A list with elements $\mathtt{\{\{a,b\},val\}}$, where {\tt val} is the value of the correlator.\\
{\bf Implementation:} We use {\tt str} and the fact that the integrand has to be proportional to the Hessian $H=\mathrm{det}(\partial_i\partial_j W)$ in order to give something non--zero. Only the non--vanishing $\omega_{ab}$ with $a\leq b$ are stored. The rest is taken care of by {\tt pickall}.

\subsection{{\tt bdry2ptinverse}}
{\bf Purpose:} Calculate the inverse $\omega^{ab}$ of the boundary metric.\\
{\bf Input:} The data for open string states and the output of {\tt corr2pt}.\\
{\bf Output:} The inverse of the boundary metric in the format of {\tt corr2pt}.
\subsection{{\tt corr3pt}}
{\bf Purpose:} Calculate the values of the cyclic representatives of the boundary three--point correlator $B_{abc}$ using the Kapustin--Li formula (\ref{eq-kapustin2}).\\
{\bf Input:}  The data for open string states and the matrix factorizations.\\
{\bf Output:} A list with elements $\mathtt{\{\{a,b,c\},val\}}$, where {\tt val} is the value of the correlator.\\
{\bf Implementation:} The routine performs the following steps:
\begin{itemize}
\item Find the cyclic representatives for $a\leq b\leq c$. In order to do this we use the {\tt ListNecklaces} function of the Mathematica package {\tt DiscreteMath`Combinatorica`}. This function works as follows:
\begin{align}
\mathtt{ListNecklaces[3,\{1,2,3\},Cyclic]}\qquad&\Rightarrow\qquad \mathtt{\{\{1,2,3\},\{1,3,2\}\}}
\end{align}
This function is very fast and efficient and uses the properties of the cyclic group.
\item Calculate the residue and write the result into a list if the value of the integral is non--zero.
\end{itemize}

\subsection{{\tt bulkbdry2pt}}
{\bf Purpose:} Calculate the bulk--boundary two--point functions $B_{a;i}$ using the Kapustin--Li formula (\ref{eq-kapustin2}).\\
{\bf Input:} The data for open string states and the matrix factorizations.\\
{\bf Output:} A list with elements $\mathtt{\{\{a,i\},val\}}$, giving the data of non--vanishing correlators.

\subsection{{\tt bulk2pt}}
{\bf Purpose:} Calculate the inverse of the topological bulk metric $\eta_{ij}$ using the formula of Vafa (\ref{eq-bulk3pt}).\\
{\bf Input:} The Landau--Ginzburg superpotential and a list containing the elements of the the bulk chiral ring.\\
{\bf Output:} A list with elements of the form $\mathtt{\{\{i,j\},val\}}$, where {\tt val} is the value of the correlator and $\mathtt{i\leq j}$.

\subsection{{\tt findindices}${}^{\ast}$}
{\bf Purpose:} Recursive procedure to build up a correlator which is allowed by the selection rules.\\
{\bf Input:} $\mathtt{findindices[q,ind,rlist,flag]}$, where {\tt q} gives the current background charge\footnote{To be precise, it is the $R$--charge minus the background charge. So, an allowed correlator has $\mathtt{q=0}$.} of the correlator, {\tt ind} is a list of labels of the current insertions, {\tt rlist} is the output of {\tt samecharge} and {\tt flag} is $0$ for the standard correlators with three unintegrated boundary insertions and $1$ for the special cases of one unintegrated boundary insertion and one unintegrated bulk insertion. The starting point is a list of unintegrated insertions and the background charge {\tt q} to cancel. The procedure calls itself, adding an integrated insertion at every step.\\
{\bf Output:} A list with elements $\mathtt{\{\{{a_0,\ldots,a_m}\},\{i_0,\ldots,i_n\}\}}$ giving the indices of a correlator which is allowed by the selection rules and where $a_0\leq a_1\leq\ldots\leq a_m$ and $i_0\leq i_1\leq\ldots\leq i_n$.\\
{\bf Implementation:} The following steps have to be implemented:
\begin{itemize}
\item If $\mathtt{q\leq 0}$, we have inserted an integrated insertion whose charge was too big. The function is stopped.
\item If $\mathtt{q=0}$, the charge selection rule is satisfied. Check if the correlator has the same suspended grade as the boundary metric. Format the output and stop the recursion.
\item If $\mathtt{q>0}$, one must insert more integrated insertions. Go through the list which encodes the $R$--charges of the integrated insertions, which is provided by {\tt samecharge}. Check that the candidate insertion is not the unit operator and make sure that the indices are in ascending order. Let the function call itself with {\tt q} being reduced by the $R$--charge of the inserted state and {\tt ind} with the label of the inserted state added.
\end{itemize}

\subsection{{\tt bulkbdryins}}
{\bf Purpose:} Produce a list of the cyclic representatives of the allowed $n$--point correlators with bulk and boundary insertions, which have three unintegrated boundary insertions.\\
{\bf Input:} The data of the cohomology elements, the integrated insertions and the output of {\tt samecharge}.\\
{\bf Output:} A list with elements of the form $\mathtt{\{\{a_0,\ldots,a_m\},\{i_0,\ldots,i_n\}\}}$ representing a correlator $B_{a_0\ldots a_m;i_0\ldots i_n}$ with at least three boundary insertions. The list contains only the cyclic representatives.\\
{\bf Implementation:} The procedure performs the following steps:
\begin{itemize}
\item Go through the combinations of three unintegrated boundary insertions and calculate the value {\tt q} by which the $R$--charges of the insertions overshoot the background charge. Make sure the there are no insertions of the unit operator.
\item Call {\tt findindices} for the given three boundary insertions and with the charge {\tt q} and {\tt flag=0}.
\item Go through the list produced by {\tt findindices} and generate the cyclic representatives for every element by using {\tt ListNecklaces}.
\item In case of multiple branes, check if the start-- and end--branes of neighboring insertions match.
\item Format the output.
\end{itemize}

\subsection{{\tt special}}
{\bf Purpose:} Produce a list of all allowed correlations with integrated insertions which only have one or two boundary insertions and at least one bulk insertion. \\
{\bf Input:} The data of the cohomology elements and the integrated insertions and the output of {\tt samecharge}.\\
{\bf Output:} A list with elements of the form $\mathtt{\{\{a_0,\ldots,a_m\},\{i_0,\ldots,i_n\}\}}$ representing a correlator $B_{a_0\ldots a_m;i_0\ldots i_n}$.\\
{\bf Implementation:} The procedure is completely analogous to {\tt bulkbdryind}. The only difference is that we call {\tt findindices} with one unintegrated boundary and one unintegrated bulk insertion and set {\tt flag=1}. 

\subsection{{\tt generateF}}
{\bf Purpose:} Generate a list of all allowed correlators $\mathcal{F}_{a_0\ldots a_m}$ as defined in (\ref{eq-fdef}) and determine their values, which may still depend on the unknown correlators $B_{a_0\ldots a_m;i_0\ldots i_n}$.\\
{\bf Input:} $\mathtt{generateF[states,bb2pt,bdry3pt,spec,npt]}$, where {\tt states} is the data of the unintegrated cohomology, {\tt bb2pt} is the output of {\tt bulkbdry2pt}, {\tt bdry3pt} is the output of {\tt corr3pt}, {\tt spec} is the output of {\tt special} and {\tt npt} is the output of {\tt bulkbdryins}.\\
{\bf Output:} A list with elements $\mathtt{\{\{a_0,\ldots,a_m\},val\}}$ specifying $\mathcal{F}_{a_0\ldots a_m}$ and its value.\\
{\bf Implementation:} The routine performs the following steps:
\begin{itemize}
\item Join the lists of various types of correlators into a list of all possible sets $a_0,\ldots,a_m$.
\item Produce datasets of the form $\mathtt{\{\{bdry\},\{\{bulk1\},\{bulk2\},\ldots \}\}}$, which collect the indices of correlators which have identical boundary insertions but different bulk insertions.
\item For every dataset, go through the different bulk insertions and use {\tt pickall} to get the coefficients of the $t$--monomials and multiply these with the appropriate powers in the $t_i$.
\item Sum up the monomials for every data set, which gives the value of $\mathcal{F}_{a_0\ldots a_m}$.
\item Format the output.
\end{itemize}
\section{Implementing the Constraint Equations}
We now come to the central task which is implementing the three sets of constraint equations: the $A_{\infty}$--relations (\ref{eq-ainf2}), the bulk--boundary crossing constraints (\ref{eq-bbcr}) and the Cardy condition (\ref{eq-cardy}). We need to use some tricks to set up these equations. The na\"ive approach would be to set up the structure of each equation at every order and then use the {\tt pickall} function to insert for the values. The problem is that the routines will be much too slow. For higher numbers of insertions an $A_{\infty}$--equation may easily have a hundred summands or more but only a handful are actually non--zero. The most efficient approach is to calculate all possible summands and assign them to the proper equations. This is the approach we will pursue here.

\subsection{{\tt allcorr}}
{\bf Purpose:} Give a list of all possible index sets for the $n$--point function, including those which belong to the same cyclic class and group the list into sublists which contain the index lists for a given number of boundary insertions.\\
{\bf Input:} The bulk--boundary two--point functions, the list of 'special' correlators, the boundary three--point functions, the list of $n$--point correlators from {\tt bulkbdryins}.\\
{\bf Output:} A nested list: $\mathtt{\{\{\{\{a_0\},\{i_0\}\},\{\{a_1\},\{i_1,i_2\}\},\ldots\},\{\{\{a_1,a_2\},\{\}\},\{\{a_2,a_1\},\{\}\},\ldots\},\ldots\}}$. 

\subsection{{\tt allF}}
{\bf Purpose:} Give a list of all possible index sets for the $\mathcal{F}_{a_0,\ldots,a_m}$, including those which belong to the same cyclic class and group the list into sublists which contain the index lists for a given number of boundary insertions.\\
{\bf Input:} The output of {\tt generateF}.\\
{\bf Output:} A list of lists whose entries are the possible indices $a_0,\ldots,a_m$ of $\mathcal{F}_{a_0,\ldots,a_m}$ for a given $m$ without taking care of the cyclicity.

\subsection{{\tt allbulk3pt}}
{\bf Purpose:} Give a list of all possible indices for the bulk threepoint functions $c_{ijk}$.\\
{\bf Input:} A list of the bulk three--point functions with elements of the form $\mathtt{\{\{i,j,k\},val\}}$, where $i\leq j\leq k$.\\
{\bf Output:} A list of indices of bulk threepoint functions with unordered indices. \\\\
Actually, the existence of these three routines render the efforts, only to store the minimum necessary amount of information by considering cyclic representatives rather obsolete. Introducing these lists, however, speeds up some parts of the procedures to set up the constraint equations.

\subsection{{\tt eqfinder}${}^{\ast}$}
{\bf Purpose:} Find all $A_{\infty}$--equations,
\begin{align}
\sum_{I\subseteq I_{0,n}}\sum_{\stackrel{k,j=0}{k\leq j}}^m(-1)^{\tilde{a}_1+\ldots\tilde{a}_k}B^{a_0}_{\phantom{a_0}a_1\ldots a_kca_{j+1}\ldots a_m;I_{0;n}\backslash I}B^c_{\phantom{c}a_{k+1}\ldots a_j;I}&=0,
\end{align}
for a given $m$ and $n$.\\
{\bf Input:} $\mathtt{eqfinder[m,n,corrs,invmet,states]}$, where {\tt corrs} is the output of {\tt allcorr} and {\tt invmet} is the inverse of the boundary metric.\\
{\bf Output:} A list with elements
\begin{align}
&\mathtt{ \Big\{\big\{\{a_0,\ldots,a_m\},\{i_0,\ldots,i_n\}\big\},\big\{\big\{\{a_1,\ldots,a_k\},\{\{a_0,\ldots,a_k,c,a_{j+1},\ldots,a_m\},\{I_{0;n}\backslash I\}\},}\nonumber \\
& \mathtt{,\{c,d\},\{\{d,a_{k+1},\ldots,a_j\},\{I\}\} \big\},\ldots\big\}\Big\}},
\end{align}
where the first list identifies the equation and the second list gives the information on the summands, which are specified by lists which determine the sign, the position of the metric insertion, and the index sets of the correlators.\\
{\bf Implementation:} The program goes through the following steps:
\begin{itemize}
\item Go through all the elements of {\tt corrs} which have less than $m+2$ boundary insertions and check if the number of bulk insertions does not exceed the number $n$, the bulk indices determine the index set $I_{0;n}\backslash I$.
\item Go through all the positions $\mathtt{c}$ of metric and split the index set into $\mathtt{a_0,\ldots,a_k}$ and $\mathtt{a_{j+1},\ldots,a_m}$. 
\item Calculate the number of boundary and bulk insertions for the second correlator of the summand and go through the elements of the corresponding sublist of {\tt corrs}.
\item Check if the first index of the correlator, $d$, together with the index $c$ of the first correlators gives a non--vanishing element of the boundary metric. If this is so, collect the index sets $\mathtt{\{d,a_{k+1},\ldots,a_j\}}$ and $\mathtt{\{I\}}$.
\item Format the equation identifier $\mathtt{\{\{a_0,\ldots,a_m\},\{i_0,\ldots,i_n\}\}}$ and check if it already exists in the list of equations. If no, generate a new element and append the data of the summand. If yes, append the information on the summand to the equation data.
\end{itemize}

\subsection{{\tt ainfty}}
{\bf Purpose:} Set up the $A_{\infty}$ relations for every $m$ and $n$ where there are non--vanishing terms.\\
{\bf Input:} The data of the unintegrated states, the inverse of the boundary metric, the bulk--boundary two--point functions, the 'special' correlators, the boundary three--point functions, the information on the cyclic representatives of the $n$--point correlators and the output of {\tt allcorr}.\\
{\bf Output:} A list of all $A_{\infty}$ relations in terms of the cyclic representatives of the unknown correlators with redundancies removed.\\
{\bf Implementation:} 
\begin{itemize}
\item Determine the maximal $m$ and $n$ for which the $A_{\infty}$--relations yield equations with non--zero summands.
\item Go through all possible values of $m$ and $n$ and call {\tt eqfinder}.
\item For every equation, determine the value of the summands using {\tt pickall} and sum up the terms.
\item Check the equation or minus the equation is already in the list\footnote{This can happen since {\tt pickall} returns the correlators in terms of cyclic representatives and different equations may give, up to signs, the same conditions on the cyclic representatives.} of equations and append it if it is not. 
\end{itemize}

\subsection{{\tt crosseqfinder}${}^{\ast}$}
{\bf Purpose:} Find all bulk--boundary crossing constraint equations
\begin{align}
\partial_i\partial_j\partial_k\mathcal{F}(t)\eta^{kl}\partial_l\mathcal{F}_{a_0\ldots a_m}(t)&=
\end{align}
\begin{align}
&=\sum_{0\leq m_1\leq\ldots m_4\leq m}(-1)^{\tilde{a}_{m+1}+\ldots+\tilde{a}_{m_3}}\mathcal{F}_{a_0\ldots a_{m_1}ba_{m_2+1}\ldots a_{m_3}ca_{m_4+1}\ldots a_m}\partial_i(t)\mathcal{F}^b_{\phantom{b}a_{m_1+1}\ldots a_{m_2}}(t)\partial_j\mathcal{F}^c_{\phantom{c}a_{m_3+1}\ldots a_{m_4}}\nonumber,
\end{align}
for a given $m$.\\
{\bf Input:} $\mathtt{crosseqfinder[m,flist,allf,bulkmet,bdrymet,bulk3pt,allbulk,states]}$, where {\tt flist} is the output of {\tt generateF}, {\tt allf} is produced by {\tt allF}. The further arguments of the function are the inverse of the bulk metric, the inverse of the boundary metric, the bulk three--point functions in ordered and complete form, where the list only contains those elements where the OPE $\phi_i\phi_j=c_{ij}^{\phantom{ij}k}\phi_k$ does not contain exact terms\footnote{Such a list can be generated easily by computing the triple derivatives of the WDVV potential \cite{Dijkgraaf:1990dj,Klemm:1991vw}. For the minimal models of type $A$ the requirement that there are no exact pieces in the OPE translates into a simple condition on the indices. For the $E_6$ it is easiest to sort the list by hand.}, and the states.\\
{\bf Output:} A list with elements $\mathtt{\{\{identifier\},\{\{lhs\},\{rhs\}\}\}}$, where
\begin{align}
\mathtt{identifier}=&\mathtt{\{a_0,\ldots,a_m,i,j\}}\nonumber \\
\mathtt{lhs}=&\mathtt{\big\{\{\{k,l\},val\},\ldots  \big\}}\nonumber \\
\mathtt{rhs}=&\mathtt{\Big\{\big\{\{\{a_0,\ldots,a_m\},\{a_{m_1+1},\ldots,a_{m_2}\},\{a_{m_2+1},\ldots,a_{m_3}\},\{a_{m_3+1},a_{m_4}\},\{a_{m_4+1},\ldots,a_m\}}\nonumber \\
&\mathtt{\{b,c,b',c'\}\},val\big\},\ldots\Big\}},
\end{align}
where {\tt val} is the value of the correlator, obtained by using the function {\tt pickF}.\\
{\bf Implementation:} The program performs the following steps:
\begin{itemize}
\item First consider the lefthand side of the equations:
\begin{itemize}
\item Check if the output of {\tt allF} contains entries for a given $m$.
\item Search the list {\tt allbulk} for possible values of $i,j,k$ and determine the index $l$ of the metric.
\item Use {\tt pickall} to calculate the partial derivative of $\mathcal{F}_{a_0\ldots a_m}$ and compute the value of the lefthand side.
\item If the value is non--zero, format the dataset and add the information to the result, i.e. produce a new entry of the equation identifier does not exist, or append the information of the summand to an existing {\tt lhs}.
\end{itemize}
\item For the righthand side of the equation we have to do the following:
\begin{itemize}
\item Start off with determining the last two factors of the summand because the derivatives with respect to the bulk parameters might vanish\footnote{The majority of the $\mathcal{F}_{a_0\ldots a_m}$ are just correlators with only boundary insertions.} and one can then avoid to produce the first factor.
\item Go through all terms in {\tt allf} which have less than $m+1$ boundary insertions and check if any of the possible $t$--derivatives is non--zero. This determines a valid second factor in a summand. Get the metric factor $b'$.
\item Given a non--vanishing second factor, go through possible candidates of the third factor in {\tt allf}. Check if any $t$--derivative is non--zero. If this is the case we have found a pair $\{i,j\}$ and have to check whether the associated bulk OPE does not contain exact terms. Get the metric factor $c'$.
\item Calculate how many boundary insertions the first factor of the summand has and go through the possible candidates in {\tt allf}. Go through the possible positions of $b$ and $c$ and check if $\omega^{b,b'}\neq 0$ and $\omega^{cc'}\neq 0$. 
\item If one gets a non--vanishing summand, format the equation identifier and the equation data and calculate the value of the summand using $\tt pickF$. Append the information of the summand to the correct equation or generate a new entry in the output list.
\end{itemize}
\end{itemize}

\subsection{{\tt sumup}}
{\bf Purpose:} Assemble the bulk--boundary crossing equations.\\
{\bf Input:} $\mathtt{sumup[states,flist,allf,bulkmet,bdrymet,bulk3pt,allbulk]}$, where the arguments are the same as for {\tt crosseqfinder}.\\
{\bf Output:} A list of bulk--boundary crossing equations with redundancies removed.\\
{\bf Implementation:}
\begin{itemize}
\item Determine the maximal $m$ for which the bulk--boundary crossing constraints can yield non--trivial information.
\item Call {\tt crosseqfinder} for every $m$.
\item Go through the equations and sum up its components.
\item Get the constraints on the correlators $B_{a_0\ldots a_m;i_0\ldots i_n}$ by extracting the coefficients of the $t_i$ using the Mathematica function {\tt CoefficientList}.
\item Add the equation to the output, if it is not already in the list.
\end{itemize}

\subsection{{\tt cardyeqfinder}${}^{\ast}$}
{\bf Purpose:} Set up the Cardy equations
\begin{align}
\partial_i\mathcal{F}_{a_0\ldots a_n}\eta^{ij}\partial_j\mathcal{F}_{b_0\ldots b_m}&=
\end{align}
\begin{align}
=\sum_{\stackrel{{0\leq n_1\leq n_2\leq n}}{0\leq m_1\leq m_2\leq m}}(-1)^{(\tilde{c}_1+\tilde{a}_0)(\tilde{c}_2+\tilde{b}_0)+\tilde{c_1}+\tilde{c_2}}\omega^{c_1d_1}\omega^{c_2d_2}\mathcal{F}_{a_0\ldots a_{n_1}d_1b_{m_1+1}\ldots m_2c_2a_{n_2+1}\ldots a_n}\mathcal{F}_{b_0\ldots b_{m_1}c_1a_{n_1+1}\ldots a_{n_2}d_2b_{m_2+1}\ldots b_m}&\nonumber
\end{align}
for a fixed $m$ and $n$.\\
{\bf Input:} $\mathtt{cardyeqfinder[n,m,flist,allf,bulkmet,bdrymet,flag]}$, where the arguments 2--6 are as for {\tt crosseqfinder} and $\mathtt{flag=0}$ gives the full set of Cardy equations and $\mathtt{flag=1}$ truncates the equations which are in the boundary changing sector.\\
{\bf Output:} A list with elements $\mathtt{\{\{identifier\},\{\{lhs\},\{rhs\}\}\}}$, where\footnote{We need not put the $a_i$ and $b_i$ into separate lists in {\tt identifier} because $n$ and $m$ are fixed.}
\begin{align}
\mathtt{identifier}=&\mathtt{\{a_1,\ldots,a_n,b_0,\ldots,b_m \}}\nonumber \\
\mathtt{lhs}=&\mathtt{\big\{\{\{i,j\},val\},\ldots\big\}}\nonumber \\
\mathtt{rhs}=&\mathtt{\Big\{\big\{\{\{a_0,\ldots,a_{n_1}\},\{a_{n_1+1},\ldots a_{n_2}\},\{a_{n_2+1},\ldots,a_n\},\{b_0,\ldots,a_{m_1}\},\{b_{m_1+1},\ldots b_{m_2}\}, }\nonumber \\
&\mathtt{,\{b_{m_2+1},\ldots,b_m\} ,\{c_1,d_1,c_2,d_2\}\},val\big\},\ldots\Big\}}
\end{align}
{\bf Implementation:}
\begin{itemize}
\item For the lefthand side we have to perform the following steps:
\begin{itemize}
\item Go through the elements of {\tt allf} which have $n+1$ entries and check if any $t$--derivative is non--vanishing.
\item Store the indices $\mathtt{a_0,\ldots,a_n}$, determine the index $i$ and determine for which $j$ the bulk metric is non--zero.
\item Go through the elements of {\tt allf} which have $m+1$ entries and check if the $t_j$--derivative is non--vanishing.
\item If {\tt flag=1} check of the summand only contains insertions of states in the boundary preserving sector.
\item Format the data of the summand and append it if the identifier already exists or otherwise generate a new entry for the equation.
\end{itemize}
\item For the righthand side, the program does the following:
\begin{itemize}
\item Go through all the elements in {\tt allf} which have less than $m+n+2$ and more than $2$ entries. 
\item Go through all the possible positions of $d_1$ and $c_2$ and separate the index sets of the first factor.
\item Calculate how many insertions the second factor of the summand has and go through the corresponding entries in {\tt allf}. Loop through the possible positions of $c_1$ and $d_2$ and check if $\omega^{c_1d_1}\neq0$ and $\omega^{c_2d_2}\neq0$.
\item Format the index sets, check if the correlator preserves the boundary, if {\tt flag=1} and compute its value using {\tt pickF}.
\item Format the data of the summand and add it to the result.
\end{itemize}
\end{itemize}

\subsection{{\tt sumup2}}
{\bf Purpose:} Assemble the Cardy equations.\\
{\bf Input:} $\mathtt{sumup2[flist,allf,bulkmet,bdrymet,flag]}$, where the arguments are the same as for {\tt cardyeqfinder}.\\
{\bf Output:} A list of all Cardy equations with redundancies removed, where the equations are in the boundary preserving sector if {\tt flag=1}.\\
{\bf Implementation:} Analogous to {\tt sumup}. 

\section{Solving the Equations and Calculation of the Effective Superpotential}
Having set up the constraint equations, we now have to solve them. Since the equations are non--linear and the algorithms described above are efficient enough to work for examples which yield up to $5000$ equations, one can in general not simply use Mathematica's {\tt Solve} routine without manipulating the equations first. The programs described below help to simplify the equations in such a way that they become tractable for {\tt Solve}. For the most complex examples which could be tackled with these programs it was however still necessary to perform some steps manually in order to obtain the solution. Having found a solution to the constraint equations it is then quite easy to sum up the correlators to get the effective superpotential.

\subsection{{\tt reduce}}
{\bf Purpose:} Insert solutions into equations, remove the variables which have been solved for and update the solutions.\\
{\bf Input:} $\mathtt{reduce[eqs,vars,sols]}$, where {\tt eqs} are the equations, {\tt vars} are the unknown variables and {\tt sols} are the solutions.\\
{\bf Output:} A list $\mathtt{\{\{eqs'\},\{vars'\},\{sols'\}\}}$ with solutions inserted, redundancies and eliminated variables removed.\\
{\bf Implementation:} 
\begin{itemize}
\item Insert {\tt sols} into {\tt eqs} and {\tt vars}.
\item Remove the zeroes and redundancies in {\tt eqs}.
\item Remove the variables which have been eliminated.
\end{itemize}

\subsection{{\tt findzero}}
{\bf Purpose:} Recursive procedure to identify the equations which are of the form $(n B_{a_0\ldots a_m;i_0\ldots i_n})^\alpha=0$, for $\alpha=1,2$ and insert their solution.\\
{\bf Input:} A list $\mathtt{\{\{eqs\},\{vars\},\{sols\}\}}$.\\
{\bf Output:} A list $\mathtt{\{\{eqs'\},\{vars'\},\{sols'\}\}}$.\\
{\bf Implementation:} 
\begin{itemize}
\item Identify the equations which set a certain variable to 0.
\item Insert the solution and remove redundancies and zeroes from the new equations, remove the eliminated variables and append the solution to $\mathtt{\{sols\}}$.
\item Call the program again with the new values.
\item Break the recursion if the number of unknowns has not reduced as compared to the previous step.
\end{itemize}

\subsection{\tt findlin}
{\bf Purpose:} Find and solve the equations which have entries linear in the variables.\\
{\bf Input:} A list $\mathtt{\{\{eqs\},\{vars\},\{sols\}\}}$ and a variable {\tt flag}.\\
{\bf Output:} A list $\mathtt{\{\{eqs'\},\{vars'\},\{sols'\}\}}$.\\
{\bf Implementation:} 
\begin{itemize}
\item If $\mathtt{flag=0}$, find equations of the form $c_1B_{a_0\ldots a_m;i_0\ldots i_n}+c_2B_{b_0\ldots b_k;j_0\ldots j_l}=0$, where $c_1,c_2$ are constants. 
\item If $\mathtt{flag=1}$ find equations which have at least one linear term at at most three summands. 
\item Solve these equations with respect to the linear variables and reduce the system using {\tt reduce}.
\end{itemize}

\subsection{{\tt solveAinf}}
{\bf Purpose:} Solve the $A_{\infty}$--relations.\\
{\bf Input:} $\mathtt{solveAinf[eqs,spec,npt]}$, where {\tt eqs} is the output of {\tt ainf}, {\tt spec} is the output of {\tt special} and {\tt npt} is the output of {\tt bulkbdryins}.\\
{\bf Output:} The complete result or a reduced set of equations.\\
{\bf Implementation:} The program performs the following steps:
\begin{itemize}
\item Build a list $\mathtt{\{\{eqs\},\{vars\},\{sols\}\}}$ for equations, unknowns and solutions.
\item Call {\tt findzero}, then {\tt findlin} with $\mathtt{flag=0}$ and finally {\tt findlin} with $\mathtt{flag=1}$.
\item If there are less than 15 equations left, try {\tt Solve} and return the result, otherwise return the remaining equations for solving them by hand.
\end{itemize}

\subsection{{\tt crossolve}}
{\bf Purpose:} Solve the bulk--boundary crossing constraints using the solution of the $A_{\infty}$--relations. It can also be used to solve the Cardy equation if the solutions of the other constraint equations are known.\\
{\bf Input:} The result of the previous equations, the equations to solve, the output of {\tt special} and the output of {\tt bulkbdryins}.\\
{\bf Output:} The solution of the equations.\\
{\bf Implementation:} 
\begin{itemize}
\item Construct a list of all unknowns and insert the known solutions. Filter out the variables which are still undetermined.
\item Insert the solution into the equations and remove redundancies.
\item Try to solve the equations using {\tt Solve} and join the new solutions with the known solutions.
\end{itemize}

\subsection{{\tt superpotential}}
{\bf Purpose:} Sum up the correlators to obtain $\mathcal{W}_{eff}$, using
\begin{align}
&\mathcal{A}_{a_0,\ldots a_m}:=(m-1)!\mathcal{F}_{(a_0,\ldots,a_m)}:=\frac{1}{m}\sum_{\sigma\in S_m}\eta(\sigma;a_1,\ldots,a_m)\mathcal{F}_{a_{\sigma(1)}\ldots a_{\sigma(m)}}\nonumber \\
&\mathcal{W}_{eff}(s;t)=\sum_{m\geq 1}\frac{1}{m}s_{a_m}\ldots s_{a_0}\mathcal{A}_{a_0\ldots a_m}(t)
\end{align}
{\bf Input:} $\mathtt{superpotential[states,flist,result]}$, where {\tt flist} if the output of {\tt generateF} and {\tt result} is the solution of the constraint equations.\\
{\bf Output:} The effective superpotential.\\
{\bf Implementation:}
\begin{itemize}
\item Generate a list {\tt fres} by inserting {\tt result} into {\tt flist}.
\item Go through {\tt fres} and compute the permutations of the given index set $\{a_0,\ldots, a_m\}$ using the Mathematica function {\tt Permutations}.
\item Get the value of the corresponding $\mathcal{F}_{a_0\ldots a_m}$ for every permutation using {\tt pickF} with target {\tt fres} and sum up the terms to obtain $\mathcal{A}_{a_0\ldots a_m}$.
\item Determine the $s$--factors and multiply with the value of $\mathcal{A}_{a_0\ldots a_m}$.
\item Sum up the terms and return $\mathcal{W}_{eff}$.
\end{itemize}

\chapter{Further Results for Minimal Models}
In this appendix we give some additional results on the $E_6$ minimal models and on effective superpotentials of type $A$ minimal models, which were obtained by solving the consistency constraints for open topological strings.

\section{The ``other'' $E_6$--model}
There is also a two--variable description for the $E_6$ minimal model:
\begin{equation}
W=x^3+y^4
\end{equation}
For completeness we list all the matrix factorizations and give
the complete spectrum. From the point of view of conformal field theory these
two incarnations of the $E_6$--model correspond to two different GSO projections
\cite{Kapustin:2003rc,Brunner:2005pq,Brunner:2005fv}.\\\\
The model has the following matrix factorizations:
\begin{align}
E_1=J_2=\left(\begin{array}{cc}x&y\\y^3&-x^2\end{array}\right)\qquad&
E_2=J_1=\left(\begin{array}{cc}x^2&y\\y^3&-x\end{array}\right)
\end{align}
\begin{align}
E_3=\left(\begin{array}{cc}x&y^2\\y^2&-x^2\end{array}\right)\qquad&
J_3=\left(\begin{array}{cc}x^2&y^2\\y^2&-x\end{array}\right)
\end{align}
\begin{align}
E_4=J_5=\left(\begin{array}{ccc}
x&y&0\\
0&x&y\\
y^2&0&x
\end{array}\right)
\qquad&
E_5=J_4=\left(\begin{array}{ccc}
x^2&-xy&y^2\\
y^3&x^2&-xy\\
-xy^2&y^3&x^2
\end{array}\right)
\end{align}
\begin{align}
E_6=\left(
\begin{array}{cccc}
x&y^2&0&0\\
y^2&-x^2&0&0\\
0&-xy&x^2&y^2\\
y&0&y^2&-x
\end{array}
\right)
\qquad&
J_6=\left(
\begin{array}{cccc}
x^2&y^2&0&0\\
y^2&-x&0&0\\
0&-y&x&y^2\\
xy&0&y^2&-x^2
\end{array}
\right)
\end{align}
Now we give the full open string spectrum of this model, including the boundary changing sector. We label the branes by $M_1,\ldots,M_6$. The numbers in the following tables are the $R$--charges multiplied by the dual Coxeter number, which is $12$. The exponents give the multiplicities.\\
The even spectrum is summarized in table \ref{tab-e6even}.
\begin{table}[h]
\begin{center}
\begin{tabular}{|c|c|c|c|c|c|c|}\hline
\vrule width 0pt height 12pt depth 6pt & $M_1$&$M_2$&$M_3$&$M_4$&$M_5$&$M_6$\\ \hline
\vrule width 0pt height 12pt depth 6pt $M_1$&$0\:10$&$4\:6$&$3\:7$&$1\:3\:5$&$5\:7\:9$&$2\:4\:6\:8$\\ \hline
\vrule width 0pt height 12pt depth 6pt $M_2$&$4\:6$&$0\:10$&$3\:7$&$5\:7\:9$&$1\:3\:5$&$2\:4\:6\:8$\\ \hline
\vrule width 0pt height 12pt depth 6pt $M_3$&$3\:7$&$3\:7$&$0\:4\:6\:10$&$2\:4\:6\:8$&$2\:4\:6\:8$&$1\:3\:5^2\:7\:9$\\ \hline
\vrule width 0pt height 12pt depth 6pt $M_4$&$5\:7\:9$&$1\:3\:5$&$2\:4\:6\:8$&$0\:2\:4\:6\:8\:10$&$2\:4^2\:6^2\:8$&$1\:3^2\:5^2\:7^2\:9$\\ \hline
\vrule width 0pt height 12pt depth 6pt $M_5$&$1\:3\:5$&$5\:7\:9$&$2\:4\:6\:8$&$2\:4^2\:6^2\:8$&$0\:2\:4\:6\:8\:10$&$1\:3^2\:5^2\:7^2\:9$\\ \hline
 \vrule width 0pt height 12pt depth 6pt $M_6$&$2\:4\:6\:8$&$2\:4\:6\:8$&$1\:3\:5^2\:7\:9$&$1\:3^2\:5^2\:7^2\:9$&$1\:3^2\:5^2\:7^2\:9$&$0\:2^2\:4^3\:6^3\:8^2\:10$ \\ \hline
\end{tabular}
\end{center}
\caption{The even spectrum for the two--variable $E_6$ model.}\label{tab-e6even}
\end{table}
The odd spectrum is given in table \ref{tab-e6odd}.
\begin{table}[h]
\begin{center}
\begin{tabular}{|c|c|c|c|c|c|c|}\hline
\vrule width 0pt height 12pt depth 6pt & $M_1$&$M_2$&$M_3$&$M_4$&$M_5$&$M_6$\\ \hline
\vrule width 0pt height 12pt depth 6pt $M_1$&$4\:6$&$0\:10$&$3\:7$&$5\:7\:9$&$1\:3\:5$&$2\:4\:6\:8$\\ \hline
\vrule width 0pt height 12pt depth 6pt $M_2$&$0\:10$&$4\:6$&$3\:7$&$1\:3\:5$&$5\:7\:9$&$2\:5\:6\:8$\\ \hline
\vrule width 0pt height 12pt depth 6pt $M_3$&$3\:7$&$3\:7$&$0\:4\:6\:10$&$2\:4\:6\:8\:$&$2\:4\:6\:8$&$1\:3\:5^2\:7\:9$\\ \hline
\vrule width 0pt height 12pt depth 6pt $M_4$&$1\:3\:5$&$5\:7\:9$&$2\:4\:6\:8$&$2\:4^2\:6^2\:8$&$0\:2\:4\:6\:8\:10$&$1\:3^5\:5^2\:7^2\:9$\\ \hline
\vrule width 0pt height 12pt depth 6pt $M_5$&$5\:7\:9$&$1\:3\:5$&$2\:4\:6\:8$&$0\:2\:4\:6\:8\:10$&$2\:4^2\:6^2\:8$&$1\:3^2\:5^2\:7^2\:9$\\ \hline
 \vrule width 0pt height 12pt depth 6pt $M_6$&$2\:4\:6\:8$&$2\:4\:6\:8$&$1\:3\:5^2\:7\:9$&$1\:3^2\:5^2\:7^2\:9$&$1\:3^2\:5^2\:7^2\:9$&$0\:2^2\:4^3\:6^3\:8^2\:10$ \\ \hline
\end{tabular}
\end{center}
\caption{The odd spectrum of the two--variable $E_6$--model.}\label{tab-e6odd}
\end{table}
By looking at the $R$--charges, one finds that for the two--variable model Serre duality pairs up bosons with bosons and fermions with fermions.

\section{Systems with more than one D--brane -- Some Examples for A--minimal Models}
\label{app-amod}
In this section we give some results for the minimal models of type $A$, whose effective superpotentials can be uniquely determined by solving the consistency constraints for open topological strings. In \cite{Herbst:2004jp} many results for such models were given, most of them however in the boundary preserving sector. In the boundary changing sector, only the case without bulk deformations was considered. Here we discuss the boundary changing sector with $t_i\neq 0$. We find that the Cardy condition is only valid in the boundary preserving sector and we give a detailed example for a case where it fails. Fortunately it turns out that it is enough to consider the Cardy condition in the boundary preserving sector in order to determine the values of all the correlators. \\
Let us first introduce some notation. The superpotential for the $A_k$--model is:
\begin{align}
W^{(k+2)}(x)=\frac{x^{k+2}}{k+2},
\end{align}
where the exponents in parentheses give the degrees of the polynomials. The matrix factorizations are:
\begin{align}
W^{(k+2)}(x)=E^{(\kappa+1)}(x)J^{(k+1-\kappa)}(x),\qquad\qquad\kappa=0,\ldots,[k/2]
\end{align}
We denote by $h$ the greatest common denominator of $E$ and $J$, i.e $E^{\kappa+1}(x)=p(x)h^{\ell+1}(x)$ and $J^{k+1-\kappa}(x)=q(x)h^{\ell+1}(x)$. $(k,\ell)$ then uniquely labels the D--brane we consider.

\subsection{The $A_3$ Model}
We consider the $A_3$ minimal model with superpotential $W=\frac{x^5}{5}$ with two D--branes. From the factorization of the superpotential we obtain the following BRST operators:
\begin{align}
Q_1=\left(
\begin{array}{cc}
0&x^2\\
\frac{x^3}{5}&0
\end{array}
\right)
\qquad&
Q_2=\left(
\begin{array}{cc}
0&x\\
\frac{x^4}{5}&0
\end{array}
\right)
\end{align}
We can associate charge matrices to these two branes \cite{Walcher:2004tx,Brunner:2003dc}:
\begin{align}
R_1=\left(
\begin{array}{cc}
\frac{1}{10}&0\\
0&-\frac{1}{10}
\end{array}
\right)
\qquad&
R_2=\left(
\begin{array}{cc}
\frac{3}{10}&0\\
0&-\frac{3}{10}
\end{array}
\right)
\end{align}
Computing the cohomology one obtains the following spectrum:
\begin{align}
\phi_1^{(0)}[1,1]=
\left(
\begin{array}{cc}
1&0\\
0&1
\end{array}
\right)
\qquad&
\phi_2^{(\frac{2}{5})}[1,1]=
\left(
\begin{array}{cc}
x&0\\
0&x
\end{array}
\right) \nonumber \\
\psi_3^{(\frac{1}{5})}[1,1]=
\left(
\begin{array}{cc}
0&1\\
-\frac{x}{5}&0
\end{array}
\right)
\qquad&
\psi_4^{(\frac{3}{5})}[1,1]=
\left(
\begin{array}{cc}
0&x\\
-\frac{x^2}{5}&0
\end{array}
\right)
\end{align}
\begin{align}
\phi_5^{(\frac{1}{5})}[1,2]=
\left(
\begin{array}{cc}
x&0\\
0&1
\end{array}
\right)
\qquad&
\psi_6^{(\frac{2}{5})}[1,2]=
\left(
\begin{array}{cc}
0&1\\
-\frac{x^2}{5}&0
\end{array}
\right)
\end{align}
\begin{align}
\phi_7^{(\frac{1}{5})}[2,1]=
\left(
\begin{array}{cc}
1&0\\
0&x
\end{array}
\right)
\qquad&
\psi_8^{(\frac{2}{5})}[2,1]=
\left(
\begin{array}{cc}
0&1\\
-\frac{x^2}{5}&0
\end{array}
\right)
\end{align}
\begin{align}
\phi_9^{(0)}[2,2]=
\left(
\begin{array}{cc}
1&0\\
0&1
\end{array}
\right)
\qquad&
\psi_{10}^{(\frac{3}{5})}[2,2]=
\left(
\begin{array}{cc}
0&1\\
-\frac{x^3}{5}&0
\end{array}
\right)
\end{align}
The numbers in brackets contain the information about the branes where the string begins and ends.\\
The deformed bulk superpotential is:
\begin{align}
\mathcal{W}(x;t)=\frac{x^5}{5}-t_4x-t_3x^2-t_2x^3+t_2^2x-t_5+t_3t_2
\end{align}
We now compute the effective superpotential for the system $(3,1)\oplus(3,0)$ by solving the constraint equations using the programs discussed in appendix \ref{app-mathematica}. We find that we have to truncate the Cardy equations to the boundary preserving sector in order to get consistent equations. In the following we give one explicit example where the Cardy constraint fails in the boundary changing sector. \\
First take the $A_{\infty}$--equation with $m=1,n=2$, where the boundary indices are $a_0=5,a_1=8$ and the bulk indices are $I=\{3,4\}$. Inserting the non--vanishing correlators one obtains:
\begin{align}
\label{eq-ainfex}
(-)^{\tilde{8}}B_{(5,8,1)}\omega^{(1,4)}B_{(4);(3,4)}+B_{(5,9,8)}\omega^{(9,10)}B_{(10);(3,4)}+B_{(5,7);(4)}\omega^{(7,6)}B_{(6,8);(3)}=0
\end{align}
Next, one observes, that the bulk--boundary crossing constraint with $m=1$, where the boundary indices are $a_0=5,a_1=7$ and the bulk indices are $i=3,j=2$, has only one non--vanishing summand:
\begin{align}
\partial_2\partial_3\partial_1\mathcal{F}\eta^{(1,4)}\partial_4\mathcal{F}_{(5,7)}=0
\end{align}
This immediately yields $B_{(5,7);(4)}=0$, which cancels the third summand in the $A_{\infty}$--equation above. Finally, we consider the Cardy equation with $m=n=0$, where $a_0=10,b_0=4$:
\begin{align}
\label{eq-cardyex}
0&=\partial_1\mathcal{F}_{10}\eta^{(1,4)}\partial_4\mathcal{F}_4+\partial_4\mathcal{F}_{10}\eta^{(4,1)}\partial_1\mathcal{F}_4 \nonumber \\
&=B_{10;1}\eta^{(1,4)}B_{(4);(3,4)}+B_{(10);(3,4)}\eta^{(4,1)}B_{4;1}
\end{align}
In the last line we cancelled a factor $t_3$ which enters both summands.  Note that the sign $(-1)^s$ in the Cardy condition (\ref{eq-cardy}) does not enter here since only the lefthand side contributes. \\
Inserting for the two-- and three--point amplitudes the $A_{\infty}$--equation yields $B_{(4);(3,4)}-B_{(10);(3,4)}=0$, whereas the Cardy equations gives $B_{(4);(3,4)}+B_{(10);(3,4)}=0$. We cannot set $B_{(4);(3,4)}=0,B_{(10);(3,4)}=0$ because this contradicts the bulk--boundary crossing relations, which explicitly assign the value $1$ to both of the correlators\footnote{These are two equations with $m=0$: The first has $a_0=4$,$i=2,j=3$, the second has $a_0=10$,$i=2,j=3$.}. So, there is obviously a contradiction. This contradiction cannot be removed by changing normalizations of the bulk and boundary metric or the states. Thus, we conclude that the Cardy equation in inconsistent with the other constraints in the boundary changing sector.\\
Truncating the Cardy equations yields, together with the $A_{\infty}$--relations and the bulk--boundary crossing constraint, a system of non--linear equations which has a unique solution. Since, in the open string sector, the effective superpotential does not contain the full information about the correlators, we also list the non--zero ones here.\\
$283$ $A_{\infty}$--equations, $36$ bulk--boundary crossing equations and $32$ Cardy constraints yield the following results for the non--vanishing correlators:
\begin{align}
&\mathcal{F}_{3}=\mathcal{F}_{(4,4)}=\mathcal{F}_{(10,10)}=t_4-t_2^2\nonumber \\
&\mathcal{F}_4=\mathcal{F}_{10}=t_5-t_2t_3 \nonumber \\
&\mathcal{F}_{(3,3)}=\mathcal{F}_{(3,4,4)}=\mathcal{F}_{(4,6,8)}=\mathcal{F}_{(6,10,8)}=\mathcal{F}_{(4,4,4,4)}=\mathcal{F}_{(10,10,10,10)}=t_2\nonumber \\
&\mathcal{F}_{(3,4)}=\mathcal{F}_{(6,8)}=\mathcal{F}_{(4,4,4)}=\mathcal{F}_{(10,10,10)}=t_3\nonumber \\
&\mathcal{F}_{(1,1,4)}=\mathcal{F}_{(1,2,3)}=-\mathcal{F}_{(1,3,2)}=\mathcal{F}_{(1,5,8)}=-\mathcal{F}_{(1,6,7)}=\mathcal{F}_{(3,5,7)}=\mathcal{F}_{(5,9,8)}=\mathcal{F}_{(6,9,7)}= \nonumber\\
&=\mathcal{F}_{(9,9,10)}=\mathcal{F}_{(2,2,6,8)}=\mathcal{F}_{(2,3,2,4)}=-\mathcal{F}_{(2,4,5,8)}=\mathcal{F}_{(2,5,10,8)}=\mathcal{F}_{(2,6,7,4)}=\nonumber \\
&=-\mathcal{F}_{(2,6,10,7)}=\mathcal{F}_{(4,5,10,7)}=-\mathcal{F}_{(5,10,10,7)}=-1 \nonumber\\
&\mathcal{F}_{(3,3,3)}=\mathcal{F}_{(3,3,4,4)}=\mathcal{F}_{(3,4,3,4)}=\mathcal{F}_{(3,4,6,8)}=\mathcal{F}_{(3,6,8,4)}=\mathcal{F}_{(3,6,10,8)}=\mathcal{F}_{(6,8,6,8)}=\mathcal{F}_{(3,4,4,4,4)}=\nonumber\\
&=\mathcal{F}_{(4,4,4,6,8)}=\mathcal{F}_{(4,4,6,10,8)}=\mathcal{F}_{(4,6,10,10,8)}=\mathcal{F}_{(6,10,10,10,8)}=\mathcal{F}_{(4,4,4,4,4,4)}=\mathcal{F}_{(10,10,10,10,10,10)}=-\frac{1}{5}
\end{align}
With that, one gets the following expression for the effective superpotential:
\begin{align}
\mathcal{W}_{eff}&=-\frac{1}{5}\bigg(\frac{u_1^6}{6}+\frac{w_1^6}{6}+u_1^4u_2+\frac{3}{2}u_2^2u_1^2+\frac{u_2^3}{3}+2u_2u_1v_{\frac{3}{2}}\tilde{v}_{\frac{3}{2}}+u_1^3v_{\frac{3}{2}}\tilde{v}_{\frac{3}{2}}+\frac{1}{2}v_{\frac{3}{2}}^2\tilde{v}_{\frac{3}{2}}^2+u_2v_{\frac{3}{2}}\tilde{v}_{\frac{3}{2}}w_1\nonumber\\
&+u_1^2v_{\frac{3}{2}}\tilde{v}_{\frac{3}{2}}w_1+u_1v_{\frac{3}{2}}\tilde{v}_{\frac{3}{2}}w_1^2+v_{\frac{3}{2}}\tilde{v}_{\frac{3}{2}}w_1^3\bigg)-t_2\left(-\frac{u_1^4}{4}-\frac{w_1^4}{4}-u_1^2u_2-\frac{u_2^2}{2}-u_1v_{\frac{3}{2}}\tilde{v}_{\frac{3}{2}}-v_{\frac{3}{2}}\tilde{v}_{\frac{3}{2}}w_1\right)\nonumber\\
&+t_3\left(\frac{u_1^3}{3}+\frac{w_1^3}{3}+u_1u_2+v_{\frac{3}{2}}\tilde{v}_{\frac{3}{2}}\right)-\left(t_4-t_2^2\right)\left(-\frac{u_1^2}{2}-\frac{w_1^2}{2}-u_2\right)+\left(t_5-t_2t_3\right)\left(-u_1-w_1\right)
\end{align}
Here we related deformation parameters $\{u,v,\tilde{v},w\}$ to the fermionic cohomology elements. Their indices indicate their homogeneous weights.

\subsection{The $A_4$ model}
We consider the $A_4$ minimal model with superpotential $W=\frac{x^6}{6}$ and three D--branes given by the following matrix factorizations:
\begin{align}
Q_1=\left(
\begin{array}{cc}
0&x^3\\
\frac{x^3}{6}&0
\end{array}
\right)
\qquad&
Q_2=\left(
\begin{array}{cc}
0&x^2\\
\frac{x^4}{6}&0
\end{array}
\right)
\qquad
Q_3=\left(
\begin{array}{cc}
0&x\\
\frac{x^5}{6}&0
\end{array}
\right)
\end{align}
The matrices that determine the R--charges are given by:
\begin{align}
R_1=\left(
\begin{array}{cc}
0&0\\
0&0
\end{array}
\right)
\qquad&
R_2=\left(
\begin{array}{cc}
\frac{1}{6}&0\\
0&-\frac{1}{6}
\end{array}
\right)
\qquad
R_3=\left(
\begin{array}{cc}
\frac{1}{3}&0\\
0&-\frac{1}{3}
\end{array}
\right)
\end{align}
The bulk chiral ring is defined by the monomials $\{1,x,x^2,x^3,x^4\}$. The three--point functions in the bulk can be calculated from the WDVV potential, given by the following expression:
\begin{align}
\mathcal{F}=&\frac{1}{2}t_5^2t_4+\frac{1}{2}t_6t_4^2+t_6t_5t_3+\frac{1}{2}t_4^2t_3^2+\frac{1}{6}t_5t_3^3+\frac{1}{2}t_6^2t_2+\frac{1}{6}t_4^3t_2+t_5t_4t_3t_2+\frac{1}{6}t_3^4t_2 \nonumber \\
&+\frac{1}{4}t_5^2t_2^2+\frac{1}{2}t_4t_3^2t_2^2+\frac{1}{6}t_4^2t_2^3+\frac{1}{8}t_3^2t_2^4+\frac{1}{210}t_2^7
\end{align}
We now calculate the effective superpotentials for some examples of systems with one and two branes.
\subsubsection{Brane 1 -- $(k,\ell)=(4,2)$}
This is the case of the self--dual brane. The spectrum is given by the following states:
\begin{align}
&\phi_1^{(0)}[1,1]=\left(
\begin{array}{cc}
1&0\\
0&1
\end{array}
\right)
\qquad
\phi_2^{(\frac{1}{3})}[1,1]=\left(
\begin{array}{cc}
x&0\\
0&x
\end{array}
\right)
\qquad
\phi_3^{(\frac{2}{3})}[1,1]=\left(
\begin{array}{cc}
x^2&0\\
0&x^2
\end{array}
\right)\nonumber \\
&\psi_4^{(0)}[1,1]=\left(
\begin{array}{cc}
0&1\\
-\frac{1}{6}&0
\end{array}
\right)
\qquad
\psi_5^{(\frac{1}{3})}[1,1]=\left(
\begin{array}{cc}
0&x\\
-\frac{x}{6}&0
\end{array}
\right)
\qquad
\psi_6^{(\frac{2}{3})}[1,1]=\left(
\begin{array}{cc}
0&x^2\\
-\frac{x^2}{6}&0
\end{array}
\right)
\end{align}
Note, that for the self--dual case bosonic and fermionic states carry the same R--charges.\\
Solving $2359$ $A_{\infty}$ relations, $306$ bulk--boundary crossing constraints and $503$ Cardy equations, one finds:
\begin{align}
&\mathcal{F}_4=\mathcal{F}_{56}=\mathcal{F}_{666}=t_4-\frac{3}{2}t_2^2\nonumber\\
&\mathcal{F}_5=\mathcal{F}_{66}=t_5-2t_2t_3\nonumber\\
&\mathcal{F}_6=t_6-\frac{1}{2}t_3^2-t_2t_4+\frac{1}{3}t_2^3\nonumber \\
&\mathcal{F}_{45}=\mathcal{F}_{466}=\mathcal{F}_{556}=\mathcal{F}_{5666}=\mathcal{F}_{66666}=t_2 \nonumber \\
&\mathcal{F}_{46}=\mathcal{F}_{55}=\mathcal{F}_{566}=\mathcal{F}_{6666}=t_3\nonumber\\
&\mathcal{F}_{116}=\mathcal{F}_{125}=\mathcal{F}_{134}=-\mathcal{F}_{143}=-\mathcal{F}_{152}=\mathcal{F}_{224}=\mathcal{F}_{2436}=\mathcal{F}_{2526}=-\mathcal{F}_{2634}=\mathcal{F}_{3435}=-1\nonumber\\
&\mathcal{F}_{446}=\mathcal{F}_{455}=\mathcal{F}_{4566}=\mathcal{F}_{4656}=\mathcal{F}_{4665}=\mathcal{F}_{5556}=\mathcal{F}_{46666}=\mathcal{F}_{55666}=\nonumber\\
&=\mathcal{F}_{56566}=\mathcal{F}_{566666}=\mathcal{F}_{6666666}=-\frac{1}{6}
\end{align} 
Summing up the contribution yields the following effective superpotential:
\begin{align}
\mathcal{W}_{eff}=&\left(t_4-\frac{3}{2}t_2^2\right)u_3
+\left(t_5-2t_3t_2\right)u_2+t_2u_3u_2+\frac{1}{2}t_3u_2^2-\frac{1}{6}u_3u_2^2+\left(t_6-\frac{1}{2}t_3^2-t_4t_2+\frac{1}{3}t_2^3\right)u_1 \nonumber \\
&+t_3u_3u_1-\frac{1}{6}u_3^2u_1+\frac{1}{2}\left(2t_4-3t_2^2\right)u_2u_1+t_2u_2^2u_1-\frac{1}{6}u_2^3u_1+\frac{1}{2}\left(t_5-2t_3t_2\right)u_1^2+t_2u_3u_1^2\nonumber \\
&+t_3u_2u_1^2-\frac{1}{2}u_3u_2u_1^2+\frac{1}{3}\left(t_4-\frac{3}{2}t_2^2\right)u_1^3+t_2u_2u_1^3-\frac{1}{3}u_2^2u_1^3+\frac{1}{4}t_3u_1^4-\frac{1}{6}u_3u_1^4+\frac{1}{5}t_2u_1^5\nonumber \\
&-\frac{1}{6}u_2u_1^5-\frac{1}{42}u_1^7
\end{align}
This expression agrees with the result given in \cite{Herbst:2004jp}.

\subsubsection{Brane 2 -- $(k,\ell)=(4,1)$}
For this case we have the following spectrum:
\begin{align}
\phi_1^{(0)}[2,2]=\left(
\begin{array}{cc}
1&0\\
0&1
\end{array}
\right)
\qquad
\phi_2^{(\frac{1}{3})}[2,2]=\left(
\begin{array}{cc}
x&0\\
0&x
\end{array}
\right)
\nonumber\\
\psi_3^{(\frac{1}{3})}[2,2]\left(
\begin{array}{cc}
0&1\\
-\frac{x^2}{6}&0
\end{array}
\right)
\qquad
\psi_4^{(\frac{2}{3})}[2,2]\left(
\begin{array}{cc}
0&x\\
-\frac{x^3}{6}&0
\end{array}
\right)
\end{align}
$79$ $A_{\infty}$--relations, $52$ bulk--boundary crossing constraints and $47$ Cardy relations yield:
\begin{align}
&\mathcal{F}_3=\mathcal{F}_{44}=t_5-2t_2t_3\nonumber\\
&\mathcal{F}_4=t_6-\frac{1}{2}t_3^2-t_2t_4+\frac{1}{3}t_2^3\nonumber\\
&\mathcal{F}_{33}=\mathcal{F}_{344}=\mathcal{F}_{4444}=t_3\nonumber\\
&\mathcal{F}_{34}=\mathcal{F}_{444}=t_3-\frac{3}{2}t_2^2\nonumber\\
&\mathcal{F}_{114}=\mathcal{F}_{123}=-\mathcal{F}_{132}=\mathcal{F}_{2324}=-1\nonumber\\
&\mathcal{F}_{334}=\mathcal{F}_{3444}=\mathcal{F}_{44444}=t_2\nonumber\\
&\mathcal{F}_{3334}=\mathcal{F}_{33444}=\mathcal{F}_{34344}=\mathcal{F}_{344444}=\mathcal{F}_{4444444}=-\frac{1}{6}
\end{align}
From this, we get the following superpotential:
\begin{align}
\mathcal{W}_{eff}=&\left(t_5-2t_3t_2\right)u_2
+\frac{1}{2}t_3u_2^2+\left(t_6-\frac{1}{2}t_3^2-t_4t_2+\frac{1}{3}t_2^3\right)u_1+\frac{1}{2}\left(2t_4-3t_2^2\right)u_2u_1 \nonumber \\
&+t_2u_2^2u_1-\frac{1}{6}u_2^3u_1+\frac{1}{2}\left(t_5-2t_3t_2\right)u_1^2+t_3u_2u_1^2+\frac{1}{3}\left(t_4-\frac{3}{2}t_2^2\right)u_1^3+t_2u_2u_1^3\nonumber \\
&-\frac{1}{3}u_2^2u_1^3+\frac{1}{4}t_3u_1^4+\frac{1}{5}t_2u_1^5-\frac{1}{6}u_2u_1^5-\frac{1}{42}u_1^7
\end{align}

\subsubsection{Brane 3 -- $(k,\ell)=(4,0)$}
The spectrum contains only two states:
\begin{align}
\phi_1^{(0)}[3,3]=\left(
\begin{array}{cc}
1&0\\
0&1
\end{array}
\right)
\qquad&
\psi_2^{(\frac{2}{3})}[3,3]=\left(
\begin{array}{cc}
0&1\\
-\frac{x^4}{6}&0
\end{array}
\right)
\end{align}
There are no $A_{\infty}$--relations, $13$ bulk boundary crossing constraints and $8$ Cardy equations and we obtain:
\begin{align}
&\mathcal{F}_{2}=t_6-\frac{1}{2}t_3^2-t_2t_4+\frac{1}{3}t_2^3\nonumber\\
&\mathcal{F}_{22}=t_5-2t_2t_3\nonumber\\
&\mathcal{F}_{112}=-1\nonumber\\
&\mathcal{F}_{222}=t_4-\frac{3}{2}t_2^2\nonumber\\
&\mathcal{F}_{2222}=t_3\nonumber\\
&\mathcal{F}_{22222}=t_2\nonumber\\
&\mathcal{F}_{222222}=-\frac{1}{6}
\end{align}
The effective superpotential is:
\begin{align}
\mathcal{W}_{eff}&=\left(t_6-\frac{1}{2}t_3^2-t_4t_2+\frac{1}{3}t_2^2\right)u_1
+\frac{1}{2}\left(t_5-2t_3t_2\right)u_1^2+\frac{1}{3}\left(t_4-\frac{3}{2}\right)u_1^3\nonumber \\
&+\frac{1}{4}t_3u_1^4+\frac{1}{5}t_2u_1^5-\frac{1}{42}u_1^7
\end{align}

\subsubsection{Brane 2 and Brane 3 -- $(4,1)\oplus(4,0)$}
Computing the cohomology yields the following spectrum:
\begin{align}
\phi_1^{(0)}[2,2]=\left(
\begin{array}{cc}
1&0\\
0&1
\end{array}
\right)
\qquad&
\phi_2^{(\frac{1}{3})}[2,2]=\left(
\begin{array}{cc}
x&0\\
0&x
\end{array}
\right)\nonumber\\
\psi_3^{(\frac{1}{3})}[2,2]=\left(
\begin{array}{cc}
0&1\\
-\frac{x^2}{6}&0
\end{array}
\right)
\qquad&
\psi_4^{(\frac{2}{3})}[2,2]=\left(
\begin{array}{cc}
0&x\\
-\frac{x^3}{6}&0
\end{array}
\right)
\end{align}
\begin{align}
\phi_5^{(\frac{1}{6})}[2,3]=\left(
\begin{array}{cc}
x&0\\
0&1
\end{array}
\right)
\qquad&
\psi_6^{(\frac{1}{2})}[2,3]=\left(
\begin{array}{cc}
0&1\\
-\frac{x^3}{6}&0
\end{array}
\right)
\end{align}
\begin{align}
\phi_7^{(\frac{1}{6})}[3,2]=\left(
\begin{array}{cc}
1&0\\
0&x
\end{array}
\right)
\qquad&
\psi_8^{(\frac{1}{2})}[3,2]=\left(
\begin{array}{cc}
0&1\\
-\frac{x^3}{6}&0
\end{array}
\right)
\end{align}
\begin{align}
\phi_9^{(0)}[3,3]=\left(
\begin{array}{cc}
1&0\\
0&1
\end{array}
\right)
\qquad&
\psi_{10}^{(\frac{2}{3})}[3,3]=\left(
\begin{array}{cc}
0&1\\
-\frac{x^4}{6}&0
\end{array}
\right)
\end{align}
Solving $449$ $A_{\infty}$ equations, $108$ bulk--boundary crossing constraints and $55$ (truncated) Cardy equations, we find:
\begin{align}
&\mathcal{F}_{3}=\mathcal{F}_{(4,4)}=\mathcal{F}_{(10,10)}=t_5-2t_2t_3\nonumber\\
&\mathcal{F}_{4}=\mathcal{F}_{10}=t_1-\frac{1}{2}t_3^2-t_2t_4+\frac{1}{3}t_2^3\nonumber\\
&\mathcal{F}_{(3,3)}=\mathcal{F}_{(3,4,4)}=\mathcal{F}_{(4,6,8)}=\mathcal{F}_{(6,10,8)}=\mathcal{F}_{(4,4,4,4)}=\mathcal{F}_{(10,10,10,10)}=t_3\nonumber\\
&\mathcal{F}_{(3,4)}=\mathcal{F}_{(6,8)}=\mathcal{F}_{(4,4,4)}=\mathcal{F}_{(10,10,10)}=t_3-\frac{3}{2}t_2^2\nonumber\\
&\mathcal{F}_{(1,1,4)}=\mathcal{F}_{(1,2,3)}=-\mathcal{F}_{(1,3,2)}=\mathcal{F}_{(1,5,8)}=-\mathcal{F}_{(1,6,7)}=\mathcal{F}_{(3,5,7)}=\mathcal{F}_{(5,9,8)}=\mathcal{F}_{(6,9,7)}=\mathcal{F}_{(9,9,10)}=\mathcal{F}_{(2,2,6,8)}\nonumber\\
&=\mathcal{F}_{(2,3,2,4)}=-\mathcal{F}_{(2,4,5,8)}=\mathcal{F}_{(2,5,10,8)}=\mathcal{F}_{(2,6,7,4)} =-\mathcal{F}_{(2,6,10,7)}=\mathcal{F}_{(4,5,10,7)}=-\mathcal{F}_{(5,10,10,7)}=-1\nonumber\\
&\mathcal{F}_{(3,3,4)}=\mathcal{F}_{(3,6,8)}=\mathcal{F}_{(3,4,4,4)}=\mathcal{F}_{(4,4,6,8)}=\mathcal{F}_{(4,6,10,8)}=\mathcal{F}_{(6,10,10,8)}=\mathcal{F}_{(4,4,4,4,4)}=\mathcal{F}_{(10,10,10,10,10)}=t_2\nonumber\\
&\mathcal{F}_{(3,3,3,4)}=\mathcal{F}_{(3,3,6,8)}=\mathcal{F}_{(3,3,4,4,4)}=\mathcal{F}_{(3,4,3,4,4)}=\mathcal{F}_{(3,4,4,6,8)}=\mathcal{F}_{(3,4,6,8,4)}=\mathcal{F}_{(3,4,6,10,8)}\nonumber\\
&=\mathcal{F}_{(3,6,8,4,4)}=\mathcal{F}_{(3,6,10,8,4)}=\mathcal{F}_{(3,6,10,10,8)}=\mathcal{F}_{(4,6,8,6,8)}=\mathcal{F}_{(6,8,6,10,8)}=\mathcal{F}_{(3,4,4,4,4,4)}=\mathcal{F}_{(4,4,4,4,6,8)}\nonumber\\
&=\mathcal{F}_{(4,4,4,6,10,8)}=\mathcal{F}_{(4,4,6,10,10,8)}=\mathcal{F}_{(4,6,10,10,10,8)}=\mathcal{F}_{(6,10,10,10,10,8)}\nonumber\\
&=\mathcal{F}_{(4,4,4,4,4,4,4)}=\mathcal{F}_{(10,10,10,10,10,10,10)}=-\frac{1}{6}
\end{align}
The effective superpotential is:
\begin{align}
\mathcal{W}_{eff}=&-\frac{1}{6}\bigg(\frac{u_1^7}{7}+\frac{w_1^7}{7}+2u_1^3u_2^2+u_1^5u_2+u_1u_2^3+u_1^4v_{\frac{3}{2}}\tilde{v}_{\frac{3}{2}}+u_1^3v_{\frac{3}{2}}\tilde{v}_{\frac{3}{2}}w_1+3u_1^2u_2v_{\frac{3}{2}}\tilde{v}_{\frac{3}{2}}+u_1^2v_{\frac{3}{2}}\tilde{v}_{\frac{3}{2}}w_1\nonumber\\
&+u_1v_{\frac{3}{2}}^2\tilde{v}_{\frac{3}{2}}^2+u_1v_{\frac{3}{2}}\tilde{v}_{\frac{3}{2}}w_1^3+u_2^2v_{\frac{3}{2}}\tilde{v}_{\frac{3}{2}}+2u_2u_1v_{\frac{3}{2}}\tilde{v}_{\frac{3}{2}}w_1+u_2v_{\frac{3}{2}}\tilde{v}_{\frac{3}{2}}w_1^2+v_{\frac{3}{2}}^2\tilde{v}_{\frac{3}{2}}w_1+v_{\frac{3}{2}}\tilde{v}_{\frac{3}{2}}w_1^4\bigg)\nonumber\\
&+t_2\left(\frac{u_1^5}{5}+\frac{w_1^5}{5}+u_1^3u_2+u_1^2v_{\frac{3}{2}}\tilde{v}_{\frac{3}{2}}+u_1u_2^2+u_1v_{\frac{3}{2}}\tilde{v}_{\frac{3}{2}}w_1+u_2v_{\frac{3}{2}}\tilde{v}_{\frac{3}{2}}+v_{\frac{3}{2}}\tilde{v}_{\frac{3}{2}}w_1^2\right)\nonumber\\
&+t_3\left(\frac{u_1^4}{4}+\frac{w_1^4}{4}+u_1^2u_2+\frac{u_2^2}{2}+u_1v_{\frac{3}{2}}\tilde{v}_{\frac{3}{2}}+v_{\frac{3}{2}}\tilde{v}_{\frac{3}{2}}w_1\right)+\left(t_4-\frac{3}{2}t_2^2\right)\left(\frac{u_1^3}{3}+\frac{w_1^3}{3}+u_1u_2+v_{\frac{3}{2}}\tilde{v}_{\frac{3}{2}}\right)\nonumber\\
&+\left(t_5-2t_2t_3\right)\left(\frac{u_1^2}{2}+\frac{w_1^2}{2}+u_2\right)+\left(t_6-\frac{1}{2}t_3^2-t_2t_4+\frac{1}{3}t_2^3\right)\left(u_1+w_1\right)
\end{align}

\subsubsection{Two Copies of Brane 3 -- $(4,0)\oplus(4,0)$}
We choose the following labelling for the states:
\begin{align}
&\phi_1^{{0}}[1,1]=\phi_3^{{0}}[1,2]=\phi_5^{{0}}[2,1]=\phi_7^{{0}}[2,2]=\left(\begin{array}{cc}
1&0\\
0&1
\end{array}
\right)
\nonumber\\
&\psi_2^{(\frac{2}{3})}[1,1]=\psi_4^{(\frac{2}{3})}[1,2]=\psi_6^{(\frac{2}{3})}[2,1]=\psi_8^{(\frac{2}{3})}[2,2]=\left(
\begin{array}{cc}
0&1\\
-\frac{x^4}{6}&0
\end{array}
\right)
\end{align}
Here brane 1 is equal to brane 2 and thus, $Q_1=Q_2=\left(\begin{array}{cc}0&x\\\frac{x^5}{6}&0\end{array}\right)$.\\
Solving the consistency conditions it turns out that one does not have to truncate the Cardy equations in this case. However, the result is also determined completely if the equations are truncated. The following result is found, after solving $78$ $A_{\infty}$--relations, $93$ bulk--boundary crossing constraints and $121$ Cardy equations:
\begin{align}
&\mathcal{F}_{2}=\mathcal{F}_{8}=t_6-\frac{1}{2}t_3^2-t_2t_4+\frac{1}{3}t_2^3\nonumber\\
&\mathcal{F}_{22}=\mathcal{F}_{46}=\mathcal{F}_{88}=t_5-2t_2t_3\nonumber\\
&\mathcal{F}_{112}=\mathcal{F}_{136}=-\mathcal{F}_{145}=\mathcal{F}_{235}=\mathcal{F}_{376}=-\mathcal{F}_{385}=\mathcal{F}_{475}=\mathcal{F}_{778}=-1\nonumber\\
&\mathcal{F}_{222}=\mathcal{F}_{246}=\mathcal{F}_{486}=\mathcal{F}_{888}=t_4-\frac{3}{2}t_4^2\nonumber\\
&\mathcal{F}_{2222}=\mathcal{F}_{2246}=\mathcal{F}_{2486}=\mathcal{F}_{4646}=\mathcal{F}_{4886}=\mathcal{F}_{8888}=t_3\nonumber\\
&\mathcal{F}_{22222}=\mathcal{F}_{22246}=\mathcal{F}_{22486}=\mathcal{F}_{24646}=\mathcal{F}_{24886}=\mathcal{F}_{46486}=\mathcal{F}_{48886}=\mathcal{F}_{88888}=t_2\nonumber\\
&\mathcal{F}_{2222222}=\mathcal{F}_{2222246}=\mathcal{F}_{2222486}=\mathcal{F}_{224646}=\mathcal{F}_{2224886}=\mathcal{F}_{2246246}=\mathcal{F}_{2246486}=\mathcal{F}_{2248646}=\mathcal{F}_{2248886}\nonumber\\
&=\mathcal{F}_{2462486}=\mathcal{F}_{2464646}=\mathcal{F}_{2464886}=\mathcal{F}_{2486486}=\mathcal{F}_{2488646}=\mathcal{F}_{2488886}=\mathcal{F}_{4646486}\nonumber\\
&=\mathcal{F}_{4648886}=\mathcal{F}_{4864886}=\mathcal{F}_{4888886}=\mathcal{F}_{8888888}=-\frac{1}{6}
\end{align}
The effective superpotential is then:
\begin{align}
\mathcal{W}_{eff}=&-\frac{1}{6}\bigg(\frac{u_1^7}{7}+\frac{w_1^7}{7}+u_1^5v_1\tilde{v}_8+u_1^4v_1\tilde{v}_8w_1+2u_1^3v_1^2\tilde{v}_8^2+u_1^3v_1\tilde{v}_8w_1^2+u_1^2v_1\tilde{v}_8w_1^3+3u_1^2v_1^2\tilde{v}_8^2w_1\nonumber\\
&+u_1v_1^3\tilde{v}_8^3+3u_1v_1^2\tilde{v}_8w_1^2+u_1v_1\tilde{v}_8w_1^4+v_1\tilde{v}_8w_1^5+2v_1^2\tilde{v}_8^2w_1^3+v_1^3\tilde{v}_8^3w_1\bigg)\nonumber\\
&+t_2\left(\frac{u_1^5}{5}+\frac{w_1^5}{5}+u_1^3v_1\tilde{v}_8+u_1^2v_1\tilde{v}_8w_1+u_1v_1^2\tilde{v}_8^2+u_1v_1\tilde{v}_8w_1^2+v_1^2\tilde{v}_8^2w_1+v_1\tilde{v}_8w_1^3\right)\nonumber\\
&+t_3\left(\frac{u_1^4}{4}+\frac{w_1^4}{4}+u_1^2v_1\tilde{v}_8+u_1v_1\tilde{v}_8w_1+\frac{1}{2}v_1^2\tilde{v}_8^2+v_1\tilde{v}_8w_1^2\right)\nonumber\\
&+\left(t_4-\frac{3}{2}t_2^2\right)\left(\frac{u_1^3}{3}+\frac{w_1^3}{3}+u_1v_1\tilde{v}_8+v_1\tilde{v}_8w_1\right)+\left(t_5-2t_2t_3\right)\left(\frac{u_1^2}{2}+\frac{w_1^2}{2}+v_1\tilde{v}_8\right)\nonumber\\
&+\left(t_6-\frac{1}{2}t_3^2-t_2t_4+\frac{1}{3}t_2^3\right)\left(u_1+w_1\right)
\end{align}

\chapter{Details on the Quartic Torus}
\label{app-torus}
In this appendix we give the explicit result for the boundary changing sector of the quartic torus. Furthermore we collect some theta function identities which are needed for calculating the B--model correlators.

\section{Boundary Changing Spectrum}
We now state the explicit results for the boundary changing operators of the quartic torus.\\
Fermions and bosons will be denoted by:
\begin{align}
&\psi_{\mathcal{A}\mathcal{B}}(i,j)=\left(\begin{array}{cc}0&\psi^{(0)}_{\mathcal{A}\mathcal{B}}(i,j)\\
\psi^{(1)}_{\mathcal{A}\mathcal{B}}(i,j)&0
\end{array}\right)\quad\textrm{and}\quad
\phi_{\mathcal{A}\mathcal{B}}(i,j)=\left(\begin{array}{cc}\phi^{(0)}_{\mathcal{A}\mathcal{B}}(i,j)&0\\
0&\psi^{(1)}_{\mathcal{A}\mathcal{B}}(i,j)
\end{array}\right),
\end{align}
respectively, where $\mathcal{A},\mathcal{B}$ denote the branes and $i,j$ denote the indices of the boundary moduli. We will use bars and tildes to make a distinction if there is more than one boundary state going from $\mathcal{A}$ to $\mathcal{B}$. Note that all states are defined only up to a normalization by a moduli--dependent factor.
\subsection{Long Branes}
From figure \ref{fig-bquartic1} one can read off that there are two fermions and two bosons stretching between two of the long branes. All of them have R--charge $1/2$. The first fermion has the following structure:
\begin{align}
{\psi}^{(0)}_{\mathcal{L}_1\mathcal{L}_2}(1,2)=\left(\begin{array}{cc}{a}(1,2)&{b}(1,2)\\
{c}(1,2)&{d}(1,2)
\end{array}\right)\quad
{\psi}^{(1)}_{\mathcal{L}_1\mathcal{L}_2}(1,2)=\left(\begin{array}{cc}
{d}(2,1)&-{b}(2,1)\\
-{c}(2,1)&{a}(2,1)
\end{array}\right),
\end{align}
where
\begin{align}
&{a}(1,2)=\left(\frac{1}{\alpha_1^1\alpha_2^1}+\frac{\alpha_3^2}{\alpha_1^2\alpha_2^2\alpha_3^1}\right)x_1\qquad {b}(1,2)=\left(-\frac{\alpha_1^2}{\alpha_2^2\alpha_3^1\alpha_3^2}+\frac{(\alpha_2^1)^2\alpha_2^2}{(\alpha_1^1)^2\alpha_1^2\alpha_3^1\alpha_3^2}\right)x_2
\nonumber \\
&{c}(1,2)=\left(\frac{\alpha_1^1}{\alpha_2^1}-\frac{\alpha_2^1(\alpha_2^2)^2}{\alpha_1^1(\alpha_1^2)^2}\right)x_2\qquad {d}(1,2)=\left(-\frac{\alpha_2^1}{\alpha_1^1(\alpha_1^2)^2}-\frac{\alpha_2^2\alpha_3^1}{(\alpha_1^1)^2\alpha_1^2\alpha_3^2}\right)x_1
\end{align}
The second fermion has the same structure with $x_1$ and $x_2$ exchanged:
\begin{align}
\bar{\psi}^{(0)}_{\mathcal{L}_1\mathcal{L}_2}(1,2)=\left(\begin{array}{cc}\bar{a}(1,2)&\bar{b}(1,2)\\
\bar{c}(1,2)&\bar{d}(1,2)
\end{array}\right)\quad
\psi^{(1)}_{\mathcal{L}_1\mathcal{L}_2}(1,2)=\left(\begin{array}{cc}
\bar{d}(2,1)&-\bar{b}(2,1)\\
-\bar{c}(2,1)&\bar{a}(2,1)
\end{array}\right),
\end{align}
where
\begin{align}
\bar{a}(1,2)=\left(\frac{\alpha_2^2}{\alpha_1^1}+\frac{\alpha_2^1\alpha_3^2}{\alpha_1^2\alpha_3^1}\right)x_2\qquad \bar{b}(1,2)=\left(-\frac{(\alpha_1^1)^2\alpha_1^2}{\alpha_2^1\alpha_3^1\alpha_3^2}+\frac{\alpha_2^1(\alpha_2^2)^2}{\alpha_1^2\alpha_3^1\alpha_3^2}\right)x_1
\nonumber \\
\bar{c}(1,2)=\left(\frac{\alpha_1^1(\alpha_1^2)^2}{\alpha_2^2}-\frac{(\alpha_2^1)^2\alpha_2^2}{\alpha_1^1}\right)x_1\qquad \bar{d}(1,2)=\left(\frac{-\alpha_1^1}{\alpha_2^2}-\frac{\alpha_1^2\alpha_3^1}{\alpha_2^1\alpha_3^2}\right)x_2
\end{align}
The first boson looks as follows:
\begin{align}
&\phi^{(0)}_{\mathcal{L}_1\mathcal{L}_2}(1,2)=\left(\begin{array}{cc}a(1,2)&b(1,2)\\
c(1,2)&d(1,2)
\end{array}\right)\quad
\phi^{(1)}_{\mathcal{L}_1\mathcal{L}_2}(1,2)=\left(\begin{array}{cc}
\frac{\alpha_3^2}{\alpha_3^1}d(1,2)&\frac{1}{\alpha_3^1\alpha_3^2}c(1,2)\\
\alpha_3^1\alpha_3^2 b(1,2)&\frac{\alpha_3^1}{\alpha_3^2}a(1,2)
\end{array}\right),
\end{align}
where
\begin{align}
a(1,2)=\left(\alpha_1^1+\frac{(\alpha_1^1)^2\alpha_2^1\alpha_3^2}{\alpha_1^2\alpha_2^2\alpha_3^1}\right)x_2\qquad b(1,2)=\left(-\frac{(\alpha_1^1)^2\alpha_1^2\alpha_2^1}{\alpha_2^2\alpha_3^1\alpha_3^2}+\frac{(\alpha_1^1)^4\alpha_2^2}{\alpha_1^2\alpha_2^1\alpha_3^1\alpha_3^2}\right)x_1
\nonumber \\
c(1,2)=\left(-\alpha_1^1(\alpha_2^1)^2+\frac{(\alpha_1^1)^3(\alpha_2^2)^2}{(\alpha_1^2)^2}\right)x_1\qquad d(1,2)=\left(\frac{(\alpha_1^1)^3}{(\alpha_1^2)^2}+\frac{(\alpha_1^1)^2\alpha_2^2\alpha_3^1}{\alpha_1^2\alpha_2^1\alpha_3^2}\right)x_2
\end{align}
For the second boson we find:
\begin{align}
\bar{\phi}^{(0)}_{\mathcal{L}_1\mathcal{L}_2}(1,2)=\left(\begin{array}{cc}\bar{a}(1,2)&\bar{b}(1,2)\\
\bar{c}(1,2)&\bar{d}(1,2)
\end{array}\right)\quad
\bar{\phi}^{(1)}_{\mathcal{L}_1\mathcal{L}_2}(1,2)\left(\begin{array}{cc}
\frac{\alpha_3^1}{\alpha_3^2}d(1,2)&\frac{1}{\alpha_3^1\alpha_3^2}c(1,2)\\
\alpha_3^1\alpha_3^2 b(1,2)&\frac{\alpha_3^1}{\alpha_3^2}a(1,2)
\end{array}\right),
\end{align}
where
\begin{align}
&\bar{a}(1,2)=\left(\frac{\alpha_1^1}{\alpha_1^2}+\frac{\alpha_2^2\alpha_3^1}{\alpha_2^1\alpha_3^2}\right)x_1\qquad\bar{b}(1,2)=\left(-\frac{\alpha_1^2(\alpha_2^1)^2}{\alpha_1^1(\alpha_3^2)^2}+\frac{\alpha_1^1(\alpha_2^2)^2}{\alpha_1^2(\alpha_3^2)^2}\right)x_2
\nonumber\\
&\bar{c}(1,2)=\left(-\frac{(\alpha_1^2)^2\alpha_2^1\alpha_3^1}{\alpha_2^2\alpha_3^2}+\frac{(\alpha_1^1)^2\alpha_2^2\alpha_3^1}{\alpha_2^1\alpha_3^2}\right)x_2\qquad\bar{d}(1,2)=\left(\frac{\alpha_1^2(\alpha_3^1)^2}{\alpha_1^1(\alpha_3^2)^2}+\frac{\alpha_2^1\alpha_3^1}{\alpha_2^2\alpha_3^2}\right)x_1
\end{align}
\subsection{Short Branes}
There is one charge $1/2$ fermion and one charge $1/2$ boson. The fermion is given by:
\begin{align}
&\psi^{(0)}_{\mathcal{S}_1\mathcal{S}_2}=\left(\begin{array}{cc}
a(1,2)&b(1,2)\\
c(1,2)&d(1,2)
\end{array}\right)\qquad
\psi^{(1)}_{\mathcal{S}_1\mathcal{S}_2}=\left(\begin{array}{cc}
-\frac{\alpha_3^2}{\alpha_3^1}d(1,2)&\alpha_3^1\alpha_3^2c(1,2)\\
\frac{1}{\alpha_3^1\alpha_3^2}b(1,2)&-\frac{\alpha_3^2}{\alpha_3^1}a(1,2)
\end{array}\right),
\end{align}
where
\begin{align}
a(1,2)&=\frac{\alpha_1^2\alpha_2^1\alpha_2^2}{\alpha_1^1}+\frac{(\alpha_2^1)^2\alpha_3^2}{\alpha_3^1}\nonumber \\
b(1,2)&=\left(\frac{\alpha_1^1\alpha_1^2\alpha_2^1\alpha_3^2}{\alpha_3^1}+\frac{(\alpha_1^2)^2(\alpha_2^1)^2\alpha_3^2}{\alpha_2^2\alpha_3^1}\right)x_2+\left(\frac{(\alpha_2^1)^3\alpha_2^2\alpha_3^2}{\alpha_1^1\alpha_3^1}+\frac{(\alpha_2^1)^2(\alpha_2^2)^2\alpha_3^2}{\alpha_1^2\alpha_3^1}\right)x_1\nonumber\\
c(1,2)&=\left(\frac{(\alpha_1^2)^2\alpha_2^1}{(\alpha_3^1)^2}+\frac{\alpha_1^1\alpha_1^2\alpha_2^2}{(\alpha_3^1)^2}\right)x_2+\left(\frac{\alpha_1^2(\alpha_2^1)^3\alpha_2^2}{(\alpha_1^1)^2(\alpha_3^1)^2}+\frac{(\alpha_2^1)^2(\alpha_2^2)^2}{\alpha_1^1(\alpha_3^1)^2}\right)x_1\nonumber \\
d(1,2)&=\left(\frac{\alpha_2^1\alpha_2^2\alpha_3^2}{(\alpha_1^1)^2\alpha_3^1}+\frac{(\alpha_2^1)^2(\alpha_3^2)^2}{\alpha_1^1\alpha_1^2(\alpha_3^1)^2}\right)x_1^2+\left(\frac{\alpha_1^2\alpha_3^2}{\alpha_1^1\alpha_3^1}+\frac{\alpha_2^1(\alpha_3^2)^2}{\alpha_2^2(\alpha_3^1)^2}\right)x_2^2
\end{align}
The boson looks as follows:
\begin{align}
\phi^{(0)}_{\mathcal{S}_1\mathcal{S}_2}=\left(\begin{array}{cc}
a(1,2)&b(1,2)\\
c(1,2)&d(1,2)
\end{array}\right)\qquad
\phi^{(1)}_{\mathcal{S}_1\mathcal{S}_2}=\left(\begin{array}{cc}
\frac{\alpha_3^1}{\alpha_3^2}d(1,2)&-\alpha_3^1\alpha_3^2c(1,2)\\
-\frac{1}{\alpha_3^1\alpha_3^2}b(1,2)&\frac{\alpha_3^2}{\alpha_3^1}a(1,2)
\end{array}\right),
\end{align}
where
\begin{align}
a(1,2)&=\left(\alpha_3^1+\frac{\alpha_1^2\alpha_2^1\alpha_3^1}{\alpha_1^1\alpha_2^2}\right)x_2+\left(\frac{(\alpha_2^1)^2\alpha_2^2\alpha_3^1}{(\alpha_1^1)^2\alpha_1^2}+\frac{\alpha_2^1(\alpha_2^2)^2\alpha_3^1}{\alpha_1^1(\alpha_1^2)^2}\right)x_1\nonumber \\
b(1,2)&=\frac{\alpha_2^2(\alpha_3^1)^2\alpha_3^2}{(\alpha_1^1)^2}-\frac{\alpha_2^1\alpha_3^1(\alpha_3^2)^2}{\alpha_1^1\alpha_1^2}\nonumber\\
c(1,2)&=\left(\frac{\alpha_2^2\alpha_3^1}{(\alpha_1^1)^3\alpha_1^2}-\frac{\alpha_2^1\alpha_3^2}{(\alpha_1^1)^2(\alpha_1^2)^2}\right)x_1^2+\left(\frac{\alpha_3^1}{(\alpha_1^1)^2\alpha_2^1}-\frac{\alpha_3^2}{\alpha_1^1\alpha_1^2\alpha_2^2}\right)x_2^2\nonumber\\
d(1,2)&=\left(\frac{\alpha_1^2\alpha_3^2}{\alpha_1^1}+\frac{\alpha_2^2\alpha_3^2}{\alpha_2^1}\right)x_2+\left(\frac{(\alpha_2^1)^2\alpha_2^2\alpha_3^2}{(\alpha_1^1)^3}+\frac{\alpha_2^1(\alpha_2^2)^2\alpha_3^2}{(\alpha_1^1)^2\alpha_1^2}\right)x_1
\end{align}

\subsection{Short to Long}
There is one fermion with charge $1/4$ and one with charge $3/4$ and, symmetrically, bosons with charges $1/4$ and $3/4$. We will denote the $3/4$--states with tildes.
The charge $1/4$ fermion reads:
\begin{align}
&\psi^{(0)}_{\mathcal{S}_1\mathcal{L}_2}=\left(\begin{array}{cc}
a(1,2)&b(1,2)\\
c(1,2)&d(1,2)
\end{array}\right)\qquad
\psi^{(1)}_{\mathcal{S}_1\mathcal{L}_2}=\left(\begin{array}{cc}
-\alpha_3^1\alpha_3^2d(1,2)&-\frac{\alpha_3^1}{\alpha_3^2}c(1,2)\\
\frac{\alpha_3^2}{\alpha_3^1}b(1,2)&\frac{1}{\alpha_3^1\alpha_3^2}a(1,2)
\end{array}\right),
\end{align}
where
\begin{align}
a(1,2)&=\frac{\alpha_1^1\alpha_3^1}{(\alpha_2^1)^2}-\frac{(\alpha_1^1)^2\alpha_3^2}{\alpha_1^2\alpha_2^1\alpha_2^2}\nonumber \\
b(1,2)&=-\frac{(\alpha_1^1)^3\alpha_1^2}{(\alpha_2^1)^2\alpha_2^2\alpha_3^2}+\frac{\alpha_1^1\alpha_2^2}{\alpha_1^2\alpha_3^2}\nonumber \\
c(1,2)&=\left(\frac{1}{\alpha_3^1}-\frac{(\alpha_1^1)^2(\alpha_1^2)^2}{(\alpha_2^1)^2(\alpha_2^2)^2\alpha_3^1}\right)x_1+\left(\frac{(\alpha_1^1)^3}{(\alpha_2^1)^3\alpha_3^1}-\frac{\alpha_1^1(\alpha_2^2)^2}{(\alpha_1^2)^2\alpha_2^1\alpha_3^1}\right)x_2\nonumber\\
d(1,2)&=\left(\frac{(\alpha_1^1)^2}{(\alpha_2^1)^2(\alpha_2^2)^2\alpha_3^1}-\frac{\alpha_1^1\alpha_1^2}{(\alpha_2^1)^3\alpha_2^2\alpha_3^2}\right)x_2+\left(-\frac{\alpha_1^1}{(\alpha_1^2)^2\alpha_2^1\alpha_3^1}+\frac{\alpha_2^2}{\alpha_1^2(\alpha_2^1)^2\alpha_3^2}\right)x_1
\end{align}
When computing the charge $3/4$ states one has to be careful with exact states. We can use those to ``gauge away'' one of the variables. We choose the convention that the charge $3/4$ states contain terms linear and quadratic in $x_1$ but not linear and quadratic in $x_2$. One mixed term $x_1x_2$ will always remain.
In this gauge the charge $3/4$ fermion reads:
\begin{align}
\tilde{\psi}^{(0)}_{\mathcal{S}_1\mathcal{L}_2}=\left(\begin{array}{cc}
\tilde{a}(1,2)&\tilde{b}(1,2)\\
\tilde{c}(1,2)&\tilde{d}(1,2)
\end{array}\right)\qquad
\tilde{\psi}^{(1)}_{\mathcal{S}_1\mathcal{L}_2}=\left(\begin{array}{cc}
-\alpha_3^1\alpha_3^2\tilde{d}(1,2)&-\frac{\alpha_3^1}{\alpha_3^2}\tilde{c}(1,2)\\
\frac{\alpha_3^2}{\alpha_3^1}\tilde{b}(1,2)&\frac{1}{\alpha_3^1\alpha_3^2}\tilde{a}(1,2)
\end{array}\right),
\end{align}
where
\begin{align}
\tilde{a}(1,2)&=\left(-(\alpha_1^1)^2\alpha_3^2+\frac{(\alpha_1^2)^2(\alpha_2^1)^2\alpha_3^2}{(\alpha_2^2)^2}\right)x_1\nonumber\\
\tilde{b}(1,2)&=\left(\frac{\alpha_1^2\alpha_3^1}{\alpha_2^2}+\frac{\alpha_1^1\alpha_2^1\alpha_3^2}{(\alpha_2^2)^2}\right)x_1\nonumber\\
\tilde{c}(1,2)&=\left(\frac{(\alpha_1^2)^2\alpha_3^2}{\alpha_1^1(\alpha_2^2)^2}+\frac{\alpha_1^2\alpha_2^1(\alpha_3^2)^2}{(\alpha_2^2)^2\alpha_3^1}\right)x_1^2+\left(-\frac{\alpha_3^2}{\alpha_2^1}-\frac{\alpha_1^1(\alpha_3^2)^2}{\alpha_1^2\alpha_2^2\alpha_3^1}\right)x_1x_2\nonumber\\
\tilde{d}(1,2)&=\left(-\frac{(\alpha_1^2)^3\alpha_2^1}{(\alpha_2^2)^3\alpha_3^1}+\frac{(\alpha_1^1)^2\alpha_1^2}{\alpha_2^1\alpha_2^2\alpha_3^1}\right)x_1x_2+\left(\frac{\alpha_1^2(\alpha_2^1)^2}{\alpha_1^1\alpha_2^2\alpha_3^1}-\frac{\alpha_1^1\alpha_2^2}{\alpha_1^2\alpha_3^1}\right)x_1^2
\end{align}
The charge $1/4$ boson looks as follows:
\begin{align}
\phi^{(0)}_{\mathcal{S}_1\mathcal{L}_2}=\left(\begin{array}{cc}
a(1,2)&b(1,2)\\
c(1,2)&d(1,2)
\end{array}\right)\qquad
\phi^{(1)}_{\mathcal{S}_1\mathcal{L}_2}=\left(\begin{array}{cc}
-\alpha_3^1\alpha_3^2d(1,2)&-\frac{\alpha_3^1}{\alpha_3^2}c(1,2)\\
\frac{\alpha_3^2}{\alpha_3^1}b(1,2)&\frac{1}{\alpha_3^1\alpha_3^2}a(1,2)
\end{array}\right),
\end{align}
where
\begin{align}
a(1,2)&=\frac{\alpha_2^1\alpha_3^1}{(\alpha_1^1)^2}-\frac{(\alpha_2^1)^2\alpha_3^2}{\alpha_1^1\alpha_1^2\alpha_2^2}\nonumber\\
b(1,2)&=-\frac{\alpha_1^2(\alpha_2^1)^2}{(\alpha_1^1)^2\alpha_2^2\alpha_3^2}+\frac{\alpha_2^1\alpha_2^2}{\alpha_1^2\alpha_3^2}\nonumber\\
c(1,2)&=\left(\frac{1}{\alpha_3^1}-\frac{(\alpha_1^2)^2(\alpha_2^1)^2}{(\alpha_1^1)^2(\alpha_2^2)^2\alpha_3^1}\right)x_2+\left(\frac{(\alpha_2^1)^3}{(\alpha_1^1)^3\alpha_3^1}-\frac{\alpha_2^1(\alpha_2^2)^2}{\alpha_1^1(\alpha_1^2)^2\alpha_3^1}\right)x_1\nonumber\\
d(1,2)&=\left(\frac{(\alpha_2^1)^2}{(\alpha_1^1)^2(\alpha_2^2)^2\alpha_3^1}-\frac{\alpha_1^2\alpha_2^1}{(\alpha_1^1)^3\alpha_2^2\alpha_3^2}\right)x_1+\left(-\frac{\alpha_2^1}{\alpha_1^1(\alpha_1^2)^2\alpha_3^1}+\frac{\alpha_2^2}{(\alpha_1^1)^2\alpha_1^2\alpha_3^2}\right)x_2
\end{align}
The charge $3/4$ boson is:
\begin{align}
\tilde{\phi}^{(0)}_{\mathcal{S}_1\mathcal{L}_2}=\left(\begin{array}{cc}
\tilde{a}(1,2)&\tilde{b}(1,2)\\
\tilde{c}(1,2)&\tilde{d}(1,2)
\end{array}\right)\qquad
\tilde{\phi}^{(1)}_{\mathcal{S}_1\mathcal{L}_2}=\left(\begin{array}{cc}
-\alpha_3^1\alpha_3^2\tilde{d}(1,2)&-\frac{\alpha_3^1}{\alpha_3^2}\tilde{c}(1,2)\\
\frac{\alpha_3^2}{\alpha_3^1}\tilde{b}(1,2)&\frac{1}{\alpha_3^1\alpha_3^2}\tilde{a}(1,2)
\end{array}\right),
\end{align}
where
\begin{align}
\tilde{a}(1,2)&=\left(-\frac{\alpha_1^1\alpha_2^2\alpha_3^1}{\alpha_1^2}+\frac{(\alpha_1^1)^2\alpha_2^1\alpha_3^2}{(\alpha_1^2)^2}\right)x_1\nonumber\\
\tilde{b}(1,2)&=\left(\frac{\alpha_1^1(\alpha_2^1)^2}{\alpha_3^2}-\frac{(\alpha_1^1)^3(\alpha_2^2)^2}{(\alpha_1^2)^2\alpha_3^2}\right)x_1\nonumber\\
\tilde{c}(1,2)&=\left(\frac{\alpha_1^1\alpha_1^2\alpha_2^1}{\alpha_2^1\alpha_3^1}-\frac{(\alpha_1^1)^3\alpha_2^2}{\alpha_1^2\alpha_2^1\alpha_3^1}\right)x_1x_2+\left(-\frac{(\alpha_2^1)^2\alpha_2^2}{\alpha_1^2\alpha_3^1}+\frac{(\alpha_1^1)^2(\alpha_2^2)^3}{(\alpha_1^2)^3\alpha_3^1}\right)x_1^2\nonumber\\
\tilde{d}(1,2)&=\left(-\frac{\alpha_1^1\alpha_2^1}{\alpha_1^2\alpha_2^2\alpha_3^1}+\frac{1}{\alpha_3^2}\right)x_1^2+\left(\frac{(\alpha_1^1)^2\alpha_2^2}{(\alpha_1^2)^3\alpha_3^1}-\frac{\alpha_1^1(\alpha_2^2)^2}{(\alpha_1^2)^2\alpha_2^1\alpha_3^2}\right)x_1x_2
\end{align}

\subsection{Long to Short}
By Serre duality, the bosons and fermions pair up with the states going from the short branes to the long branes.
The charge $1/4$ fermion reads:
\begin{align}
\psi^{(0)}_{\mathcal{L}_1\mathcal{S}_2}=\left(\begin{array}{cc}
a(1,2)&b(1,2)\\
c(1,2)&d(1,2)
\end{array}\right)\qquad
\psi^{(1)}_{\mathcal{L}_1\mathcal{S}_2}=\left(\begin{array}{cc}
\frac{1}{\alpha_3^1\alpha_3^2}d(1,2)&-\frac{\alpha_3^2}{\alpha_3^1}c(1,2)\\
\frac{\alpha_3^1}{\alpha_3^2}b(1,2)&-{\alpha_3^1\alpha_3^2}a(1,2)
\end{array}\right),
\end{align}
where
\begin{align}
a(1,2)&=\frac{1}{\alpha_3^1}-\frac{\alpha_1^1\alpha_2^1\alpha_3^2}{\alpha_1^2\alpha_2^2(\alpha_3^1)^2}\nonumber\\
b(1,2)&=\left(\frac{(\alpha_2^1)^3\alpha_3^2}{\alpha_1^1\alpha_1^2(\alpha_3^1)^2}-\frac{\alpha_1^1\alpha_1^2\alpha_2^1\alpha_3^2}{(\alpha_2^2)^2(\alpha_3^1)^2}\right)x_2+\left(\frac{(\alpha_1^1)^3\alpha_3^2}{\alpha_2^1\alpha_2^2(\alpha_3^1)^2}-\frac{\alpha_1^1\alpha_2^1\alpha_2^2\alpha_3^2}{(\alpha_1^2)^2(\alpha_3^1)^2}\right)x_1\nonumber\\
c(1,2)&=-\frac{(\alpha_1^1)^2\alpha_1^2}{\alpha_2^2\alpha_3^1}+\frac{(\alpha_2^1)^2\alpha_2^2}{\alpha_1^2\alpha_3^1}\nonumber\\
d(1,2)&=\left(\frac{\alpha_1^1\alpha_3^2}{\alpha_2^1\alpha_2^2}-\frac{(\alpha_1^1)^2(\alpha_3^2)^2}{\alpha_1^2(\alpha_2^2)^2\alpha_3^1}\right)x_2+\left(-\frac{\alpha_2^1\alpha_3^2}{\alpha_1^1\alpha_1^2}+\frac{(\alpha_2^1)^2(\alpha_3^2)^2}{(\alpha_1^2)^2\alpha_2^2\alpha_3^1}\right)x_1
\end{align}
For the $3/4$ fermion we find:
\begin{align}
\tilde{\psi}^{(0)}_{\mathcal{L}_1\mathcal{S}_2}=\left(\begin{array}{cc}
\tilde{a}(1,2)&\tilde{b}(1,2)\\
\tilde{c}(1,2)&\tilde{d}(1,2)
\end{array}\right)\qquad
&\tilde{\psi}^{(1)}_{\mathcal{L}_1\mathcal{S}_2}=\left(\begin{array}{cc}
\frac{1}{\alpha_3^1\alpha_3^2}\tilde{d}(1,2)&-\frac{\alpha_3^2}{\alpha_3^1}\tilde{c}(1,2)\\
\frac{\alpha_3^1}{\alpha_3^2}\tilde{b}(1,2)&-{\alpha_3^1\alpha_3^2}\tilde{a}(1,2)
\end{array}\right),
\end{align}
where
\begin{align}
\tilde{a}(1,2)&=\left(-\frac{(\alpha_2^1)^2}{\alpha_1^2\alpha_3^1}+\frac{\alpha_2^1\alpha_2^2}{\alpha_1^1\alpha_3^2}\right)x_1\nonumber\\
\tilde{b}(1,2)&=\left(-\frac{\alpha_1^2(\alpha_2^1)^2}{\alpha_2^2\alpha_3^1}+\frac{(\alpha_2^1)^4\alpha_2^2}{(\alpha_1^1)^2\alpha_1^2\alpha_3^1}\right)x_1x_2+\left(\frac{(\alpha_1^1)^2}{\alpha_3^1}-\frac{(\alpha_2^1)^2(\alpha_2^2)^2}{(\alpha_1^2)^2\alpha_3^1}\right)x_1^2\nonumber\\
\tilde{c}(1,2)&=\left(-\frac{\alpha_1^1\alpha_1^2\alpha_2^1}{\alpha_3^2}+\frac{(\alpha_2^1)^3(\alpha_2^2)^2}{\alpha_1^1\alpha_1^2\alpha_3^2}\right)x_1\nonumber\\
\tilde{d}(1,2)&=\left(-\frac{(\alpha_2^1)^2\alpha_2^2\alpha_3^1}{(\alpha_1^1)^2\alpha_1^2}+\frac{(\alpha_2^1)^3\alpha_3^2}{\alpha_1^1(\alpha_1^2)^2}\right)x_1^2+\left(\alpha_3^1-\frac{\alpha_1^1\alpha_2^1\alpha_3^2}{\alpha_1^2\alpha_2^2}\right)x_1x_2
\end{align}
The charge $1/4$ boson is given by the following expression:
\begin{align}
&\phi^{(0)}_{\mathcal{L}_1\mathcal{S}_2}=\left(\begin{array}{cc}
a(1,2)&b(1,2)\\
c(1,2)&d(1,2)
\end{array}\right)\qquad
\phi^{(1)}_{\mathcal{L}_1\mathcal{S}_2}=\left(\begin{array}{cc}
-\frac{1}{\alpha_3^1\alpha_3^2}d(1,2)&\frac{\alpha_3^2}{\alpha_3^1}c(1,2)\\
-\frac{\alpha_3^1}{\alpha_3^2}b(1,2)&{\alpha_3^1\alpha_3^2}a(1,2)
\end{array}\right),
\end{align}
where
\begin{align}
a(1,2)&=\left(\frac{1}{(\alpha_1^2)^2\alpha_2^2\alpha_3^1}-\frac{1}{\alpha_1^1\alpha_1^2\alpha_2^1\alpha_3^2}\right)x_2+\left(-\frac{(\alpha_1^1)^2}{(\alpha_1^2)^3(\alpha_2^1)^2\alpha_3^1}+\frac{\alpha_1^1\alpha_2^2}{(\alpha_1^2)^2(\alpha_2^1)^3\alpha_3^2}\right)x_1\nonumber\\
b(1,2)&=-\frac{1}{\alpha_3^1}+\frac{(\alpha_1^1)^2(\alpha_2^2)^2}{(\alpha_1^2)^2(\alpha_2^1)^2\alpha_3^1}\nonumber\\
c(1,2)&=\left(-\frac{\alpha_1^1}{\alpha_2^1\alpha_2^2\alpha_3^2}+\frac{(\alpha_1^1)^3\alpha_2^2}{(\alpha_1^2)^2(\alpha_2^1)^3\alpha_3^2}\right)x_2+\left(\frac{\alpha_2^1}{\alpha_1^1\alpha_1^2\alpha_3^2}-\frac{\alpha_1^1(\alpha_2^2)^2}{(\alpha_1^2)^3\alpha_2^1\alpha_3^2}\right)x_1\nonumber\\
d(1,2)&=-\frac{\alpha_2^2\alpha_3^1}{\alpha_1^2(\alpha_2^1)^2}+\frac{\alpha_1^1\alpha_3^2}{(\alpha_1^2)^2\alpha_2^1}
\end{align}
Finally, the charge $3/4$ boson is:
\begin{align}
&\tilde{\phi}^{(0)}_{\mathcal{L}_1\mathcal{S}_2}=\left(\begin{array}{cc}
\tilde{a}(1,2)&\tilde{b}(1,2)\\
\tilde{c}(1,2)&\tilde{d}(1,2)
\end{array}\right)\qquad
\tilde{\phi}^{(1)}_{\mathcal{L}_1\mathcal{S}_2}=\left(\begin{array}{cc}
-\frac{1}{\alpha_3^1\alpha_3^2}\tilde{d}(1,2)&\frac{\alpha_3^2}{\alpha_3^1}\tilde{c}(1,2)\\
-\frac{\alpha_3^1}{\alpha_3^2}\tilde{b}(1,2)&{\alpha_3^1\alpha_3^2}\tilde{a}(1,2)\end{array}\right),
\end{align}
where
\begin{align}
\tilde{a}(1,2)&=\left(-\frac{(\alpha_1^1)^2\alpha_2^2}{\alpha_1^2(\alpha_2^1)^2\alpha_3^1}+\frac{(\alpha_1^1)^3\alpha_3^2}{(\alpha_1^2)^2\alpha_2^1(\alpha_3^1)^2}\right)x_1^2+\left(\frac{1}{\alpha_3^1}-\frac{\alpha_1^1\alpha_2^1\alpha_3^2}{\alpha_1^2\alpha_2^2(\alpha_3^1)^2}\right)x_1x_2\nonumber\\
\tilde{b}(1,2)&=\left(\frac{\alpha_1^1\alpha_1^2\alpha_2^1\alpha_3^2}{(\alpha_3^1)^2}-\frac{(\alpha_1^1)^3(\alpha_2^2)^2\alpha_3^2}{\alpha_1^2\alpha_2^1(\alpha_3^1)^2}\right)x_1\nonumber\\
\tilde{c}(1,2)&=\left(\frac{(\alpha_1^1)^2\alpha_1^2}{\alpha_2^2\alpha_3^1}-\frac{(\alpha_1^1)^4\alpha_2^2}{\alpha_1^2(\alpha_2^1)^2\alpha_3^1}\right)x_1x_2+\left(-\frac{(\alpha_2^1)^2}{\alpha_3^1}+\frac{(\alpha_1^1)^2(\alpha_2^2)^2}{(\alpha_1^2)^2\alpha_3^1}\right)x_1^2\nonumber\\
\tilde{d}(1,2)&=\left(\frac{\alpha_1^1\alpha_2^2\alpha_3^2}{\alpha_2^1}-\frac{(\alpha_1^1)^2(\alpha_3^2)^2}{\alpha_1^2\alpha_3^1}\right)x_1
\end{align}
\section{Theta Function Identities}
\label{app-theta}
In this appendix we collect definitions and useful identities for theta functions. Standard references are for instance \cite{Mumford1,FarkasKra}.
The theta functions with characteristics are defined as follows:
 \begin{align}
\Theta\left[\begin{array}{c}c_1\\c_2\end{array}\right](u,\tau)&=\sum_{m\in\mathbb{Z}}q^{(m+c_1)^2/2}e^{2\pi i(u+c_2)(m+c_1)},
\end{align}
where $q=e^{2\pi i \tau}$. \\
For our purpose we need the Jacobi theta functions:
\begin{align}
&\Theta_1(u,\tau)\equiv\Theta\left[\begin{array}{c}\frac{1}{2}\\\frac{1}{2}\end{array}\right](u,\tau)\qquad \Theta_2(u,\tau)\equiv\Theta\left[\begin{array}{c}\frac{1}{2}\\0\end{array}\right](u,\tau)
\nonumber\\
&\Theta_3(u,\tau)\equiv\Theta\left[\begin{array}{c}0\\0\end{array}\right](u,\tau)\qquad\Theta_4(u,\tau)\equiv\Theta\left[\begin{array}{c}0\\\frac{1}{2}\end{array}\right](u,\tau)
\end{align}
We write $\Theta_i(0,\tau)\equiv \Theta_i(\tau)$.\\
These theta functions are symmetric in the $u$--argument:
\begin{align}
\Theta_1(-u,\tau)=-\Theta_1(u,\tau)\quad\Theta_2(-u,\tau)=\Theta_2(u,\tau)\quad\Theta_3(-u,\tau)=\Theta_3(u,\tau)\quad\Theta_4(-u,\tau)=\Theta_4(u,\tau)
\end{align}
In particular, one sees that $\Theta_1(0,\tau)=0$.
To uniformize the $\alpha_i$ (\ref{eq-uniformization}), we used the identities
\begin{align}
\Theta_3^2(u,\tau)\Theta_4^2(\tau)&=\Theta_4^2(u,\tau)\Theta_3^2(\tau)-\Theta_1^2(u,\tau)\Theta_2^2(\tau)\nonumber\\
\Theta_2^2(u,\tau)\Theta_4^2(\tau)&=\Theta_4^2(u,\tau)\Theta_2^2(\tau)-\Theta_1^2(u,\tau)\Theta_3^2(\tau).
\end{align}
In order to simplify the cohomology elements we applied the following addition rules:
\begin{align}
\label{eq-simp1}
\Theta_4(u_1+u_2,\tau)\Theta_4(u_1-u_2,\tau)\Theta_4(0,\tau)^2&=\Theta_4(u_1,\tau)^2\Theta_4(u_2,\tau)^2-\Theta_1(u_1,\tau)^2\Theta_1(u_2,\tau)^2\nonumber\\
\Theta_1(u_1+u_2,\tau)\Theta_1(u_1-u_2,\tau)\Theta_3(0,\tau)^2&=\Theta_1(u_1,\tau)^2\Theta_3(u_2,\tau)^2-\Theta_3(u_1,\tau)^2\Theta_1(u_2,\tau)^2\nonumber\\
&=\Theta_4(u_1,\tau)^2\Theta_2(u_2,\tau)^2-\Theta_2(u_1,\tau)^2\Theta_4(u_2,\tau)^2\nonumber\\
\Theta_1(u_1+u_2,\tau)\Theta_2(u_1-u_2,\tau)\Theta_4(0,\tau)^2&=\Theta_1(u_1,\tau)^2\Theta_4(u_2,\tau)^2-\Theta_4(u_1,\tau)^2\Theta_1(u_2,\tau)^2
\end{align}
\begin{align}
\label{eq-simp2}
\Theta_1(u_1+u_2,\tau)\Theta_4(u_1-u_2,\tau)\Theta_2(0,\tau)\Theta_3(0,\tau)&=\Theta_1(u_1,\tau)\Theta_2(u_2,\tau)\Theta_3(u_2,\tau)\Theta_4(u_1,\tau)+(u_1\leftrightarrow u_2)\nonumber\\
\Theta_4(u_1+u_2,\tau)\Theta_1(u_1-u_2,\tau)\Theta_2(0,\tau)\Theta_3(0,\tau)&=\Theta_1(u_1,\tau)\Theta_2(u_2,\tau)\Theta_3(u_2,\tau)\Theta_4(u_1,\tau)-(u_1\leftrightarrow u_2)\
\end{align}
These identities are actually just special cases of more general identities. In order to determine the correlators we need the most general addition theorems. 
For this, we introduce some more notation \cite{Mumford1}:
\begin{align}
&x_1=\frac{1}{2}(x+y+u+v)\quad y_1=\frac{1}{2}(x+y-u-v)\quad u_1=\frac{1}{2}(x-y+u-v)\quad v_1=\frac{1}{2}(x-y-u+v)
\end{align}
Furthermore we define $\Theta_i^u\equiv\Theta_i(u,\tau)$. For our calculations we can make use of the following formulas:
\begin{align}
\label{eq-bigtheta1}
-\Theta_1^x\Theta_1^y\Theta_1^u\Theta_1^v-\Theta_2^x\Theta_2^y\Theta_2^u\Theta_2^v+\Theta_3^x\Theta_3^y\Theta_3^u\Theta_3^v+\Theta_4^x\Theta_4^y\Theta_4^u\Theta_4^v&=2\Theta_1^{x_1}\Theta_1^{y_1}\Theta_1^{u_1}\Theta_1^{v_1}\nonumber \\
\Theta_1^x\Theta_1^y\Theta_1^u\Theta_1^v-\Theta_2^x\Theta_2^y\Theta_2^u\Theta_2^v+\Theta_3^x\Theta_3^y\Theta_3^u\Theta_3^v-\Theta_4^x\Theta_4^y\Theta_4^u\Theta_4^v&=2\Theta_4^{x_1}\Theta_4^{y_1}\Theta_4^{u_1}\Theta_4^{v_1}
\end{align}
\begin{align}
\label{eq-bigtheta2}
\Theta_3^x\Theta_3^y\Theta_2^u\Theta_2^v+\Theta_4^x\Theta_4^y\Theta_1^u\Theta_1^v-\Theta_2^x\Theta_2^y\Theta_3^u\Theta_3^v-\Theta_1^x\Theta_1^y\Theta_4^u\Theta_4^v&=2\Theta_4^{x_1}\Theta_4^{y_1}\Theta_1^{u_1}\Theta_1^{v_1}\nonumber \\
\Theta_3^x\Theta_3^y\Theta_2^u\Theta_2^v-\Theta_4^x\Theta_4^y\Theta_1^u\Theta_1^v-\Theta_2^x\Theta_2^y\Theta_3^u\Theta_3^v+\Theta_1^x\Theta_1^y\Theta_4^u\Theta_4^v&=2\Theta_1^{x_1}\Theta_1^{y_1}\Theta_4^{u_1}\Theta_4^{v_1}
\end{align}
What we need for our calculations are the differences of the two relations (\ref{eq-bigtheta1}) and (\ref{eq-bigtheta2}):
\begin{align}
\label{eq-mytheta}
\Theta_1^x\Theta_1^y\Theta_1^u\Theta_1^v-\Theta_4^x\Theta_4^y\Theta_4^u\Theta_4^v&=\Theta_4^{x_1}\Theta_4^{y_1}\Theta_4^{u_1}\Theta_4^{v_1}-\Theta_1^{x_1}\Theta_1^{y_1}\Theta_1^{u_1}\Theta_1^{v_1}\nonumber\\
\Theta_1^x\Theta_1^y\Theta_4^u\Theta_4^v-\Theta_4^x\Theta_4^y\Theta_1^u\Theta_1^v&=\Theta_1^{x_1}\Theta_1^{y_1}\Theta_4^{u_1}\Theta_4^{v_1}-\Theta_4^{x_1}\Theta_4^{y_1}\Theta_1^{u_1}\Theta_1^{v_1}
\end{align}

\bibliographystyle{fullsort}
\bibliography{thesisbib}

\providecommand{\href}[2]{#2}\begingroup\raggedright\begin{thebibliography}{10%
0}

\bibitem{ausgang}
``{P}rayer to {O}ur {L}ord {W}itten.''
\newblock performed at the TH Chrismas Play at Cern in December 2006; {\tt
  http://resonaances.blogspot.com/2007/01/cern-th-2007.html}.

\bibitem{Knapp:2006rd}
J.~Knapp and H.~Omer, ``{M}atrix {F}actorizations, {M}inimal {M}odels and
  {M}assey {P}roducts,'' {\em JHEP} {\bf 05} (2006) 064,
\href{http://www.arXiv.org/abs/hep-th/0604189}{{\tt hep-th/0604189}}.

\bibitem{Knapp:2007kq}
J.~Knapp and H.~Omer, ``{M}atrix {F}actorizations and {H}omological {M}irror
  {S}ymmetry on the {T}orus,'' {\em JHEP} {\bf 03} (2007) 088,
\href{http://www.arXiv.org/abs/hep-th/0701269}{{\tt hep-th/0701269}}.

\bibitem{mirrorsymmetry}
K.~Hori, S.~Katz, A.~Klemm, R.~Pandharipande, R.~Thomas, C.~Vafa, R.~Vakil, and
  E.~Zaslow, {\em {M}irror {S}ymmetry}.
\newblock American Mathematical Society, 2003.

\bibitem{Greene:1996cy}
B.~R. Greene, ``{S}tring {T}heory on {C}alabi-{Y}au {M}anifolds,''
\href{http://www.arXiv.org/abs/hep-th/9702155}{{\tt hep-th/9702155}}.

\bibitem{Neitzke:2004ni}
A.~Neitzke and C.~Vafa, ``{T}opological {S}trings and their {P}hysical
  {A}pplications,''
\href{http://www.arXiv.org/abs/hep-th/0410178}{{\tt hep-th/0410178}}.

\bibitem{Vonk:2005yv}
M.~Vonk, ``{A} {M}ini-{C}ourse on {T}opological {S}trings,''
\href{http://www.arXiv.org/abs/hep-th/0504147}{{\tt hep-th/0504147}}.

\bibitem{polchinski1}
J.~Polchinski, {\em {S}tring {T}heory {V}olume {I} - {A}n {I}ntroduction to the
  {B}osonic {S}tring}.
\newblock Cambridge University Press, Cambridge, 1998.

\bibitem{polchinski2}
J.~Polchinski, {\em {S}tring {T}heory {V}olume {II} - {S}uperstring {T}heory
  and {B}eyond}.
\newblock Cambridge University Press, Cambridge, 1998.

\bibitem{Sharpe:2003dr}
E.~Sharpe, ``{L}ectures on {D}-branes and {S}heaves,''
\href{http://www.arXiv.org/abs/hep-th/0307245}{{\tt hep-th/0307245}}.

\bibitem{Aspinwall:2004jr}
P.~S. Aspinwall, ``{D}-branes on {C}alabi-{Y}au {M}anifolds,''
\href{http://www.arXiv.org/abs/hep-th/0403166}{{\tt hep-th/0403166}}.

\bibitem{Lerche:1989uy}
W.~Lerche, C.~Vafa, and N.~P. Warner, ``{C}hiral {R}ings in {N=2}
  {S}uperconformal {T}heories,'' {\em Nucl. Phys.} {\bf B324} (1989)
427.

\bibitem{Witten:1991zz}
E.~Witten, ``{M}irror {M}anifolds and {T}opological {F}ield {T}heory,''
\href{http://www.arXiv.org/abs/hep-th/9112056}{{\tt hep-th/9112056}}.

\bibitem{Witten:1993yc}
E.~Witten, ``{P}hases of {N} = 2 {T}heories in {T}wo {D}imensions,'' {\em Nucl.
  Phys.} {\bf B403} (1993) 159--222,
\href{http://www.arXiv.org/abs/hep-th/9301042}{{\tt hep-th/9301042}}.

\bibitem{Birmingham:1991ty}
D.~Birmingham, M.~Blau, M.~Rakowski, and G.~Thompson, ``{T}opological {F}ield
  {T}heory,'' {\em Phys. Rept.} {\bf 209} (1991)
129--340.

\bibitem{Ooguri:1996ck}
H.~Ooguri, Y.~Oz, and Z.~Yin, ``{D}-branes on {C}alabi-{Y}au {S}paces and their
  {M}irrors,'' {\em Nucl. Phys.} {\bf B477} (1996) 407--430,
\href{http://www.arXiv.org/abs/hep-th/9606112}{{\tt hep-th/9606112}}.

\bibitem{Polishchuk:1998db}
A.~Polishchuk and E.~Zaslow, ``{C}ategorical {M}irror {S}ymmetry: {T}he
  {E}lliptic {C}urve,'' {\em Adv. Theor. Math. Phys.} {\bf 2} (1998) 443--470,
\href{http://www.arXiv.org/abs/math.ag/9801119}{{\tt math.ag/9801119}}.

\bibitem{alg-geom/9411018}
M.~Kontsevich, ``{H}omological {A}lgebra of {M}irror {S}ymmetry,''
  \href{http://www.arXiv.org/abs/alg-geom/9411018}{{\tt alg-geom/9411018}}.

\bibitem{Orlov:2003yp}
D.~Orlov, ``{T}riangulated {C}ategories of {S}ingularities and {D}-branes in
  {L}andau-{G}inzburg {M}odels,''
\href{http://www.arXiv.org/abs/math.ag/0302304}{{\tt math.ag/0302304}}.

\bibitem{eisenbud}
D.~Eisenbud, ``{H}omological {A}lgebra on a {C}omplete {I}ntersection with an
  {A}pplication to {G}roup {R}epresentations,'' {\em Trans. Amer. Math. Soc}
  {\bf 260} (1980) 35.

\bibitem{Kapustin:2002bi}
A.~Kapustin and Y.~Li, ``{D}-branes in {L}andau-{G}inzburg {M}odels and
  {A}lgebraic {G}eometry,'' {\em JHEP} {\bf 12} (2003) 005,
\href{http://www.arXiv.org/abs/hep-th/0210296}{{\tt hep-th/0210296}}.

\bibitem{Brunner:2003dc}
I.~Brunner, M.~Herbst, W.~Lerche, and B.~Scheuner, ``{L}andau-{G}inzburg
  {R}ealization of {O}pen {S}tring {TFT},'' {\em JHEP} {\bf 11} (2006) 043,
\href{http://www.arXiv.org/abs/hep-th/0305133}{{\tt hep-th/0305133}}.

\bibitem{Lazaroiu:2003zi}
C.~I. Lazaroiu, ``{O}n the {B}oundary {C}oupling of {T}opological
  {L}andau-{G}inzburg {M}odels,'' {\em JHEP} {\bf 05} (2005) 037,
\href{http://www.arXiv.org/abs/hep-th/0312286}{{\tt hep-th/0312286}}.

\bibitem{Herbst:2004ax}
M.~Herbst and C.-I. Lazaroiu, ``{L}ocalization and {T}races in {O}pen-{C}losed
  {T}opological {L}andau- {G}inzburg {M}odels,'' {\em JHEP} {\bf 05} (2005)
  044,
\href{http://www.arXiv.org/abs/hep-th/0404184}{{\tt hep-th/0404184}}.

\bibitem{Warner:1995ay}
N.~P. Warner, ``{S}upersymmetry in {B}oundary {I}ntegrable {M}odels,'' {\em
  Nucl. Phys.} {\bf B450} (1995) 663--694,
\href{http://www.arXiv.org/abs/hep-th/9506064}{{\tt hep-th/9506064}}.

\bibitem{Aspinwall:2006ib}
P.~S. Aspinwall, ``{T}he {L}andau-{G}inzburg to {C}alabi-{Y}au {D}ictionary for
  {D}-branes,''
\href{http://www.arXiv.org/abs/hep-th/0610209}{{\tt hep-th/0610209}}.

\bibitem{Aspinwall:2007cs}
P.~S. Aspinwall, ``{T}opological {D}-{B}ranes and {C}ommutative {A}lgebra,''
\href{http://www.arXiv.org/abs/hep-th/0703279}{{\tt hep-th/0703279}}.

\bibitem{hhp}
M.~Herbst, K.~Hori, and D.~C. Page to appear.

\bibitem{Witten:1998cd}
E.~Witten, ``{D}-branes and {K}-theory,'' {\em JHEP} {\bf 12} (1998) 019,
\href{http://www.arXiv.org/abs/hep-th/9810188}{{\tt hep-th/9810188}}.

\bibitem{Walcher:2004tx}
J.~Walcher, ``{S}tability of {L}andau-{G}inzburg {B}ranes,'' {\em J. Math.
  Phys.} {\bf 46} (2005) 082305,
\href{http://www.arXiv.org/abs/hep-th/0412274}{{\tt hep-th/0412274}}.

\bibitem{Brunner:2005fv}
I.~Brunner and M.~R. Gaberdiel, ``{M}atrix {F}actorisations and {P}ermutation
  {B}ranes,'' {\em JHEP} {\bf 07} (2005) 012,
\href{http://www.arXiv.org/abs/hep-th/0503207}{{\tt hep-th/0503207}}.

\bibitem{Brunner:2005pq}
I.~Brunner and M.~R. Gaberdiel, ``{T}he {M}atrix {F}actorisations of the
  {D}-model,'' {\em J. Phys.} {\bf A38} (2005) 7901--7920,
\href{http://www.arXiv.org/abs/hep-th/0506208}{{\tt hep-th/0506208}}.

\bibitem{Enger:2005jk}
H.~Enger, A.~Recknagel, and D.~Roggenkamp, ``{P}ermutation {B}ranes and
  {L}inear {M}atrix {F}actorisations,'' {\em JHEP} {\bf 01} (2006) 087,
\href{http://www.arXiv.org/abs/hep-th/0508053}{{\tt hep-th/0508053}}.

\bibitem{Brunner:2006tc}
I.~Brunner, M.~R. Gaberdiel, and C.~A. Keller, ``{M}atrix {F}actorisations and
  {D}-branes on {K3},'' {\em JHEP} {\bf 06} (2006) 015,
\href{http://www.arXiv.org/abs/hep-th/0603196}{{\tt hep-th/0603196}}.

\bibitem{Fredenhagen:2006qw}
S.~Fredenhagen and M.~R. Gaberdiel, ``{G}eneralised {N} = 2 {P}ermutation
  {B}ranes,'' {\em JHEP} {\bf 11} (2006) 041,
\href{http://www.arXiv.org/abs/hep-th/0607095}{{\tt hep-th/0607095}}.

\bibitem{Keller:2006tf}
C.~A. Keller and S.~Rossi, ``{B}oundary {S}tates, {M}atrix {F}actorisations and
  {C}orrelation {F}unctions for the {E}-models,'' {\em JHEP} {\bf 03} (2007)
  038,
\href{http://www.arXiv.org/abs/hep-th/0610175}{{\tt hep-th/0610175}}.

\bibitem{Schmidt-Colinet:2007vi}
C.~Schmidt-Colinet, ``{T}ensor {P}roduct and {P}ermutation {B}ranes on the
  {T}orus,'' {\em JHEP} {\bf 07} (2007) 044,
\href{http://www.arXiv.org/abs/hep-th/0701128}{{\tt hep-th/0701128}}.

\bibitem{greuel-knoerrer}
G.~M. Greuel and K.~Kn{\"o}rrer, ``{E}infache {K}urvensingularit{\"a}ten und
  {T}orsionsfreie {M}oduln,'' {\em Math. Ann.} {\bf 270} (1985) 417--425.

\bibitem{Brunner:2004mt}
I.~Brunner, M.~Herbst, W.~Lerche, and J.~Walcher, ``{M}atrix {F}actorizations
  and {M}irror {S}ymmetry: {T}he {C}ubic {C}urve,'' {\em JHEP} {\bf 11} (2006)
  006,
\href{http://www.arXiv.org/abs/hep-th/0408243}{{\tt hep-th/0408243}}.

\bibitem{Govindarajan:2005im}
S.~Govindarajan, H.~Jockers, W.~Lerche, and N.~P. Warner, ``{T}achyon
  {C}ondensation on the {E}lliptic {C}urve,'' {\em Nucl. Phys.} {\bf B765}
  (2007) 240--286,
\href{http://www.arXiv.org/abs/hep-th/0512208}{{\tt hep-th/0512208}}.

\bibitem{Kapustin:2003ga}
A.~Kapustin and Y.~Li, ``{T}opological {C}orrelators in {L}andau-{G}inzburg
  {M}odels with {B}oundaries,'' {\em Adv. Theor. Math. Phys.} {\bf 7} (2004)
  727--749,
\href{http://www.arXiv.org/abs/hep-th/0305136}{{\tt hep-th/0305136}}.

\bibitem{Kapustin:2003rc}
A.~Kapustin and Y.~Li, ``{D}-branes in {T}opological {M}inimal {M}odels: {T}he
  {L}andau-{G}inzburg {A}pproach,'' {\em JHEP} {\bf 07} (2004) 045,
\href{http://www.arXiv.org/abs/hep-th/0306001}{{\tt hep-th/0306001}}.

\bibitem{Ashok:2004zb}
S.~K. Ashok, E.~Dell'Aquila, and D.-E. Diaconescu, ``{F}ractional {B}ranes in
  {L}andau-{G}inzburg {O}rbifolds,'' {\em Adv. Theor. Math. Phys.} {\bf 8}
  (2004) 461--513,
\href{http://www.arXiv.org/abs/hep-th/0401135}{{\tt hep-th/0401135}}.

\bibitem{Herbst:2004jp}
M.~Herbst, C.-I. Lazaroiu, and W.~Lerche, ``{S}uperpotentials, {A}(infinity)
  {R}elations and {WDVV} {E}quations for {O}pen {T}opological {S}trings,'' {\em
  JHEP} {\bf 02} (2005) 071,
\href{http://www.arXiv.org/abs/hep-th/0402110}{{\tt hep-th/0402110}}.

\bibitem{Ashok:2004xq}
S.~K. Ashok, E.~Dell'Aquila, D.-E. Diaconescu, and B.~Florea, ``{O}bstructed
  {D}-branes in {L}andau-{G}inzburg {O}rbifolds,'' {\em Adv. Theor. Math.
  Phys.} {\bf 8} (2004) 427--472,
\href{http://www.arXiv.org/abs/hep-th/0404167}{{\tt hep-th/0404167}}.

\bibitem{Hori:2004ja}
K.~Hori and J.~Walcher, ``{F}-term {E}quations near {G}epner {P}oints,'' {\em
  JHEP} {\bf 01} (2005) 008,
\href{http://www.arXiv.org/abs/hep-th/0404196}{{\tt hep-th/0404196}}.

\bibitem{Herbst:2004zm}
M.~Herbst, C.-I. Lazaroiu, and W.~Lerche, ``{D}-brane {E}ffective {A}ction and
  {T}achyon {C}ondensation in {T}opological {M}inimal {M}odels,'' {\em JHEP}
  {\bf 03} (2005) 078,
\href{http://www.arXiv.org/abs/hep-th/0405138}{{\tt hep-th/0405138}}.

\bibitem{Herbst:2006kt}
M.~Herbst, ``{Q}uantum {A}-infinity {S}tructures for {O}pen-{C}losed
  {T}opological {S}trings,''
\href{http://www.arXiv.org/abs/hep-th/0602018}{{\tt hep-th/0602018}}.

\bibitem{Hori:2006ic}
K.~Hori and J.~Walcher, ``{D}-brane {C}ategories for {O}rientifolds: {T}he
  {L}andau-{G}inzburg {C}ase,''
\href{http://www.arXiv.org/abs/hep-th/0606179}{{\tt hep-th/0606179}}.

\bibitem{Diaconescu:2006id}
D.-E. Diaconescu, A.~Garcia-Raboso, R.~L. Karp, and K.~Sinha, ``{D}-brane
  {S}uperpotentials in {C}alabi-{Y}au {O}rientifolds (projection),''
\href{http://www.arXiv.org/abs/hep-th/0606180}{{\tt hep-th/0606180}}.

\bibitem{Jockers:2006sm}
H.~Jockers, ``{D}-brane {M}onodromies from a {M}atrix-{F}actorization
  perspective,'' {\em JHEP} {\bf 02} (2007) 006,
\href{http://www.arXiv.org/abs/hep-th/0612095}{{\tt hep-th/0612095}}.

\bibitem{Hori:2004zd}
K.~Hori and J.~Walcher, ``{D}-branes from {M}atrix {F}actorizations,'' {\em
  Comptes Rendus Physique} {\bf 5} (2004) 1061--1070,
\href{http://www.arXiv.org/abs/hep-th/0409204}{{\tt hep-th/0409204}}.

\bibitem{Cargese}
H.~Jockers and W.~Lerche, ``{M}atrix {F}actorizations, {D}-{B}ranes and their
  {D}eformations,''
\href{http://www.arXiv.org/abs/arXiv:0708.0157 [hep-th]}{{\tt arXiv:0708.0157
  [hep-th]}}.

\bibitem{Vafa:1990mu}
C.~Vafa, ``{T}opological {L}andau-{G}inzburg {M}odels,'' {\em Mod. Phys. Lett.}
  {\bf A6} (1991)
337--346.

\bibitem{Vafa:1989xc}
C.~Vafa, ``{S}tring {V}acua and {O}rbifoldized {L}-{G} {M}odels,'' {\em Mod.
  Phys. Lett.} {\bf A4} (1989)
1169.

\bibitem{klo}
M.~Kontsevich unpublished.

\bibitem{Knorrer}
H.~{Kn\"orrer}, ``{C}ohen--{M}acaulay {M}odules on {H}ypersurface
  {S}ingularities {I},'' {\em Invent. Math.} {\bf 88} (1987) 153--164.

\bibitem{Kajiura:2005yu}
H.~Kajiura, K.~Saito, and A.~Takahashi, ``{M}atrix {F}actorizations and
  {R}epresentations of {Q}uivers {II}: {T}ype {ADE} {C}ase,''
\href{http://www.arXiv.org/abs/math.ag/0511155}{{\tt math.ag/0511155}}.

\bibitem{Siqveland1}
A.~Siqveland, ``{T}he {M}ethod of {C}omputing {F}ormal {M}oduli,'' {\em J.
  Alg.} {\bf 241} (2001) 292--327.

\bibitem{yoshino}
Y.~Yoshino, {\em Cohen--Macaulay Modules over Cohen--Macaulay Rings}.
\newblock Cambridge University Press, Cambridge, 1990.

\bibitem{Laza}
R.~Laza, G.~Pfister, and D.~Popescu, ``{M}aximal {C}ohen-{M}acaulay {M}odules
  over the {C}one of an {E}lliptic {C}urve,'' {\em Journal of Algebra} {\bf
  253} (2003) 209--236.

\bibitem{Recknagel:1997sb}
A.~Recknagel and V.~Schomerus, ``{D}-branes in {G}epner {M}odels,'' {\em Nucl.
  Phys.} {\bf B531} (1998) 185--225,
\href{http://www.arXiv.org/abs/hep-th/9712186}{{\tt hep-th/9712186}}.

\bibitem{Recknagel:2002qq}
A.~Recknagel, ``{P}ermutation {B}ranes,'' {\em JHEP} {\bf 04} (2003) 041,
\href{http://www.arXiv.org/abs/hep-th/0208119}{{\tt hep-th/0208119}}.

\bibitem{Caviezel:2005th}
C.~Caviezel, S.~Fredenhagen, and M.~R. Gaberdiel, ``{T}he {RR} {C}harges of
  {A}-type {G}epner {M}odels,'' {\em JHEP} {\bf 01} (2006) 111,
\href{http://www.arXiv.org/abs/hep-th/0511078}{{\tt hep-th/0511078}}.

\bibitem{curtoPHD}
C.~Curto, ``{M}atrix {M}odel {S}uperpotentials and {C}alabi--{Y}au {S}paces:
  {A}n {ADE} {C}lassification.'' Phd thesis, 2005.

\bibitem{Curto:2006}
C.~Curto and R.~Morrison, David, ``{T}hreefold {F}lops via {M}atrix
  {F}actorization,'' \href{http://www.arXiv.org/abs/math/0611014}{{\tt
  math/0611014}}.

\bibitem{McKay}
J.~McKay, ``{G}raphs, {S}ingularities and {F}inite {G}roups,'' {\em Proc.
  Sympos. Pure Math.} {\bf 37} (1980) 183--186.

\bibitem{khovanov1}
M.~Khovanov and L.~Rozansky, ``{M}atrix {F}actorizations and {L}ink
  {H}omology,'' \href{http://www.arXiv.org/abs/math.QA/0401268}{{\tt
  math.QA/0401268}}.

\bibitem{khovanov2}
M.~Khovanov and L.~Rozansky, ``{M}atrix {F}actorizations and {L}ink {H}omology
  {II},'' \href{http://www.arXiv.org/abs/math.QA/0505056}{{\tt
  math.QA/0505056}}.

\bibitem{Gukov:2005qp}
S.~Gukov and J.~Walcher, ``{M}atrix {F}actorizations and {K}auffman
  {H}omology,''
\href{http://www.arXiv.org/abs/hep-th/0512298}{{\tt hep-th/0512298}}.

\bibitem{Becker:2006ks}
K.~Becker, M.~Becker, C.~Vafa, and J.~Walcher, ``{M}oduli {S}tabilization in
  non-geometric {B}ackgrounds,'' {\em Nucl. Phys.} {\bf B770} (2007) 1--46,
\href{http://www.arXiv.org/abs/hep-th/0611001}{{\tt hep-th/0611001}}.

\bibitem{Lazaroiu:2001nm}
C.~I. Lazaroiu, ``{S}tring {F}ield {T}heory and {B}rane {S}uperpotentials,''
  {\em JHEP} {\bf 10} (2001) 018,
\href{http://www.arXiv.org/abs/hep-th/0107162}{{\tt hep-th/0107162}}.

\bibitem{Witten:1992fb}
E.~Witten, ``{C}hern-{S}imons {G}auge {T}heory as a {S}tring {T}heory,'' {\em
  Prog. Math.} {\bf 133} (1995) 637--678,
\href{http://www.arXiv.org/abs/hep-th/9207094}{{\tt hep-th/9207094}}.

\bibitem{Merkulov}
S.~A. Merkulov, ``{S}trong {H}omotopy {A}lgebras of a k{\"a}hler manifold,''
  {\em Int. Math. Res. Notices} {\bf 3} (1999) 153,
  \href{http://www.arXiv.org/abs/math/9809172}{{\tt math/9809172}}.

\bibitem{Kajiura:2005sn}
H.~Kajiura and J.~Stasheff, ``{O}pen-{C}losed {H}omotopy {A}lgebra in
  {M}athematical {P}hysics,'' {\em J. Math. Phys.} {\bf 47} (2006) 023506,
\href{http://www.arXiv.org/abs/hep-th/0510118}{{\tt hep-th/0510118}}.

\bibitem{siqvelandPHD}
A.~Siqveland, ``{M}atric {M}assey {P}roducts and {F}ormal {M}oduli {L}ocal and
  {G}lobal.'' Phd thesis, University of Oslo, Dept. of Mathematics, 1995.

\bibitem{Hofman:2000ce}
C.~Hofman and W.-K. Ma, ``{D}eformations of {T}opological {O}pen {S}trings,''
  {\em JHEP} {\bf 01} (2001) 035,
\href{http://www.arXiv.org/abs/hep-th/0006120}{{\tt hep-th/0006120}}.

\bibitem{Dijkgraaf:1990dj}
R.~Dijkgraaf, H.~L. Verlinde, and E.~P. Verlinde, ``{T}opological {S}trings in
  d $<$ 1,'' {\em Nucl. Phys.} {\bf B352} (1991)
59--86.

\bibitem{Herbst:2006nn}
M.~Herbst, W.~Lerche, and D.~Nemeschansky, ``{I}nstanton {G}eometry and
  {Q}uantum {A}(infinity) {S}tructure on the {E}lliptic {C}urve,''
\href{http://www.arXiv.org/abs/hep-th/0603085}{{\tt hep-th/0603085}}.

\bibitem{Siqveland2}
A.~Siqveland, ``{G}lobal {M}atric {M}assey {P}roducts and the {C}ompactified
  {J}acobian of the {E6}--{S}ingularity,'' {\em J. Alg.} {\bf 241} (2001)
  259--291.

\bibitem{laudal1}
O.~A. Laudal, ``{F}ormal {M}oduli of {A}lgebraic {S}tructures,'' {\em Lecture
  Notes in Mathematics} {\bf 754} (1978).

\bibitem{laudal2}
O.~A. Laudal, ``{M}atric {M}assey {P}roducts and {F}ormal {M}oduli {I},'' {\em
  Lecture Notes in Mathematics} {\bf 1183} (1986) 218--240.

\bibitem{Siqveland3}
A.~Siqveland, ``{G}eneralized {M}atric {M}assey {P}roducts for {G}raded
  {M}odules,'' \href{http://www.arXiv.org/abs/math.AG/0603425}{{\tt
  math.AG/0603425}}.

\bibitem{eriksen}
E.~Eriksen, ``{C}onnections and {M}onodromy on {M}odules.'' Technical Report.
  University of Oslo. {\tt }, 1995.

\bibitem{math.AG/0303166}
E.~Eriksen, ``{An introduction to noncommutative deformations of modules},''
  \href{http://www.arXiv.org/abs/math.AG/0303166}{{\tt math.AG/0303166}}.

\bibitem{eriksen2}
E.~Eriksen and T.~S. Gustavsen, ``{C}omputing {C}onnections on {M}odules,''
  \href{http://www.arXiv.org/abs/math.AG/0602616}{{\tt math.AG/0602616}}.

\bibitem{eriksen3}
E.~Eriksen, ``{C}onnections on {M}odules over {S}imple {C}urve
  {S}ingularities,'' \href{http://www.arXiv.org/abs/math.AG/0603259}{{\tt
  math.AG/0603259}}.

\bibitem{wiki:artin}
Wikipedia, ``{A}rtinian {R}ing --- {W}ikipedia{,} {T}he {F}ree
  {E}ncyclopedia,'' 2007.

\bibitem{Schlessinger}
M.~Schlessinger, ``{F}unctors of {A}rtin {R}ings,'' {\em Trans. Am. Math. Soc.}
  {\bf 130} (1968) 208--222.

\bibitem{wiki:phs}
Wikipedia, ``{P}rincipal {H}omogeneous {S}pace --- {W}ikipedia{,} {T}he {F}ree
  {E}ncyclopedia,'' 2007.

\bibitem{Lazaroiu:2000rk}
C.~I. Lazaroiu, ``{O}n the {S}tructure of {O}pen-{C}losed {T}opological {F}ield
  {T}heory in two {D}imensions,'' {\em Nucl. Phys.} {\bf B603} (2001) 497--530,
\href{http://www.arXiv.org/abs/hep-th/0010269}{{\tt hep-th/0010269}}.

\bibitem{Lerche:1991re}
W.~Lerche and N.~P. Warner, ``{P}olytopes and {S}olitons in {I}ntegrable, {N=2}
  {S}upersymmetric {L}andau-{G}inzburg {T}heories,'' {\em Nucl. Phys.} {\bf
  B358} (1991)
571--599.

\bibitem{Eguchi:1996nh}
T.~Eguchi and S.-K. Yang, ``{A} new {D}escription of the {E(6)}
  {S}ingularity,'' {\em Phys. Lett.} {\bf B394} (1997) 315--322,
\href{http://www.arXiv.org/abs/hep-th/9612086}{{\tt hep-th/9612086}}.

\bibitem{Lerche:1996an}
W.~Lerche and N.~P. Warner, ``{E}xceptional {SW} {G}eometry from {ALE}
  {F}ibrations,'' {\em Phys. Lett.} {\bf B423} (1998) 79--86,
\href{http://www.arXiv.org/abs/hep-th/9608183}{{\tt hep-th/9608183}}.

\bibitem{Eguchi:2001fm}
T.~Eguchi, N.~P. Warner, and S.-K. Yang, ``{ADE} {S}ingularities and {C}oset
  {M}odels,'' {\em Nucl. Phys.} {\bf B607} (2001) 3--37,
\href{http://www.arXiv.org/abs/hep-th/0105194}{{\tt hep-th/0105194}}.

\bibitem{Zuber:2000ia}
J.-B. Zuber, ``{CFT}, {BCFT}, {ADE} and all that,''
\href{http://www.arXiv.org/abs/hep-th/0006151}{{\tt hep-th/0006151}}.

\bibitem{Klemm:1991vw}
A.~Klemm, S.~Theisen, and M.~G. Schmidt, ``{C}orrelation {F}unctions for
  {T}opological {L}andau-{G}inzburg {M}odels with {$c \leq 3$},'' {\em Int. J.
  Mod. Phys.} {\bf A7} (1992)
6215--6244.

\bibitem{getzlerpriv}
E.~Getzler private communication.

\bibitem{GPS05}
G.-M. Greuel, G.~Pfister, and H.~Sch\"onemann, ``{\sc Singular} 3.0,'' {A
  Computer Algebra System for Polynomial Computations}, Centre for Computer
  Algebra, University of Kaiserslautern, 2005.
\newblock {\tt http://www.singular.uni-kl.de}.

\bibitem{GPdeform}
G.-M. Greuel and G.~Pfister, {\em {\tt deform.lib}. {A} {\sc {S}ingular} 3.0
  {L}ibrry for {C}omputing {D}eformations}, 2006.

\bibitem{Lerche:1989cs}
W.~Lerche, D.~Lust, and N.~P. Warner, ``{D}uality {S}ymmetries in {N=2}
  {L}andau-{G}inzburg {M}odels,'' {\em Phys. Lett.} {\bf B231} (1989)
417.

\bibitem{atiyah}
M.~Atiyah, ``{V}ector {B}undles over an {E}lliptic {C}urve,'' {\em Proc. London
  Math. Soc.} {\bf VII (3)} (1957) 414--452.

\bibitem{PolishchukFukaya}
A.~Polishchuk, ``{M}assey and {F}ukaya {P}roducts on {E}lliptic {C}urves,''
  {\em Adv.Theor.Math.Phys.} {\bf 4} (2000) 1187--1207,
  \href{http://www.arXiv.org/abs/math.ag/9803017}{{\tt math.ag/9803017}}.

\bibitem{PolishchukAppell}
A.~Polishchuk, ``{M.P.} {A}ppell's {F}unctions and {V}ector {B}undles of {R}ank
  $2$ on {E}lliptic {C}urves,''
  \href{http://www.arXiv.org/abs/math.AG/9810084}{{\tt math.AG/9810084}}.

\bibitem{PolishchukMirror}
A.~Polishchuk, ``{H}omological {M}irror {S}ymmetry with {H}igher {P}roducts,''
  \href{http://www.arXiv.org/abs/math.ag/9901025}{{\tt math.ag/9901025}}.

\bibitem{Polishchuk:2000kx}
A.~Polishchuk, ``${A}_{\infty}$-structures on an {E}lliptic {C}urve,'' {\em
  Commun. Math. Phys.} {\bf 247} (2004) 527--551,
\href{http://www.arXiv.org/abs/math.ag/0001048}{{\tt math.ag/0001048}}.

\bibitem{PolishchukTheta}
A.~Polishchuk, ``{I}ndefinite {T}heta {S}eries of {S}ignature (1,1) from the
  point of view of {H}omological {M}irror {S}ymmetry,''
  \href{http://www.arXiv.org/abs/math.ag/0003076}{{\tt math.ag/0003076}}.

\bibitem{math.AG/0012018}
B.~Kreussler, ``{Homological Mirror Symmetry in Dimension One},''
  \href{http://www.arXiv.org/abs/math.AG/0012018}{{\tt math.AG/0012018}}.

\bibitem{Kajiura:2007zz}
H.~Kajiura, ``An {$A_\infty$}-structure for {L}ines in a {P}lane,''
\href{http://www.arXiv.org/abs/math.QA/0703164}{{\tt math.QA/0703164}}.

\bibitem{Cremades:2003qj}
D.~Cremades, L.~E. Ibanez, and F.~Marchesano, ``{Y}ukawa {C}ouplings in
  {I}ntersecting {D}-brane {M}odels,'' {\em JHEP} {\bf 07} (2003) 038,
\href{http://www.arXiv.org/abs/hep-th/0302105}{{\tt hep-th/0302105}}.

\bibitem{Govindarajan:2006uy}
S.~Govindarajan and H.~Jockers, ``{E}ffective {S}uperpotentials for {B}-branes
  in {L}andau-{G}inzburg {M}odels,'' {\em JHEP} {\bf 10} (2006) 060,
\href{http://www.arXiv.org/abs/hep-th/0608027}{{\tt hep-th/0608027}}.

\bibitem{Dell'Aquila:2005jg}
E.~Dell'Aquila, ``{D}-branes in {T}oroidal {O}rbifolds and {M}irror
  {S}ymmetry,'' {\em JHEP} {\bf 04} (2006) 035,
\href{http://www.arXiv.org/abs/hep-th/0512051}{{\tt hep-th/0512051}}.

\bibitem{Lerche:1991wm}
W.~Lerche, D.~J. Smit, and N.~P. Warner, ``{D}ifferential {E}quations for
  {P}eriods and {F}lat {C}oordinates in two-dimensional {T}opological {M}atter
  {T}heories,'' {\em Nucl. Phys.} {\bf B372} (1992) 87--112,
\href{http://www.arXiv.org/abs/hep-th/9108013}{{\tt hep-th/9108013}}.

\bibitem{Harvey:1988ur}
J.~A. Harvey and S.~G. Naculich, ``{C}onformal {F}ield {T}heory and {G}enus
  {Z}ero {F}unction {F}ields,'' {\em Nucl. Phys.} {\bf B305} (1988)
417.

\bibitem{Giveon:1990ay}
A.~Giveon and D.-J. Smit, ``{S}ymmetries on the {M}oduli {S}pace of (2,2)
  {S}uperstring {V}acua,'' {\em Nucl. Phys.} {\bf B349} (1991)
168--206.

\bibitem{FarkasKra}
H.~M. Farkas and I.~Kra, {\em Theta Constants, Riemann Surfaces and the Modular
  Group}.
\newblock American Mathematical Society, Providence, Rhode Island, 2001.

\bibitem{Mumford1}
D.~Mumford, {\em Tata Lectures on Theta, Vol. 1}.
\newblock Birkhauser, 1983.

\bibitem{Bershadsky:1993cx}
M.~Bershadsky, S.~Cecotti, H.~Ooguri, and C.~Vafa, ``{K}odaira-{S}pencer
  {T}heory of {G}ravity and {E}xact {R}esults for {Q}uantum {S}tring
  {A}mplitudes,'' {\em Commun. Math. Phys.} {\bf 165} (1994) 311--428,
\href{http://www.arXiv.org/abs/hep-th/9309140}{{\tt hep-th/9309140}}.

\bibitem{Eynard:2007hf}
B.~Eynard, M.~Marino, and N.~Orantin, ``{H}olomorphic {A}nomaly and {M}atrix
  {M}odels,'' {\em JHEP} {\bf 06} (2007) 058,
\href{http://www.arXiv.org/abs/hep-th/0702110}{{\tt hep-th/0702110}}.

\bibitem{Walcher:2007tp}
J.~Walcher, ``{Extended Holomorphic Anomaly and Loop Amplitudes in Open
  Topological String},''
\href{http://www.arXiv.org/abs/arXiv:0705.4098 [hep-th]}{{\tt arXiv:0705.4098
  [hep-th]}}.

\bibitem{Walcher:2006rs}
J.~Walcher, ``{O}pening {M}irror {S}ymmetry on the {Q}uintic,''
\href{http://www.arXiv.org/abs/hep-th/0605162}{{\tt hep-th/0605162}}.

\bibitem{Marino:2006hs}
M.~Marino, ``{O}pen {S}tring {A}mplitudes and {L}arge {O}rder {B}ehavior in
  {T}opological {S}tring {T}heory,''
\href{http://www.arXiv.org/abs/hep-th/0612127}{{\tt hep-th/0612127}}.

\bibitem{Mayr:2001xk}
P.~Mayr, ``{N = 1} {M}irror {S}ymmetry and open/closed {S}tring {D}uality,''
  {\em Adv. Theor. Math. Phys.} {\bf 5} (2002) 213--242,
\href{http://www.arXiv.org/abs/hep-th/0108229}{{\tt hep-th/0108229}}.

\bibitem{Lerche:2001cw}
W.~Lerche and P.~Mayr, ``On {N = 1} {M}irror {S}ymmetry for {O}pen {T}ype {II}
  {S}trings,''
\href{http://www.arXiv.org/abs/hep-th/0111113}{{\tt hep-th/0111113}}.

\end{thebibliography}\endgroup
\newpage
\thispagestyle{empty}
\begin{center}
{\bf\Large Curriculum Vitae} \\
\end{center}

\subsection*{Personal Data}
\begin{itemize}
\item {\bf Name:} Johanna Knapp 
\item {\bf Date of Birth:} 10 July 1981
\item {\bf Place of Birth:} Krems a. d. Donau, Austria
\item {\bf Nationality:} Austrian 
\end{itemize}

 \subsection*{Education}
\begin{itemize}
\item February 2005 -- July 2007: PhD studies at CERN via a grant of the ``Austrian CERN doctoral program'', under the supervision of Prof. Wolfgang Lerche.
\item since 2005: PhD studies at the University of Technology, Vienna, Austria, under the supervision of Prof. Maximilian Kreuzer.
\item November 2004: Master of Science (``Diplomingenieur'') in Physics.
\item 1999-2004: Studies of ``Technische Physik'' (technical
  physics) at the Vienna University of Technology, Vienna, Austria.
\item June 1999: ``Matura'' -- graduation from high school with distinction at the ``Bundesgymnasium Krems, Piaristengasse 2'', Krems a. d. Donau, Austria.
\end{itemize}

\subsection*{Grants and Awards}
\begin{itemize}
\item February 2005 - July 2007: Scholarship from the Austrian CERN Doctoral Program.
\item 2004: Grant for excellent achievements during the studies from the University of Technology, Vienna.
\item 2003: ``TOP Stipendium des Landes Nieder\"osterreich'': special grant by the region of Lower Austria for students in technical and natural sciences
\end{itemize}

\subsection*{Publications}
\begin{itemize}
\item J.Knapp, H. Omer, `` Matrix Factorizations and Homological Mirror Symmetry on the Torus'', {\em JHEP}, {\bf 0703} (2007) 088, {\tt hep-th/0701269}
\item J. Knapp, H. Omer, ``Matrix Factorizations, Minimal Models and Massey Products'', {\em JHEP}, {\bf 0605} (2006) 064, {\tt hep-th/0604189}
\item S. Guttenberg, J. Knapp, M. Kreuzer, ``On the Covariant Quantization of Type II Superstrings'', {\em JHEP} {\bf 0406} (2004) 030, {\tt hep-th/0405007} 
\item F. Aigner, M. Hillbrand, J. Knapp, G. Milovanovic, V. Putz, R. Sch\"ofbeck, M. Schweda, ``Technical Remarks and Comments on the UV/IR-Mixing Problem of a Noncommutative Scalar Quantum Field Theory'', {\em Czech. J. Phys.} {\bf 54} (2004) 711-719, {\tt hep-th/0302038}
\end{itemize}

\end{document}